\documentclass[11pt,letterpaper]{article}
\pdfoutput=1
\usepackage{jheppub}
\usepackage{amsmath,amssymb,amsfonts,bm}
\usepackage{graphicx,wrapfig}
\usepackage{multirow}
\usepackage{verbatim}
\usepackage{appendix}
\usepackage{rotating}
\usepackage{multirow}
\usepackage{subcaption}

\numberwithin{equation}{section}

\newtheorem{theorem}{Theorem}[section]
\newtheorem{definition}{Definition}[section]
\newtheorem{prop}{Proposition}[section]
\newtheorem{conj}{Conjecture}[section]
\newtheorem{cor}{Corollary}[section]

\graphicspath{{./images/}}

\newcommand{\IR}{\mathcal{I}\!\!\mathcal{R}}

\newcommand{\mg}[1]{\textbf{\color{blue} [{\sc mg}: {#1}]}}

\newcommand{\dd}{{\mathrm{d}}}

\newcommand{\bea}{\begin{eqnarray}}
\newcommand{\eea}{\end{eqnarray}}
\newcommand{\be}{\begin{equation}}
\newcommand{\ee}{\end{equation}}
\newcommand{\bd}{\begin{definition}}
\newcommand{\ed}{\end{definition}}
\newcommand{\mb}{\mathbf}

\newcommand{\wt}{\widetilde}
\newcommand{\wh}{\widehat}
\newcommand{\ol}{\overline}
\newcommand{\ds}{\displaystyle}
\newcommand{\oct}{{\raisebox{-.035cm}{\begin{turn}{45}$\square$\end{turn}}}}
\newcommand{\soct}{{\raisebox{-.035cm}{\begin{turn}{0}$\diamond$\end{turn}}}}

\newcommand{\eg}{\emph{e.g.}}
\newcommand{\ie}{\emph{i.e.}}
\newcommand{\cf}{\emph{cf.}}

\newcommand{\Z}{{\mathbb Z}}
\newcommand{\R}{{\mathbb R}}
\newcommand{\C}{{\mathbb C}}
\newcommand{\Q}{{\mathbb Q}}

\newcommand{\Li}{{\rm Li}}

\newcommand{\Tr}{{\rm Tr \,}}
\renewcommand{\Re}{{\rm Re}}
\renewcommand{\Im}{{\rm Im}}
\newcommand{\bs}{\backslash}
\newcommand{\pd}{\partial}

\newcommand{\lra}{\longrightarrow}
\newcommand{\CA}{\mathcal{A}}
    \newcommand{\CB}{\mathcal{B}}
\newcommand{\CC}{\mathcal{C}}

\newcommand{\CF}{\mathcal{F}}

\newcommand{\CI}{\mathcal{I}}

\newcommand{\CK}{\mathcal{K}}
\newcommand{\CL}{\mathcal{L}}
\newcommand{\CM}{\mathcal{M}}
\newcommand{\CN}{\mathcal{N}}
\newcommand{\CO}{\mathcal{O}}
\newcommand{\CP}{\mathcal{P}}

\newcommand{\CR}{\mathcal{R}}
\newcommand{\CS}{\mathcal{S}}

\newcommand{\CV}{\mathcal{V}}
\newcommand{\CW}{\mathcal{W}}
\newcommand{\CX}{\mathcal{X}}

\newcommand{\CZ}{\mathcal{Z}}

\newcommand{\arrowsL}{ \underset{\longrightarrow}{\overset{\longrightarrow}{\cdots}}}
\newcommand{\arrowsD}{	\downarrow\! \vdots \! \downarrow}

\title{K-Decompositions and 3d Gauge Theories}

\author[1]{Tudor Dimofte}
\author[1,2]{Maxime Gabella}
\author[3]{Alexander B. Goncharov}

\affiliation[1]{Institute for Advanced Study, Einstein Dr., Princeton, NJ 08540, USA}
\affiliation[2]{Institut de Physique Th\'eorique, CEA/Saclay, 91191 Gif-sur-Yvette, France}
\affiliation[3]{Yale University Mathematics Dept., New Haven, CT 06520, USA}

\abstract{This paper combines several new constructions in mathematics and physics. Mathematically, we study \emph{framed} flat $PGL(K,\C)$-connections on a large class of 3-manifolds $M$ with boundary. We introduce a moduli space $\CL_K(M)$ of framed flat connections on the boundary $\pd M$ that extend to $M$.
Our goal is to understand an open part of $\CL_K(M)$ as a Lagrangian subvariety in the symplectic moduli space $\CX^{\rm un}_K(\pd M)$ of framed flat 
connections on the boundary --- and more so, as a ``$\mathrm{K}_2$-Lagrangian,'' meaning that the $\mathrm{K}_2$-avatar of the symplectic form restricts to zero.
We construct an open part of $\CL_K(M)$ from elementary data associated with the {\it hypersimplicial $K$-decomposition} of an ideal triangulation of $M$, in a way that generalizes (and combines) both Thurston's gluing equations in 3d hyperbolic geometry and the cluster coordinates for framed flat $PGL(K,\C)$-connections on surfaces.
By using a canonical map from the complex of configurations of decorated flags to the Bloch complex, we prove that any generic component of $\CL_K(M)$ is $\mathrm{K}_2$-isotropic as long as $\pd M$ satisfies certain topological constraints (Theorem \ref{thm:K2}). In some cases this easily implies that $\CL_K(M)$ is $\mathrm{K}_2$-Lagrangian.
For general $M$, we extend a classic result of Neumann and Zagier on symplectic properties of $PGL(2)$ gluing equations to reduce the $\mathrm{K}_2$-Lagrangian property to a combinatorial statement.

Physically, we translate the $K$-decomposition of an ideal triangulation of $M$ and its symplectic properties to produce an explicit construction of 3d $\CN=2$ superconformal field theories $T_K[M]$ resulting (conjecturally) from the compactification of $K$ M5-branes on $M$. This extends known constructions for $K=2$. Just as for $K=2$, the theories $T_K[M]$ are described as IR fixed points of abelian Chern-Simons-matter theories. Changes of triangulation (2--3 moves) lead to abelian mirror symmetries that are all generated by the elementary duality between $N_f=1$ SQED and the XYZ model. In the large $K$ limit, we find evidence that the degrees of freedom of $T_K[M]$ grow cubically in $K$.
}

\begin{document}

\maketitle

\section{Introduction}

This paper presents a combination of mathematical and physical results. Its main goal is a physical one: to algorithmically define three-dimensional supersymmetric $\CN=2$ quantum field theories $T_K[M]$ labeled by an oriented topological 3-manifold $M$ and an integer $K\geq 2$. The theories $T_K[M]$ are meant to coincide with the compactification of the six-dimensional $\CN=(2,0)$ superconformal theory with symmetry algebra $A_{K-1}$ on 3-manifolds $M$. This implies that the theories $T_K[M]$ should possess several important properties, relating their observables to the topology and geometry of $M$ --- in particular, to the geometry of the moduli space of flat $SL(K,\C)$-connections on $M$ and its quantization.  These properties are summarized in Table \ref{tab:math-phys} below (page \pageref{tab:math-phys}). 
The combinatorial definition we give of $T_K[M]$ makes many of its expected properties manifest.

For $K=2$, and a particular class of 3-manifolds $M$ with boundary, theories $T_2[M]$ were defined in \cite{DGG}. The key idea of \cite{DGG} was to decompose $M$ into topological ideal tetrahedra (\ie\ to choose a \emph{triangulation} of $M$). Then, after assigning a canonical ``tetrahedron theory'' $T_2[\Delta]$ to each tetrahedron,
% that satisfied the desired geometric properties,
$T_2[M]$ was constructed by ``gluing'' together the tetrahedron theories
\be \textstyle M = \bigcup_{i=1}^N\Delta_i\quad\Rightarrow\quad T_2[M] = \big(T_2[\Delta_1]\otimes \cdots\otimes T_2[\Delta_N]\big)/\sim\,. \label{glueT2} \ee
Physically, each $T_2[\Delta_i]$ contained a 3d chiral multiplet, and the gluing operation `$\sim$' added superpotential interactions and gauged some global symmetries, producing an abelian Chern-Simons-matter theory with an explicit Lagrangian. The final step in the gluing was to flow to the infrared, defining $T_2[M]$ as the infrared limit of the Chern-Simons-matter theory.

\begin{figure}[htb]
\centering
\includegraphics[width=4in]{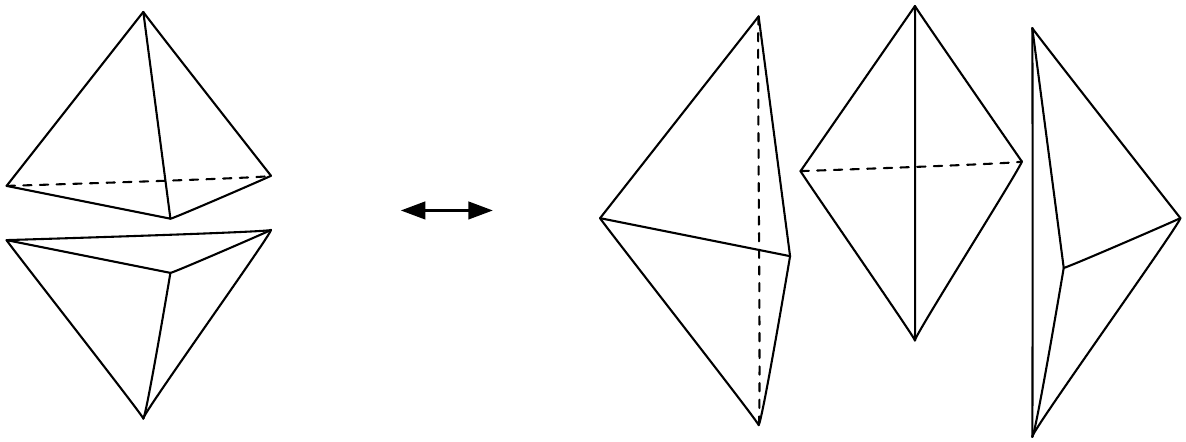}
\caption{The 2--3 move: decomposing a bipyramid into two or three tetrahedra.}
\label{fig:23-intro}
\end{figure}

The gluing of \eqref{glueT2} was done in such a way that, generically, the definition of $T_2[M]$ would be independent of the choice of triangulation. Topologically, any two ideal triangulations can be related by local 2--3 Pachner moves as depicted in Figure \ref{fig:23-intro} \cite{Matveev-spines, Piergallini}. The basic theory $T_2[\text{bipyramid}]$ associated with a triangular bipyramid and constructed from the triangulation on the LHS of Figure \ref{fig:23-intro} was 3d $\CN=2$ quantum electrodynamics (SQED) with two chiral multiplets. The theory constructed from the triangulation on the RHS was the ``XYZ model,'' consisting of three chiral multiplets and a cubic interaction. These two 3d theories are equivalent (dual) in the infrared \cite{AHISS}, ensuring a local triangulation-independence. Lifting the local independence to a global independence of any theory $T_2[M]$ turned out to be subtle for two reasons: 1) it required interchanging infrared limits for different pieces of a Chern-Simons-matter Lagrangian; and 2) not all triangulations could be used to produce sensible Chern-Simons-matter Lagrangians, and the ``good'' triangulations that work are not all known to be related by sequences of 2--3 moves. Thus, strictly speaking, the triangulation-independence of $T_2[M]$ was only conjectural.
% a conjecture, albeit a well motivated one.

The observables of theories $T_2[M]$ were related to classical and quantum hyperbolic geometry, because flat $SL(2,\C)$-connections on $M$ are (roughly speaking) hyperbolic metrics.%
\footnote{Precisely, hyperbolic metrics on $M$ are in 1-1 correspondence with flat $PGL(2,\C)$-connections  whose holonomy representation $\rho:\pi_1(M)\to PGL(2,\C)$ is discrete, faithful, and torsion-free, \cf\ \cite{thurston-1980}. The $\rho$ is  the holonomy representation of the hyperbolic metric. 
%In general there exist flat $PGL(2,\C)$-connections that do not come from hyperbolic metrics. 
The relation between hyperbolic geometry and $PGL(2,\C)$-connections was heavily exploited in \cite{Witten-gravCS} and subsequent works to study quantum gravity in three dimensions, and in \cite{gukov-2003} and subsequent works to understand aspects of quantum Chern-Simons theory.} %
The properties that allowed $T_2[M]$ to be defined as in \eqref{glueT2} were a direct generalization of the symplectic properties that Neumann and Zagier \cite{NZ} observed in Thurston's gluing equations \cite{thurston-1980} for ideal hyperbolic tetrahedra. These same symplectic properties allowed Thurston's gluing equations to be quantized in \cite{Dimofte-QRS} (following \cite{hikami-2006, DGLZ}). The lift of Thurston's gluing equations to theories $T_2[M]$ amounts to a categorification of hyperbolic geometry --- though many details of this categorification remain to be worked out.

In this paper, we extend the triangulation methods of \cite{DGG} to general $K\geq 2$. This requires developing some new mathematics. At the classical level, we need to describe flat $PGL(K,\C)$-connections on triangulated 3-manifolds in a manner analogous to Thurston's description of hyperbolic metrics. 
To achieve this, we enhance moduli spaces of flat connections on a 3-manifold $M$ with additional \emph{framing data}, much as in the 2d constructions of \cite{FG-Teich}. Namely, we consider flat connections together with a choice of invariant flags along certain loci on the boundary $\pd M$. This requires the introduction of extra topological data on $\pd M$.%
\footnote{Examples of this ``extra structure'' include the laminations discussed in early versions of \cite{FG-qdl-cluster}, and expanded on in \cite{FG-laminations}. The slightly more general notion of framing data that we present in Section \ref{sec:basics} arose in collaboration with the authors of \cite{DGV-hybrid}.} %
Then, for admissible 3-manifolds $M$, we construct two algebraic varieties
\be \begin{array}{rl}  \wt\CL_K(M) &\simeq \{\text{framed flat $PGL(K,\C)$-connections on $M$}\}\,, \\[.1cm]
 \CX^{\rm un}_K(\pd M) & \simeq \{\text{framed unipotent flat $PGL(K,\C)$-connections on $\pd M$}\}\,. \end{array} \label{framed-intro}
\ee
Although we do not give precise definitions here, let us present a basic example. 

Let $M$ be a polyhedron. Then the  moduli space 
 $\CX^{\rm un}_K(\pd M)$ parametrizes flat $PGL(K, \C)$-connections on the sphere $\pd M$ punctured at the vertices of the polyhedron, with unipotent monodromies 
 around the vertices, plus a choice of an invariant flag near each of the vertices. Turning to $\wt\CL_K(M)$,   it makes little sense to talk about a moduli space of just flat connections on a polyhedron, since a polyhedron  is simply connected, and so any flat connection is trivial.  However, the \emph{framed} moduli space of a polyhedron  parameterizes configurations of 
  flags in $\C^K$, labelled by the vertices of the polyhedron. For example, in the most fundamental example when $M$ is a tetrahedron we get a configuration space of 
  four flags in $\C^K$. 
 
The space $\CX^{\rm un}_K(\pd M)$ is a singular complex symplectic space, which carries a canonical  symplectic form on its 
non-singular part, and more so,  a $\mathrm{K}_2$-avatar of the symplectic form~\cite{FG-Teich}. 
We consider  the image of the natural projection
\be \CL_K(M) := {\rm im}(\wt \CL_K(M)\to \CX^{\rm un}_K(\pd M)) \,.
\ee
When $M$ is a knot complement and $K=2$, the subvariety $\CL_K(M)$ is a curve defined by the A-polynomial of the knot \cite{cooper-1994}.   
The spaces $\CL_K(M)$ and $\widetilde \CL_K(M)$ may have several irreducible components. We introduce a notion of  a generic component, and conjecture that 
any generic component of $\CL_K(M)$ is a $\mathrm{K}_2$-Lagrangian subvariety of $\CX^{\rm un}_K(\pd M)$. This implies that 
it is Lagrangian for the symplectic form on $\CX^{\rm un}_K(\pd M)$.
We prove that, assuming that $M$ has no toric components on the boundary,  any generic component of $\CL_K(M)$ is $\mathrm{K}_2$-isotropic.%
\footnote{The graph of any cluster transformation 
is $\mathrm{K}_2$-isotropic, and hence $\mathrm{K}_2$-Lagrangian: this follows immediately from the basic 
relation of cluster transformations to the Bloch complex \cite[Sec 6]{FG-cluster}, and provided first general examples of $\mathrm{K}_2$-Lagrangians.  In particular, when $M$ is a cobordism of triangulated 2d surfaces generated by flips of the triangulation, this implies  that $\CL_K(M)$ is a $\mathrm{K}_2$-Lagrangian. The $\mathrm{K}_2$-Lagrangian property 
of the cluster transformations is crucial for their quantization. The $\mathrm{K}_2$-Lagrangian property of A-polynomials was discussed in \cite{Dunfield-mahler, Champ-hypA}, and argued to be necessary for quantization in \cite{GS-quant}.}

The moduli space $\CX^{\rm un}_K(\pd M)$, as well as a larger moduli space $\CX_K(\pd M)$ where the unipotence condition is dropped, 
were introduced and studied in \cite{FG-Teich},
generalizing the classical Teichm\"uller theory. The moduli space 
$\CX^{\rm un}_K(\pd M)$ is birationally isomorphic to Hitchin's moduli space $\CM_H$ for $\pd M$ in any fixed complex structure. Hitchin's moduli space, as a hyperk\"ahler manifold, has played a major role in the study of 4d supersymmetric gauge theories, \cf\ \cite{Kapustin-Witten, GMN}. 
In applications it is often important to consider a hyperk\"ahler resolution of singularities in $\CM_H$. The framing data of $\CX^{\rm un}_K(\pd M)$ in \eqref{framed-intro} 
partially resolves the space of flat $PGL(K, \C)$-connections on $\partial M$ and  can be thought of as an algebraic counterpart for the hyperk\"ahler resolution of $\CM_H$.

For $K=3$, the variety $\wt \CL_K(M)$ was described in \cite{BFG-sl3}. For general $K> 2$, the space $\wt \CL_K(M)$ was studied in \cite{GGZ-slN}, following \cite{GTZ-slN, Zickert-rep, Zickert-sl3}. We will comment on the relation between these works and the current paper at the end of Section \ref{sec:mathperspective}.

\begin{figure}[htb]
\centering
\includegraphics[width=5in]{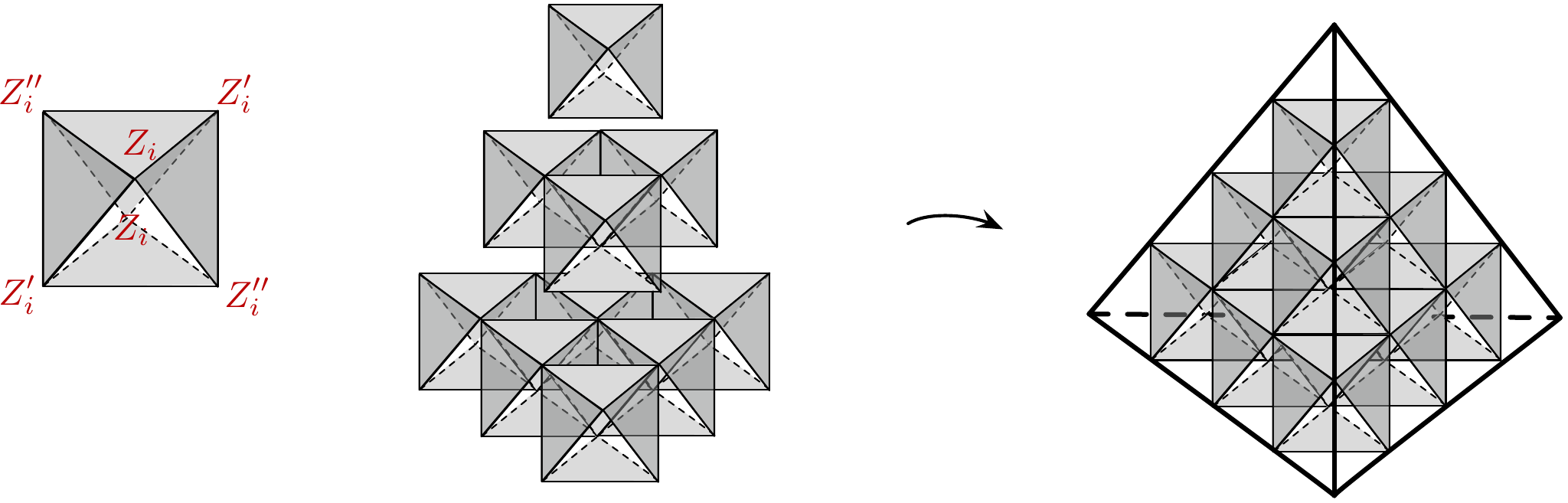}
\caption{Stacking $\frac 16K(K^2-1)$ octahedra to form an $A_{K-1}$ tetrahedron (for $K=4$).}
\label{fig:octs-intro}
\end{figure}

In \cite{FG-Teich} it was shown that an open part of $\CX_K(\pd M)$ is covered by affine charts $\CX_K(\pd M,\mb t)$ of specific type, the cluster Poisson coordinate charts,  labeled by 2d ideal triangulations $\mb t$ of $\pd M$. Here we similarly glue together $\wt \CL_K(M)$ from affine varieties $\wt \CL_K(M,\mb t_{3d})$, labeled by 3d ideal triangulations $\mb t_{3d}$. To define $\wt \CL_K(M,\mb t_{3d})$, we use an important tool: the {\it hypersimplicial $K$-decomposition} of an ideal triangulation, or \emph{K-decomposition} for short. Given an ideal triangulation $\mb t_{3d}=\{\Delta_i\}_{i=1}^N$ of $M$, we further divide each tetrahedron $\Delta_i$ into $\frac16 K(K^2-1)$ octahedra $\oct_j$ (Figure \ref{fig:octs-intro}). To each octahedron we assign a triple of ``cross-ratio parameters'' $z,z',z''$ with $zz'z''=-1$, and the variety $\wt\CL_K[M,\mb t_{3d}]$ is cut out from $(\C^*)^{2\cdot\#\text{octahedra}}=(\C^*)^{\frac 13 NK(K^2-1)}$ by \emph{gluing equations} that include
\be z'' + z^{-1} = 1 \label{oct-eq-intro} ~.\ee
These gluing equations are  analogues of Thurston's gluing equations from hyperbolic geometry. Symplectic properties of the  gluing equations (which we prove in some cases) provide  one of the ways to prove that $\CL_K(M)\subset \CX^{\rm un}_K(\pd M)$ is $\mathrm{K}_2$-isotropic.

Some features of the cluster coordinate charts for $\CX_K(\pd M)$ were recently generalized by the spectral networks of \cite{GMN-spectral, GMN-snakes}. The generalization uses the geometry of $K$-fold spectral covers of the surface $\pd M$. It would be very interesting to describe the variety $\wt \CL_K(M)$ in a similar way, using $K$-fold covers of the 3-manifold $M$, perhaps along the lines of \cite{CCV, Cordova-tangles}.

Coming back to 3d gauge theories: we use the structure of the space $\wt \CL_K(M)$, and in particular the symplectic properties of gluing equations, to formulate a definition of $T_K[M]$. Namely, given a triangulation $M=\bigcup_{i=1}^N\Delta_i$ and a further $K$-decomposition into $\frac16 NK(K^2-1)$ octahedra $\oct_j$, we propose that
\be T_K[M] \;=\; \left(\bigotimes_{i=1}^N T_K[\Delta_i]\right)\bigg/\!\!\sim \;\;\;=\; \Bigg(\bigotimes_{j=1}^{\frac16NK(K^2-1)} T[\oct_j] \Bigg)\Big/\!\!\sim\;, \label{glueTK}\ee
in direct analogy to \eqref{glueT2}. Here $T[\oct_j]$ is the canonical theory of a single chiral multiplet (identical to $T_2[\Delta]$), and gluing is again implemented by superpotentials and gauging. Thus $T_K[M]$ acquires a description as the infrared limit of an abelian Chern-Simons-matter theory. The invariance of $T_K[M]$ under 2--3 Pachner moves follows (again, conjecturally) from basic 2--3 moves for the octahedra.

The theory $T_K[\Delta]$ associated with a single tetrahedron at higher $K$ is no longer so simple or canonical as for $K=2$.  The definition \eqref{glueTK} implies that its UV Lagrangian contains $\frac16K(K^2-1)$ chiral multiplets. For the first few $K$, we find in Section \ref{sec:Tpoly}:
\be \label{TKthys-intro}
\begin{array}{c|c|c}
 K & T_K[\Delta] & \text{flavor sym.} \\\hline
 2 & \text{1 free chiral} & U(1) \\
 3 & \text{4 free chirals} & U(1)^{4} \\
 4 & \text{10 chirals + degree-six superpotential} & U(1)^9 \\
 5 & \;\text{20 chirals, $U(1)^2$ gauge group, $W=\sum$(four monopole ops)}\; & U(1)^{16}
\end{array} \ee
In general, for all $K\geq 5$, the gluing rules force $T_K[\Delta]$ to have both a nontrivial superpotential and a nontrivial gauge group.

Constructing theories $T_K[M]$ for all $K\geq2$ allows us to study some features of the $K\to \infty$ limit. In particular, by relating degrees of freedom of theories $T_K[M, \Pi ]$ to the volume of $PGL(K,\C)$-connections on $M$, we will find evidence of the large-$K$ scaling
\be \text{$\#$ degrees of freedom of $T_K[M]$} \;\sim\; \text{$\#$ octahedra} \;\sim\;  K(K^2-1)\,. \ee
This agrees beautifully with predictions from M-theory and holography \cite{HenningsonSkenderis, HMM-anomalies}.
We note, however, that $K$ does not appear as a continuously tunable parameter in $T_K[M]$, such as the rank of a gauge group. (This is obvious, for example, in \eqref{TKthys-intro}.) Recently a number of 3d theories that do allow analytic continuation in $K$ were studied in \cite{FGSS-AD}; it would be very interesting to relate these to the present constructions.

As we review momentarily, part of the 3d-3d correspondence relates partition functions of $T_K[M]$ on spheres (and more generally lens spaces) to partition functions of Chern-Simons theory on $M$ itself, with complex gauge group $PGL(K,\C)$. Thus, the combinatorial construction of UV Lagrangians for theories $T_K[M]$ \emph{implies} a combinatorial construction of partition functions for $PGL(K,\C)$ Chern-Simons theory on $M$.
Alternatively, one may simply say that the symplectic properties of $PGL(K,\C)$ gluing equations allow a systematic quantization of the pair of spaces $\CL_K(M)\subset \CX^{\rm un}_K(\pd M)$, generalizing \cite{Dimofte-QRS}. 
The result is a construction of a subsector of $PGL(K,\C)$ Chern-Simons theory that generalizes a circle of ideas initiated in \cite{gukov-2003, DGLZ} for $K=2$.%
\footnote{The $PGL(K,\C)$ Chern-Simons partitions constructed with the current methods have limited TQFT-like properties under cutting-and-gluing operations that preserve a triangulation and some additional structures. It is nevertheless an open problem to define $PGL(K,\C)$ Chern-Simons theory (for any $K\geq 2$) as a full TQFT, with complete cutting-and-gluing rules. This has recently been emphasized in \cite{CDGS, GukovPei, PeiKe}, along with new proposals for resolving the problem.} %

In the remainder of this introduction, we review some basic features of the 3d-3d correspondence between observables of $T_K[M]$ and the geometry of flat $PGL(K,\C)$-connections on $M$; then we provide a more detailed summary of our main mathematical results.

\subsection{The 3d-3d correspondence}
\label{sec:intro3d}

The 3d-3d correspondence was first conjectured in \cite{DGH}, and has since been studied in a multitude of papers. Some of its fundamental ideas were developed in \cite{Yamazaki-3d, DGG, CCV}, and the first physical proofs of the correspondence appeared in \cite{CJ-S3, LY-S2}.
(The recent review \cite{D-volume} contains further discussion and references.) New and interesting aspects of the correspondence are still being developed, \cf\ \cite{CDGS, GukovPei, PeiKe, Yamazaki-defects}, which appeared after the first version of this paper.

The basic idea behind the 3d-3d correspondence is that the compactification of the 6d $(2,0)$ theory of type $A_{K-1}$ on $ \R^3 \times M$, with a topological twist along $M$ that preserves four supercharges, should lead to a 3d $\CN=2$ theory $T_K[M]$ in $\R^3$. Alternatively, $T_K[M]$ may be defined as the effective field theory on a stack of $K$ M5-branes wrapping $M\times \R^3$ in the M-theory background $ \R^5\times T^*M$.
More generally, compactification of the 6d $(2,0)$ theory (or of 5-branes) on a $d$-manifold $X^d$ should produce an effective theory $T_K[X^d]$ in $\R^{6-d}$. Notable examples include the 2d-4d correspondence developed in \cite{Witten-M, Gaiotto-dualities, GMN, AGT} and many other works, and the 4d-2d correspondence introduced in \cite{GGP-4d}.

Various properties of the parent $(2,0)$ theory imply relations between observables in the theory $T_K[M]$ and the geometry of flat $G_\C$ connections on $M$, where $G_\C$ is a complex group with Lie algebra $A_{K-1}$.%
\footnote{Some subtle discrete choices can modify the theory $T_K[M]$ and the appropriate form of the group $G_\C$. See, for example, \cite{Tachi-discrete}.} %
Some of these relations are summarized in Table \ref{tab:math-phys}.

\begin{table}[htb]
\centering
\be \notag %\label{geom-phys}
\begin{array}{c@{\quad}c@{\quad}c}
    \underline{\;T_K[M, \Pi]\;} && \underline{\;M,\;\text{$G_\C$ connections}\;} \\[.2cm]
    \text{coupling to $T_K[\pd M]$} &\longleftrightarrow& \text{polarization $\Pi$ for $\CX^{\rm un}_K(\pd M)$} \\
    \text{rank of flavor symmetry group} &\longleftrightarrow& \text{$\frac12\dim_\C \CX^{\rm un}_K(\pd M)$}\\
    \text{twisted superpotential on $\C\times S^1$} &\longleftrightarrow& \text{classical $G_\C$ Chern-Simons functional} \\
    \text{SUSY parameter space on $\C\times S^1$} &\longleftrightarrow& %\text{flat connections extending to $M$},
    \CL_K(M) \\[.05cm]
    \text{flavor line operators} &\longleftrightarrow& \text{quantized algebra of functions on $\CX^{\rm un}_K(\pd M)$}\\
    \text{identities for line operators} & \longleftrightarrow & \text{quantization of $\CL_K(M)$} \\
    \text{partition function on lens space $L(k,1)$} &\longleftrightarrow& \text{$G_\C$ Chern-Simons theory on $M$ at level $k$} \\
    \text{index on $S^2\times_q S^1$} & \longleftrightarrow & \text{$G_\C$ Chern-Simons theory on $M$ at level $k=0$} \\
    \text{holomorphic blocks on $\C\times_q S^1$} & \longleftrightarrow & \text{analytically cont'd CS wavefunctions on $M$}
    \end{array}
\ee
\caption{A sampling of relations in the 3d-3d correspondence.}
\label{tab:math-phys}
\end{table}

An important aspect to mention is that when $M$ has a boundary (which will be true throughout this paper), the theory $T_K[M]$ is not truly an isolated 3d theory, but rather a boundary condition for a 4d theory $T_K[\pd M]$. By the 2d-4d correspondence, the rank of the gauge group of $T_K[\pd M]$ is half the dimension of the space of flat connections on the boundary $\CX^{\rm un}_K(\pd M)$. To obtain an isolated 3d theory, we must choose a weak-coupling limit%
\footnote{More precisely, we mean here a weak-coupling limit for the abelian Seiberg-Witten theory that describes $T_K[\pd M]$ on its Coulomb branch. Such a weak-coupling limit always exists.} %
for the theory $T_K[\pd M]$ \cite{DGV-hybrid}. This choice amounts to a polarization  $\Pi$ of one of affine charts of the complex symplectic space $\CX_K(\pd M)$. The resulting 3d theory should be denoted $T_K[M,\Pi]$. It acquires a flavor symmetry of rank $\frac12\dim\CX^{\rm un}_K(\pd M)$.

By the 2d-4d correspondence, putting the 4d theory $T_K[\pd M]$ in a background with angular momentum quantizes its algebra of line operators (see \cite{GMNIII} and similar ideas in \cite{GW-surface}).
An example of such a background is $(\C\times_q S^1)\times \R_+$, where $(\C\times_q S^1):= \{(w,t)\in \C\times [0,1]\}/(w,1)\sim(qw,0)$.
The resulting quantum algebra $\hat \CX^{\rm un}_K(\pd M)$ is a quantization of $\CX^{\rm un}_K(\pd M)$ --- it is the same quantization presented in \cite{FockChekhov, Kash-Teich} for $K=2$ and \cite{FG-Teich, FG-cluster, FG-qdl-cluster} for $K>2$.
When acting on a 3d boundary theory $T_K[M,\Pi]$, line operators of $T_K[\pd M]$ satisfy additional relations. These relations define an $\hat \CX^{\rm un}_K(\pd M)$-module $\hat \CL_K(M)$ whose characteristic variety (or ``classical limit'') should be $\CL_K(M)$. In other words, the module $\hat \CL_K(M)$ is a quantization of the Lagrangian $\CL_K(M)$.

Some of the more interesting observables of a 3d $\CN=2$ theory $T_K[M,\Pi]$ are partition functions on squashed lens spaces $L(k,1)_b$.
For $k=0$, $k=1$, and $k>1$, respectively, it was conjectured in \cite{DGG-index}, \cite{DGG}, and (after the first version of this paper) in  \cite{D-levelk} that these partition functions are equivalent to partition functions of $G_\C$ Chern-Simons theory at level $k$ on $M$ itself. This was proven for $k=0,1$ in \cite{LY-S2, CJ-S3}.
The quantum algebra of line operators (in fact, two commuting copies of it) acts on the $L(k,1)_b$ partition function of $T_K[M,\Pi]$ in such a way that the relations of the module $\hat \CL_K(M)$ are satisfied. Correspondingly, a quantization of the Lagrangian variety $\CL_K(M)$ should annihilate a $G_\C$ Chern-Simons partition function on $M$~\cite{gukov-2003}.

A somewhat simpler observable of $T_K[M,\Pi]$ is the effective twisted superpotential of the theory on $\C\times S^1$ (at $q=1$). It is the generating function for the supersymmetric parameter space that also appears on the LHS of Table \ref{tab:math-phys}.
As a concrete example, let's review how this works for a standard $K=2$ tetrahedron and for the 2--3 Pachner move.

The tetrahedron theory $T_2[\Delta]=T[\oct]$ contains a single free chiral multiplet $\phi$, with a $U(1)$ flavor symmetry (see Section \ref{sec:Toct}). Its effective twisted superpotential on $\C\times S^1$ is
\be \wt W(z) = \Li_2(z^{-1}) ~.\ee
The dilogarithm here is a standard 1-loop contribution to the twisted superpotential coming from Kaluza-Klein modes of the 3d chiral $\phi$~\cite{NS-I}. The complex parameter $z$ is the twisted mass of $\phi$, associated with its $U(1)$ symmetry. The supersymmetric parameter space is then defined by
\be \CL_2[\Delta] := \Big\{ \exp \frac{d\wt W}{d\log z}=z''\Big\} = \big\{ z''+z^{-1}=1\big\}\,. \label{LD-physics} \ee
By the 3d-3d correspondence, this should correspond to the Lagrangian $\CL_2[\Delta]$. Indeed, \eqref{LD-physics} is identical to \eqref{oct-eq-intro}, and we recognize it as the canonical $\mathrm{K}_2$-Lagrangian subvariety of $\CX_2[\pd\Delta] \simeq (\C^*)^2$,
with symplectic form $d\log z''\wedge d\log z$ and $\mathrm{K}_2$ form $z''\wedge z$.
The superpotential $\wt W(z)$ itself is interpreted as the complexified hyperbolic volume of an ideal hyperbolic tetrahedron.

Now consider a bipyramid as in Figure \ref{fig:23-intro}. Triangulating by two tetrahedra (with a suitable choice of boundary polarization) leads to a description of $T_2$[bipyramid] as 3d $\CN=2$ SQED, \ie\ a $U(1)$ gauge theory with two chiral multiplets. Its twisted superpotential is
\be \wt W_{(2)}(x,y) = \Big[ \Li_2(s^{-1}) + \Li_2(s/x) + \tfrac12\log(s)^2+\log s(\log y-i\pi)  \Big]_{d/ds = 0}\,.  \ee
Each dilogarithm comes from a chiral multiplet of SQED (geometrically: from a tetrahedron),
and the effect of the $U(1)$ gauge group is to extremize the superpotential with respect to a dynamical variable $s$. In contrast, triangulation by three tetrahedra gives the XYZ model, whose twisted superpotential is
\be \wt W_{(3)}(x,y) = \Li_2(x^{-1}) + \Li_2(y^{-1}) + \Li_2(xy) + \tfrac12(\log(xy)-i\pi)^2\,. \ee
This time there is no minimization (since there is no dynamical gauge group), but the arguments of the dilogarithms are constrained by the superpotential of the XYZ model, so that their product is one. Geometrically, the constraint is associated with the internal edge of the triangulation.

Twisted superpotentials of a 3d $\CN=2$ theory are insensitive to renormalization-group flow. (Supersymmetry protects them from quantum corrections beyond one-loop.) Then the infrared equivalence of SQED and the XYZ model implies that we should have
\be  \wt W_{(2)}(x,y) = \wt W_{(3)}(x,y)\,, \label{W23-intro} \ee
with appropriate choices of branch cuts.
Mathematically \eqref{W23-intro} is the 5-term identity for the dilogarithm. The superpotentials $\wt W_{(2)}$ and $\wt W_{(3)}$ simply compute the complexified hyperbolic volume of the bipyramid in two different ways. The SUSY parameter space of $T_2[\text{bipyramid}]$ is $\CL_2[\text{bipyramid}]=\{ d\wt W(x,y)/d\log x = p_x,\, d\wt W(x,y)/d\log y = p_y\}$; it is an algebraic variety that may equivalently be computed from either $\wt W_{(2)}$ or $\wt W_{(3)}$.

Heuristically, the theory $T_K[M]$ should be thought of as a categorical analogue of the motivic $PSL(K,\C)$-volume of $M$, or (equivalently) the $PSL(K,\C)$-class of $M$ in the Bloch group (see pages \pageref{hdflbl}--\pageref{mot-vol}). For $K=2$, this perspective was advocated in the introduction to \cite{DGG}. In brief, the Bloch group $B_2(F)$ of a field $F$ is the abelian group generated by elements $\{z\}_2$ with $z\in F^*-\{1\}=F-\{0,1\}$, modulo five-term relations
\be [x]-[y]+\Big[\frac yx\Big] -\Big[\frac{1-x^{-1}}{1-y^{-1}}\Big] + \Big[\frac{1-x}{1-y}\Big] = 0\,,\qquad x,y\in F^*-\{1\}\,.\ee
For a three-manifold $M$, we take $F=\C(\wt {\cal L}_K(\pd M))$ to be the field of functions on (a regular component of) the moduli space of framed flat $PGL(K,\C)$-connections on $M$. Then the motivic $PGL(K,\C)$-volume of $M$ is an element $\{M\}_2= \sum_{i=1}^{\#\,\text{octahedra}} \{z_i\}_2 \in B_2(F)$, obtained by summing the parameters $z_i$ associated to each octahedron in a $K$-decomposition of $M$, viewing these parameters as functions on $\wt {\cal L}_K(\pd M)$. The five-term relations ensure that the element $\{M\}_2$ is invariant under (generic) 2--3 moves. The real $PGL(K,\C)$-volume of $M$ is obtained by applying the Bloch-Wigner dilogarithm map \eqref{defBW} to the motivic volume. We may compare this to the theory $T_K[M, \Pi]$. It is defined by taking a tensor product of octahedron theories and adding some interactions to enforce the gluing --- similar to the formal sum of octahedron parameters defining $\{M\}_2$. Just like $\{M\}_2$, the theory $T_K[M,\Pi]$ is invariant under (generic) 2--3 moves. The real $PGL(K,\C)$-volume of $M$ may be recovered (roughly) by evaluating the real part of the twisted superpotential of the theory on $\C\times S^1$; but by instead evaluating partition functions of $T_K[M,\Pi]$ one also recovers many quantum invariants, as in Table \ref{tab:math-phys}.

\subsection{A mathematical perspective}
\label{sec:mathperspective}

Given a 3-manifold $M$ with boundary, 
consider the moduli space ${\rm Loc}_K(\partial M)$ of flat $PGL(K, \C)$-connections on the boundary $\partial M$. 
This space is symplectic, with the symplectic form at a generic point 
 given by the Atiyah-Bott / Goldman  
 construction. 
The moduli space of flat connections on the boundary $\partial M$ that can be 
extended to $M$ is expected to be a Lagrangian subvariety ${\cal L}_K(M) \subset 
{\rm Loc}_K(\partial M)$. 

One can  quantize the symplectic space 
${\rm Loc}_K(\partial M)$ by defining a non-commutative $q$-deformation   
of the $\ast$-algebra of regular functions 
on ${\rm Loc}_K(\partial M)$, and constructing its $\ast$-representation in an infinite-dimensional 
 Hilbert space ${\cal H}_K({\partial M})$. 
This has nothing to do with a 3-manifold: the problem makes sense for
 any oriented 2d surface $\CC$. 
Assuming that the surface is hyperbolic, that is $\chi(\CC) <0$,  and 
has at least one hole, the quantization 
was done  
in \cite{FG-Teich, FG-cluster, FG-qdl-cluster}.  
It generalizes the quantization of Teichm\"uller spaces 
 \cite{Kash-Teich, FockChekhov}, related to the $SL(2)$ case.

 The next goal is to quantize the Lagrangian subvariety ${\cal L}_K(M)$. By this we mean 
defining a line in 
${\cal H}_K({\partial M})$ that must be 
annihilated by the $q$-deformations of the equations defining the 
subvariety ${\cal L}_K(M)$. In the $SL(2)$ case, this was discussed in a series of physics papers \cite{gukov-2003}, \cite{hikami-2006, DGLZ, Dimofte-QRS, DGG-index}, \cite{EO, GS-quant, BorotEynard}, and in the closely related mathematical works \cite{KashAnd, AK-new, Gar-index, GHRS-index}. (An alternative mathematical approach to the quantization of ${\cal L}_K(M)$ has been proposed using the theory of skein modules, \cf\ \cite{Frohman-Gelca, Sikora-quant}.)
Subsequent to the initial version of this paper, further constraints on a consistent quantization were uncovered in \cite{D-levelk, AK-complexCS}.

The following problem motivated this project. We would like to have a local procedure for the quantization of the moduli space of flat $PGL(K,\C)$-connections on 3-manifolds, by decomposing the manifold into tetrahedra, quantizing the tetrahedra, and then gluing the quantized tetrahedra. 
This approach for $SL(2)$ was implemented  in \cite{Dimofte-QRS} by introducing the phase space and the Lagrangian subvariety related to a hyperbolic tetrahedron. However, already for $SL(2)$, an attempt to understand even the Lagrangian subvariety itself as a moduli space of flat connections immediately faces a serious problem: 

\vskip 2mm
\noindent {\it Any flat connection on a tetrahedron is trivial, so the corresponding moduli space 
is just a point, and thus cannot 
produce a Lagrangian subvariety in the phase space assigned to 
the boundary, which in this case is a four-holed sphere}. 

\vskip 2mm
\noindent The problem is that the tetrahedron has trivial topology, while the moduli space of flat connections 
is a topological invariant, and hence also becomes trivial.

We suggest a solution  to this problem based on the following idea. We consider moduli spaces of flat connections 
on 3-manifolds with {\it framings}. A framing amounts to introducing invariant flags on 
each of the so-called 
{\it small boundary components}, which we define below, invariant under the holonomy around the component.  
This, remarkably, allows one to produce the missing Lagrangian subvariety for the tetrahedron. 
The corresponding moduli spaces are defined for arbitrary admissible manifolds, and 
 can be ``symplectically'' glued 
from the ones assigned to the tetrahedra. So we use the invariant flags to {\it localize} 
flat connections to tetrahedra. 

This notion of framing generalizes the key idea used in \cite{FG-Teich} to introduce cluster coordinates 
on the moduli space ${\cal X}_{K}(\CC)$ of framed flat $PGL(K)$-connections on a surface $\CC$ with punctures: invariant flags were invoked to localize flat connections on ideal triangles. In the three-dimensional case, a related notion of ``decorations'' was used by \cite{GTZ-slN, Zickert-rep, Zickert-sl3} to study representations of $\pi_1(M)$.

In Section \ref{sec:admissible} we start with a careful discussion of a class of 3-manifolds with boundary, 
which we call {\it admissible 3-manifolds}, which are glued from truncated tetrahedra. 
Since the boundary faces of truncated tetrahedra are of two different types, triangles and hexagons, 
the boundary  of the manifold obtained by gluing them along the hexagonal faces 
also has two kinds of boundary components, {\it big} (formed by the unglued hexagons) and 
{\it small} (formed by the triangles).
We say that such a manifold is admissible if 1) the fundamental groups of the small boundary components are abelian; 2) the small boundary components are not spheres; and 3) the big boundary components have negative Euler character.

The second and third conditions are technical, and simply allow us to avoid stacks in our constructions. The first condition, however, is dictated by the notion of framing: \emph{every} vector bundle with flat connection on an admissible 3-manifold admits at least one choice of framing.
Indeed, if the fundamental group of the small boundary is abelian, then the holonomies of a flat connection on the small boundary all commute with each other, and a family of commuting operators 
in a vector space $V_K$ always has an invariant flag. (Typically there are just $K!$  invariant flags; in a basis in which all operators are diagonal with different eigenvalues, choosing a flag amounts to ordering the basis.)
So by adding a framing to a flat vector bundle we enlarge the moduli space by taking its cover 
and partially resolving its singularities, rather than cutting it down.

In section \ref{sec:spaces} we discuss  the moduli spaces  presented in \eqref{framed-intro}: the space $\wt \CL_K(M)$ of framed flat $PGL(K,\C)$-connections on $M$, the space $\CX^{\rm un}_K(\pd M)$ of framed unipotent flat $PGL(K,\C)$-connections on the boundary, and the prospective Lagrangian $\CL_K(M) = \text{im}\big(\wt \CL_K(M)\to\CX^{\rm un}_K(\pd M)\big)$.

The simplest example of an admissible 3-manifold is the tetrahedron $\Delta$, 
whose boundary is understood as a sphere with four punctures. We associate to the boundary the moduli space ${\cal X}_2^{\rm un}(\partial \Delta)$ of framed flat $PGL(2,\C)$-connections on the four-punctured sphere with unipotent monodromy around the punctures.
It is a two-dimensional symplectic space. 
Its Lagrangian subspace ${\cal L}_2(\Delta)$ 
consists of the connections that can be extended to the bulk with framings at the 
four vertices at the boundary. Since any connection on the ball is trivial, the only data left is the four flags, 
which in this case amounts to a configuration of four lines in a two-dimensional space $V_2$. 
The resulting Lagrangian pair
\be \label{ellagp}
{\cal L}_2(\Delta) \subset {\cal X}_2^{\rm un}(\partial \Delta)
\ee
is our main building block. 
It has a natural Zariski open part, which deserves a special notation $\CL_\soct\subset \CP_{\pd\soct}$,
 described in coordinates as follows: 
\be \label{ellagpoct}
\CP_{\pd\soct} := \{z, z', z'' \in \C^*~|~ zz'z'' =-1\}, ~~~~ \CL_\soct:= \{z, z', z'' \in \CP_{\pd\soct} ~|~  z''+z^{-1}-1=0\}.
\ee
The natural compactification of the symplectic space $\CP_{\pd\soct} = \C^*\times \C^*$ 
given as a moduli space (\ie\ a stack) ${\cal X}_2^{\rm un}(\partial \Delta)$ should help to deal with non-generic framings.

\subsubsection*{Framed flat connections from octahedra}

Our first major goal is to build the moduli space $\wt \CL_K(M)$
of framed flat $PGL(K,\C)$-connections on an admissible 3-manifold $M$ out of these building blocks.
To achieve this, we choose an ideal triangulation $\mb t_{3d}$ of $M$, and a further \emph{hypersimplicial $K$-decomposition} of each tetrahedron in $\mb t_{3d}$ (Figure \ref{fig:octs-intro}).  We show in Section \ref{sec:FFC} that this $K$-decomposition has precisely the combinatorial and geometric 
data that we need to describe $\wt \CL_K(M)$ and its projection $\CL_K(M)$ to the boundary moduli space.

We  use the framing data on a flat connection to assign to each tetrahedron in the triangulation $\mb t_{3d}$ a \emph{configuration} of four flags at its vertices. One can think of these as four flags in the $K$-dimensional space $V_K$, considered modulo the action of $PGL(K,\C)$.%
\footnote{It is important to notice that the sets of configurations of objects of any kind associated with a vector space depend only on the dimension of the space and not on the choice of vector space itself; thus configurations assigned to isomorphic vector spaces are \emph{canonically} isomorphic.} %
The $K$-decomposition of tetrahedra is used to construct various generalized cross-ratios that determine the configuration of four flags. These generalized cross-ratios correspond precisely to the parameters $z_i,z_i',z_i''$ assigned to the vertices of each octahedron $\oct_i$ (as on the left of Figure \ref{fig:octs-intro}). Once the parameters are identified with cross-ratios, they naturally satisfy certain monomial relations of the form $c_v=1$, stating that the product of all octahedron parameters sitting at any vertex $v$ of the $K$-decomposition is trivial. These gluing constraints generalize Thurston's gluing equations for an ideal hyperbolic triangulation.

These projective geometry constructions can be restated as follows.
Let $\mb t_{3d}$ be an ideal triangulation of $M$, inducing an ideal triangulation $\mb t$ of the big boundary. 
We  construct a Zariski-open  subset 
\be \CX^{\rm un}_K(\pd M,\mb t) \simeq (\C^*)^{2d}\; \subset\; \CX^{\rm un}_K(\pd M)\,,\qquad 2d=\text{dim}_\C\CX^{\rm un}_K(\pd M)  \ee 
of the space of framed unipotent flat connections on $\pd M$, generalizing the cluster coordinate charts defined by \cite{FG-Teich}. 
We write the $\C^*$-coordinates of $\CX^{\rm un}_K(\pd M,\mb t)$ as monomials of the octahedron parameters $z_i,z_i'$ (where $i$ ranges over the octahedra). 
This just means that we get a  projection
\be \label{Xt-intro}
\textstyle \widetilde \pi:\prod_{i=1}^N\CP_{\pd\soct_i} \simeq (\C^*)^{2N}  \lra \CX^{\rm un}_K(\pd M,\mb t) \simeq (\C^*)^{2d}\,,
\ee
where $i=1,...,N$ indexes the octahedra in the $K$-decomposition of $\mb t_{3d}$.
The projection is not quite canonical in the presence of small-torus boundary components, but canonical if they are absent.

We define an open subset of the space $\wt \CL(M)$ by intersecting the product of octahedron Lagrangians $\CL_{\soct_i}$ with the gluing constraints
\be 
\textstyle   \wt\CL_K(M,\mb t_{3d}) = \Big( \prod_{i=1}^N \CL_{\soct_i}\Big) \cap \{c_v=1\}_{\text{vertices $v$}}\,. \label{tL3d-intro} 
\ee
By showing that a set of octahedron parameters that satisfies the gluing equations can be used to uniquely reconstruct a framed flat connection on $M$ (Section \ref{sec:projbases}), we prove that

\medskip
\noindent{\bf Theorem \ref{thm:XM}} (page \pageref{thm:XM}) \textit{The intersection \eqref{tL3d-intro}  parameterizes a Zariski-open subset of the space of framed flat connections on $M$.}

\medskip
\noindent
The restriction of  the map (\ref{Xt-intro}) to the subvariety $\{c_v=1\}$ is a canonical map
    \be \label{mappi}
     \pi: \prod_{i=1}^N \CP_{\pd\soct_i} \,\cap\, \{c_v=1\} \lra \CX^{\rm un}_K(\pd M,\mb t).
    \ee      
Let  $\CL_K(M,\mb t_{3d}) := \wt\pi (\wt\CL_K(M,\mb t_{3d})) = \pi(\wt\CL_K(M,\mb t_{3d}))$ be the image of $\wt \CL_K(M,\mb t_{3d})$  under the projections $\wt \pi$ or $\pi$\,:
\be   \CL_K(M,\mb t_{3d}) \subset \CX^{\rm un}_K(\pd M,\mb t)\,.\ee
It is an open subset of the space $\CL_K(M)$. 

Theorem \ref{thm:XM} implies that changing the bulk ($\mb t_{3d}$) and boundary ($\mb t$) triangulations amounts to  birational transformations of the spaces $\wt \CL_K(M,\mb t_{3d})$, $\CL_K(M,\mb t_{3d})$ and $\CX^{\rm un}_K(\pd M,\mb t)$.

Changes of bulk triangulation are generated by 2--3 moves, which can be decomposed into elementary 2--3 moves acting on octahedra in a $K$-decomposition, as described in Section~\ref{sec:23}. Changes of boundary triangulation correspond to cluster transformations on $\CX^{\rm un}_K(\pd M,\mb t)$~\cite{FG-Teich}.
 Taking the union of the spaces assigned to a given bulk triangulation ${\bf t}_{3d}$ over the set of  the bulk triangulations  compatible with a fixed boundary triangulation $\mb t$, we obtain a triple 
 $$
 \wt \CL_K(M,\mb t) \to \CL_K(M,\mb t) \subset \CX^{\rm un}_K(\pd M,\mb t).
 $$
 It depends only on $\mb t$. One may then vary $\mb t$ to eliminate the dependence on boundary triangulation. We emphasize that even  then  we  do not cover 
 the whole moduli spaces $\wt \CL_K(M)$ and $ \CX^{\rm un}_K(\pd M)$. In particular, only components of $\wt \CL_K(M)$ corresponding to irreducible flat connections on $M$
 will be detected.

\subsubsection*{Symplectic gluing}

Our next major goal is to understand the symplectic properties of $\CL_K(M)$ and $\CX^{\rm un}_K(\pd M)$. They are summarized in the Symplectic Gluing Conjecture (Conj. \ref{conj:symp}, page \pageref{conj:symp}): for any admissible 3-manifold $M$ with bulk triangulation $\mb t_{3d}$ and corresponding big-boundary triangulation $\mb t$, we expect that
\begin{itemize}
\item The moduli space $\CX^{\rm un}_K(\pd M,\mb t)$, equipped with the canonical complex symplectic form, is isomorphic to  a holomorphic symplectic quotient of the product of octahedron spaces $\prod_{i=1}^N \CP_{\pd\soct_i}$, equipped with the product symplectic structure $\sum_{i=1}^N d\log z_i\wedge d\log z_i'$. 
Precisely, it is the symplectic quotient for the Hamiltonian $(\C^*)^{N-d}$ action  whose Hamiltonians are given by  the gluing monomials $c_v$. Thus
\be \textstyle \CX^{\rm un}_K(\pd M,\mb t) \simeq \big(\prod_{i=1}^N \CP_{\pd\soct_i}\,\cap\, \{c_v=1\}\big)\big/(\C^*)^{N-d}\,.\label{Xred-intro} \ee
\item $\CL_K(M,\mb t_{3d})\subset \CX^{\rm un}_K(\pd M,\mb t)$ is a Lagrangian subvariety; it coincides with the image of the product Lagrangian $\prod_{i=1}^N\CL_{\soct_i}\subset \prod_{i=1}^N\CP_{\pd\soct_i}$ under the symplectic quotient \eqref{Xred-intro}.
\end{itemize}

 Recall the projection 
$\textstyle \widetilde \pi:\prod_{i=1}^N\CP_{\pd\soct_i}   \lra \CX^{\rm un}_K(\pd M,\mb t)$ from (\ref{Xt-intro}).  We prove that the gluing monomials $c_v$ Poisson commute, and that they Poisson commute with the pullbacks by the  map $\wt \pi^*$ 
of the  cluster coordinates on $\CX^{\rm un}_K(\pd M,\mb t)$, given by the monomials in the octahedron parameters.
Thus the first claim (\ref{Xred-intro}) of the 
Symplectic Gluing Conjecture reduces to the claim that exactly $N-d$ of the equations $c_v=1$ are independent.
An easy Euler-characteristic count shows that in the presence of $t$ small-torus boundaries, 
the total number of gluing monomials $c_v$ equals $N-d+(K-1)t$; therefore, it remains to show that there are exactly $(K-1)t$ relations among the gluing monomials.

Also recall the map $\pi$ from (\ref{mappi}). 
The claim that ($\CX^{\rm un}_K(\pd M,\mb t),\Omega)$, as a holomorphic symplectic space, is the reduction of the product of octahedron spaces means that 
\be \label{symplred}
    \textstyle \pi^*\Omega = \sum_{i=1}^N d\log z_i \wedge d\log z_i'\big|_{c_v=1}\,. 
\ee

Since the number of monomials  $c_v$ is $N-d+(K-1)t$,  the dimension 
  of $\wt \CL(M; {\bf t}_{3d})$ is at least $d-(K-1)t$, see (\ref{tL3d-intro}).  
The second claim of the Symplectic Gluing Conjecture is that the image of $\wt \CL(M; {\bf t}_{3d})$ under the projection by the Hamiltonian flows 
  of the Hamiltonians $c_v$  is Lagrangian.
When the only small-boundary components of $\pd M$ are discs, Theorem \ref{thm:K2} guarantees that the image of $\wt \CL(M; {\bf t}_{3d})$ is isotropic; thus it would suffice to show that its dimension is $\geq d$.
           
We refer to the relationship between the pair $\CL_K(M,\mb t_{3d})\subset \CX^{\rm un}_K(\pd M,\mb t)$ and the elementary octahedron pairs $\CL_{\soct_i}\subset \CP_{\pd\soct_i}$ as {\it symplectic gluing.}
Since our parameters are assigned to the octahedra 
of the $K$-decomposition of $M$, the 
Symplectic Gluing Conjecture says that 
\begin{itemize} 
\item {\it the Lagrangian pair
${\cal L}_K(M) \subset {\cal X}^{\rm un}_K(\pd M)$ is obtained by symplectic gluing 
of the elementary Lagrangian pairs (\ref{ellagp}) parametrized by the octahedra of the $K$-decomposition of $M$ corresponding to an ideal triangulation $\mb t_{3d}$ of $M$.}
\end{itemize} 
 
 The symplectic gluing is studied in Sections \ref{sec:Bloch} and \ref{sec:combi}  from two different perspectives.
In both settings, it is very natural to promote the notion of a Lagrangian subvariety to the much stronger notion of a \emph{$\mathrm{K}_2$-Lagrangian}. Let us explain what this means.

\subsubsection*{$\mathrm{K}_2$-Lagrangians}

Let $F^*:= F -\{0\}$ 
be the multiplicative group of a field $F$. Recall that $F^*\wedge F^*=\Lambda^2 F^*$ 
is the abelian group generated by elements of the form $a\wedge b$, $a,b\in F^*$, 
with $a\wedge b=-b\wedge a$ and $(a_1a_2)\wedge b=a_1\wedge b+a_1\wedge b$. The group $\mathrm{K}_2(F)$ is the quotient 
of $F^*\wedge F^*$ by the subgroup generated by Steinberg relations $(1-z)\wedge z$, 
where $z\in F^* -\{1\}$:
\be \label{defK2}
\mathrm{K}_2(F) = F^* \wedge F^*/\{(1-z)\wedge z\}. 
\ee
Next, let $X$ be a complex algebraic variety. Denote by $\C(X)$ the field of rational functions on $X$. 
Then  there is a homomorphism to the space $\Omega_{\rm log}^2(X)$ of holomorphic 2-forms with logarithmic singularities 
on $X$:
\be
d\log: \C(X)^* \wedge \C(X)^* \longrightarrow \Omega_{\rm log}^2(X), ~~~~ f\wedge g 
\longmapsto d\log f \wedge d\log g\,.
\ee
The map kills elements $(1-f) \wedge f$, and so the 
image of an element $W \in \C(X)^* \wedge \C(X)^*$ depends only on its class in $\mathrm{K}_2(\C(X))$.

It was proved in \cite{FG-Teich} that the symplectic form on the 
moduli space ${\cal X}^{\rm un}_K(\CC)$ of unipotent flat connections on a 2d surface $\CC$ 
can be upgraded to its motivic avatar, a class ${\cal W}$ in $\mathrm{K}_2$ of ${\cal X}^{\rm un}_K(\CC)$. 
The symplectic form is recovered as $d\log({\cal W})$.
From our current perspective, the construction of \cite{FG-Teich} applies directly to the big boundaries of admissible 3-manifolds, and we explain the simple generalization to small boundaries in Sections \ref{sec:K2lift}.
(An even wider class of examples is provided by cluster ${\cal A}$-varieties in \cite{FG-cluster}.)
This motivates the following definition. 

\bd \label{def:K2} Let ${X}$ be a complex variety with a class ${\cal W}$ in $\mathrm{K}_2(\C(X))$ such that 
$d\log({\cal W})$ is a symplectic form at the generic part of $X$.  
A subvariety ${L} \subset X$ is called a $\mathrm{K}_2$-Lagrangian subvariety if 
${\cal W}$ restricts to zero in $\mathrm{K}_2(\C(L))$, and $2\,{\rm dim}\,L = {\rm dim}\,X$.
\ed

\noindent{\bf Examples}: 1. The space $\CP_{\pd\soct}$ has the symplectic form $d\log z\wedge d\log z'$. It lifts to a symbol 
$z\wedge z'$, which is invariant, up to $2$-torsion, under the cyclic shift $z, z', z'' \longmapsto z', z'', z$.  
The curve $\CL_{\soct}$ is a $\mathrm{K}_2$-Lagrangian, since $z\wedge z'$ restricts to $(1-z)\wedge z$ on $\CL_{\soct}$. 

2. The graph of any cluster transformation ${\cal A} \lra {\cal A}$ of a cluster ${\cal A}$-variety ${\cal A}$ is 
 a $\mathrm{K}_2$-Lagrangian subvariety of the product ${\cal A}\times {\cal A}$, see Section 6 of  \cite{FG-cluster}.
\vskip 3mm

\medskip
\noindent{\bf Theorem \ref{thm:K2}} i) \textit{If $\pd M$ has only big boundary and small discs, any generic component of 
$\CL_K(M)$ is a $\mathrm{K}_2$-isotropic subspace of $\CX^{\rm un}_K(\pd M)$, \ie\ the restriction of the $\mathrm{K}_2$-class $\CW$ is zero.}

ii) \textit{If $M$ is a convex polyhedron, then $\CL_K(M)$ is a $\mathrm{K}_2$-Lagrangian subvariety of $\CX^{\rm un}_K(\pd M)$.}
\medskip

\noindent Part ii) of Theorem \ref{thm:K2} follows from  part i) and an easy  dimension count. Indeed, $\CL_K(M)$ is just 
 the configuration space of flags parametrized by  vertices of the polyhedron $M$:
$$
\CL_K(M) = G \backslash {\cal B}^{\{\mbox{\rm vertices of $M$}\}}, ~~~~{\rm dim}\,\CL_K(M)  = v(M)\cdot {\rm dim}\,{\cal B} - {\rm dim}\, G.
$$
Here $G = PGL_K$ and $v(M)$ is the number of vertices of $M$. On the other hand, 
$$
{\rm dim}\,{\cal X}^{\rm un}_K(M) = {\rm dim}(G)(v(M)-2) - {\rm rk}(G)v(M) = 2 {\rm dim}\,\CL_K(M).
$$

 Since  the cluster coordinates on $\CX^{\rm un}_K(\pd M, \mb t)$ and  the gluing constraints $c_v$ are monomials of the octahedron parameters $z_i,z_i'$, 
  a proof of Conjecture \ref{conj:symp} implies a $\mathrm{K}_2$-analog of (\ref{symplred}): 
\be 
\textstyle \pi^*\CW =  \sum_{i=1}^N z_i\wedge z_i' \big|_{c_v=1}\qquad \text{(mod torsion)}\,. 
\ee
It would immediately imply  that $\CL_K(M,\mb t_{3d})\subset \CX^{\rm un}_K(\pd M,\mb t)$ is $\mathrm{K}_2$-isotropic.

\vskip 3mm
In Section \ref{sec:Bloch} we prove Theorem \ref{thm:K2} by using the canonical map of complexes \cite{G93}:
\be \label{hdflbl}
\alpha_\bullet: ~\mbox{\it Complex of generic configurations of decorated flags in $V_K$} \lra \mbox{\it Bloch complex}. 
\ee  
To define it, we define first a closely related  homomorphism of complexes, where the notation will be explained momentarily:
\be \label{flags2Bloch01}
\begin{array}{ccccccc}
\cdots & \stackrel{\dd}{\lra} &A^{(K)}_4 & \stackrel{\dd}{\lra} 
 &A^{(K)}_3 & \stackrel{\dd}{\lra}  &A^{(K)}_2 \\
&&\downarrow \mbox{\footnotesize $\gamma_5$} &&\downarrow \mbox{\footnotesize $\gamma_4 $}&&\downarrow \mbox{\footnotesize $\gamma_3 $} \\
\cdots & \lra&R_2(F)& \stackrel{i}{\lra} &\Z[F^*-\{1\}]& \stackrel{\delta}{\lra} & \wedge^2(F^*)~.
\end{array} 
\ee

Here the abelian group $A^{(K)}_n$ is given by formal integral linear combinations of the configurations of 
$n+1$ generic decorated flags in $V_K$. The differential ${\rm d}$ is the standard simplicial differential. 
This way we get   the complex of generic configurations of decorated flags in $V_K$. 

Let us define the bottom complex. We denote by   $\Z[F^*-\{1\}]$  the free abelian group with a basis 
$\{z\}$, where $z$ runs through the elements of $F^*-\{1\}$. The homomorphism $\delta$ in (\ref{flags2Bloch01}) is defined by setting $\delta \{z\}:=(1-z)\wedge z$. So 
its cockerel  is the group $\mathrm{K}_2(F)$ in (\ref{defK2}).  

  Let $R_2(F) \subset \Z[F^*-\{1\}]$ be the subgroup  generated by the 
five-term relations
\be
\sum_{i=0}^4(-1)^i\{r(z_0, \ldots, \widehat z_i, \ldots, z_4)\}\subset \Z[F^*-\{1\}]\,,
\ee
where  $(z_0, \ldots, z_4)$ run through   generic configurations of $5$ points 
on ${\mathbb P}^1(F)$, and $r(-)$ is the cross-ratio. One shows that the restriction of the map $\delta$ to the subgroup $R_2(F)$ is zero. 
So we get the bottom complex, where $i$ is the natural embedding. 

The 
 {\it Bloch group} $B_2(F)$ is  the quotient of the  group $\Z[F^*-\{1\}]$ 
 by the five-term relations:
$$
B_2(F):= \frac{\Z[F^*-\{1\}]}{R_2(F)}\,.
$$
Since $\delta(R_2(F))=0$, the map $\delta$ descends to the Bloch group, and we get the {\it Bloch  complex}:
\be \label{blcmap}
B_2(F) \stackrel{\delta}{\lra} \wedge^2(F^*).
\ee
So the map of complexes (\ref{flags2Bloch01}) induces a map of complexes (\ref{hdflbl}).

Let us assume now that $F=\C$. Recall  the Bloch-Wigner version of the dilogarithm 
\be \label{defBW-intro}
{\cal L}i_2(z):= {\rm Im}\big(\Li_2(z) + \log(1-z)\log|z|\big)\,. 
\ee
It is a single-valued function, well defined for all $z \in \C$. So it defines a group homomorphism 
$$
{\cal L}i_2: \Z[(\C)] \lra \R, ~~~ \{z\}_2 \to {\cal L}i_2(z)\,. 
$$
It satisfies the Abel five-term relation, i.e. its restriction to the subgroup $R_2(\C)$ is zero. Therefore it gives rise to   a group homomorphism
\be \label{bgrouph}
{\cal L}i_2: B_2(\C) \lra \R, ~~~ \{z\}_2 \to {\cal L}i_2(z)\,. 
\ee

The Bloch group is famously related to 3d hyperbolic geometry, in the following way. Consider the 
scissor congruence group ${\cal P}_2$ of ideal hyperbolic polyhedra. It is an abelian group with the generators  assigned to 
ideal oriented hyperbolic tetrahedra ${\rm I}(z_1, z_2, z_3, z_4)$ with the vertices $z_1, ..., z_4$. The generators  $[{\rm I}(z_1, z_2, z_3, z_4)]$ satisfy two kinds of relations. 
First, the cutting and gluing  relation: cutting an ideal hyperbolic bipyramid in
two different ways into $2$ or $3$ ideal tetrahedra amounts to the same element of  ${\cal P}_2$. 
Second, changing the orientation of a tetrahedron amounts to a sign change:  $[{\rm I}(0, 1, \infty, z)] = - [{\rm I}(0, 1, \infty, \overline z)]$. 
Denote by $B_2(\C)^-$ the aniinvariants of the action of  complex conjugation on the group $B_2(\C)$. 
Then there is a canonical group isomorphism  
$$
B_2(\C)^- \lra {\cal P}_2, ~~~~\{z\}_2 \to [{\rm I}(0, 1, \infty, z)].
$$
Finally, 
the map $\delta$ shows up in the formula for the differential of the dilogarithm:
\be \label{diffdilog}
d{\cal L}i_2(z) =   \log|1-z| d{\rm arg}(z) -   \log|z| d{\rm arg}(1-z).
\ee

Let us return finally to 
the  homomorphism of complexes \eqref{flags2Bloch01}. To define it,  one uses a key construction of \cite{G93} relating configurations of flags to configurations of vectors:
\be \label{hdflbc}
\mbox{\it Complex of generic configurations of decorated flags in $V_K$} ~\lra ~ 
\mbox{\it biGrassmannian 
complex}.
\ee
Given a single generic configuration of $(n+1)$ decorated flags in $V_K$, 
one assigns to it a collection of points in the 
Grassmannians ${\bf G}_{p}^q$ where $p+q+2=n+1$. 
The $K$-decomposition of simplices is crucial here:  viewing the $(n+1)$ flags as assigned to the vertices of an $n$-dimensional simplex $\Delta$, each hypersimplex in the $K$-decomposition of $\Delta$ gives rise to a single point in a Grassmannian that matches the type of the hypersimplex.
Combining \eqref{hdflbc} with the homomorphism of complexes, defined in~\cite{G95}, which we review in Section \ref{sec:Bloch-cx}:
 \be \label{hdflblc}
\mbox{\it biGrassmannian 
complex} ~\lra ~ 
\mbox{\it Bloch complex}, 
\ee 
we arrive at the homomorphism of complexes \eqref{flags2Bloch01}, and hence \eqref{hdflbl}.

The homomorphism of complexes \eqref{hdflbl} controls a number of features of the geometry of framed flat connections on an admissible three-manifold $M$ and its boundary. 
Let us elaborate on this.

First, choosing compatible triangulations $\mb t_{3d}$ and $\mb t$ of $M$ and its big boundary, we assign to 
points in the moduli spaces $\CX^{\rm un}_K(\pd M,\mb t)$ and $\wt\CL_K(M,\mb t_{3d})$  elements in the complex of generic configurations of decorated flags in $V_K$. 
 For example, we may start from a generic point of $\wt\CL_K(M,\mb t_{3d})$, representing a framed flat connection on a triangulated manifold $M$. 
We pick  a decorated flag representative for each  flag of the framing: we can do this  thanks to the unipotence condition. 
 Then we assign to each tetrahedron 
of the triangulation the configuration of four decorated flags obtained by the restriction of the framed connection to the tetrahedron. The formal sum of these 
configurations over all tetrahedra of the triangulation is an element of the group $A^{(K)}_3$ which we assign to the framed flat connection. 
Similarly,  we assign to  a generic point of $\CX^{\rm un}_K(M,{\mb t})$ an element of $A^{(K)}_2$. Then we find:

1) The component $\gamma_3$ of the map (\ref{flags2Bloch01}) was used in \cite[Section 15]{FG-Teich} to define 
the $\mathrm{K}_2$-class ${\cal W}$ on the  space  ${\cal X}^{\rm un}_K(\CC)$ for a\ 2d surface $\CC$. 
It provides a $\mathrm{K}_2$-class  on the space ${\cal X}^{\rm un}_K(\pd M)$. 

2) Theorem \ref{thm:K2} tells that the restriction of the $\mathrm{K}_2$-class  ${\cal W}$ on ${\cal X}^{\rm un}_K(\pd M)$ to ${\cal L}_K(M)$ is zero. 
 The component  $\gamma_4$ of  map (\ref{flags2Bloch01}) tells how exactly it becomes zero. 
Precisely, given an ideal triangulation ${\bf t}$   of $\pd M$, the class ${\cal W}$  has a natural lift to $\Lambda^2\C^*$ \cite[Section 15]{FG-Teich}. 
The map $\gamma_4$ presents its restriction to ${\cal L}_K(M)$ 
as a sum of Steinberg relations $(1-z_i)\wedge z_i$. 
These $z_i$'s are just our octahedron parameters. They are precisely the terms of the map   $\gamma_4$ in (\ref{flags2Bloch01}).

The very existence of the map of complexes (\ref{flags2Bloch01}) implies that 
the image $\sum\{z_i\}_2$ of the map $\gamma_4$ in the Bloch group $B_2(\C)$ is independent of the choice of the balk triangulation ${\bf t}_{3d}$. 
We call it the {\it motivic volume}.\label{mot-vol}  Applying the dilogarithm homomorphism \eqref{bgrouph} to it we define 
 in Section \ref{sec3.3} the volume of a generic framed flat $PGL(K, \C)$-connection 
on $M$. 
We stress that although the motivic volume of a framed flat connection does not depend 
on the balk triangulation ${\bf t}_{3d}$, it does depend on the boundary triangulation ${\bf t}$. 

Using  formula \eqref{diffdilog} and  the commutativity of the last square in \eqref{flags2Bloch01}, we get  
a formula for the variation of the volume  of generic framed flat connections on $M$. 
It generalizes the Neumann-Zagier formula 
for variation of volumes of hyperbolic 3-manifolds with toric boundary \cite{NZ}, and the work of Bonahon \cite{Bonahon-vol} on hyperbolic 3-manifolds with geodesic boundary. The  $PGL(3,\C)$-analog of the variation formula 
was beautifully established by Bergeron-Falbel-Guilloux in \cite{BFG-sl3}.

If the big boundary is absent, 
the motivic volume lies in the kernel of the Bloch complex map (\ref{blcmap}), and thus defines an element 
of $\mathrm{K}_3^{\rm ind}(\C)\otimes \Q$ due to a theorem of Suslin \cite{Sus1}. 
This follows immediately from the construction and the fact that 
the map (\ref{hdflbl}) is a map of complexes. 
The value of the regulator on it is the Chern-Simons invariant of the connection.

3) What exactly happens under the 2--3 moves changing the  triangulation ${\bf t}_{3d}$? 
In Section \ref{sec:23} we  prove that the $PGL(K)$ pentagon relation can be reduced to a sequence of 2--3 Pachner moves in the $K=2$ case, 
which we call elementary pentagon relations. 
The elementary pentagon relations match the 
terms of the component  $\alpha_5$ of the map (\ref{hdflbl}).

The 2--3 move on a 3-manifold can be viewed as a cobordism that amounts to attaching a 4-simplex 
to the 3-manifold. We believe that the $K$-decomposition of this 4-simplex 
should play an important role 
in construction of invariants of 4-manifolds. The fact that the elementary pentagons 
which appear in the decomposition of the 2--3 $PGL(K)$-move match the hypersimplices 
of type $\Delta^{1,2}$ and $\Delta^{2,1}$ 
of the $K$-decomposition of the 4-simplex, which are the 4d analogs of the octahedron,  agrees nicely with this idea.

\subsubsection*{Combinatorics}

In Section \ref{sec:combi}, we generalize to $K> 2$ a combinatorial analysis of ideal triangulations (and their symplectic properties) pioneered by Neumann-Zagier \cite{NZ} and extended by Neumann \cite{Neumann-combinatorics} for $K=2$. We introduce an algebra of paths on slices in the $K$-decomposition of a triangulation of an admissible 3-manifold $M$. These paths provide a simple, geometric interpretation of the monomial map relating coordinates on $\CX_K(\pd M,\mb t)$, gluing constraints $c_v$, and octahedron parameters $z_i,z_i',z_i''$. 

We use the geometry of paths to prove a weak version of the Symplectic Gluing Conjecture, essentially saying that the symplectic quotient on the RHS of \eqref{Xred-intro} makes sense and contains $\CL_K(\pd M,\mb t)$ as a subspace (Proposition \ref{prop:symp}).

The geometry of paths in Section \ref{sec:combi} have another application: they allow us to consistently define logarithmic lifts of octahedron parameters and coordinates on $\CX^{\rm un}_K(\pd M,\mb t)$ --- thus lifting the symplectic quotient \eqref{Xred-intro} to universal covers. As initially discussed in \cite{Dimofte-QRS}, this is a necessary requirement for quantization of the pair $\CL_K(M,\mb t_{3d})\subset \CX^{\rm un}_K(\pd M,\mb t)$. It is also a necessary requirement for the definition of gauge theories $T_K[M]$ in later sections of the paper. This lift was introduced by Neumann in \cite{Neumann-combinatorics} in order to calculate the Chern-Simons invariant of a flat $PGL(2,\C)$-connection.

\subsubsection*{Related mathematical works}

There are several recent papers closely related to our work. Bergeron-Falbel-Guilloux~\cite{BFG-sl3} were the first to study parametrizations of spaces of flat connections on 3-manifolds for $K>2$, considering $PGL(3,\C)$-connections. The authors localized flat connections to tetrahedra, defined the equivalent of our octahedron parameters, and generalized the classic formulas for variations of the volume.
While this paper was in preparation, Garoufalidis, Goerner, and Zickert \cite{GGZ-slN} (following \cite{GTZ-slN, Zickert-rep, Zickert-sl3}) explained how to localize flat $PGL(K,\C)$-connections to tetrahedra by considering ``decorated representations'' of $\pi_1(M)$. The ``decorations'' of \cite{GGZ-slN} are identical to our notion of ``framing,'' and lead to the same parameters $z_i,z_i',z_i''$ and gluing equations for the space $\wt \CL_K(M,\mb t_{3d})$ of framed flat connections on $M$ that we discuss. A major difference between \cite{GGZ-slN} and the present work includes our simultaneous study of both bulk and boundary moduli spaces $\CL_K(M)\subset \CX^{\rm un}_K(\pd M)$ as $\mathrm{K}_2$-Lagrangian pairs obtained from symplectic gluing (which necessitates the introduction of admissible manifolds, with big and small boundaries). This allows us to easily obtain (\eg) the formula for variation of the volume.
Another major difference is our use of geometric $K$-decompositions.

Since the first version of this paper, a proof of an important result for manifolds with boundary consisting entirely of small tori, such as knot complements,   
generalizing the work of Neuman \cite{Neumann-combinatorics} for $K=2$ has been proposed by \cite{GZ-gluing}. 
It implies  
 the Symplectic Gluing Conjecture.   The proof  is extremely technical. It would be very satisfying to find a simpler, more fundamental proof. 
  For the case $K=2$, a simple topological perspective on symplectic gluing was given in \cite{DV-NZ}.

Methods for quantizing the pair $\CL_K(M)\subset \CX^{\rm un}_K(\pd M)$ have been significantly improved since the first version of this paper. In particular, it was realized that in order to quantize $\CL_K(M,\mb t_{3d})\subset \CX^{\rm un}_K(\pd M, \mb t)$ for a fixed triangulation $\mb t_{3d}$ --- to obtain both a Chern-Simons wavefunction and a set of difference operators that annihilate it --- it is necessary for the triangulation $\mb t_{3d}$ to admit a \emph{positive angle structure} \cite{KashAnd, AK-new, AK-complexCS, Gar-index, GHRS-index, D-levelk}. This condition is also necessary physically in order for the gauge theory $T_K[M]$ to flow to a well-defined conformal field theory in the infrared. The positive angle structure seems to provide an intrinsically three-dimensional substitute for the notion of positivity in the study of cluster varieties such as $\CX_K(\pd M,\mb t)$. This relationship will be explored elsewhere.

\subsection{Organization}

To recap, the paper is organized as follows.

In Section \ref{sec:basics} we introduce 
the notion of a triangulated admissible 3-manifold (Section~\ref{sec:admissible}),  
 different moduli spaces with framed flat connections (Section~\ref{sec:spaces}), 
and explain 
how the framing allows a localization of framed flat connections to configurations of flags (Section~\ref{sec:configs}). We then define the hypersimplicial $K$-decomposition (Section \ref{sec:Kdec}).
 
In Section \ref{sec:FFC} we analyze how octahedron parameters are used to describe framed flat connections on an admissible $M$ and its boundary, and begin to discuss the symplectic properties of the moduli spaces $\CL_K(M)\subset \CX^{\rm un}_K(\pd M)$. We review how 
the $K$-decompositions of the 2d simplices, called $K$-triangulations, 
were used in \cite{FG-Teich} to define cluster coordinates on $\CX_K^{\rm un}(\pd M)$. 
Then we revisit the octahedron parameters related to a $K$-decomposition of $M$, and show (Theorem \ref{thm:XM}) 
how the elementary symplectic pairs $\CL_\soct\subset \CP_{\pd\soct}$ for octahedra are glued together to construct a Zariski open part for the pair $\CL_K(M)\subset \CX^{\rm un}_K(\pd M)$. 
We also study 2--3 moves, and decompose 
them into sequences of elementary ones.

In Section \ref{sec:Bloch}, we  formulate the Symplectic Gluing Conjecture (Section \ref{sec:perspective}). Then we introduce all the ingredients necessary to understand the 
homomorphism of complexes  \eqref{hdflbl}. This allows us to prove  the main new result,  Theorem \ref{thm:K2}, telling 
 that ${\cal L}_K({\cal M})$ 
is $\mathrm{K}_2$-isotropic in the case that $\pd M$ consists entirely of big boundary, with
holes filled in by small discs. 
We discuss how to generalize this to arbitrary admissible~$M$.

In Section \ref{sec:combi} we follow the combinatorial approach to symplectic gluing.
We discuss the combinatorics of octahedron parameters, the Poisson brackets that they induce on $\CX_K(\pd M)$, and the abstract data needed for quantization. We defer to Appendix \ref{app:NZ} 
the proof of the most nontrivial result about Poisson brackets for eigenvalue coordinates on $\CX_K(\pd M)$.

In Sections \ref{sec:Tpoly} and \ref{sec:Tcusp} we return to the physical motivations of this paper. We use the combinatorial data of Section \ref{sec:combi} to construct simple theories $T_K[M]$ associated with polyhedra and show how 2--3 moves encode mirror symmetries. We then discuss general properties of theories $T_K[M]$ when $M$ is a knot or link complement, including the large-$K$ scaling behavior. (At the end of Section \ref{sec:Tcusp} we provide a few examples of the moduli spaces associated with simple knot complements, and how their coordinates are computed from paths in the $K$-decomposition.)

The complete dictionary between the symplectic data of Section \ref{sec:combi}, class-$\CR$ gauge theories, and other quantum invariants is reviewed in Appendix \ref{app:symp}.

%%%%%%%%%%%%%%%%%%%%%%%%%%%%%%%%%%%%
%%%%%%%%%%%%%%%%%%%%%%%%%%%%%%%%%%%%
%%%%%%%%%%%%%%%%%%%%%%%%%%%%%%%%%%%%
\section{Basic tools and definitions}
\label{sec:basics}

We carefully introduce the basic objects studied throughout the paper.

\subsection{Gluing admissible 3-manifolds from truncated tetrahedra} \label{sec:admissible}

\begin{figure}[tbh]
\centering
\includegraphics[width=6in]{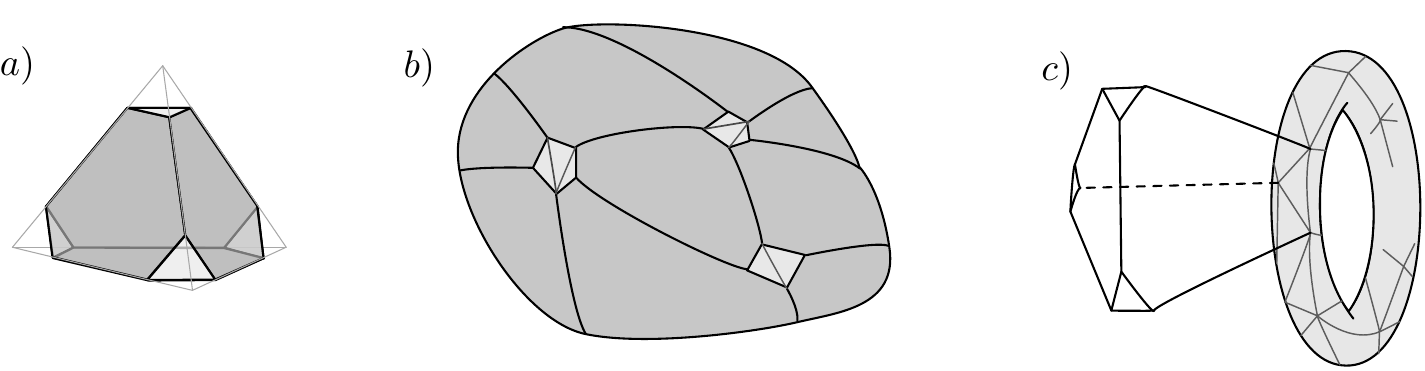}
\caption{($a$) truncated tetrahedra, glued together to form 3-manifolds that have either ($b$) big boundaries tiled by hexagonal faces with holes filled in by small vertex triangles; or ($c$) small torus boundaries tiled exclusively by vertex triangles.}
\label{fig:trunc-intro}
\end{figure}

Let us truncate a tetrahedron by cutting off its vertices. The resulting {\it truncated tetrahedron} 
has two kinds of faces: (small) triangles replacing the original vertices, and (big) hexagons.  
Let us glue truncated tetrahedra into an oriented manifold with boundary by allowing pairs of hexagonal faces to be glued, 
but not triangular ones (Figure~\ref{fig:trunc-intro}). 
The  boundary of the resulting manifold $M$ is tiled by triangles and unglued hexagons. 
We call the part tiled by triangles the {\it small boundary}, and 
the part tiled by hexagons the {\it big boundary}.

A small boundary component could have topology of any oriented surface. 
However we impose two additional constraints. We say that a gluing is {\it admissible} if it satisfies the following topological conditions:
\begin{itemize}
\item The small boundary components have abelian fundamental groups.
\item The small boundary components are not spheres.
\item The big boundary components have negative Euler character.
\end{itemize}
The reason for the first condition is to ensure that any flat connection admits a framing (Section \ref{sec:spaces}).
The second condition is ultimately necessary for the existence of logarithmic path coordinates in Section \ref{sec:combi} and subsequent quantization and definition of $T_K[M]$, though it could be relaxed classically. Mathematically, the second and third conditions eliminate stacks that should be assigned to 
spheres with less than three punctures.

A 3-manifold with boundary is {\it admissible} if it is homeomorphic to a manifold 
with boundary obtained by an admissible gluing 
of truncated tetrahedra. 
We consider only admissible manifolds $M$. 
An ideal triangulation of an admissible manifold is an admissible tiling of the manifold by truncated tetrahedra.

A small boundary component must be a {\it closed torus}, an {\it annulus}, or
a {\it disc}. A big boundary component is a surface with $h\geq 1$ holes (see Figure \ref{fig:trunc-intro}b), provided by the edges
of the truncated tetrahedra shared by triangles and hexagons. 
The hexagonal tiling is equivalent to a 2d ideal triangulation of this surface --- shrinking small discs to points we 
get an ideal triangulation. 
The small and big boundary components are connected as follows. 

\begin{itemize}
\item
The small tori are closed, disjoint boundary components.
\item
Boundary circles of small annuli must be  glued to 
boundary circles of big surfaces. 
\item
The boundary circle of a small disc is glued to a boundary circle 
of a big boundary surface. 
So each disc fills a hole of a big boundary. 
\end{itemize}
Here are some useful examples of admissible 3-manifolds to keep in mind:
\vspace{-.1cm}
\begin{enumerate}\setlength{\itemsep}{-.05cm}
\item The most basic example is given by 
a  tetrahedron. Its  big boundary is a four-holed sphere,  
and each hole is filled by a small boundary disc. 
More generally, any convex polyhedron 
gives rise to an admissible 3-manifold. Indeed, consider a triangulation of the polyhedron 
into tetrahedra. Its big boundary is a sphere with 
the holes matching the vertices of the polyhedron. 

\item Another fundamental example is a  hyperbolic 3-manifold, triangulated into ideal hyperbolic tetrahedra. 
The big boundary is then the geodesic boundary. It is a union of  geodesic surfaces. 
The small torus boundaries are cusps or deformed cusps. 
The cusps can be regularized by taking horoball neighborhoods of all the ideal hyperbolic tetrahedra vertices 
ending on them. This truncates the vertices of the ideal hyperbolic tetrahedra. 
In each tetrahedron, the truncated-vertex triangles are Euclidean. 
Moving from tetrahedron to tetrahedron going around the cusp, we sweep out a Euclidean surface. 
Metrically, the surface closes up into a Euclidean torus if the cusp holonomy is parabolic. Otherwise it keeps spiraling.

\item Our last example is given by a link complement.  
It has only small torus boundaries, one for every excised link component.

\end{enumerate}

\begin{figure}[htb]
\centering
\includegraphics[width=5.7in]{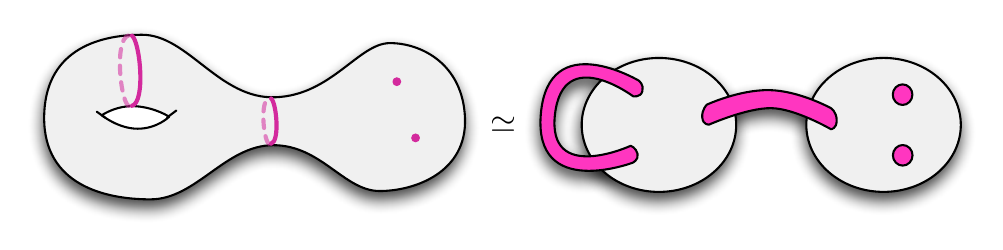}
\caption{A topological surface with lamination and marked points is equivalent to a big boundary with small annuli and small discs connecting its holes. Here, two loops of a lamination are replaced by two annuli (depicted as long tubes); and two marked points are replaced by discs. The remaining big boundary consists of two 3-punctured spheres.}
\label{fig:lam}
\end{figure}

Topologically, we can describe the boundary $\pd M$ of an admissible 3-manifold in terms of laminations.
A {\it simple lamination} on a surface $\CC$ with marked points $\{p_1,p_2,...\}$ is a collection of non-intersecting simple unoriented 
non-isotopic loops on $\CC\bs\{p_1,p_2,...\}$ 
modulo isotopy.
The boundary $\partial M$ is a disjoint union of tori and  
closed oriented surfaces $\CC$ 
with simple laminations and marked points. The small annuli in $\pd M$ are neighborhoods of the loops of the laminations, and the small discs are neighborhoods of the marked points (Figure \ref{fig:lam}). The remainder of $\pd M$ is big boundary.

There is also a nice physical interpretation of the admissibility conditions. We ultimately want to study 3-manifolds $M$ on which the 6d $(2,0)$ theory can be compactified. The big boundaries we have just described are asymptotic regions of $M$ in six dimensions (\ie\ of $M\times \R^3$), of the form $\CC\times \R_+$. The small boundaries come from regularizing codimension-two defects of the 6d theory. The defects may either be closed loops in $M$ (hence small torus boundaries), infinite lines connecting two asymptotic regions (hence small annuli), or in special cases \emph{half}-infinite lines attached to one asymptotic region (hence small discs). This is explained further in \cite{DGV-hybrid}. Note that it is impossible to obtain small spheres --- there are no codimension-three defects in the 6d theory, which could appear point-like on $M$.

\subsection{Configurations of flags}
\label{sec:flags}

To define the correct phase space assigned to the boundary $\partial M$ of an admissible 3-manifold $M$, 
and especially to define its Lagrangian subvariety corresponding to $M$,  
we start with a small digression on flags and configurations of flags.

A flag $F_\bullet$ in a $K$-dimensional complex vector 
space $V_K$ is a collection of nested subspaces
\be  F_\bullet:\quad   \varnothing   =F_0 \subset F_1 \subset F_2 \subset\cdots \subset F_K =  V_K\,, 
~~~{\rm dim}\,F_i=i.
\ee
We also use the notation 
$F^i :=  F_{K-i}$ 
for the codimension $i$ subspace of the flag. Then a flag is denoted 
\be \label{flag} F^\bullet:\quad   \varnothing   =F^K \subset F^{K-1} \subset F^{K-2} \subset\cdots \subset F^0 =  V_K\,, ~~~{\rm codim}\,F^i=i.
\ee
In fact later on we will mostly use the second convention. 
We say that a basis $\{f_1,f_2,\ldots, f_K\}$ in $V$ is adjusted for a flag $F_\bullet$ if 
$F_{i}$ is spanned by $\{f_1,f_2,\ldots, f_i\}$. 

A flag $F_\bullet$ in $V_K$ induces a canonical dual flag 
$F_\bullet^*$ in the dual vector space $V_K^*$:
\be \label{dualflag} 
F_\bullet^*:\quad \varnothing = F_0^* \subset F_1^* \subset \cdots \subset F_K^* = V_K^*\,,~~~~F_i^*:= (V_K/F^i)^*.
\ee
Indeed, 
the projection $V_K/F^{i+1}\to V_K/F^i$ induces a dual injection $(V_K/F^{i})^*\hookrightarrow (V_K/F^{i+1})^*$. 

A collection of $m$ flags $(F_{1, \bullet}, F_{2, \bullet}, \ldots, F_{m, \bullet})$ in $V_K$ is said to be {\it generic} if 
for any collection of integers $a_1, \ldots, a_m$ that sum to $K$ one has an isomorphism
$$
F_{1, a_1} \oplus \ldots \oplus F_{m, a_m} = V_K. 
$$

The space of flags is denoted by ${\cal B}$, the 
moduli space of  configurations of $m$ flags is denoted by ${\rm Conf}_m({\cal B})$, and 
the variety of  generic configurations of $m$ flags is denoted by ${\rm Conf}^*_m({\cal B})$.
Recall that if we have a set $X$ with an action of a group $G$, the \emph{configurations} of $n$ elements of $X$ are the orbits of $G$ acting on $X^n$. In this case, a configuration of flags consists of the $PGL(V_K)$-orbit of a tuple of flags.
For  $PGL(2, \C)$, the space of flags is ${\mathbb C\mathbb P}^1$. The 
cross-ratio provides an isomorphism 
\be
{\rm Conf}^*_4({\mathbb C\mathbb P}^1) \stackrel{=}{\longrightarrow} {\mathbb C\mathbb P}^1-\{0, 1, \infty\}. 
\ee

\subsection{The basic moduli spaces} \label{sec:spaces}

A vector bundle ${\cal V}$ with a flat connection
gives rise to a new bundle with a flat connection on the same space, the flag bundle ${\cal V}_{\cal B}$, 
whose fiber at a point $x$ is the space of all flags at the fiber of ${\cal V}$ at $x$. 

\bd
Let $M$ be an admissible 3-manifold. A {\rm framing} on a vector bundle $\CV$ with a flat 
connection on $M$  
is a choice of a flat section of the flag bundle ${\cal V}_{\cal B}$ 
on each of the small components of $\partial M$.

Let $\pd M^*$ be the total boundary $\pd M$ with all small discs removed. A framing on a vector bundle $\CV$ with flat connection on $\pd M^*$ is a choice of flat section of ${\cal V}_{\cal B}$ on each remaining small component of $\pd M^*$ and on each $S^1$ boundary of $\pd M^*$.
\ed 
To define a framing we need to define it for each of the small boundary components. 
A framing at a boundary component can be thought of as an invariant flag, that is
a  flag in the fiber at a point of the 
component, preserved by the holonomies around the component.

A collection of commuting linear transformations in a finite-dimensional complex vector space
 always admits  an invariant flag, \ie\ a flag preserved by all the transformations. 
Therefore, since the fundamental group of each small boundary components is abelian, 
a framing is an additional structure on a vector bundle with connection --- it always exists. 
This is why we consider only small boundary components with the abelian fundamental group. Otherwise 
the very existence of an invariant flag is a severe restriction on a flat connection.

\vskip 2mm
If the holonomy around a puncture is unipotent, and  consists of a single Jordan block, 
then there is a unique invariant flag there.
Otherwise additional freedom arises. For example, if the holonomy is trivial, 
then the choice of invariant flag is completely arbitrary. 
\vskip 2mm

Using the notion of framing, we can define now the moduli spaces needed in the paper.  

\bd \hspace{1in}\\ \indent i) The moduli space 
${\cal X}_K(\partial M)$ parametrizes framed $PGL(K,\C)$-bundles with flat connections on $\partial M^*$. 

ii) The subspace ${\cal X}_K^{\rm un}(\partial M)\subset {\cal X}_K(\partial M)$ parametrizes 
the framed flat connections on $\partial M^*$ with unipotent holonomy around the holes ($S^1$ boundaries) where small discs were removed. 

iii) The moduli space 
${\cal L}_K(M)$ 
parametrizes framed flat $PGL(K,\C)$-connections on $\partial M^*$   
that can be extended to framed flat connections on $M$.

iv) The moduli space $\wt \CL_K(M)$ parametrizes framed flat $PGL(K,\C)$-connections on $M$. 
\ed

Any loop on $\pd M^*$ around a hole is contractible in $M$. So 
the holonomy of a flat connection in $M$ around any such loop in $\pd M$ is trivial, and thus 
the invariant flags near every puncture are completely unrestricted. So
${\cal L}_K(M)$ embeds into the moduli space of connections with trivial holonomies around the holes. 
The moduli spaces are related as follows: 
\be
{\cal L}_K(M) \subset {\cal X}_K^{\rm un}(\partial M)\subset {\cal X}_K(\partial M).
\ee
Moreover, $\CL_K(M)$ is the image of the projection $\wt \CL_K(M)\to \CX_K(\pd M)$.

The moduli space ${\cal X}_K(\partial M)$ has a canonical Poisson structure. 
The moduli space ${\cal X}_K^{\rm un}(\partial M)$ is symplectic. It is  realized as a closure of 
a symplectic leaf 
of ${\cal X}_K(\partial M)$. It serves the role of the phase space.  
The subspace  ${\cal L}_K(M)$ is supposed to be a Lagrangian subspace.

Zariski open parts of these moduli spaces can be understood 
by introducing coordinates. However  the coordinates are not everywhere defined. 
The moduli spaces themselves are of fundamental importance.

\vskip 3mm
The moduli space ${\cal X}_K(\partial M)$  is naturally  decomposed into a product of the 
moduli spaces assigned to the components of $\partial M$. To state this precisely, let 
us discuss the moduli spaces assigned to the small boundary components. 

\paragraph{1. The phase space for a surface with a simple lamination.} 
Let $\CC$ be an oriented surface with $h>0$ holes and $\chi(\CC)<0$. 
Let $\gamma$ be a simple lamination on $\CC$. (Thus, we have in mind that $\CC$ is the union of the big boundary of $M$ and the small annular boundaries that connect some pairs of holes; the small annuli are collapsed to lamination curves.) A framing on a flat vector bundle with connection on $(\CC, \gamma)$ is a choice 
of a flat section of the flag bundle over each component of the lamination $\gamma$, and near 
each of the holes. Then 
\be {\cal X}_K(\CC, \gamma):= \left\{\begin{array}{l}
\text{framed flat $PGL(K,\C)$-connections on $(\CC, \gamma)$} 
 \end{array}\right\}/{\rm isomorphisms} \,.
\ee
This is a finite-dimensional complex space. If $g$ is the genus of $\CC$, its dimension is
$-\chi(\CC)~{\rm dim}(PGL(K))$. If we further fix the eigenvalues of the $PGL(K,\C)$ holonomies
around each of the holes, the dimension is cut down to
\be \label{dKgh}
d_K(g,h) = -\chi(\CC)~{\rm dim}(PGL(K))  - h~{\rm rk}(PGL(K)) = 2(K^2-1)(g-1)+K(K-1)h\,.
\ee

Let ${\rm Loc}_K(\CC)$ be the traditional  moduli space 
of vector bundles with flat connections on $\CC$. Forgetting the 
framing, we get a projection, which is a finite cover over the generic point: 
\be \pi: {\cal X}_K(\CC) \longrightarrow {\rm Loc}_K(\CC)\,. \ee
Over the locus of connections with unipotent holonomies around the punctures it 
is generically one to one, and partially resolves the  singularities of the traditional unipotent moduli space.

\paragraph{2. The phase space and the coordinate phase space for a surface.}
The moduli space ${\cal X}_K(\CC)$ was introduced in \cite{FG-Teich}, for the case with no lamination. It was proved there that 
any 2d ideal triangulation ${\bf t}$ gives rise to a rational coordinate system $\{x_i^{\bf t}\}$ on 
the moduli space ${\cal X}_K(\CC)$. Moreover, it was shown that 
there is a Zariski open subset 
\be
{\cal X}_K(\CC; {\bf t}) \subset {\cal X}_K(\CC) 
\ee
on which the coordinates $\{x_i^{\bf t}\}$ are well defined, and 
which is identified with a complex torus:
\be
{\cal X}_K(\CC; {\bf t}) \stackrel{=}{\longrightarrow} (\C^*)^d, ~~~~ p \longmapsto x_i^{\bf t}(p)\,.
\ee
We call the complex torus ${\cal X}_K(\CC; {\bf t})$ the 
{\it coordinate phase space associated with an ideal triangulation}, or just the {\it coordinate phase space}.

\vskip 2mm
There is a holomorphic Poisson structure on ${\cal X}_K(\CC)$,
which in any of the coordinate systems $\{x^{\bf t}_i\}$ is given by 
\be
\{x^{\bf t}_i, x^{\bf t}_j\} = \varepsilon^{\bf t}_{ij}x^{\bf t}_i x^{\bf t}_j, ~~~\varepsilon^{\bf t}_{ij}\in \Z.
\ee
The Poisson tensor $\varepsilon^{\bf t}_{ij}$ depends on the choice of the triangulation. The eigenvalues of holonomies around holes in $\C$ generate the center of the Poisson algebra; after fixing them, the Poisson tensor can be inverted to define a holomorphic symplectic form.

With non-empty lamination, the corresponding moduli space 
${\cal X}_K(\CC, \gamma)$ is studied and coordinatized in \cite{FG-laminations}; 
see also \cite{DGV-hybrid, Kabaya-pants} for the case $K=2$.

\paragraph{3. The phase space for a torus $T^2$.} 
The phase space ${\cal X}_K(T^2)$ is the moduli space of flat $PGL(K, \C)$-connections 
on $T^2$ together with a framing, given by a flat section of the associated flag bundle on $T^2$. 
Alternatively, a framing is a choice of a flag in a fiber invariant under both A- and B-cycle holonomies, or 
 a reduction of the structure group to a Borel subgroup. 
There is a birational equivalence, that is 
an isomorphism at the generic point
\be  \label{PT2intro}
 {\cal X}_K(T^2) \stackrel{\sim}{=} ({\Bbb C}^*\times {\Bbb C}^*)^{K-1}.
\ee
It assigns to a framed flat connection on $T^2$ the   {\it ordered} collections $\{\ell_a\}$ and $\{m_b\}$, where
\be \label{ABholonomies}
{\rm diag}(1, \ell_1, \ell_1\ell_2, \ldots, \ell_1 \ldots\ell_{K-1}) ~~\mbox{and}~~ 
{\rm diag}(1, m_1, m_1m_2, \ldots, m_1 \ldots m_{K-1})
\ee are  the 
diagonal parts of the $PGL(K, \C)$-holonomies $\text{Hol}_A,\,\text{Hol}_B$ around the A- and B-cycles of the torus. 
Let $\kappa_{ab}$ 
be the Cartan matrix.  
Then the holomorphic symplectic form is 
\be  \label{torusS}
\ds \Omega_{T^2} = \text{Tr}\,d\log\Big(\frac{\text{Hol}_A}{\det\text{Hol}_A{}^{\frac1K}}\Big)\wedge d\log\Big(\frac{\text{Hol}_B}{\det\text{Hol}_B{}^{\frac1K}}\Big) = \sum_{a,b=1}^{K-1} (\kappa^{-1})_{ab}\,
d\log \ell_a \wedge d\log m_b\,,
\ee
corresponding to Poisson brackets $\{\ell_a,m_b\} = \kappa_{ab}\ell_am_b$.
The traditional moduli space of flat connections on the torus is birationally isomorphic to the quotient 
of (\ref{PT2intro}) by the symmetric group  
$S_K$. It makes it more difficult to 
introduce the canonical coordinates, since they must be invariants of the symmetric group. 
 The choice of invariant flag orders the eigenvalues. 
Notice that for the torus there is no extra unipotency condition.

\vskip 3mm
Since $\partial M$ is a disjoint union of tori $T_j^2$ and punctured surfaces with simple laminations  
$(\CC_i, \gamma_i)$, the moduli space ${\cal X}_K^{\rm un}(\partial M)$ is a product:
$$
{\cal X}_K^{\rm un}(\partial M) = \prod_i {\cal X}_K(T_i^2) \times \prod_j{\cal X}_K(\CC_j, \gamma_j).
$$

\paragraph{4. The moduli space ${\cal L}_K(M, {\bf t})$.}
Let us consider now an admissible 3-manifold $M$ whose boundary $\partial M$ does not have small annuli. 
Choose a triangulation ${\bf t}$ of the big boundary of $\partial M$. 
We define 
\be
{\cal L}_K(M, {\bf t}):= {\cal L}_K(M) \cap \bigg(\prod_i {\cal X}_K(T_i^2) \times \prod_j{\cal X}_K(\CC_j; {\bf t}_j)\bigg)\,.
\ee

Let us see now what we get in our two running examples.

\paragraph{Examples:} 1. Let $M= B_m$ be a 3d ball with $m\geq 3$ small discs on the boundary. It 
can be approached combinatorially as a convex polyhedron with $m$ vertices. 
Then ${\cal X}^{\rm un}_K(\partial M)$ 
is the moduli space of framed flat $PGL(K)$-bundles with connections on $S^2$ minus $m ~\mbox{discs}$, and 
unipotent monodromies around the discs. 
Now let us figure out the Lagrangian subspace. Since any flat connection on a ball is trivial, 
the invariant flags are unrestricted, and are the only non-trivial part of the data. 
So ${\cal L}_K(B_m)$ is the configuration space of $m$ flags, 
realized as a framing data in a trivialized vector bundle with connection:
\be
{\cal L}_K(B_m) = {\rm Conf}_m({\cal B}) \subset {\cal X}^{\rm un}_K(S^2 - m ~\mbox{discs}). 
\ee
The statement that this is a Lagrangian subvariety is non-trivial even in this case. It follows from the 
result of  \cite{FG-Teich} that a flip of a 2d triangulation is a birational transformation that preserves the Poisson/sympectic structure. 

2. Let $M= S^3 \bs {\cal K}$ be the complement of a knot ${\cal K}$ in $S^3$. 
Then the small boundary is a torus $T^2$. 
The $SL(2,\C)$ version of ${\cal L}_2(M)$ is a complex curve.%
\footnote{More precisely, the top-dimensional components of ${\cal L}_2(M)$ are a complex curve. Exceptional zero-dimensional components can also arise, but they are usually excluded by additional stability conditions.} %
It is usually called the A-polynomial curve of the knot \cite{cooper-1994}.  
Therefore for $SL(K,\C)$, we arrive at a natural generalization of the A-polynomial as a Lagrangian subvariety in the phase space (\ref{PT2intro}). Upon quantization, the operators $\hat \CL_{K}(M)$ quantizing the equations defining the Lagrangian subvariety $\CL_{K}(M)$
are expected to provide recursion relations for the colored HOMFLY polynomials of a knot, just as the quantized A-polynomial (conjecturally) 
provides a recursion relation for the colored Jones polynomial \cite{Gar-Le, garoufalidis-2004, gukov-2003}. (See \eg\ \cite{Gar-qSL3, FGS-superA, Gar-a} for quantizations of $\CL_{K}(M)$ in several special cases.)

\vskip 3mm

\paragraph{Conclusion:} At first glance,  
the moduli space ${\cal X}^{\rm un}_K(\pd M)$ of framed unipotent flat connections 
on the boundary $\pd M$ of an admissible 3-manifold $M$ is just a modification 
of the traditional moduli space ${\rm Loc}^{\rm un}_K(\pd M)$ 
of unipotent flat connections on $\pd M$, making it more accessible. 
It turns out however that the moduli space ${\cal X}^{\rm un}_K(\pd M)$
is really indispensable  in construction of the Lagrangian subvariety 
assigned to $M$: 
$$
\mbox{\it Quite often, there is simply no room in  ${\rm Loc}^{\rm un}_K(\pd M)$ for a 
Lagrangian assigned to $M$!}
$$
Indeed, the canonical projection 
$
\pi: {\cal X}^{\rm un}_K(\pd M)\longrightarrow {\rm Loc}^{\rm un}_K(\pd M)
$ 
can map ${\cal L}_K(M)$ to a subvariety 
whose dimension is too small to be Lagrangian.

\subsection{From framed flat bundles to configurations of flags} \label{sec:configs}

\begin{wrapfigure}{r}{1.5in}
\vspace{-.5cm}
\includegraphics[width=1.5in]{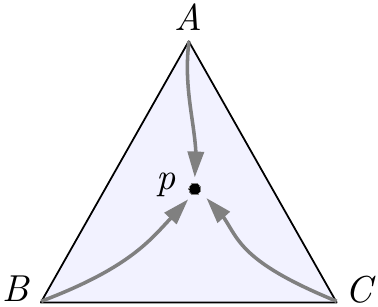}
\caption{Parallel transport of flags to the common point $p$.}
\label{fig:ABCparalleltransp}
\end{wrapfigure}

Given a framed $PGL(K, \C)$-vector bundle ${\cal V}$ with a flat connection on a 2d surface
$\CC$ with punctures, and given an ideal triangulation  of $\CC$, 
\ie\ a triangulation with vertices at the punctures, any ideal {\it triangle} 
$t$ of the triangulation gives rise to a configuration of three flags $(A,B,C)$ 
in a $K$-dimensional complex vector space $V_K$. 
Namely, consider the restriction of the flat bundle ${\cal V}$ to the triangle $t$. 
Then the flat sections defining the framing provide flat sections of the associated flag bundle 
${\cal V}_{\cal B}$ at each punctured disc on $\CC$ in the vicinity of the vertices of
the triangle.
Using the connection, we parallel transport each of the invariant flags to a point $p$ inside of the triangle $t$. 
Since the connection is flat, and the triangle is contractible, the resulting collection of flags 
in the fiber ${\cal V}_p$ of the bundle at $p$ does not depend on the 
paths from the small triangles to $p$. The resulting configuration of three flags $(A,B,C)$ 
does not depend on any choices involved. More generally, given any ideal polygon on $\CC$ 
we can run the same construction and construct a configuration of flags in $V_K$ 
corresponding to the flat sections near the vertices of the polygon. 
This construction goes back to \cite{FG-Teich}, and leads to a 
construction of the coordinates associated with the triangulation in {\it loc. cit}. 
We will recall it later on, in Section \ref{sec:bdyflat}.

Literally the same construction with minimal adjustments
 can be applied to framed flat vector bundles 
 on an admissible 3-manifold with a given ideal triangulation. Here is how it goes. 
Consider a framed $PGL(K, \C)$-bundle with a flat connection ${\cal V}$ on an admissible 3-manifold $M$, 
and an ideal triangulation of $M$. Then an ideal tetrahedron $T$ of the triangulation 
gives rise to a configuration of four flags $(A,B,C,D)$ 
in a $K$-dimensional complex vector space $V_K$. 
Indeed, let us restrict ${\cal V}$ to the truncated tetrahedron $T$. 
The flat sections defining the framing provide flat sections of the flag bundle 
${\cal V}_{\cal B}$ at each of the four small triangles assigned to the vertices of the original tetrahedra. 
Using the connection, we parallel transport the corresponding invariant flags 
to a point $p$ inside of the tetrahedra $T$. 
Since the connection is flat, and the tetrahedron is contractible, the resulting collection of flags 
in the fiber ${\cal V}_p$ does not depend on the 
paths from the small triangles to $p$. So the resulting configuration of four flags $(A,B,C,D)$ 
is well defined.  Similarly,  any ideal polyhedron in $M$, and in particular 
an ideal triangle or an ideal edge of the triangulation, 
leads to a configuration of flags labeled by the vertices of the polyhedron.

The very notions of a {\it framed} bundle with flat connection and an {\it admissible} manifold 
were designed in such a way that we can run the above constructions. 
They deliver a collection of configurations of flags assigned to the 
ideal tetrahedra (in 3d) or triangles/rectangles (in 2d). 
So the next important question is how to deal with configurations of flags.

\subsection{Hypersimplicial $K$-decomposition} \label{sec:Kdec}

\begin{wrapfigure}[8]{r}{1.5in}
\vspace{-.2in}
\includegraphics[width=1.5in]{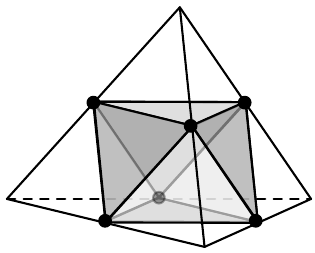}
\caption{Forming $\Delta^{1,1}$.}
\label{fig:D11}
\end{wrapfigure}

We now describe a construction that plays a crucial role in this paper, namely the hypersimplical $K$-decomposition of an ideal triangulation in dimension $n$.

Let $p$ and $q$ be two non-negative integers satisfying $p+q=n-1$. 
An $n$-dimensional hypersimplex $\Delta^{p,q}$ is defined  \cite{GGL} as the intersection of 
the $n+1$-dimensional cube $0 \leq x_i \leq 1$ with the hyperplane 
$\sum_{i=0}^n  x_i = q+1$. Combinatorially, it is 
isomorphic to the convex hull of the centers of $p$-dimensional faces 
of an $n$-dimensional simplex. 

\vskip 2mm
The hypersimplices $\Delta^{p, 0}$ and $\Delta^{0, q}$ are just simplices. 
The simplest hypersimplex different from a simplex 
is the octahedron $\Delta^{1,1}$. 
It is the convex hull of the centers of the edges of a tetrahedron (Figure \ref{fig:D11}).
\vskip 2mm

The boundary of a hypersimplex $\Delta^{p,q}$ is a union of $n+1$ hypersimplices 
$\Delta^{p-1,q}$ and $n+1$ hypersimplices $\Delta^{p,q-1}$. 
Indeed, they are given by the intersections with the hyperplanes $x_i=0$ and $x_i=1$ of the unit cube. 
For example, the boundary of the octahedron 
$\Delta^{1,1}$ consists of four $\Delta^{1, 0}$-triangles  and four $\Delta^{0, 1}$-triangles. 
The boundary of $\Delta^{2,1}$ is given by five octahedra $\Delta^{1,1}$ and five $\Delta^{2,0}$-tetrahedra , etc. 
\vskip 2mm

Consider the standard coordinate space $\R^{n+1}$. It contains the integral lattice $\Z^n \subset \R^n$. 
The hyperplanes  $x_i =s$,
 where $s\in \Z$ and 
$i=0, \ldots, n$, cut the space into unit cubes with vertices at the integral points. 
Given a positive integer $K$,  consider the $n$-dimensional 
simplex $\Delta^n_K$ given by the intersection of the 
hyperplane $\sum x_i =K$ with the positive orthant $x_i \geq 0$:
\be
\Delta^n_K = \{(x_0, \ldots, x_n) ~|~ \sum_{i=0}^n x_i =K\}.
\ee
The hyperplanes $x_i =s$, where $s\in \Z$, cut the simplex $\Delta^n_K$ 
into a union of hypersimplices. 
Indeed, the hyperplane 
$\sum x_i =K$ intersect each of the standard unit  lattice cubes either by an empty set, or by a 
hypersimplex. 

\bd \cite{FG-Teich}
A hypersimplicial $K$-decomposition (or $K$-decomposition, for short) of an $n$-dimensional simplex is a decomposition 
of the simplex $\Delta^n_K$ into hypersimplices provided by 
the hyperplanes $x_i =s$, $s\in \Z$. 
\ed

The polyhedra of a hypersimplicial $K$-decomposition have vertices at the {\it lattice points} 
\be
p_{b_0, \ldots, b_n}, \qquad b_0+ \ldots + b_n=K, \qquad b_i \in \Z.
\ee
The hypersimplices of $K$-decomposition are the closures of  
connected components of the complement to the hyperplanes $x_i =s$. The hypersimplices 
$\Delta^{p,q}$ of the $K$-decomposition match the $ {K+q\choose n}$ solutions of the equation 
\be \label{a-equation}
a_0 + \ldots + a_n= K-(p+1), \qquad a_i \in \Z_{\geq 0}.
\ee
Finally, a $K$-decomposition of a simplex induces $K$-decom\-po\-si\-tions of 
any faces of the simplex.

\paragraph{Examples:} 
\begin{enumerate}
\item
A $K$-decomposition of a segment is its decomposition into $K$ equal little segments. 

\item
When $n=2$ we get a $K$-triangulation of a triangle. 
It is given by three families of parallel lines, each consisting of $K$ lines including one of the sides, 
with triple point intersections, 
as shown on the Figure \ref{fig:triangleSL3}. They induce decompositions of the sides of the triangle into 
$K$ little segments. 
 
\begin{figure}[htb]
\centering
\includegraphics[width=2in]{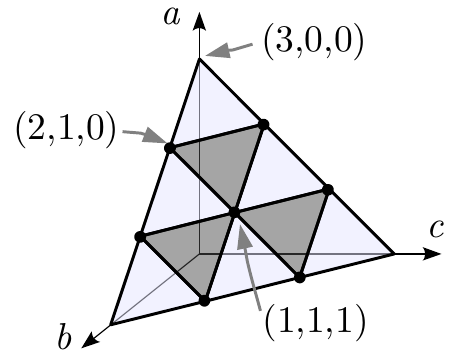}
\caption{2-simplex with lattice points $(a,b,c)$, $a+b+c=3$.}
\label{fig:triangleSL3}
\end{figure}

A $K$-triangulated triangle carries  ${K+2\choose 2}$ lattice points $p_{abc}$ 
indexed by triples of non-negative integers $(a,b,c)$ with $a+b+c=K$. 
The original triangle is decomposed into $K^2$ small triangles. 
They are of two types: ${K+1\choose 2}$ of them are ``upright'' $\Delta^{0,1}$-triangles,
 shaded white, and labeled by the triples $(a,b,c)$ with $a+b+c=K-1$; ${K\choose 2}$ of them are ``upside-down'' $\Delta^{1,0}$-triangles, shaded black 
and labeled by the triples $(a,b,c)$ with $a+b+c=K-2$.

\item
When $n=3$ we get a $K$-decomposition of a tetrahedron. It consists of 
${K+2\choose 3}$ upright $\Delta^{0,2}$-tetrahedra, ${K+1\choose 3}$ octahedra $\Delta^{1,1}$, and 
${K\choose 3}$ upside-down $\Delta^{2,0}$-tetrahedra. 
It induces a $K$-triangulation of each of the four faces of the tetrahedron.  
The lattice points $p_{abcd}$ are labeled by quadruples of non-negative integers that sum to $K$, see Figure~\ref{fig:hypersimplices}.

\begin{figure}[htb]
\centering
\includegraphics[scale=.8]{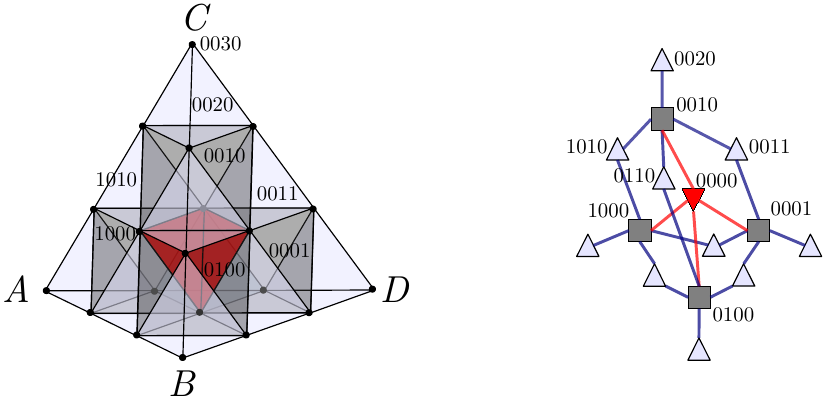}
\caption{The $3$-decomposition of the tetrahedron. The upright $\Delta^{0,2}$-tetrahedra (white) correspond to lines, the octahedra (black) to planes, and the upside-down $\Delta^{2,0}$-tetrahedron (red) to a 3-space. 
Adjacent hypersimplices obey incidence relations, which can be represented as a hypergraph, shown on the right.}
\label{fig:hypersimplices}
\end{figure}

\item
When $n=4$ we get a $K$-decomposition of a four-dimensional simplex into 
hypersimplices of type $\Delta^{0,3}$, $\Delta^{1,2}$, $\Delta^{2,1}$, $\Delta^{3,0}$. 

\end{enumerate}

%%%%%%%%%%%%%%%%%%%%%%%%%%%%%%%
\subsection{Configuration of flags and hypersimplices} \label{sec:config-basic}
Consider a generic configuration of $n+1$ flags $(F^\bullet_{0}, \ldots, F^\bullet_{n})$ in a $K$-dimensional complex vector space $V_K$.
We assign them to the vertices of an $n$-dimensional simplex $\Delta^n_K$. 

These flags define a collection of $(p+1)$-dimensional 
linear subspaces  in $V_K$, assigned to the $\Delta^{p,q}$-hypersimplices 
of the hypersimplicial  $K$-decomposition of the simplex $\Delta^n_K$. 
Namely, let $\Delta^{p,q}_{\bf a}$ be the $\Delta^{p,q}$-hypersimplex corresponding to a given partition 
\be \label{pcf}
{\bf a} = (a_0, \ldots , a_n ), \qquad a_0 + \ldots + a_n=K-(p+1), \qquad  a_i\geq 0~.%p+1=K-q, q=d
\ee

\bd \label{def2.4} Given a generic configuration of $n+1$ flags $(F^\bullet_{0}, \ldots, F^\bullet_{n})$ in $V_K$, 
we assign to $\Delta^{p,q}_{\bf a}$ a $(p+1)$-dimensional  subspace $\mathfrak F_{\bf a}$ 
of $V_K$ given by intersection of the flag subspaces $F_{i}^{a_i}$: 
\be \Delta^{p,q}_{\bf a}
 \quad\leadsto\quad \mathfrak F_{\bf a} = F^{a_0}_{0}\cap F^{a_1}_{1}\cap \ldots \cap F^{a_n}_{n}, \qquad 
{\bf a} = (a_0, \ldots, a_n)~.
\ee
\ed
Each hypersimplex $\Delta^{p,q}_{\bf a}$ is surrounded by $n+1$ hypersimplices $\Delta^{p-1,q+1}_{\bf b}$, 
where ${\bf b}$ is obtained from ${\bf a}$ by adding $1$ to one of the coordinates 
$(a_0, \ldots, a_n)$. So the collection of ${\bf b}$'s is
$$
(a_0+1, a_1, \ldots, a_n), ~(a_0, a_1+1, \ldots, a_n), ~~\ldots ~~,  ~
(a_0, a_1, \ldots, a_n+1).
$$
Therefore we get a collection of $n+1$ codimension-one subspaces $\mathfrak F_{\bf b}\subset  \mathfrak F_{\bf a}$.

One way to organize this data is to assign to the hypersimplex $\Delta^{p,q}_{\bf a}$ 
the configuration of $n+1$ hyperplanes $\mathfrak F_{\bf b}$ in the space $\mathfrak F_{\bf a}$, which we denote by 
${\cal P}{\cal C}_{\bf a}(F^{a_0}_{0},  F^{a_1}_{1},  \ldots , F^{a_n}_{n})$. 
\bd \label{def2.5} Given a generic configuration of $n+1$ flags $(F^\bullet_{0}, \ldots, F^\bullet_{n})$ in $V_K$, 
we assign to $\Delta^{p,q}_{\bf a}$ a configuration of $n+1$ hyperplanes 
in a $p$-dimensional projective space:
\be \label{congph} \Delta^{p,q}_{\bf a}
 \quad\longmapsto \quad 
{\cal P}{\cal C}_{\bf a}(F^{a_0}_{0},  F^{a_1}_{1},  \ldots , F^{a_n}_{n}) \in {\rm Conf}_{n+1}({\Bbb P}^p), \qquad 
{\bf a} = (a_0, \ldots, a_n)~.
\ee
\ed
Equivalently, it is a configuration of $n+1$ points in the $p$-dimensional 
projective space of hyperplanes in $\mathfrak F_{\bf a}$. 
Here are two basic examples (explained in much greater detail in Section~\ref{sec:FFC}):
\begin{itemize}
\item 
Let $n=3$. Then the octahedron $\Delta^{1,1}_{\bf a}$ gives rise to a configuration of four lines $\mathfrak F_{\bf b}$ in a 
$2$-plane $\mathfrak F_{\bf a}$.
The cross-ratio of these four lines is the coordinate $z_{\bf a}$ assigned to the 
octahedron $\Delta^{1,1}_{\bf a}$. 

\item 
Let $n=4$. Then the hypersimplex $\Delta^{1,2}_{\bf a}$ gives rise to a configuration of five lines $\mathfrak F_{\bf b}$ in a 
$2$-plane $\mathfrak F_{\bf a}$.
The hypersimplex $\Delta^{2, 1}_{\bf a}$ gives rise to a configuration of five planes $\mathfrak F_{\bf b}$ in a 
$3$-space $\mathfrak F_{\bf a}$.

Notice that generic configurations of five planes in a 3-space can be identified with 
generic configurations of five lines in a 2-plane. A nice way to see this is to pass to 
the corresponding projective configuration of five points in the projective plane, and 
notice that given five generic points in 
the projective plane, there is a unique conic containing them. 
This conic is isomorphic to a projective line, thus providing an isomorphism. 
(Another way to see it is to combine the two canonical isomorphisms (\ref{gale0}) and (\ref{gale}) from Section \ref{sec:BiGr}.)

This way we get two kinds of ``pentagons'' related to the octahedral coordinates. 
We will return to these two pentagons in Section \ref{sec:23}, and explore the relevant geometry.  
\end{itemize}

\paragraph{Remark.} The constructions   in Definitions \ref{def2.4}, \ref{def2.5} are dual to the one used in  \cite{G93}. 
Indeed, a flag in a vector space $V$ determines the dual  flag in the dual space $V^*$.  
A configuration of   flags in $V$ gives rise to  configuration of the dual  flags; 
the intersection of flag subspaces corresponds to the 
quotients by direct sums of flag subspaces in \cite{G93}.

%%%%%%%%%%%%%%%%%%%%%%%%%%%%%%%%%%%%%%%%%
%%%%%%%%%%%%%%%%%%%%%%%%%%%%%%%%%%%%%%%%%
%%%%%%%%%%%%%%%%%%%%%%%%%%%%%%%%%%%%%%%%%

\section{Localization of framed flat connections} 
\label{sec:FFC}

Given a 2d ideal triangulation $\mb t$ of the big boundary of an admissible 3-manifold $M$, we know from \cite{FG-Teich} how to build coordinates on a Zariski open subset $\CX^{\rm un}_K(\pd M,\mb t)$ of the space of framed flat connections on $\pd M$ --- the coordinate phase space associated with $\pd M$ and $\mb t$. We will review the construction in Section \ref{sec:bdyflat}. (We also explained how to extend the definition of $\CX^{\rm un}_K(\pd M,\mb t)$ to small torus boundary components in Section \ref{sec:spaces}.)
Now let $\mb t_{\rm 3d}$ be a 3d ideal triangulation of $M$ compatible with a triangulation $\mb t$ of the big boundary.
We will construct coordinates associated with the octahedra of the $K$-decomposition of $M$.
We will prove in Sections \ref{sec:bulkflat}--\ref{sec:projbases} that octahedron coordinates in the bulk of $M$ parametrize an open subset $\wt \CL_K(M,\mb t_{\rm 3d})$ of the space $\wt \CL_K(M)$ of framed flat connections on $M$. The space $\wt \CL_K(M,\mb t_{\rm 3d})$ projects to a submanifold $\CL_K(M,\mb t_{\rm 3d}) \subset \CX^{\rm un}_K(\pd M,\mb t)$ that parametrizes the flat connections in the boundary phase space that extend to the bulk. By using sufficiently refined 3d triangulations, or taking a union over all 3d triangulations, we obtain a submanifold $\CL_K(M,\mb t)$ that only depends on $\mb t$.

Our main \emph{conceptual} goal is to understand the nature of $\CL_K(M,\mb t)$ as a \emph{Lagrangian submanifold} of $\CX^{\rm un}_K(\pd M,\mb t)$, with its Atiyah-Bott-Goldman symplectic structure. We will do this by expressing $\CL_K(M,\mb t) \subset \CX^{\rm un}_K(\pd M,\mb t)$ as a symplectic gluing of elementary symplectic pairs $\CL_\soct\subset \CP_{\pd \soct}$ associated with octahedra. We start explaining how this should work in this section, but defer a full treatment to Sections \ref{sec:Bloch} and \ref{sec:combi}.

In Section \ref{sec:23}, we will investigate how 2--3 moves act on $\wt \CL_K(M,\mb t_{\rm 3d})$ and $\CL_K(M,\mb t_{\rm 3d})$, and interpreted 2--3 moves in terms of the hypersimplicial $K$-decompositions of 4-simplices.

\subsection{Boundary phase spaces} 
\label{sec:bdyflat}

Let $\CC$ be an oriented surface with  at least one puncture and $\chi(\CC)<0$ --- \eg\ a component of the big boundary of an admissible 3-manifold $M$. 
Our first item of business is to review the ${\cal X}$-coordinates 
on the moduli space ${\cal X}_K(\CC)$ of framed flat $PGL(K,\C)$-connections on $\CC$ \cite{FG-Teich}.%
\footnote{These $\CX$-coordinates were recently reviewed and generalized in \cite{GMN-spectral, GMN-snakes} in the context of the physical 6d (2,0) theory compactified on surfaces $\CC$. The work of \cite{GMN-spectral, GMN-snakes} should tie in beautifully with our 3d constructions, though many details remain to be explored.}

\subsubsection{Coordinates from the $K$-triangulation}\label{sec4.1.1}
\label{sec:2dlines}

\begin{figure}[htb]
\centering
\includegraphics[scale=.8]{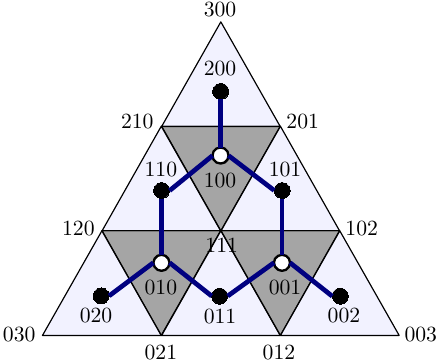}
\caption{3-triangulation of a triangle, with triples of integers corresponding to points $p_{abc}$ with $a+b+c=3$, lines $\mathfrak L_{abc}$ with $a+b+c=2$ (black dots), and planes $\mathfrak P_{abc}$ with $a+b+c=1$ (white dots). The blue connectors between black and white dots stand for incidence relations between lines and planes.}
\label{fig:linesplanes}
\end{figure}

Let us fix an ideal triangulation of $\CC$. 
Then, given  a $PGL(K)$-vector bundle with a framed flat connection, each ideal triangle  
gives rise to a configuration of three flags $(A,B,C)$ in $V_K$, assigned to the vertices of the triangle (as discussed in Section \ref{sec:configs}).
We assume that it is a generic configuration of flags.   
Then it gives rise to configurations  
 of lines and planes in ${V}_K$, associated with the white $\Delta^{0,1}$- and black $\Delta^{1,0}$-triangles 
in the $K$-triangulation:
\be 
\Delta^{0,1}_{abc}  \quad\leadsto\quad \mathfrak L_{abc} = A^a\cap B^b\cap C^c\,, ~~~~a+b+c=K-1\,,
\ee
\be 
\Delta^{1,0}_{abc} \quad \leadsto\quad \mathfrak P_{abc} = A^a\cap B^b\cap C^c\,,~~~~a+b+c=K-2\,,
\ee
as in Figure \ref{fig:linesplanes}.
The plane $\mathfrak P_{abc}$ on a black triangle contains all three lines $\mathfrak L_{a+1,b,c}$, $\mathfrak L_{a,b+1,c}$, and $\mathfrak L_{a,b,c+1}$ on the white triangles surrounding it. If the configuration of flags $(A,B,C)$ is generic, 
these are the only relations among the lines.

Next, every \emph{internal} lattice point $p_{abc}$ of the $K$-triangulation, which is labeled by three \emph{strictly positive} integers that add to $K$, gives rise to a 3-space $$
\mathfrak V_{a-1,b-1,c-1} = A^{a-1}\cap B^{b-1}\cap C^{c-1}.
$$
 This space contains all three planes and all six lines on the black and white triangles surrounding it. This data allows us to define a coordinate $x_{abc}$ associated with the internal point. It is a triple-ratio of the collection of lines and planes contained in $\mathfrak V_{a-1,b-1,c-1}$, and there are several ways to describe it. For example, choosing any six vectors $v_{a'b'c'}$ that generate the six respective lines $\mathfrak L_{a'b'c'}$ 
that surround $p_{abc}$, and setting
\bea
&l_1:=  v_{a+1,b-1,c-1}, ~~l_2:=  v_{a-1,b+1,c-1}, ~~l_3:=  v_{a-1,b-1,c+1},&  \nonumber \\ 
&l_4:=  v_{a-1,b,c}, \qquad l_5:=  v_{a,b-1,c},\qquad l_6:=  v_{a,b,c-1},  &  \nonumber
\eea
we can define the triple-ratio as
\be \label{deftrip}
 x_{abc} = \frac{
 \langle   l_1\wedge   l_6\wedge   l_4 \rangle
 \langle   l_2\wedge   l_4\wedge   l_5\rangle \langle   l_3\wedge   l_5\wedge   l_6\rangle }
 {\langle   l_1\wedge   l_5\wedge   l_4 \rangle
 \langle   l_2\wedge   l_6\wedge   l_5\rangle\langle   l_3\wedge   l_4\wedge l_6\rangle }
\,.
\ee%
Here $\langle v\wedge v'\wedge v''\rangle$ is defined as follows. 
Choose a volume form $\Omega_3$ in a three-dimensional vector space $V_3$. Given a triple of vectors 
$v, v', v''$ in $V_3$, we set $\langle v\wedge v'\wedge v''\rangle:= (\Omega_3, v\wedge v'\wedge v'')$. 
In any unimodular basis in $V_3$,  it is the determinant of the matrix 
expressing $v, v', v''$ in this basis. 

The triple-ratio is independent of the choice of the vectors $v$, since every vector occurs an equal number of times in the numerator and denominator. It is also independent of the choice of the volume form $\Omega_3$. The triple-ratio is the only invariant of a configuration of three flags in $V_3$. More generally, the ${K-1\choose 2}$ triple-ratios $x_{abc}$ assigned to interior points of a $K$-triangulation 
parametrize the space of configurations of three flags $(A,B,C)$ in ${V}_K$.

\begin{figure}[htb]
\centering
\includegraphics[scale=.63]{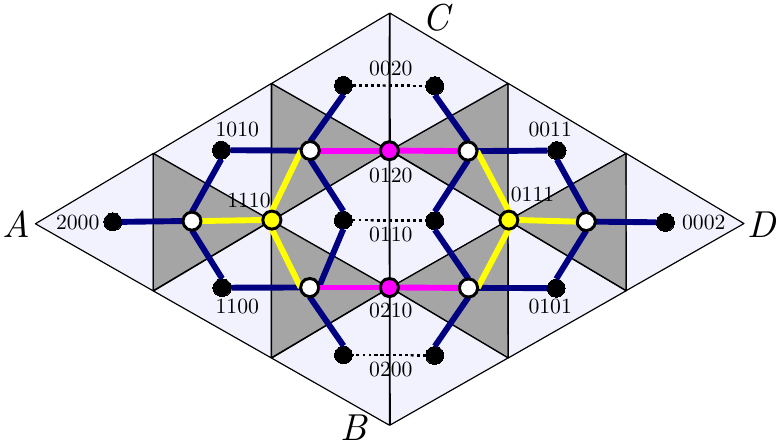}
\caption{Two triangles glued along an edge. The points $p_{0bc0}$ on the edge are shown in magenta, together with the incidence relations with the neighboring planes (white dots). The lines (black dots) on adjacent white triangles are identified. Internal lattice points are shown in yellow.}  
\label{fig:crossratio}
\end{figure}

To describe framed flat connections on the surface $\CC$, the triple-ratio coordinates in the interior of each triangle are supplemented by cross-ratio coordinates on edges. Every edge of the ideal triangulation belongs to a distinguished quadrilateral, formed by the two triangles sharing the edge. We parallel-transport the four flags $A,B,C,D$ at its vertices to a point $p$ in its interior. 
Then we can simultaneously decorate the $K$-triangulations of both triangles with a collection of lines, planes, and 3-spaces. We label the objects on the triangle $ABC$ with quadruples of integers $(a,b,c,0)$ and those on the triangle $BCD$ with quadruples $(0,b,c,d)$. See Figure~\ref{fig:crossratio}.

The $K-1$ lattice points $p_{abcd}$ on the edge have labels $(0,b,c,0)$ with $b,c>0$ and $b+c=K$. They belong to the $K$-decomposition of both adjacent triangles. The lines on the white triangles adjacent to the edge are identified. The four lines surrounding $p_{0bc0}$ have labels $\mathfrak L_{1,b-1,c-1,0},\,\mathfrak L_{0,b,c-1,0},\,\mathfrak L_{0,b-1,c,0},$ and $\mathfrak L_{0,b-1,c-1,1}$, and belong to the same copy of $V_2 = B^{b-1}\cap C^{c-1}\subset V_K$.
To define a cross-ratio coordinate, we choose four vectors $ a, b, c, d$ inside each respective line, and set
\be x_{0bc0} = \frac{\langle  a\wedge  b\rangle\langle  c\wedge  d\rangle}{\langle  a\wedge  c\rangle\langle  b\wedge  d\rangle}.  \label{eq:crossratiox0bc0}
\ee
Here $\langle  v\wedge  v' \rangle$ is defined using a volume form in $V_2$.

\begin{figure}[tbh]
\centering
\includegraphics[width=1.8in]{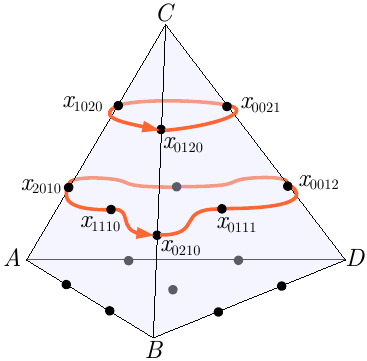}
\caption{Unipotent holonomy constraints around a puncture.}
\label{fig:tetHolonomy}
\end{figure}

A surface $\CC$ of genus $g$ with $h$ punctures has $-2\chi(\CC)$ triangles and $-3\chi(\CC)$ edges in any ideal triangulation. Therefore, the cross-ratios and triple-ratios form a total of
\be -((K-1)(K-2)+3(K-1))\chi(\CC) = -(K^2-1)\chi(\CC) \ee
local coordinates for the space ${\cal X}_K(\CC)$. The requirement that the holonomies at every puncture are unipotent imposes $(K-1)h$ independent constraints on the coordinates. 
Explicitly, the constraints at a given puncture say that the product of \emph{minus} the coordinates around each of the $K-1$ loops surrounding the puncture is equal to one. This is illustrated in Figure \ref{fig:tetHolonomy}, in the case that $\CC$ is the boundary of an ideal tetrahedron, and $K=3$; the
%\begin{wrapfigure}{r}{1.8in}
%\end{wrapfigure}
products of three coordinates $(-x_\bullet)$ around the first loop and six coordinates $(-x_\bullet)$ around the second loop must both equal one.
 (The origin of these constraints is explained in \cite{FG-Teich}; we will 
re-derive the constraints in Section \ref{sec:projbases}.)

Altogether, we find that cross-ratios and triple-ratios give
\be d_K(g,h) = -(K^2-1)\chi(\CC) - (K-1)h \ee
independent coordinates for the moduli space ${\cal X}^{\rm un}_K(\CC)$. This is precisely the expected complex dimension. The coordinates are $\C^*$-valued and cover an open patch. Notice that the cross-ratios and triple-ratios never equal zero or infinity, as long as all of the invariant flags at the punctures are generic.

\paragraph{Dual coordinates:}
In \cite{FG-Teich}, the coordinates for the moduli space of framed flat connections on a punctured surface were defined slightly differently than we have presented above. Rather than defining lines $\mathfrak L_{abc}$ and $\mathfrak L_{abcd}$ and their invariants by intersecting subspaces of flags, \cite{FG-Teich} used quotients of flags. The two approaches are very simply related by the operation of dualizing flags, as in \eqref{dualflag}. That is, our cross-ratio and triple-ratio coordinates defined from a collection of flags $(A,B,C,\ldots)$ (and a flat connection) on $\CC$ as above are equivalent to (minus) the  coordinates used in \cite{FG-Teich} defined using the dual flags $(A^*,B^*,C^*,\ldots)$. 
The cross-ratio defined in \cite{FG-Teich} differs from ours by an additional minus sign.

\subsubsection{Symplectic structure}
\label{sec4.1.2}

\begin{figure}[htb]
\centering
\includegraphics[scale=.63]{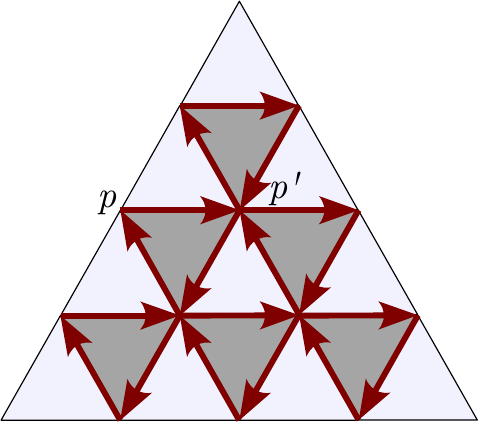}
\caption{Poisson structure on a 4-triangulated triangle. The coordinates $x$ and $x'$ at the lattice points $p$ and $p'$ have the Poisson bracket $\{ x, x' \} = + x x'$ since there is one positively oriented edge from $p$ to $p'$.}
\label{fig:Poisson}
\end{figure}

The Poisson structure on an open patch of ${\cal X}_K(\CC)$ corresponding to an ideal triangulation is described  by drawing arrows circulating clockwise around every small black triangle in the $K$-decomposition of the triangulated surface $\CC$ (see Figure~\ref{fig:Poisson}). Then the Poisson bracket of two cross-ratio or triple-ratio coordinates $x,x'$ at lattice points $p,p'$ is
\be \label{flatPoisson}
 \{x,x'\} = (\text{\# edges from $p$ to $p'$ $-$ \# edges from $p'$ to $p$})\cdot xx'\,.\ee

After projection to the traditional moduli space ${\rm Loc}_K(\CC)$, the symplectic and Poisson structures agree 
with the standard Goldman symplectic form on the non-singular part of the moduli space. 
The latter is expressed simply in terms of complex flat connections $\CA$ as $\Omega = \int_\CC \Tr(\delta \CA\wedge \delta \CA)$.

\subsection{The Lagrangian pair assigned to a 3d triangulation of $M$}
\label{sec:bulkflat}

Having described coordinates for the moduli space of framed flat connections on oriented 2d 
punctured surfaces $\CC$, we now proceed to study the 
framed flat connections on admissible 3-manifolds $M$. 
In the process of building coordinates for flat 3d connections, we will find a natural way to 
understand the spaces $\CL_{K}(M)$ of framed flat $PGL(K)$-connections on the boundary $\pd M$ that extend to the bulk. 

Precisely, we fix an ideal triangulation ${\bf t}$ of the big boundary of $\partial M$. 
Then, as we discussed in Section~\ref{sec:spaces}, there is a complex torus ${\cal X}^{\rm un}_K(\partial M, {\bf t})$, 
called the coordinate phase space,  
realized as a Zariski open part of the symplectic moduli space  ${\cal X}^{\rm un}_K(\partial M)$. 
Pick a 
3d triangulation $\mb t_{\rm 3d}$ of $M$ inducing the 2d triangulation ${\bf t}$. We determine 
explicitly a Zariski open subset $\wt \CL_K(M,\mb t_{\rm 3d})$ of framed flat connections on $M$ and a corresponding Lagrangian subvariety
\be \label{clagrphs}
{\cal L}_K( M, \mb t_{\rm 3d})\subset {\cal X}^{\rm un}_K(\partial M, {\bf t})\,, 
\ee
which parametrizes the framed flat connections from ${\cal X}^{\rm un}_K(\partial M, {\bf t})$ 
that extend to the bulk. Specifically, ${\cal L}_K( M, \mb t_{\rm 3d})$ is the (closure of) the image of the projection of $\wt \CL_K(M,\mb t_{\rm 3d})$ to ${\cal X}^{\rm un}_K(\partial M, {\bf t})$.
We call (\ref{clagrphs}) a {\it Lagrangian pair}, emphasizing both 
the complex symplectic torus and its Lagrangian subvariety.

Generically, $\wt \CL_K(M,\mb t_{\rm 3d})$ and $\CL_K(M, \mb t_{\rm 3d})$ will turn out to be independent of $\mb t_{\rm 3d}$, depending only on the fixed boundary triangulation $\mb t$. For certain ``degenerate'' choices of $\mb t_{\rm 3d}$, these spaces may lose components. By taking a union over all 3d triangulations, we can obtain spaces $\wt \CL_K(M,\mb t)$ and $\CL_K(M,\mb t)$ that depend only on the boundary triangulation $\mb t$.

The boundary of a tetrahedron, understood as an admissible 3-manifold, 
 is a four-punctured sphere with a canonical triangulation. 
This triangulation, for $K=2$, provides the most basic Lagrangian pair (\ref{clagrphs}), 
which we call the {\it elementary Lagrangian pair} --- the Lagrangian pair 
for the $PGL(2)$-tetrahedron. It is our basic building block, and deserves a special notation:
\be
{\cal L}_{ \soct} \subset \CP_{\pd \soct}\,. 
\ee
We assign 
the elementary Lagrangian pair
 to each of the octahedrons of the $K$-decomposition of $M$ provided by the 3d triangulation $\mb t_{\rm 3d}$.

\subsubsection{The Lagrangian pair for the $PGL(2)$-tetrahedron}\label{sec4.2.2}

\begin{figure}[htb]
\centering
\includegraphics[width=3.8in]{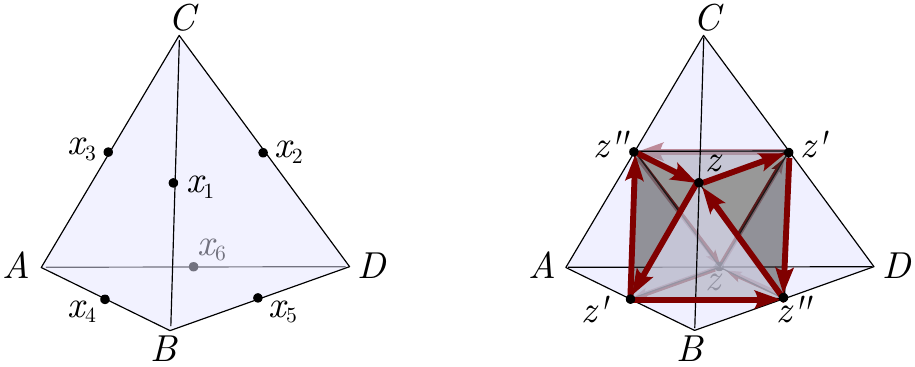}
\caption{Coordinates for framed $PGL(2,\C)$-connections on an ideal tetrahedron.}
\label{fig:SL2tet}
\end{figure}

Let us review the framed flat $PGL(2,\C)$-connections on a tetrahedron. We denote by ${\bf t}$ the canonical triangulation of the tetrahedron. 
Let us choose framing flags $A,B,C,D$ at the four punctures, as in Figure~\ref{fig:SL2tet}. Each flag is just a line. 
Abusing notation, we denote the line defining  
the flag $A$ by $A$, etc. 
Just as in Section \ref{sec:bdyflat}, we parametrize ${\cal X}_2(\pd \Delta, {\bf t})$ by six coordinates $x_1,\ldots,x_6$ assigned to the edges. Each $x_i$ is obtained by parallel-transporting the four flags to a common point $p_i$ 
near the edge, and then taking a cross-ratio. For example,
\be \label{crossx1}
x_1 = \frac{\langle a\wedge b\rangle\langle c\wedge d\rangle}
{\langle a\wedge c\rangle\langle b\wedge d\rangle}\bigg|_{p_1}\,,
\ee
where we emphasize the evaluation of flags at $p_1$. 
Computing \eqref{crossx1}, we choose any four non-zero vectors $(a, b, c, d)$ in the lines $(A,B,C,D)$.
 There are four relations among the $x$'s, reflecting the unipotent holonomy: 
the product of $(-x_i)$ around every vertex equals one, \ie
\be x_1x_2x_3= x_1x_4x_5= x_3x_4x_6 = x_2x_5x_6 = -1\,. \label{tet2rels}\ee
The corresponding four monomials are central with respect to the Poisson bracket \eqref{flatPoisson}.

Relations \eqref{tet2rels} imply that coordinates on opposite edges are equal. It is then convenient to 
re-label the $x$'s as $z,z',z''$ (as on the right of Figure~\ref{fig:SL2tet}), with $zz'z''=-1$. 
We therefore find a two-dimensional complex torus sitting as a Zariski open part of 
 the phase space ${\cal X}^{\rm un}_2(\pd \Delta)$:  
\be 
{\cal P}_{\pd \soct}:= {\cal X}^{\rm un}_2(\pd \Delta, {\bf t}) \subset {\cal X}^{\rm un}_2(\pd \Delta)\,.
\label{Poct-flat}
\ee
The notation emphasizes the octahedron $\oct$. We call it the {\it octahedron phase space}. 
Explicitly, 
\be 
{\cal P}_{\pd \soct}=
\{z,z',z''\in \C^*\,|\, zz'z''=-1\}~, \label{Ptet2} 
\ee
with Poisson brackets 
\be 
\{z,z'\} = zz'\,, ~~\{z',z''\} = z'z''\,, ~~\{z'',z\} = z''z\,. %\label{phspace3}
\ee
Now let's fill in the bulk of the tetrahedron. Since the bulk is contractible, any bundle 
with a flat connection on the bulk is trivial. 
So the choice of invariant flags $A,B,C,D$ becomes absolutely crucial --- the flags are unrestricted, 
and they carry the only nontrivial degrees of freedom of framed flat connections in the bulk. 

After filling in the tetrahedron we can parallel-transport all flags to a common point $p$ of the tetrahedron. 
The previous three cross-ratios $z,z',z''$ can be computed at this point:
\be \label{3dSL2z}
z = \frac{\langle a \wedge b \rangle\langle c \wedge d\rangle}
{\langle a\wedge c \rangle\langle b\wedge d\rangle}\bigg|_{p},\quad
z' = \frac{\langle b\wedge d\rangle\langle c \wedge a\rangle}
{\langle b\wedge c \rangle\langle d\wedge a\rangle}\bigg|_{p},\quad
z'' = \frac{\langle d\wedge a\rangle\langle c \wedge b\rangle}
{\langle d\wedge c\rangle\langle a\wedge b\rangle}\bigg|_{p}.
\ee
They are no longer independent. In addition to the relation $zz'z''=-1$, there is a new Pl\"ucker relation
\be {\cal L}_{ \soct}:=\CL_2(\Delta,\mb t_{3d})=\wt \CL_2(\Delta,\mb t_{3d}):\quad z''+z^{-1}-1=0\,. \label{Plucker2}\ee
The curve defined by \eqref{Plucker2} parametrizes framed flat connections in the bulk (with the canonical triangulation $\mb t_{3d}$) for which the configuration of framing flags is generic. It is isomorphic to its image $\CL_2(\Delta,\mb t_{3d}) \subset {\cal X}^{\rm un}_2(\pd \Delta, {\bf t})$, the subset of connections on the boundary that extend to the bulk. 
So we have defined the elementary Lagrangian pair ${\cal L}_{ \soct}\subset  {\cal P}_{\pd \soct}$.

\subsubsection{The Lagrangian pair for the $PGL(K)$-tetrahedron}
\label{sec:SLKtet}

\begin{figure}
\centering
\begin{subfigure}{.5\textwidth}
  \centering
  \includegraphics[width=.7\linewidth]{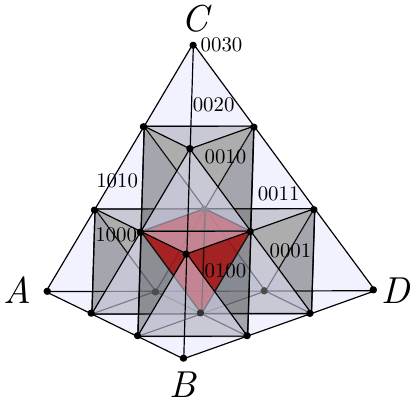}
\end{subfigure}%
\begin{subfigure}{.5\textwidth}
  \centering
  \includegraphics[width=.7\linewidth]{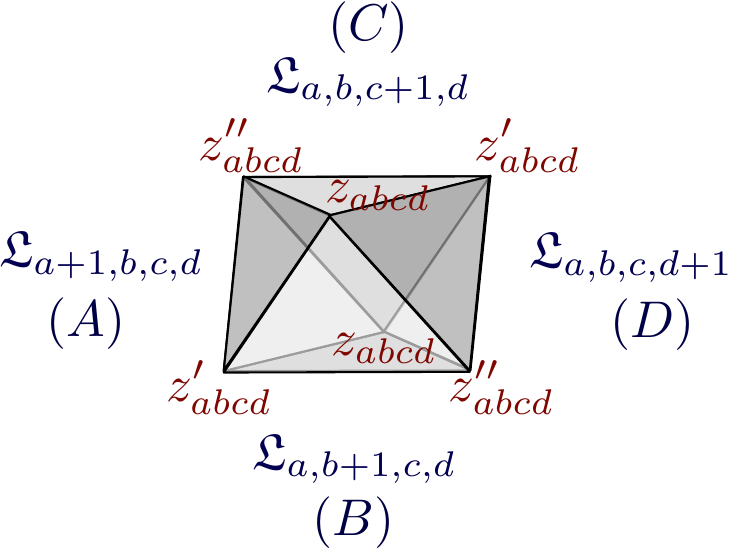}
\end{subfigure}
\caption{\emph{Left}: 3-decomposition of a tetrahedron. \emph{Right}: Cross-ratios for an octahedron.}
\label{fig:hyper3}
\end{figure}

In order to generalize to $PGL(K,\C)$-connections, we consider a $K$-decomposition of the 3d tetrahedron as in Figure \ref{fig:hypersimplices}, repeated for convenience on the left of Figure \ref{fig:hyper3}.

%\begin{wrapfigure}{l}{2in}
%\includegraphics[width=2in]{hyper3}
%\caption{3-decomposition of a tetrahedron.}
%\label{fig:hyper3}
%\end{wrapfigure}

Framed flat connections on the tetrahedron come with a choice of four invariant 
flags $A,B,C,D$ near the vertices. Working on the boundary of the tetrahedron, we parallel-transport the flags to the midpoint of any of the six edges, by restricting the connection to the quadrilateral 
defined by the faces adjacent to the edge. The resulting configuration of four flags at this midpoint gives rise to  a system of lines $\mathfrak L_{abcd}$ on the white triangles lying on the two tetrahedron faces adjacent to the edge. These lines then give us cross-ratio and triple-ratio coordinates for the lattice points on these faces and on the edge, just as in Figure~\ref{fig:crossratio}.

Repeating this for each of the six edges we construct all the cross-ratio and triple-ratio coordinates 
on ${\cal X}^{\rm un}_K(\pd \Delta,\mb t)$ assigned to the canonical triangulation of the tetrahedron boundary. There are $4\times {K-1\choose 2}+6\times (K-1)$ coordinates altogether, with $4\times (K-1)$ relations coming from the unipotent-holonomy constraints at the four vertices. This leaves exactly $2(K-1)^2$ independent coordinates on ${\cal X}^{\rm un}_K(\pd \Delta,\mb t)$.

Just as in the $SL(2)$ case, 
if we fill in the bulk of the 3d tetrahedron, the flat $PGL(K,\C)$-connection must be trivial. The choice of flags is the only data remaining. Now the coordinates obey additional Pl\"ucker relations. The simplest and most beautiful way to describe the relations is to use octahedral coordinates.

A generic  configuration of four flags $(A,B,C,D)$ in $V_K$
defines a configuration of lines in $V_K$ assigned to the 
$\Delta^{0,2}$-tetrahedra, and 2-planes assigned to the octahedra $\oct_{abcd}:=\Delta^{1,1}_{abcd}$, 
of  the $K$-decomposition:
\be \Delta^{0,2}_{abcd} \quad\leadsto\quad \mathfrak L_{abcd} = A^a\cap B^b\cap C^c \cap D^d\, ,~~~~a+b+c+d=K-1.
\ee
\be \hspace{-.2in}\oct_{abcd}=\Delta^{1,1}_{abcd} \quad \leadsto\quad \mathfrak P_{abcd} = A^a\cap B^b\cap C^c\cap D^d\,,~~~~a+b+c+d=K-2.\ee
The four lines sitting on the $\Delta^{0,2}$-tetrahedra surrounding an octahedron $\oct_{abcd}$ are the lines
\be \label{4lines}
\mathfrak L_{a+1,b,c,d},~~\mathfrak L_{a,b+1,c,d},~~\mathfrak L_{a,b,c+1,d},~~\mathfrak L_{a,b,c,d+1}, ~~~~a+b+c+d=K-2.
\ee
 They belong the 2-plane $\mathfrak P_{abcd}$. They provide the
 three cross-ratios \emph{exactly} as in \eqref{3dSL2z}, replacing the $(A,B,C,D)$ lines there with the four lines here:
\be
z_{abcd}, ~~z_{abcd}',~~z_{abcd}'', ~~~~a+b+c+d=K-2.
\ee
%\begin{wrapfigure}{r}{2.2in}
%\includegraphics[width=2.2in]{oct-z}
%\caption{Cross-ratios for an octahedron.}
%\label{fig:octz}
%\end{wrapfigure}
These cross-ratios are associated with the six vertices of the octahedron $\oct_{abcd}$. 
The cross-ratios assigned to opposite vertices are equal, 
as on the right of Figure~\ref{fig:hyper3} (\cf\ Figure~\ref{fig:SL2tet}).   They satisfy the 
standard monomial relation
\be z_{abcd}z_{abcd}'z_{abcd}'' = -1\,, \label{Poct-exp} \ee
as well as the Pl\"ucker relation
\be z_{abcd}''+ z_{abcd}^{-1}-1 = 0\,. \label{Loct-exp} \ee
These relations encode the space of framed flat connections in the bulk of the tetrahedron and (isomorphically) the Lagrangian submanifold $\CL_K({\Delta})$ of framed connections on the boundary that extend to the bulk. To see this, 
 we must relate octahedron parameters to the cross-ratio and triple-ratio coordinates on the boundary.

The key to relating bulk and boundary coordinates is the realization that the 3d array of ${K+2\choose 3}$ lines in the $K$-decomposition of the tetrahedron restricts to the standard 2d arrays of lines in the $K$-decomposition of the boundary.

Consider, for example, a lattice point $p_{abcd}$ on an edge of the boundary. It has a label induced by the $K$-decomposition of the bulk, with exactly two of $a,b,c,d$ vanishing. A unique octahedron vertex touches this point, and the cross-ratio $x_{abcd}$ that we would assign from the boundary perspective is nothing but the bulk cross-ratio at the octahedron vertex. Explicitly, the $6\times (K-1)$ edge coordinates on the boundary of the tetrahedron are

\be \label{crossz} \text{edges:}\quad
\begin{array}{c} x_{0bc0}=z_{0b'c'0}\,,\quad  x_{ab00}=z'_{a'b'00}\,,\quad x_{a0c0} = z''_{a'0c'0}\,, \\[.2cm]
x_{a00d}=z_{a'00d'}\,,\quad x_{00cd}=z'_{00c'd'}\,,\quad x_{0b0d}=z''_{0b'0d'}\,, \end{array}\qquad \ee
where $(a',b',c',d')=(a-1,b-1,c-1,d-1)$.

\begin{figure}[htb]
\centering
\includegraphics[width=5.5in]{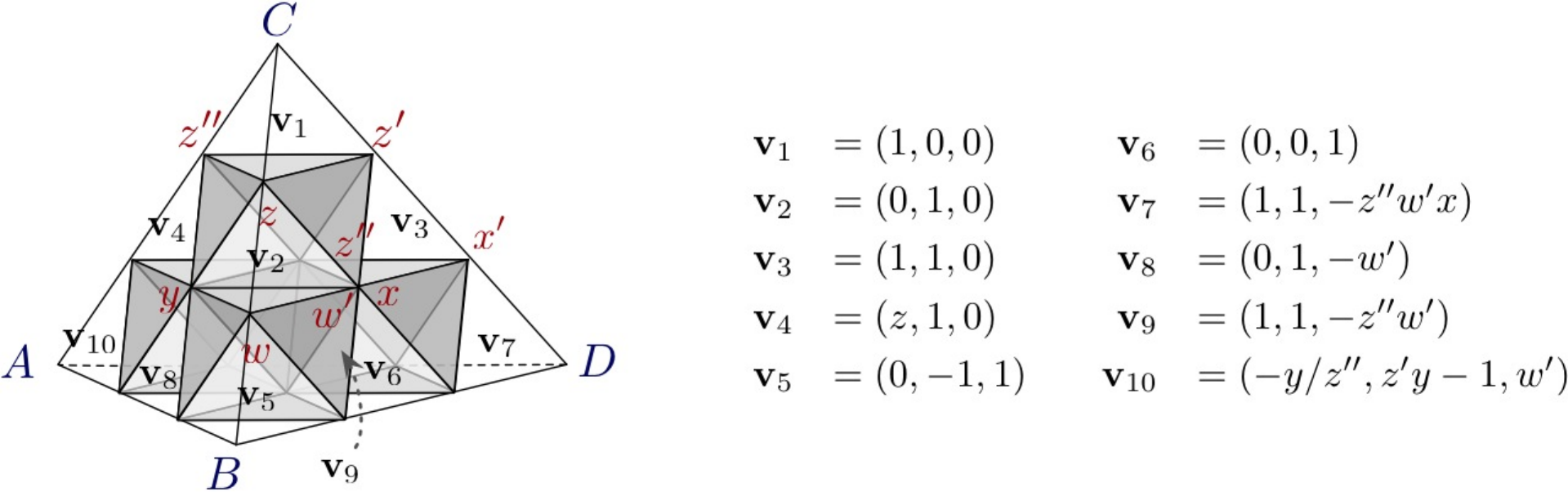}
\caption{Ten vectors parametrizing the ten lines on the $K=3$ tetrahedron. We have used a $GL(3,\C)$ transformation to put the vectors in the given form; otherwise they are completely generic. We also show the corresponding octahedron parameters $(z,w,x,y)$, on their corresponding vertices. Here $zz'z''=-1$ and $z''+z^{-1}-1=0$, so that $z'=(1-z)^{-1},\, z''=1-z^{-1}$, and similarly for $w,x,y$.}
\label{fig:SL3-lines}
\end{figure}

Similarly, every triple-ratio $x_{abcd}$ on the boundary, with exactly one of $a,b,c,d$ vanishing, is the product of the three octahedron parameters at the vertices of the three octahedra that touch $p_{abcd}$. For example, on face $BCD$, we have triple-ratio coordinates
\be \text{faces:}\qquad \label{triplez} x_{0bcd} = z_{0,b-1,c-1,d}\, z'_{0,b,c-1,d-1}\,z''_{0,b-1,c,d-1}\,,\qquad\quad\ee
and similarly for the other faces. There are several ways to demonstrate this relation.
It is sufficient to do the case $K=3$, using an explicit parametrization for the ten lines in the 3d $K$-decomposition, in a chosen basis for $V_3$. (Specifying ten lines with the standard incidence relations is equivalent to specifying the configuration of four flags.) Such a parametrization is shown in Figure \ref{fig:SL3-lines}, conveniently chosen to encode the four octahedron cross-ratios, called $z,w,x,y$. Thus, for example
\be z'' = \frac{\langle  v_1\wedge  v_2 \rangle \langle  v_3\wedge  v_4\rangle}{\langle  v_1\wedge  v_3 \rangle \langle  v_2\wedge  v_4\rangle}\,,\quad
 w' = \frac{\langle  v_5\wedge  v_6 \rangle \langle  v_2\wedge  v_8\rangle}{\langle  v_5\wedge  v_2 \rangle \langle  v_6\wedge  v_8\rangle}\,,\quad
 x = \frac{\langle  v_9\wedge  v_6 \rangle \langle  v_3\wedge  v_7\rangle}{\langle  v_9\wedge  v_3 \rangle \langle  v_6\wedge  v_7\rangle}\,.
 \ee
We can then check that the expected triple-ratio coordinate in the middle of (say) the $BCD$ face, namely
\be x_{0111} = \frac{ \langle  v_1 \wedge  v_2\wedge  v_6\rangle \langle  v_5\wedge  v_6 \wedge v_3\rangle \langle  v_7\wedge  v_3\wedge  v_2\rangle}
{ \langle  v_1\wedge  v_3\wedge  v_6\rangle \langle  v_5 \wedge v_2\wedge  v_3\rangle \langle  v_7\wedge  v_6\wedge  v_2\rangle}\,, \ee
is equal to the product $xw'z''$.

Finally, let us look at the internal lattice points of the $K$-decomposition of a tetrahedron: points $p_{abcd}$ with none of $a,b,c,d$ vanishing. (This first occurs for $K=4$.) Each such point coincides with six octahedron vertices, and we could try to assign to the point a coordinate $c_{abcd}$ that is the product of the six surrounding octahedron parameters. This product, however, is automatically trivial, due to a cancellation in the cross-ratios involved. Explicitly:
\be \label{internalz}
 c_{abcd} = z_{a,b-1,c-1,d}\,z_{a-1,b,c,d-1}\, z'_{a,b,c-1,d-1}\,z'_{a-1,b-1,c,d}\,z''_{a,b-1,c,d-1}\,z''_{a-1,b,c-1,d} \equiv 1\,.\ee

We have shown that a generic configuration of four flags $(A,B,C,D)$ induces octahedron parameters $z_{abcd},z_{abcd}',z_{abcd}''$ 
in the bulk of a tetrahedron that satisfy relations \eqref{Poct-exp}--\eqref{Loct-exp} and \eqref{internalz}, and are related to the standard boundary coordinates via \eqref{crossz}--\eqref{triplez}. Conversely, if we start with triples of octahedron parameters satisfying these properties, it is possible to reconstruct uniquely the configuration of four flags. One way to show this is to use the coordinates to reconstruct the array of ${K+2\choose 3}$ lines on the upright simplices $\Delta^{0,2}$. (We will see explicitly that this can be done in Section \ref{sec:projbases}.) 
Identifying the lines is equivalent to specifying the configuration of flags.

Summarizing, there is a canonical bijection between the following two spaces:

\begin{enumerate}

\item \emph{The space of octahedron parameters}, given by arbitrary collections of the numbers 
\be
z_{abcd},z_{abcd}',z_{abcd}'' \in \C^*-\{1\}\,, ~~~~a+b+c+d=K-2
\ee
satisfying the relations \eqref{Poct-exp}--\eqref{Loct-exp} and \eqref{internalz},

\item \emph{The subspace of generic configuration of four flags
\begin{equation} \label{lpair}
{\rm Conf}^*_4({\cal B})=\wt \CL_K(\Delta,\mb t_{\rm 3d})=\CL_K(\Delta,\mb t_{3d})=\CL_K(\Delta, \mb t)\;\; \subset \CX^{\rm un}_K(\pd \Delta, {\bf t})\,.
\ee}
\end{enumerate}
The Lagrangian pair for the $PGL(K)$-tetrahedron is given by (\ref{lpair}). Note that in this case we again find that the projection to the boundary $\wt \CL_K(\Delta,\mb t_{\rm 3d})\to \CL_K(\Delta,\mb t_{3d})$ is one-to-one.

\subsubsection{The coordinate phase space for a tetrahedron via symplectic reduction}
\label{sec:tet-symp}

The coordinate symplectic phase space related to the canonical triangulation ${\bf t}$ of a tetrahedron is a complex torus 
${\cal X}^{\rm un}_K(\pd \Delta, {\bf t})$ realized as a Zariski open subset in $\CX^{\rm un}_K(\pd \Delta)$. 
We can obtain the symplectic coordinate phase space ${\cal X}^{\rm un}_K(\pd \Delta, {\bf t})$ as a symplectic reduction of a product of 
the elementary {\it octahedron phase spaces} $\CP_{\pd \soct}$ defined in \eqref{Ptet2}. 

Consider the product of the octahedron phase spaces over the octahedrons of the $K$-decomposition of 
a tetrahedron, equipped with the product symplectic structure: 
\be
{\cal P}_\times := \prod_{\soct_i\in \Delta} \CP_{\pd \soct_i}\,.
\ee
The  edge \eqref{crossz} and face \eqref{triplez} coordinates on the boundary of the tetrahedron, written in the 
octahedron parameters, can be viewed as a monomials on the complex torus $\CP_\times $. 
We call them the edge and face coordinate monomials. 
The Poisson brackets between them coincide with the Poisson brackets 
of the corresponding edge and face coordinates for the 
 Atiyah-Bott-Goldman Poisson structure \eqref{flatPoisson} on $\CX^{\rm un}_K(\pd \Delta)$. 
The octahedron relations $zz'z''=-1$ imply that they satisfy monomial relations given by 
the unipotent holonomy constraints around each puncture, as in Figure \ref{fig:tetHolonomy}. 

The products \eqref{internalz} of six octahedron parameters around the ${K-1\choose 3}$ internal points of the $K$-decomposition are no longer equal to one. They are monomial functions $c_k$ on $\CP_\times$. 
The $c_k$ Poisson-commute with each other, as well as with the boundary coordinates. 
A nice graphical way to see that the boundary monomials and the $c_k$ have the expected commutation relations is to draw arrows on each (oriented) octahedron, as on the right of Figure~\ref{fig:SL2tet}. After gluing together the octahedra, the Poisson bracket between product coordinates on any two points of the $K$-decomposition is proportional to the net number of arrows going from one to the other. All arrows in the bulk of the tetrahedron cancel out in pairs, while the arrows on the boundary reproduce the configuration in Figure \ref{fig:Poisson}.

It is also easy to see that the ${K-1\choose 3}$ monomial functions $c_k$ are all independent. Indeed, since $\text{Conf}_4^*(\CB)$ has complex dimension $(K-1)^2$ and (from Section \ref{sec:SLKtet}) $\text{Conf}_4^*(\CB)=\wt\CL_K(\Delta,\mb t_{3d}) = \big(\prod_{\soct_i}\CL_{\soct_i}\big)\cap \{c_k\}_{\text{vertices $k$}}$, there must be at least $\text{($\#$ octahedra)}-\text{dim}\,\text{Conf}_4^*(\CB)={K+1\choose 3}-(K-1)^2={K-1\choose 3}$ independent $c_k$'s.

It follows that there is a canonical isomorphism of symplectic spaces
\be {\cal X}^{\rm un}_K({\pd \Delta}, {\bf t}) = \CP_\times \big/\!\!\big/(c_k-1)\,.\label{redtet} \ee
The notation on the right denotes the symplectic reduction corresponding to the family of commuting moment maps $c_k$ at the level set 
$c_k=1$. Each moment map $c_k$ provides an independent Hamiltonian flow that amounts to an action 
of a copy of the group $\C^*$. We take the quotient by the resulting action of the complex torus,  
and then set $c_k=1$. Note that ${\rm dim}~\CP_\times \big/\!\!\big/(c_k-1) = 2(K-1)^2$, as expected.

We identify $\CL_K(\Delta, {\bf t}_{\rm 3d}) $, the space of generic configurations of four flags from \eqref{lpair}, as the image of the product Lagrangian $\prod_{\soct_i\in\Delta}\CL_{\soct_i}\subset  \CP_\times$ under the symplectic reduction \eqref{redtet}. It follows that
\be
\CL_K(\Delta, {\bf t}_{\rm 3d})  \subset {\cal X}^{\rm un}_K({\pd\Delta}, {\bf t})
\ee 
is Lagrangian as well.

\subsubsection{Gluing constraints for 3-manifolds}
\label{sec:SLKpoly}

Now let $M$ be an admissible 3-manifold with an ideal triangulation $\mb t_{3d}$ and a framed flat $PGL(K,\C)$-connection. 
Then we can define the parameters $z_i,z_i',z_i''$ for every octahedron in the $K$-decomposition of $M$.

\begin{wrapfigure}{r}{1.6in}
\includegraphics[width=1.6in]{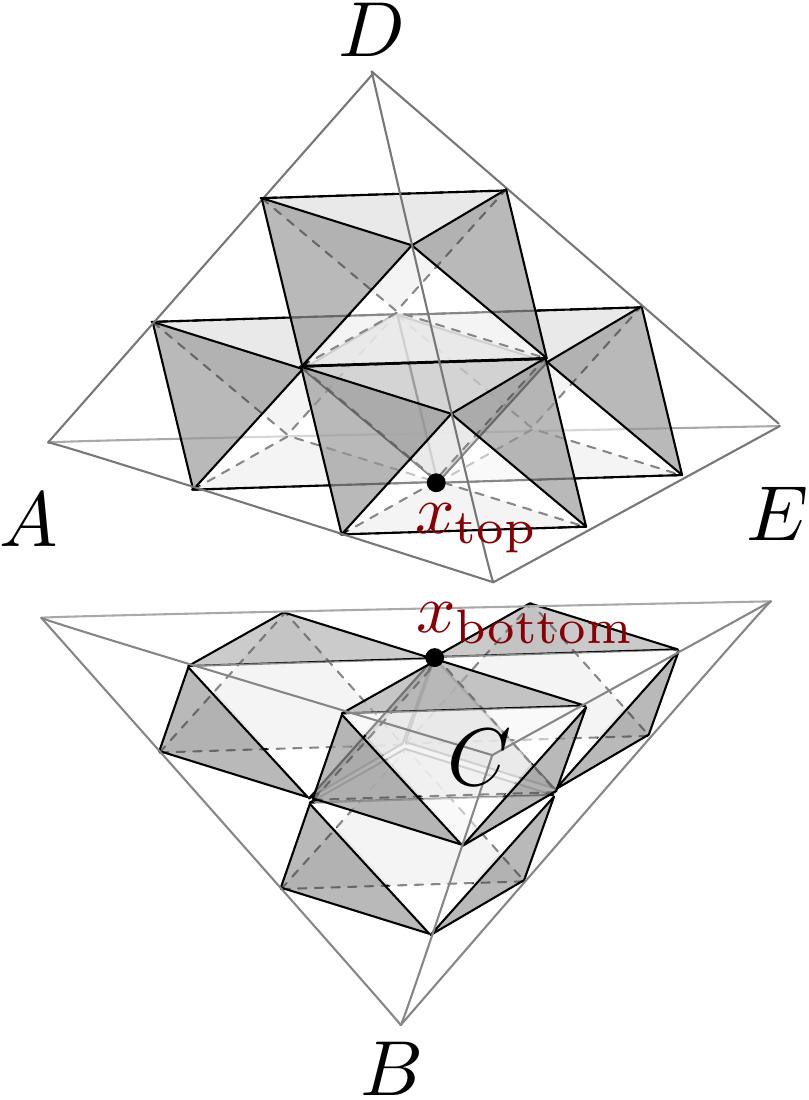}
\caption{Lattice point on a pair of glued faces.}
\label{fig:faceglue}
\end{wrapfigure}

In Section \ref{sec:SLKtet} we found that at every internal point in the $K$-decomposition of a tetrahedron the product of octahedron parameters is one. We also saw that the product of parameters on a boundary point is equal to the cross-ratio or triple-ratio coordinate in the boundary phase space. Both of these facts generalize. Now the product at every internal point in the $K$-decomposition of a 3-manifold --- including points on faces and edges of glued tetrahedra --- is one; while products at external points are equal to big-boundary coordinates. We explain briefly why this is so. (The situation for small torus boundaries is slightly more subtle, and will be deferred to Section \ref{sec:torusbdy}.)

First consider a point $p$ on a big internal face in $M$, shared by two tetrahedra, as in Figure \ref{fig:faceglue}. As a boundary point for the top tetrahedron, $p$ is assigned a triple-ratio $x_{\rm top}$ formed from flags $A,C,E$. As a boundary of the bottom tetrahedron, $p$ is also assigned a triple-ratio $x_{\rm bottom}$ formed from $A,C,E$, but in the opposite orientation. Therefore, $x_{\rm top}x_{\rm bottom}=1$. Moreover, the quantity $x_{\rm top}x_{\rm bottom}$ is itself the product of the six octahedron parameters at vertices that touch $p$, so the product of these octahedron parameters must be one.

\begin{figure}[htb]
\centering
\includegraphics[scale=.4]{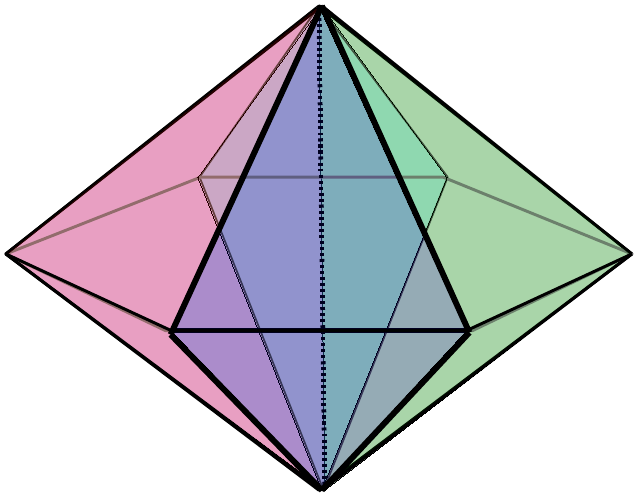}
\caption{Hexagonal bipyramid obtained by gluing six tetrahedra along a common edge.}
\label{fig:hexabipyramid}
\end{figure}

%\begin{wrapfigure}{r}{2.5in}
%\includegraphics[width=2.5in]{trivedge}
%\caption{Configuration of octahedra around a point on a trivalent internal edge.}
%\label{fig:trivedge}
%\end{wrapfigure}

\begin{figure}
\centering
\begin{subfigure}{.5\textwidth}
  \centering
  \includegraphics[width=.75\linewidth]{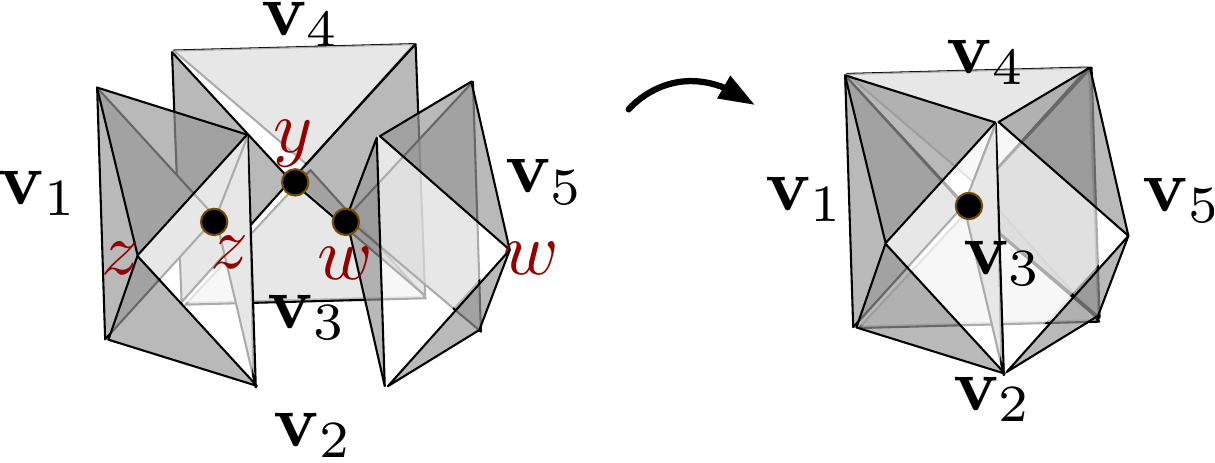}
\end{subfigure}%
\begin{subfigure}{.5\textwidth}
  \centering
  \includegraphics[width=.7\linewidth]{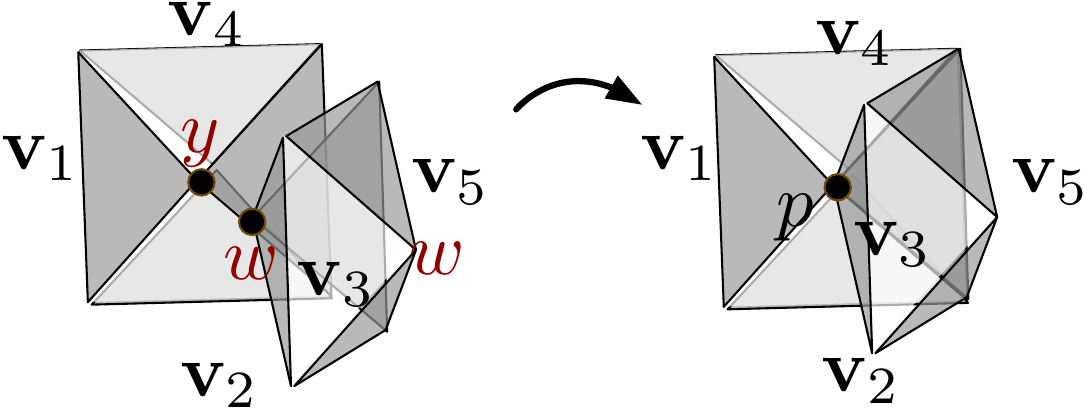}
\end{subfigure}
\caption{\emph{Left}: Configuration of octahedra around a point on a trivalent internal edge. \emph{Right}: Point on an external edge.}
\label{fig:trivedge}
\end{figure}

Next, consider points on an internal edge $E$ in $M$ shared by $N$ tetrahedra. We picture this arrangement as an $N$-gonal bipyramid, as in Figure \ref{fig:hexabipyramid}, and use the flags at the $N+2$ vertices to define octahedron parameters throughout the bipyramid.%
\footnote{Note that the $N$ tetrahedra forming this bipyramid need not all be distinct. Multiple edges of a single tetrahedron might be glued to $E$. This does \mg{not} alter the conclusion of the current argument; one just needs to be careful about identifying octahedron cross-ratios with the proper orientations.} %
Each of the $K-1$ lattice points on $E$ is touched by $N$ octahedron vertices. The product of the $N$ octahedron cross-ratios at each lattice point involves numerators and denominators that cancel in pairs, and so automatically equals one. For example, a situation with $N=3$ is shown on the left of Figure \ref{fig:trivedge}. The three octahedron parameters touching a point on the internal edge are
\be z=\frac{\langle  v_4\wedge  v_1\rangle \langle  v_3\wedge v_2 \rangle}
{\langle  v_4\wedge  v_3\rangle \langle  v_1\wedge v_2 \rangle}\,,\quad
w=\frac{\langle  v_4\wedge  v_3\rangle \langle  v_5\wedge v_2 \rangle}
{\langle  v_4\wedge  v_5\rangle \langle  v_3\wedge v_2 \rangle}\,, \quad
y=\frac{\langle  v_4\wedge  v_5\rangle \langle  v_1\wedge v_2 \rangle}
{\langle  v_4\wedge  v_1\rangle \langle  v_5\wedge v_2 \rangle}\,,
\ee
so that $zwy=1$.

%\begin{wrapfigure}{r}{2.3in}
%\includegraphics[width=2.3in]{bivedge}
%\caption{Point on an external edge.}
%\label{fig:bivedge}
%\end{wrapfigure}

Finally, let us consider external lattice points on a big boundary. For a point on a boundary face, we already know from Section \ref{sec:SLKtet} that its triple-ratio coordinate is a product of three octahedron parameters. For a boundary edge $E$, let us again suppose that $E$ is shared by $N$ tetrahedra in the bulk. The boundary cross-ratio parameter $x$ for a point $p$ on $E$ can be computed as a product of octahedron parameters that touch $p$. Just like in the case of internal edges above, there will be a cancellation between numerators and denominators, but the cancellation is incomplete, and simply leaves behind the single expected cross-ratio $x$. For example, in the situation depicted on the right of Figure \ref{fig:trivedge} (for $N=2$), the octahedron product is
\be y\cdot w = \frac{\langle  v_4\wedge  v_5\rangle \langle  v_1\wedge v_2 \rangle}
{\langle  v_4\wedge  v_1\rangle \langle  v_5\wedge v_2 \rangle} \cdot 
\frac{\langle  v_4\wedge  v_3\rangle \langle  v_5\wedge v_2 \rangle}
{\langle  v_4\wedge  v_5\rangle \langle  v_3\wedge v_2 \rangle}
=
\frac{\langle  v_4\wedge  v_3\rangle \langle  v_1\wedge v_2 \rangle}
{\langle  v_4\wedge  v_1\rangle \langle  v_3\wedge v_2 \rangle}\,,
\ee
which is independent of $v_5$ and agrees with the boundary cross-ratio.

Notice that the octahedron parameters constructed from a framed flat connection on $M$ are well-defined 
and  take values in $\C^*-\{1\}$ as long as the configurations of four flags 
parallel-transported to the interior of every tetrahedron are generic. Conversely, we claim that starting with a collection of 
octahedron parameters $z_i,z_i',z_i'' \in \C^*-\{1\}$ in the $K$-decomposition of $M$ that satisfy the octahedron equations \eqref{Poct-exp}--\eqref{Loct-exp} and the condition that the product at any internal point is one, we can reconstruct a unique framed flat connection on $M$. So:

\begin{theorem} \label{thm:XM}
The octahedron parameters satisfying the equations \eqref{Poct-exp}--\eqref{Loct-exp} and the monomial gluing relations
parametrize an open subset $\wt \CL_K(M,\mb t_{3d})$ of the space of framed flat connections $\wt \CL_K(M)$.
\end{theorem}

\noindent We will prove this in Section \ref{sec:projbases}. \\

We would also like to show that the phase space $\CX^{\rm un}_K({\pd M}, {\bf t})$ is a symplectic reduction of octahedron phase spaces \eqref{Poct-flat}. Observe that if we define gluing monomials $c_k$ on the product phase space
\be
\CP_\times=\prod_{\soct_i\in M} \CP_{\pd \soct_i}
\ee 
as the product of octahedron parameters at every internal lattice point of the $K$-decomposition of $M$, then the $c_k$ all Poisson-commute with each other. This follows easily from a cancellation-of-arrows argument as in Section \ref{sec:tet-symp}. Moreover, for the lattice points on the big boundary of $M$, we define functions $x_j$ to be the products of octahedron parameters there. Then they satisfy the standard boundary commutation relations and commute with all the $c_k$, because while octahedron arrows cancel out in the bulk they leave behind the standard arrangements of Figure \ref{fig:Poisson} on the boundary. We then expect
that there is an equivalence 
\be \CX^{\rm un}_K({\pd M}, {\bf t}) = \CP_\times \big/\!\!\big/(c_k-1)\,. \label{redbig} \ee
Moreover, the Lagrangian submanifold $\CL_K(M, \mb t_{\rm 3d})\subset \CX^{\rm un}_K({\pd M}, {\bf t})$ should be obtained as the image of a product of octahedron Lagrangians $\CL_\times =\prod_{\soct_i\in M}\CL_{\soct_i}$ under the reduction \eqref{redbig}.
We will formalize this in Section \ref{sec:perspective}.

\subsection{Holonomy representation}
\label{sec:projbases}

We would like to show that, given octahedron parameters in the $K$-decomposition of an admissible manifold $M$ that satisfy the octahedron equations \eqref{Poct-exp}--\eqref{Loct-exp} as well as the gluing equations at every internal lattice point of the $K$-decomposition, we can reconstruct a unique framed flat $PGL(K,\C)$-connection on $M$ --- and thus prove Theorem \ref{thm:XM}. To do so, we will extend the 2d notion of local projective bases labeled by ``snakes'' \cite{FG-Teich} to three dimensions.

We will begin with a framed flat connection on $M$, and derive the rules for constructing projective bases and the transformations from one projective basis to another (which all involve octahedron parameters). Then we will run the argument in reverse, starting from abstract octahedron parameters and using 3d snakes to reconstruct projective bases and a framed flat connection.

\subsubsection{3d snakes in a tetrahedron}
\label{sec:dragontet}

We first consider a single ideal tetrahedron and its $K$-decomposition. 

We start with a generic configuration of four flags in $V_K$.  
It gives rise to ${K+2\choose 3}$ lines $\mathfrak L_{abcd}$ labeled by upright tetrahedra $\Delta^{0,2}_{abcd}$. 
We are going to construct projective bases in $V_K$ labeled by geometric objects that we call \emph{3d snakes}.

\begin{wrapfigure}{r}{2.2in}
\includegraphics[width=2.2in]{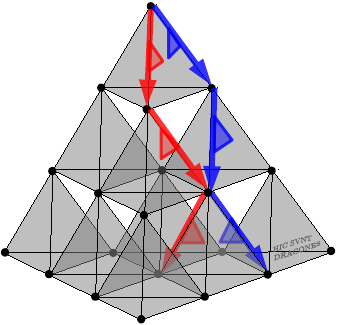}
\caption{3d snakes in the $(K-1)$-decomposition of a tetrahedron, from the top vertex to the opposite face (here for $K=4$).}
\label{fig:3dsnake}
\end{wrapfigure}

To define a 3d snake, we begin with
 an oriented path on the edges of the $(K-1)$-decomposition of a tetrahedron, which
 starts at any of the  tetrahedron vertices and ends at a lattice point on the opposite face, 
consisting of $K-1$ segments (the small edges it traverses). We orient the segments 
toward the opposite face. We decorate each segment of the path by a \emph{fin}. 
To define a fin observe that 
a path segment is an edge of a unique $\Delta^{0,2}$ tetrahedron;
 the fin is a face of that $\Delta^{0,2}$ tetrahedron containing the edge. 
We visualize it by saying that a fin points from the midpoint of the segment to one of the two unoccupied vertices of the $\Delta^{0,2}$ tetrahedron 
(Figure \ref{fig:3dsnake}). We define a 3d snake as a path as above whose segments are decorated by fins.

It may help to observe that every $\Delta^{0,2}_{abcd}$ tetrahedron in the $K$-decomposition corresponds to a lattice point of the $(K-1)$-decomposition; thus the vertices of snakes are labeled by lines $\mathfrak L_{abcd}$. Moreover, every octahedron in the $K$-decomposition corresponds to a black $\Delta^{0,2}$ tetrahedron in the $(K-1)$-decomposition, and octahedron parameters $z,z',z''$ label the \emph{edges} of these black $\Delta^{0,2}$ tetrahedra.

\begin{figure}[htb]
\centering
\includegraphics[width=4.3in]{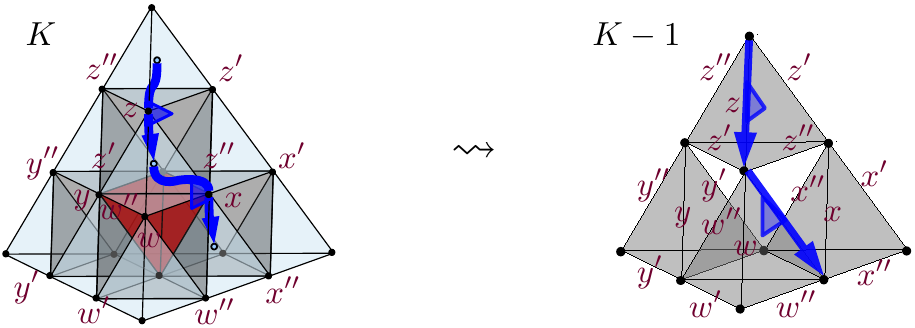}
\caption{Correspondence between the $K$ and $(K-1)$-decompositions, for $K=3$. (It is much more natural to draw snakes on the latter.)}
\label{fig:KK1}
\end{figure}

\begin{figure}[htb]
\centering
\includegraphics[width=3.4in]{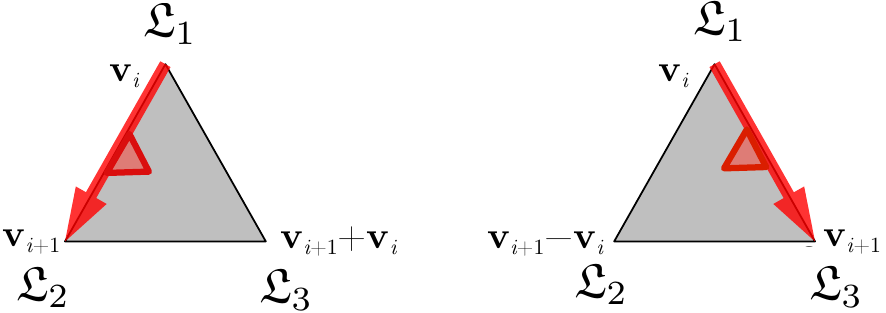}
\caption{Using the $i$-th segment of a snake to select part of a projective basis. Here there are three lines $\mathfrak L_1,\mathfrak L_2,\mathfrak L_3$ associated with the vertices of a small black tetrahedron. The indicated vectors must belong to these lines.}
\label{fig:2dbases-L}
\end{figure}

The $K$ lines $\mathfrak L_{abcd}$ that label the vertices of a 3d snake  
are generic since 
the original configuration of four flags is generic. Let us associate a basis $( v_1,\ldots, v_K)$ 
to the snake so that the vector $ v_i$  lies in the line at
 the $i$-th vertex of the snake (numbered from tail to head), and the vector $ v_i\pm  v_{i+1}$ 
belongs to the line at the vertex to which the $i$-th fin points to. Here we take a sum if the fin points to the right and a difference if the fin points to the left, as in Figure \ref{fig:2dbases-L}. 
These conditions determine the basis uniquely, up to overall scaling; thus a snake defines a projective basis. 
To construct the basis we can choose any $ v_1$ at the tail of the snake, and use the fin relations to 
determine every subsequent $ v_i$.

\begin{figure}[htb]
\centering
\includegraphics[width=5in]{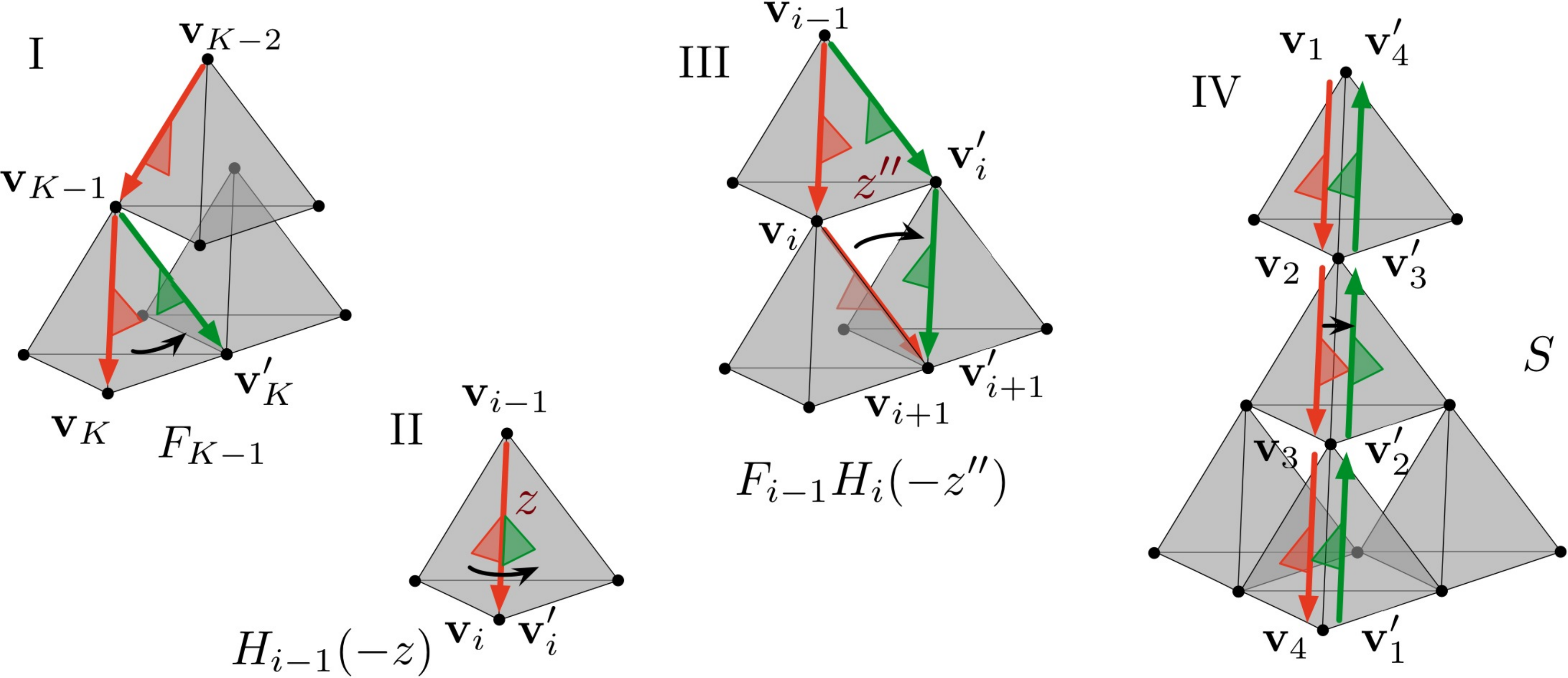}
\caption{The four fundamental snake moves. In each case we go from the orange snake to the green snake. Move IV is shown for $K=4$.}
\label{fig:dragonmoves}
\end{figure}

There are four elementary moves that can be used to move a 3d snake to any position within a tetrahedron. They are shown in Figure \ref{fig:dragonmoves}. Each move acts as a simple $PGL(K,\C)$ transformation on the projective basis associated with a 3d snake. The moves are expressed in terms of two kinds of $GL(K,\C)$ matrices. The first are the 
elementary unipotent matrices $F_i := {\rm Id} +E_{i+1,i}$ (for $i=1,\ldots,K-1$), where $E_{n,k}$ is the matrix with entry $1$ on the $n$-th row and $k$-th column and zero elsewhere. The second are the diagonal matrices $H_i(x) = \text{diag}(1,\ldots,1,x,\ldots,x)$, with the $(i+1)$-st through $K$-th entries equal to $x$.
Their arguments are always the negatives of octahedron parameters. Observe that $F_i$ and $H_j(x)$ commute unless $i=j$.

Move I can only occur at the head of a snake, adjacent to the face where it ends. It moves the head across a small black tetrahedron, keeping the fins on the same tetrahedron face, and acts as $F_{K-1}$, sending
\be \begin{pmatrix}  v_1 \\ \vdots \\  v_{K-1}\\  v_{K} \end{pmatrix} \;\mapsto\;  
F_{K-1} \begin{pmatrix}  v_1 \\ \vdots \\  v_{K-1}\\  v_{K} \end{pmatrix} = 
\begin{pmatrix}  v_1 \\ \vdots \\  v_{K-1}\\  v_{K-1} +  v_K \end{pmatrix}\,.
\ee
(Here we act on the left on column vectors containing the $ v_i$.)
Move II is a fin-flip that can occur at any segment along the snake; it simply multiplies $ v_i$ (at the head of the segment) and all subsequent vectors by (minus) the octahedron parameter associated with the segment's edge. Move III can act on any two consecutive segments of a snake in the position shown; note that the fins on the second segments are pointing inwards. Move III is a sort of combination of Moves I and II, and involves the octahedron parameter on the edge that the snake slides across.%
\footnote{This transformation is perhaps the most nontrivial to derive. It suffices to apply it to snakes on a $K=3$ tetrahedron, in which case the parametrizations of Figure \ref{fig:SL3-lines} easily lead to the answer.} %
Finally, Move IV is a reversal of orientation that makes sense only when a snake lies entirely along an edge of a big tetrahedron; the fins are unmodified. Move IV reverses the order of the basis vectors, and negates every even-numbered vector, thus multiplying by an anti-diagonal matrix
\be S = \begin{pmatrix} \cdots & 0&0&1 \\
 \cdots & 0 &-1&0 \\
 \cdots &1&0&0 \\  
  \!\!\text{\rotatebox{90}{$\ddots$}} & \;\vdots\; & \;\vdots\; & \;\vdots\;  \end{pmatrix}\,.
\ee

\begin{figure}[htb]
\hspace{-.25in}\includegraphics[width=6.5in]{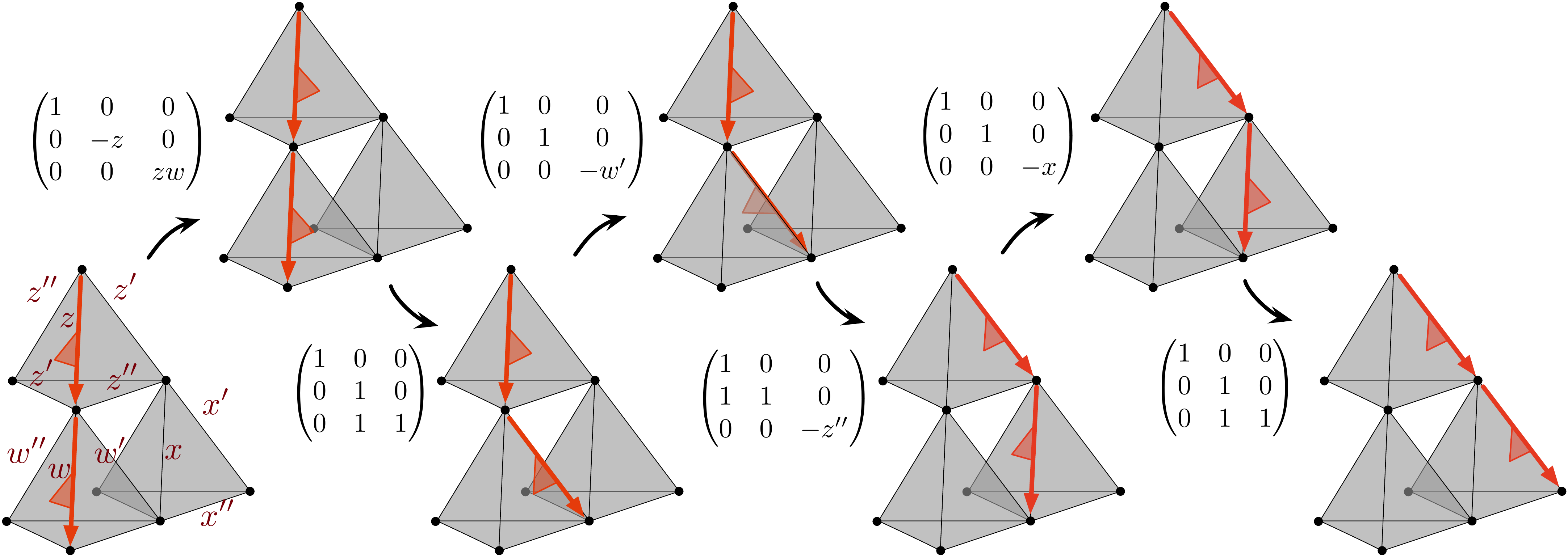}
\caption{Moving a snake from one edge to another, for $K=3$.}
\label{fig:dragonface3}
\end{figure}

The first three moves preserve a flag; \ie\ they are lower-triangular matrices. Indeed, any move keeping the tail of a snake attached to a specific vertex of the big tetrahedron preserves the flag associated with that vertex. An example of the sequence of moves that can be used to move a snake from one edge of a big tetrahedron to another (with fins oriented the same way) appears in Figure \ref{fig:dragonface3}. Multiplying the matrices in this example shows that the transformations of 3d snakes along a 2d surface are equivalent to the snake transformations of \cite{FG-Teich} --- given the usual identification of octahedron parameters and boundary cross- and triple-ratios.

The single most important thing to observe about snakes on a tetrahedron is that the projective bases associated with them are well-defined --- \ie\ the $PGL(K,\C)$ holonomy inside the tetrahedron can be trivialized --- \emph{if and only if the octahedron relations $zz'z''=-1$, $z''+z^{-1}-1=0$, and the sextic internal gluing constraints are satisfied.}

%\begin{wrapfigure}{r}{2.7in}
\begin{figure}[tbh]
\centering
\includegraphics[width=2.7in]{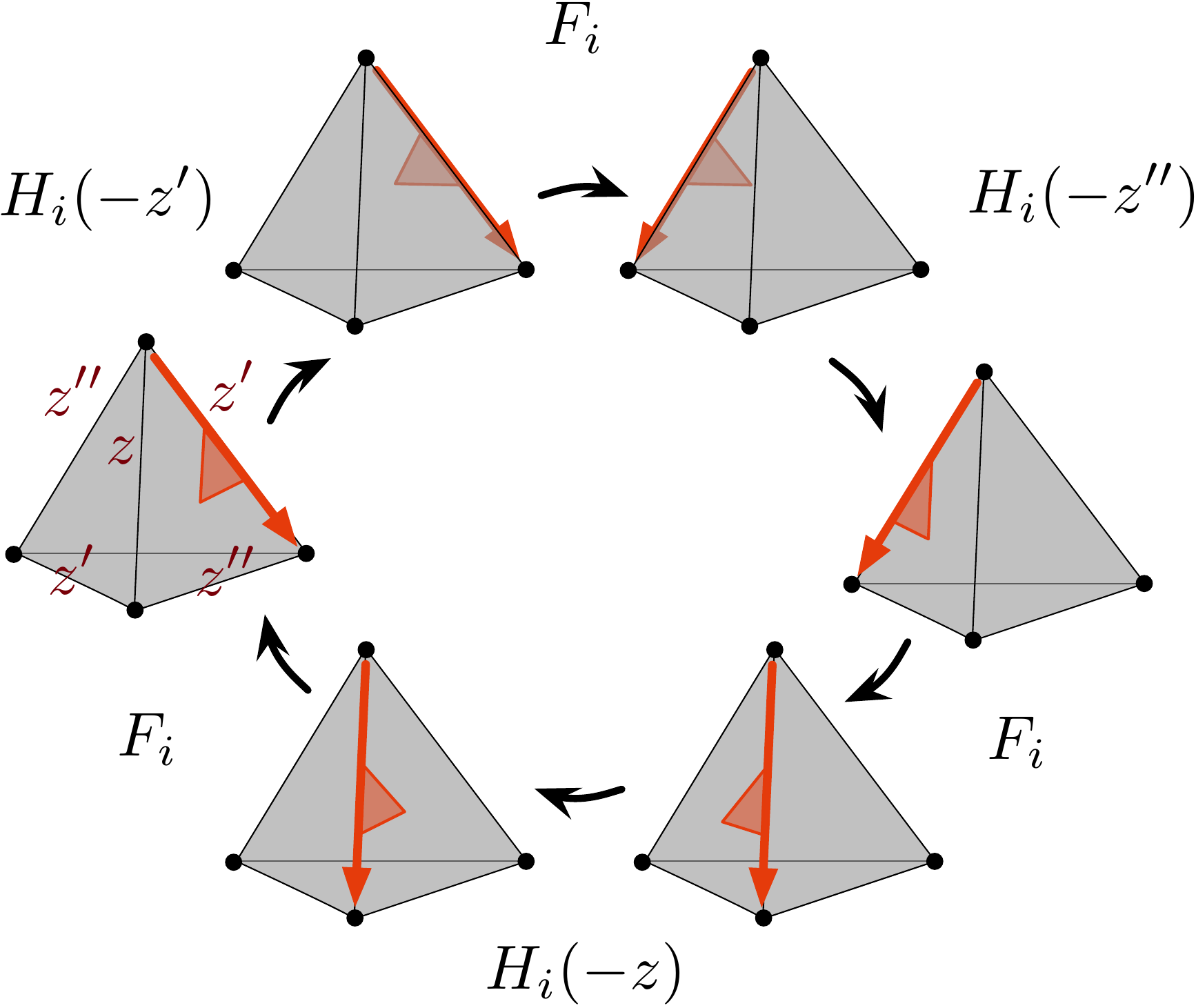}
\caption{Moving a snake around a black tetrahedron to obtain coordinate constraints.}
\label{fig:dragoncycle}
\end{figure}
%\end{wrapfigure}

On one hand, if we derive both octahedron parameters and projective bases by using a configuration of four flags, then the bases are automatically well defined and the relations are automatically satisfied. Conversely, if we simply place a triple of unconstrained coordinates $z_{abcd}, z_{abcd}',z_{abcd}'' \in \C^* -\{1\}$ on every octahedron and \emph{define} the $PGL(K,\C)$ transformations from one snake to another using the rules of Figure \ref{fig:dragonmoves}, we will find that the holonomy around a closed path in the space of snakes is trivial if and only if the standard relations are obeyed.

For example, to move a snake all the way around a black tetrahedron with parameters $z,z',z''$, as shown in Figure \ref{fig:dragoncycle}, we can use a combination of Moves II (fin-flips) and either Moves I or III. Either way, the total holonomy is diagonal aside from a $2\times 2$ block conjugate to
\be  \begin{pmatrix} 1 & 0 \\ -zz'(z''+z^{-1}-1) &\; -zz'z'' \end{pmatrix}\,. \ee
Having trivial eigenvalues requires $zz'z''=-1$, and having a trivial holonomy requires $z''+z^{-1}-1=0$. If $K\geq 4$, then the $(K-1)$-decomposition of a tetrahedron contains ${K-1\choose 3}$ upside-down $\Delta^{2,0}$ tetrahedra  in addition to the upright black ones. We may move snakes around these upside-down tetrahedra in a similar way, and find that the holonomy is trivial if and only if the sextic gluing constraints are satisfied --- \ie\ the product of six octahedron vertices at corresponding internal points in the $K$-triangulation is one.
 
Once we know that the $PGL(K,\C)$ holonomy inside a big tetrahedron is trivial, we can use octahedron parameters and snake moves to reconstruct the configuration of framing flags. We first choose \emph{any} projective basis $(v_1,\ldots,v_K)$ for $\C^K$ associated with a snake along one of the edges of the tetrahedron. The basis is unique up to $PGL(K,\C)$ isomorphism.
Then snake moves determine vectors (hence lines $\mathfrak L_{abcd}$) on every other lattice point of the $(K-1)$-decomposition, which uniquely reconstruct the configuration of flags.

\subsubsection{3d snakes in a 3-manifold}
\label{sec:dragonM}

Now we extend 3d snakes and projective bases to an entire triangulated 3-manifold $M$, and prove that octahedron parameters are coordinates for an open subset of framed flat connections.

%\begin{wrapfigure}[12]{r}{3in}
\begin{figure}[tbh]
\centering
\includegraphics[width=3in]{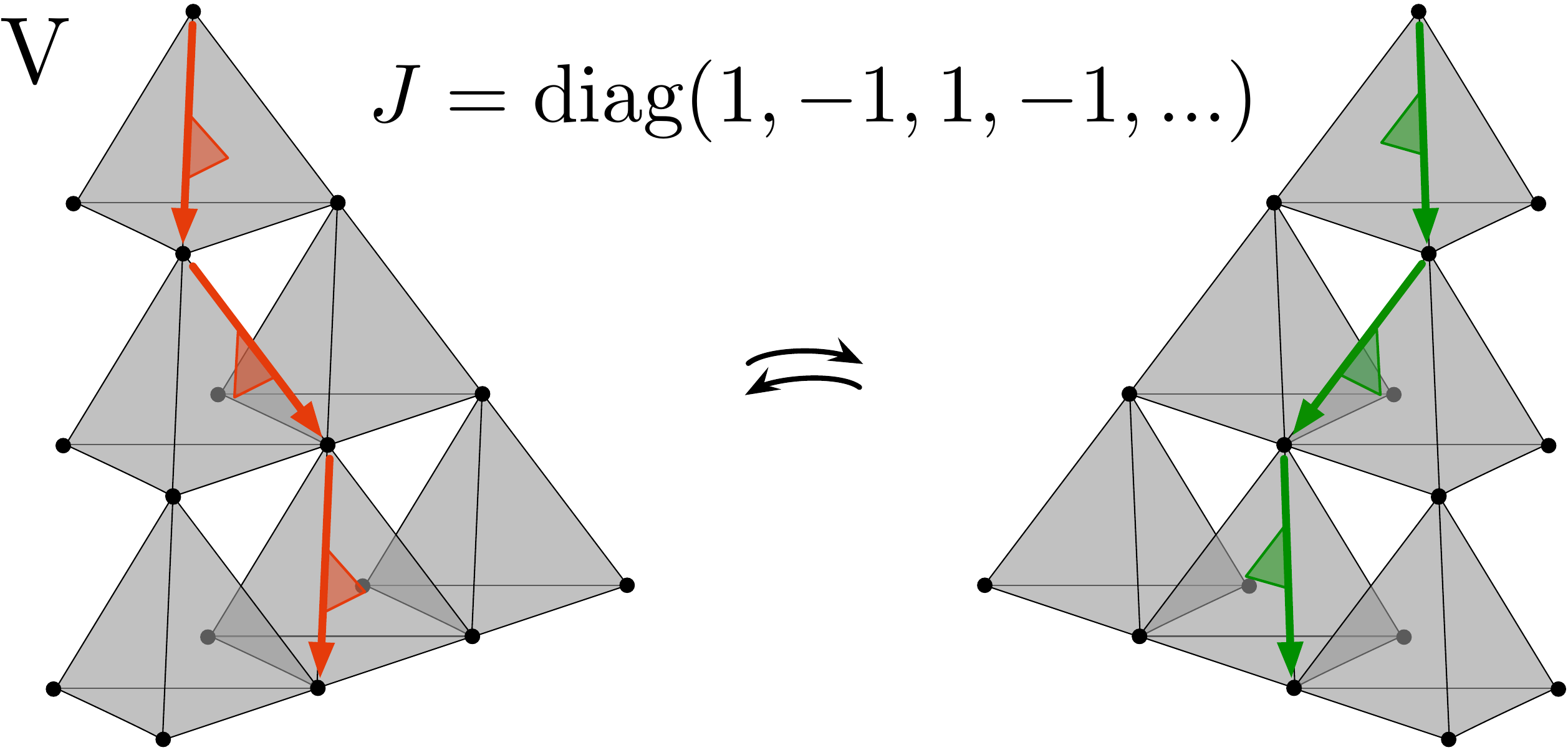}
\caption{The fifth snake move: jumping between glued tetrahedra, fins along the common face.}
\label{fig:dragonjump}
\end{figure}
%\end{wrapfigure}

We simply need one more elementary move to transport a snake from one tetrahedron to another, across a common face (Figure \ref{fig:dragonjump}).
It can be derived by looking at a bipyramid, as in Section \ref{sec:SLKpoly} (see Figure \ref{fig:faceglue}).
We assume that all fins lie along the common glued face, and don't change position during the jump between tetrahedra.
The lines determining the snakes' projective bases are all built from the three flags at the vertices of the common glued face (in Figure \ref{fig:faceglue}, these are flags $A,C,E$), but they come with opposite orientation on the two sides. Therefore, the snake move is implemented by a diagonal matrix $J$ of alternating $\pm 1$'s.

We can now run the argument backwards. Suppose that we assign arbitrary parameters
\be \label{octacoord}
z,z',z'' \in \C^*-\{1\}\,,\quad\text{with}\quad zz'z''=-1\,, \quad z''+z^{-1}-1=0
\ee
to 
each octahedron of the $K$-decomposition of a triangulated 3-manifold $M$. 
Consider the set of all snakes in the manifold. We assign to each snake $s$ a standard 
$K$-dimensional vector space 
$\C^K_{s}$ with a basis. We treat the latter as a projective basis, \ie\ consider bases that differ by common $\C^*$ rescaling as equivalent. If two snakes $s_1$ and $s_2$ are related by an elementary move of type I--V 
from Figures \ref{fig:dragonmoves} and \ref{fig:dragonjump}, we define a linear map 
$\C^K_{s_1} \to \C^K_{s_2}$ using the snake move matrices and the 
octahedral coordinates (\ref{octacoord}). The collection of the vector spaces $\C^k_{s}$ and the maps between them 
determines a $PGL(K,\C)$-connection. The holonomy of this connection inside of a tetrahedra 
is trivial if and only if the octahedron parameters satisfy the two relations in (\ref{octacoord}), and the sextic gluing constraints for internal lattice points of tetrahedra. Now by moving snakes among different tetrahedra we can show that the $PGL(K,\C)$ holonomy around any contractible cycle for the snakes in $M$ is trivial if and only if the face and edge gluing constraints of Section \ref{sec:SLKpoly} are also obeyed.

\begin{figure}[htb]
\centering
\includegraphics[width=5.4in]{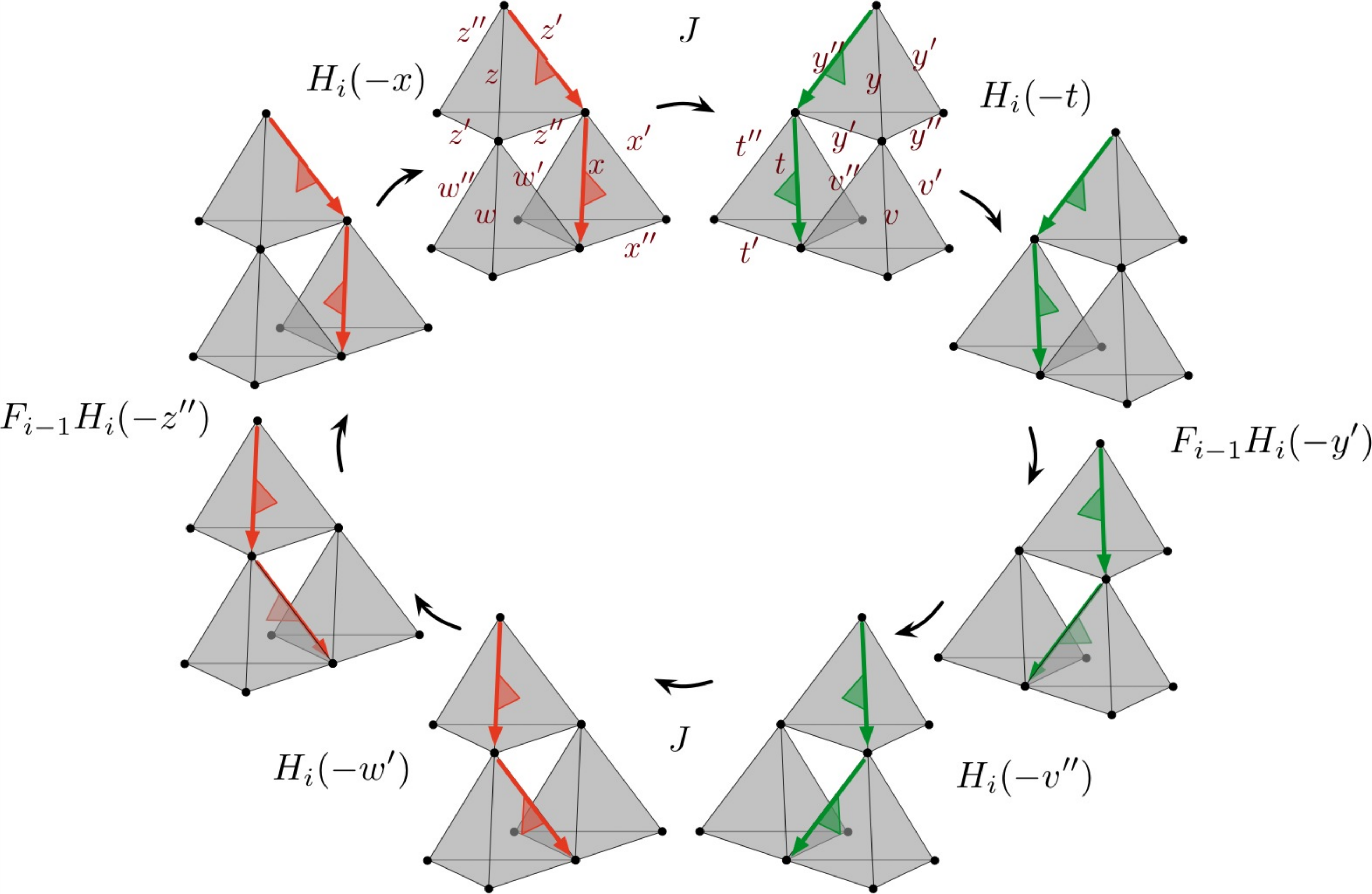}
\caption{Moving a snake around a lattice point of the $K$-triangulation that lies on a pair of glued faces. The four images on the left (orange snakes) are from one tetrahedron, and the images on the right (green snakes) are from a second.}
\label{fig:dragonfaceglue}
\end{figure}

For example, to find the face constraints we can compute the holonomy around the loop shown in Figure \ref{fig:dragonfaceglue}. The loop encircles a lattice point $p$ of the $K$-decomposition that lies on the common glued face (corresponding to a white triangle in the $(K-1)$-triangulation of the face). It is easy to see that corresponding holonomy matrix is $\text{diag}(1,\ldots,1,z''w'xy'v''t,\ldots,z''w'xy'v''t)$, with $K-i$ nontrivial entries. For a trivial holonomy the product $z''w'xy'v''t$ of the six octahedron parameters that touch $p$ must equal one --- \ie\ the face constraint must be satisfied.

In a similar way, we may analyze the internal edge constraints by moving snakes in a circle around that edge. The corresponding holonomy matrix is a product of diagonal fin-flips and $J$-matrices. Requiring trivial holonomy forces the products of octahedra coordinates at every point on the edge to equal one. To make this explicit, suppose that $N$ tetrahedra are glued around an edge and in each tetrahedron there are $K-1$ octahedra touching the edge, with parameters $z_i^{(I)}$ (where $1\leq i\leq K-1$, $1\leq I\leq N$). Define $D(x_1,\ldots,x_{K-1}) := H_1(x_1)H_2(x_2)\cdots H_{K-1}(x_{K-1}) = \text{diag}(1,x_1,x_1x_2,\ldots,x_1\cdots x_{K-1})$. Then the total snake holonomy around the edge is
\be \prod_{I=1}^N \big[D(-z_1^{(I)},\ldots,-z_{K-1}^{(I)})\cdot J\big] = \prod_{I=1}^N D\big(z_1^{(I)},\ldots,z_{K-1}^{(I)}\big)\,. \ee
The matrix is trivial if and only if $\prod_{I=1}^N z_i^{(I)}=1$ for all $i$. The minus signs that accompany octahedron parameters are cancelled out by the signs in $J$-matrices!

\begin{figure}[htb]
\centering
\includegraphics[width=5in]{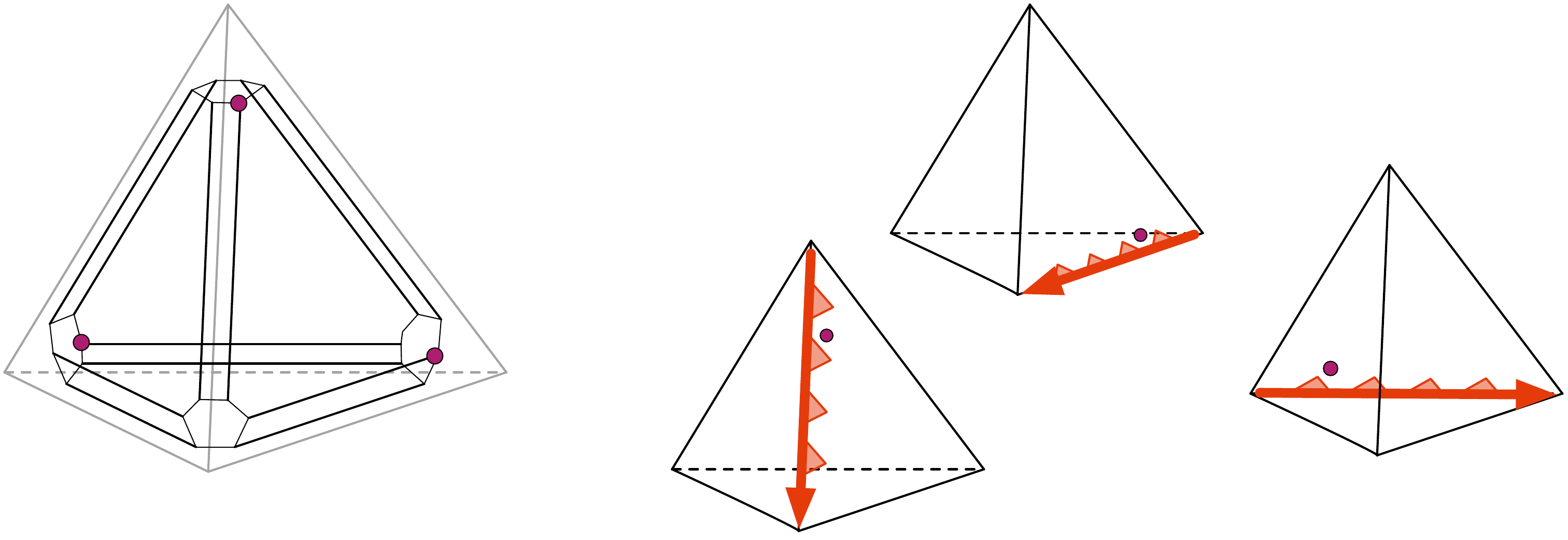}
\caption{The associahedron, and canonical snakes corresponding to three of its 24 vertices.}
\label{fig:marks}
\end{figure}

\begin{wrapfigure}{r}{1.6in}
\vspace{-.5cm}
\centering
\includegraphics[width=1.5in]{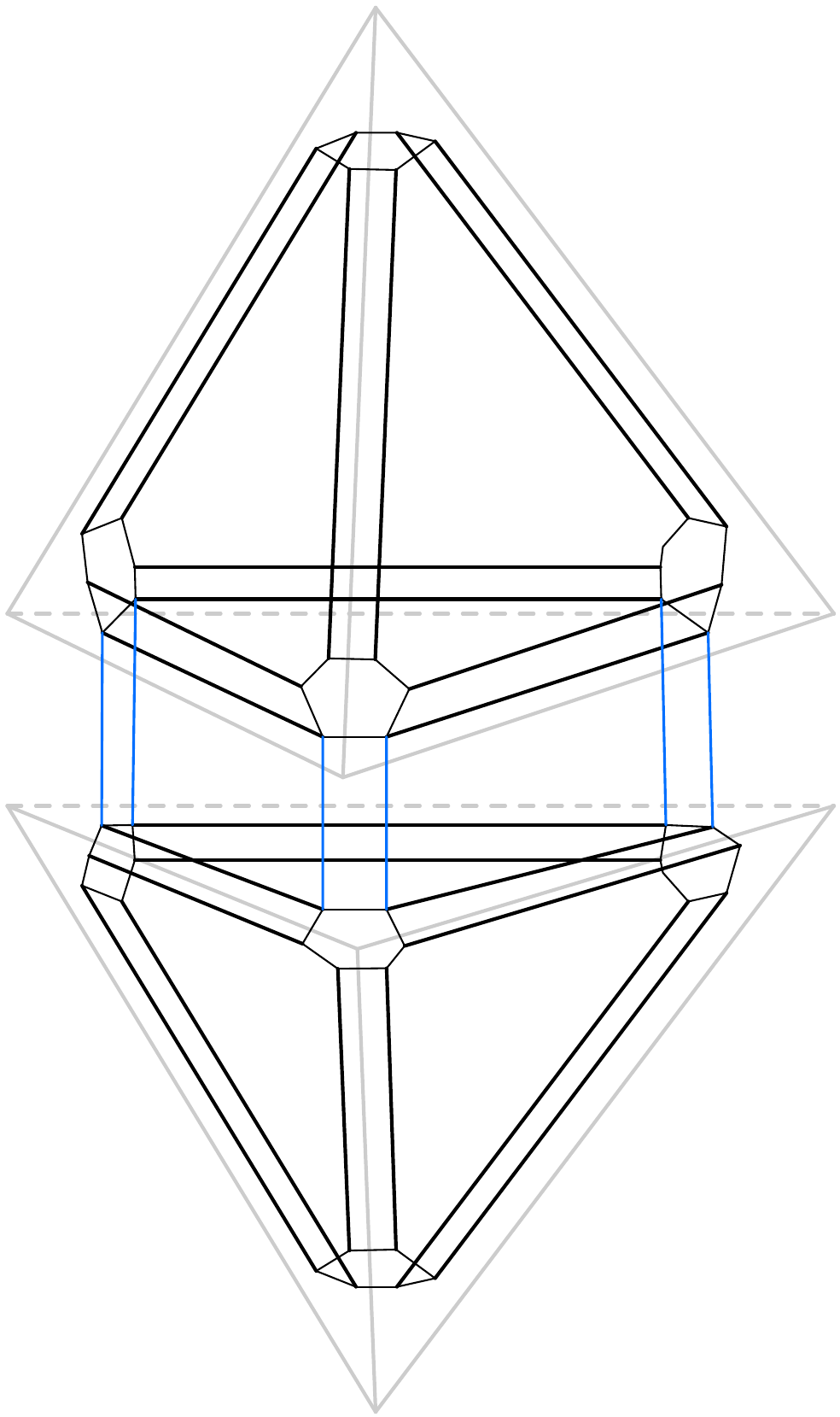}
\caption{Connecting glued associahedra.}
\label{fig:mark-graph}
\end{wrapfigure}

In combination with loops inside tetrahedra, the face and edge loops just discussed are enough to generate all the homotopically trivial transformations of snakes. Thus, we find that imposing all the standard octahedron gluing constraints allows to define a flat $PGL(K)$-connection on the space of snakes, which induces a flat $PGL(K)$-connection on the 3-manifold $M$.

To be more explicit about the last step, we may ``doubly truncate'' every big tetrahedron $\Delta$ in the triangulation of $M$ (Figure \ref{fig:marks}). This polyhedron has 24 vertices. It has 
hexagons in places of the original vertices and faces, and rectangles in place of the original edges. It is isomorphic to the convex hull of 
the orbit of a generic point in the space $\R^3$, realized as a hyperplane $x_1+x_2+x_3+x_4=1$ 
in $\R^4$, under the action of the group $S_4$ permuting the coordinates $(x_1, \ldots, x_4)$. 
It is nothing else but the 3d associahedron.%
\footnote{A similar ``cutting'' procedure in 
$\R^d$ leads to a $d$-dimensional associahedron with $(d+1)!$ vertices. It is isomorphic to 
 the convex hull 
of the $S_{d+1}$-orbit of a generic point in the hyperplane $\R^d \subset \R^{d+1}$. 
For example, for $d=2$ we get a hexagon.}

Now, we mark the 24 vertices of each doubly-truncated tetrahedron in the triangulation of $M$. Each marked point labels one of 24 canonical snakes on the edges of a given tetrahedron, with all fins 
going along the same tetrahedron face. The location of the point indicates: 1) the edge choice; 2) 
the snake orientation; and 3) the fins.

At every marked point $s$ in $M$ we fix a standard 
$K$-dimensional vector space 
$\C^K_{s}$ with a projective basis. The snake holonomy matrices then define a $PGL(K,\C)$ transformation 
between these vector spaces, determined by the octahedron parameters. Within a single tetrahedron we use Moves I--IV to obtain holonomies along the edges of the associahedron,
and we know that any closed cycle in the graph has trivial holonomy.
We can represent this by filling in the closed cycles with 2-cells. We then extend the graph from one tetrahedron to another by using Move V, and fill in all the closed cycles corresponding to face and edge gluing constraints with 2-cells as well, since these cycles also have trivial holonomy. In the end, we obtain a 2-complex that has the same fundamental group as $M$, so we have produced a flat $PGL(K,\C)$-connection on $M$. Finally, as at the end of Section \ref{sec:dragontet}, we use the projective bases assigned to snakes within each tetrahedron to reconstruct the framing flags; and, by construction, the flags at all small boundaries of $M$ will be preserved by the flat $PGL(K,\C)$-connection. So we arrive at a framed flat $PGL(K,\C)$-connection on $M$.  \\

Notice again that the space $\wt \CL_K(M,\mb t_{\rm 3d})$ of octahedron parameters parametrizes framed flat connections for which the 4-tuples of framing flags for each tetrahedron in $\mb t_{\rm 3d}$ are generic. For certain choices of $\mb t_{\rm 3d}$, the space $\wt \CL_K(M,\mb t_{\rm 3d})$ may lose components. For example, if $\mb t_{\rm 3d}$ includes a tetrahedron glued to itself along two faces, forming a univalent internal edge, the space $\wt \CL_K(M,\mb t_{\rm 3d})$ will be empty. (It is impossible to satisfy the gluing constraints along the univalent edge with octahedron parameters in $\C^*-\{1\}$.) By taking a union over all 3d triangulation $\mb t_{\rm 3d}$ we can define a space $\wt \CL_K(M,\mb t)$ that depends only on the boundary triangulation. For any sufficiently refined $\mb t_{3d}$, the algebraic closures of $\wt \CL_K(M,\mb t_{\rm 3d})$ and $\wt \CL_K(M,\mb t)$ must coincide.

\subsection{Small torus boundaries}
\label{sec:torusbdy}

Let us take a moment to spell out how the boundary phase space $\CX_K^{\rm un}(\pd M,\mb t)$ gets enriched in the presence of small torus boundaries.

\begin{figure}[htb]
\centering
\includegraphics[width=2.6in]{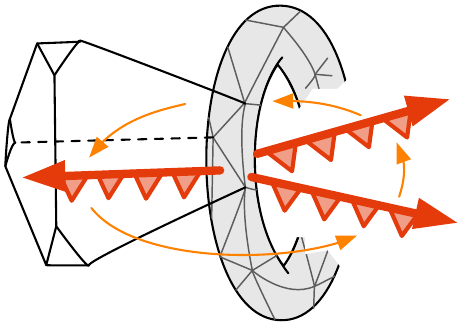}
\caption{Computing the holonomy around a cycle of the torus boundary.}
\label{fig:torus-snake}
\end{figure}

Suppose that $\pd M$ consists of a single small torus boundary, and choose A and B cycles for it. The 3d snakes of Section \ref{sec:projbases} show us how to express the holonomy eigenvalues around the A and B cycles in terms of octahedron parameters. We compute the full holonomy around a cycle by fixing the tail of a snake on the torus boundary, and swinging the head around the desired cycle (Figure \ref{fig:torus-snake}). Only snake moves I, II, III, and V are used, which leads to a lower-triangular holonomy matrix; this must be so, since the holonomy must preserve the flag on the torus. Therefore, the eigenvalues can simply be read off from the diagonal of the matrix, and take the form of Laurent monomials in the octahedron parameters. A more intuitive and constructive method for finding these monomials will be described in Section \ref{sec:combi}.

For now, we make the following claims. As coordinates on the product phase space $\CP_\times = \prod_{\soct_i\in M}\CP_{\pd\soct_i}$, the holonomy eigenvalues Poisson-commute with all of the internal gluing functions $c_k$. Moreover, they satisfy the expected Atiyah-Bott-Goldman Poisson bracket for the torus boundary. Explicitly, let us parametrize the $PGL(K,\C)$ eigenvalues as
\be \label{bdyAB}
 \text{Hol}(\text{A})= 
\begin{pmatrix}
   1 &0&0& \cdots & 0 \\
    & m_1 &0&\cdots& 0 \\
    && m_1m_2 & \cdots &0 \\
    &*&&\ddots & \\
    &&&& m_1m_2\cdots m_{K-1}
  \end{pmatrix}, \quad \text{Hol}(\text{B})= 
\begin{pmatrix}
   1 &0&0& \cdots & 0 \\
    & \ell_1 &0&\cdots& 0 \\
    && \ell_1\ell_2 & \cdots &0 \\
    &*&&\ddots & \\
    &&&& \ell_1\ell_2\cdots \ell_{K-1}
  \end{pmatrix},
\ee
The functions $l_a$ and $m_b$ are the characters of the Cartan group 
of $PGL(K)$ provided by the simple positive roots.  
The Poisson brackets between them are described by the Cartan matrix $\kappa_{ab}$:
\be \{\ell_a,m_b\} = \kappa_{ab}\,\ell_a m_b\,, ~~~~\{\ell_a,\ell_b\} = \{m_a,m_b\}=0. \ee
\be \kappa_{ab} = \begin{cases} 2 & a=b \\
 -1 & |a-b|=1 \\
 0 & |a-b|>1 \end{cases}
\ee
Here we assumed that the intersection number of the A and B cycles on the torus $\langle A, B\rangle =1$, 
with the orientation of the torus induced from that of $M$. 
Otherwise the formula reads $\{\ell_a,m_b\} = \langle A, B\rangle\kappa_{ab}\,\ell_a m_b$. 

We again expect that the boundary phase space associated to a small torus%
\footnote{Although we continue to use the notation ``$\CX_K^{\rm un}(\pd M)$'' to discuss boundary phase spaces associated with small tori, we emphasize again that we \emph{never} require unipotent holonomy around the cycles of a torus. It is only a condition at the holes in the big boundary of $M$.} %
can be expressed as a symplectic quotient of a product of octahedron spaces:
\be
%\CX_K({T^2})\simeq
\CX^{\rm un}_K({\pd M}) \simeq  \CP_\times\big/\!\!\big/(c_k-1)\,.
\ee
Notice that here the result of the (linear) symplectic reduction is a complex torus. Thus, we hope to obtain the smooth part $\CX^{\rm un}_K({\pd M})\subset \CX_K(T^2)$, isomorphic to $(\C^*)^{2K-2}$, discussed in Section \ref{sec:spaces}.

We may motivate the symplectic reduction by counting degrees of freedom. An ideal triangulation of $M$ with $N$ tetrahedra has exactly $N$ internal edges (this can be verified by computing the Euler character) and $2N$ internal faces. Therefore, the number of internal gluing functions is
\be \#\,c_k = N {K-1\choose 3} + (2N)  {K-1\choose 2} + N  {K-1\choose 1}  = N {K+1\choose 3}\,.\ee
This is the same as the number of octahedra in the $K$-decomposition of $M$. Thus, if all the $c_k$'s were independent moment maps, the quotient $\CP_\times\big/\!\!\big/(c_k-1)$ would be zero-dimensional. However, in the presence of a small torus boundary, it turns out that there are $K-1$ automatic relations among the $c_k$. We will demonstrate this in Section \ref{sec:combi}. Then --- assuming that the remaining $c_k$ are independent --- $\CP_\times\big/\!\!\big/(c_k-1)$ does have dimension $2(K-1)$, as required.

If we take the image of the product octahedron Lagrangian $\CL_\times$ under the reduction $\CP_\times\big/\!\!\big/(c_k-1)$, 
then we will obtain a Lagrangian submanifold $\CL_K(M, \mb t_{3d})\subset \CX_K(T^2)$. It is typically just a component of the variety of framed flat connections on the boundary that extend to the bulk. In particular, the requirement that the framing flags for each big tetrahedron in $M$ be generic (so that octahedron parameters are well defined) implies that $\CL_K(M,\mb t_{3d})$ only parametrizes flat connections whose boundary holonomy has finite stabilizer. Whenever the boundary holonomies have coincident eigenvalues, these eigenvalues must fit in a single Jordan block.

Finally, recall that if $\pd M$ consists of multiple big components $\CC_i$ and small torus components $T^2_i$, the boundary phase space is just a straightforward product $\CX^{\rm un}({\pd M},\mb t) =  \prod_{i=1}^b\CX^{\rm un}_K({\CC_i},\mb t)\times \prod_{i=1}^t \CX_K({T^2_i})$. The expected equivalence $\CX_K(\pd M,\mb t)\simeq \CP_\times \big/\!\!\big/(c_k-1)$ should generalize in the obvious way.

\subsection{2--3 moves via 4d cobordism}
\label{sec:23}

As a final application of the ideas in this section, we describe how 2--3 moves act on $K$-decompositions and octahedron parameters.

Recall that a 2--3 Pachner move (shown in Figure \ref{fig:23-intro} on page \pageref{fig:23-intro}) 
affects a triangular bipyramid in the triangulation of a 3-manifold $M$. It changes the bulk triangulation 
of the bipyramid, leaving its boundary intact. Any two bulk triangulations of $M$ can be connected
  a sequence of 2--3 moves \cite{Matveev-spines, Piergallini}. 
So if $M$ is an admissible 3-manifold, a 2--3 Pachner move leaves the big-boundary triangulation fixed. 

In general, alterations of $d$-dimensional triangulations (Pachner moves) can be thought of as elementary cobordisms that amount to 
gluing $(d+1)$-dimensional simplices to  $d$-triangulations. For example,  in two dimensions, changing a triangulation of a quadrilateral by performing a \emph{flip} is equivalent to gluing a 3d tetrahedron onto the quadrilateral. 
We should think about the 2-3 move as a cobordism provided by a 4-simplex.

Studying framed flat $PGL(K)$-connections on a $d$-manifold,   
 it is essential to realize that $d$-simplices are no longer 
the most elementary pieces of the manifold: we rather have to 
consider $K$-decompositions of simplices in a $d$-dimensional triangulation. Therefore Pachner moves, understood as cobordisms provided by $(d+1)$-simplices, should no longer 
be the most elementary cobordisms. Instead, the elementary cobordisms 
should reflect the hypersimplices of the $K$-decomposition of a $(d+1)$-simplex. 
The Pachner move needs to be decomposed into a sequence of such elementary cobordisms. 

In two dimensions, this led to the decomposition of a flip into ${K+1\choose 3}$ elementary mutations. 
Each mutation corresponded to one of the octahedra $\Delta^{1,1}$ in a 3d tetrahedron \cite{FG-Teich}. 
Now we would like to proceed similarly, and decompose a 2--3 move 
as a sequence of elementary moves that reflect the 
 hypersimplicial $K$-decomposition of a 4-simplex.

\begin{figure}[htb]
\centering
\includegraphics[width=5.5in]{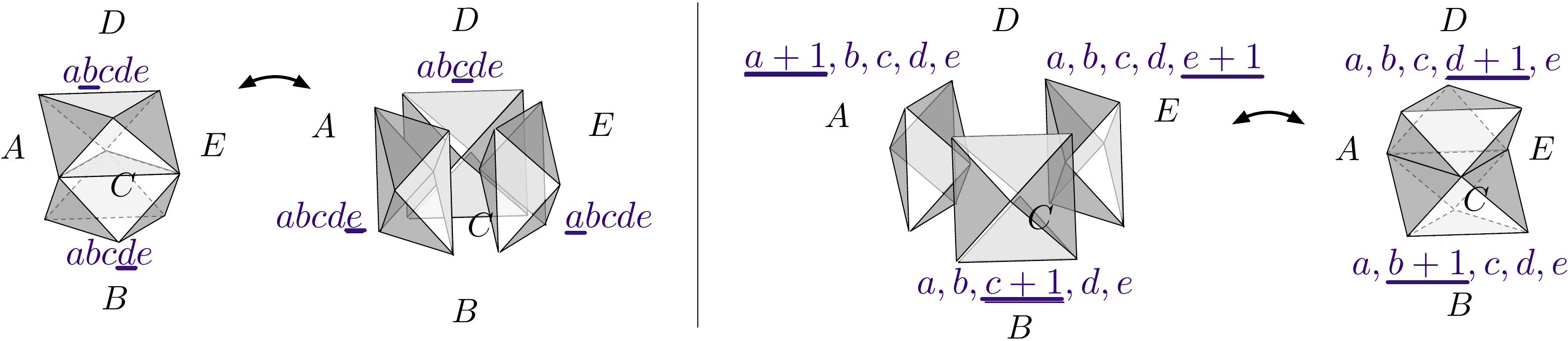}
\caption{The two elementary 2--3 moves for octahedra, corresponding to cobordisms through $\Delta^{1,2}$ hypersimplices (left) and $\Delta^{2,1}$ hypersimplices (right).}
\label{fig:23-oct-abc}
\end{figure}

The proof of the invariance of the motivic volume map under the 2--3 moves 
suggests that the effect of the Pachner  2--3 move on the framed flat $PGL(K)$-connections 
should be presented as a composition of the elementary moves of two types, 
parametrized by the set of all  $\Delta^{1,2}$- and $\Delta^{2,1}$-hypersimplices 
in the $K$-decomposition of a 4-simplex. This includes

\begin{itemize}  \item the ${K+2\choose 4}$ elementary moves via cobordisms through the $\Delta^{1,2}$-hypersimplices; and 

\item the ${K+1\choose 4}$ elementary moves from cobordisms through $\Delta^{2,1}$-hypersimplices. 

\end{itemize}
Altogether, these are 
$\frac12 K^2(K^2-1)$ elementary ``2--3 moves'' on octahedra $\Delta^{1,1}$ in the $K$-decomposition. 

This is exactly what we find. 
Topologically, the two types of moves (Figure \ref{fig:23-oct-abc}) look identical. Indeed, the simplices $\Delta^{1,2}$ and $\Delta^{2,1}$ are isomorphic, both having five 3d octahedra on their boundary. Geometrically, we will see (momentarily) that the first elementary move changes coordinates on the configuration space of five lines in a 2-dimensional space $V_2$, while the second move changes coordinates on the space of five planes in a 3-dimensional space $V_3$,
\be \begin{array}{c@{\;\;\leftrightarrow\;\;}l}
  \Delta^{1,2} & \text{Configurations of 5 lines in $V_2$} \\[.1cm]
  \Delta^{2,1} & \text{Configurations of 5 planes in $V_3$}\,.
  \end{array}
\ee
These two moduli spaces are canonically isomorphic.

\begin{figure}[htb]
\hspace{-.5in}
\includegraphics[width=7in]{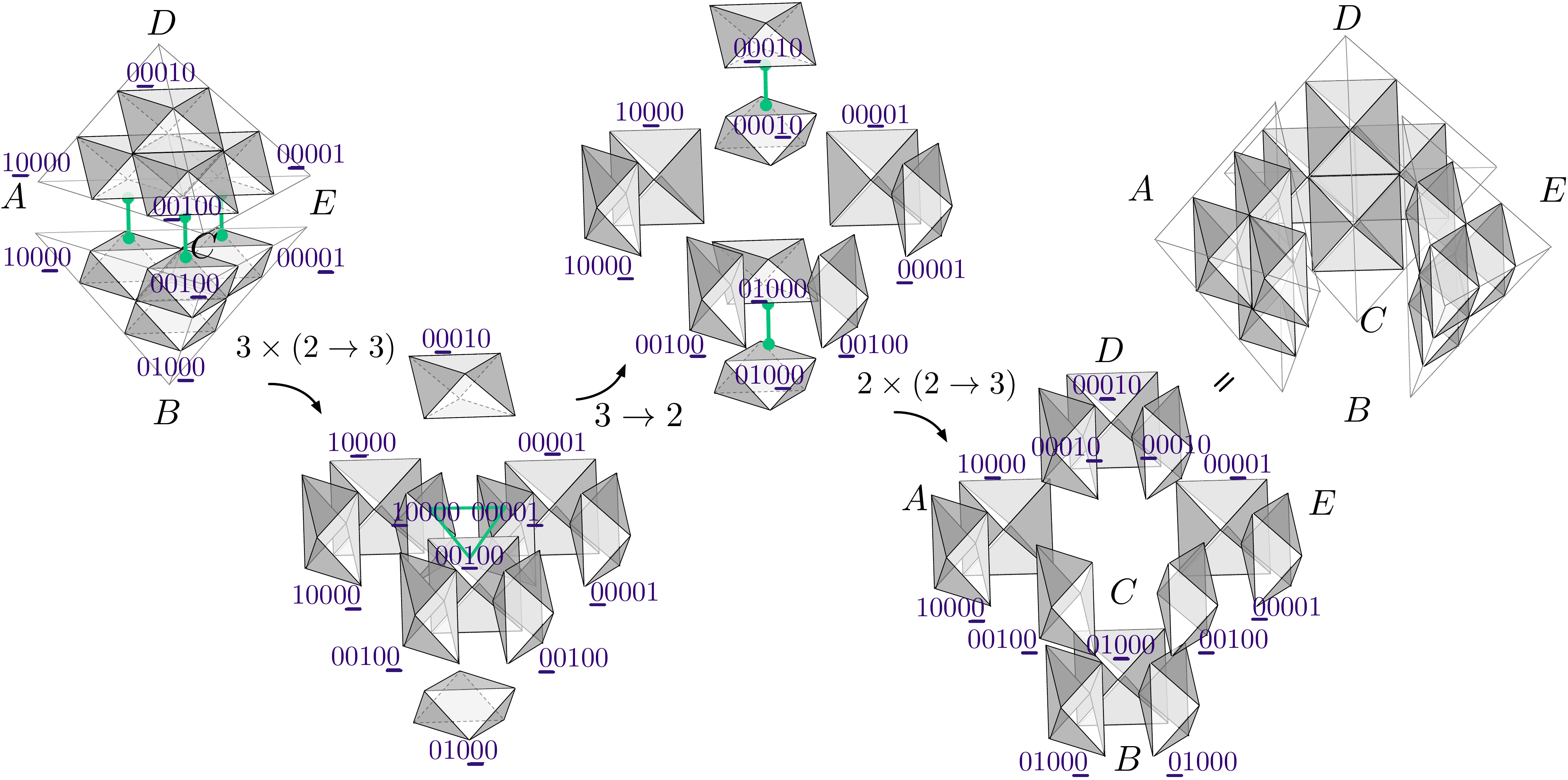}
\caption{The sequence of elementary 2--3 moves for $K=3$ (labels are for future reference).}
\label{fig:23-3-abc}
\end{figure}

\begin{figure}[htb]
\centering
\includegraphics[width=6in]{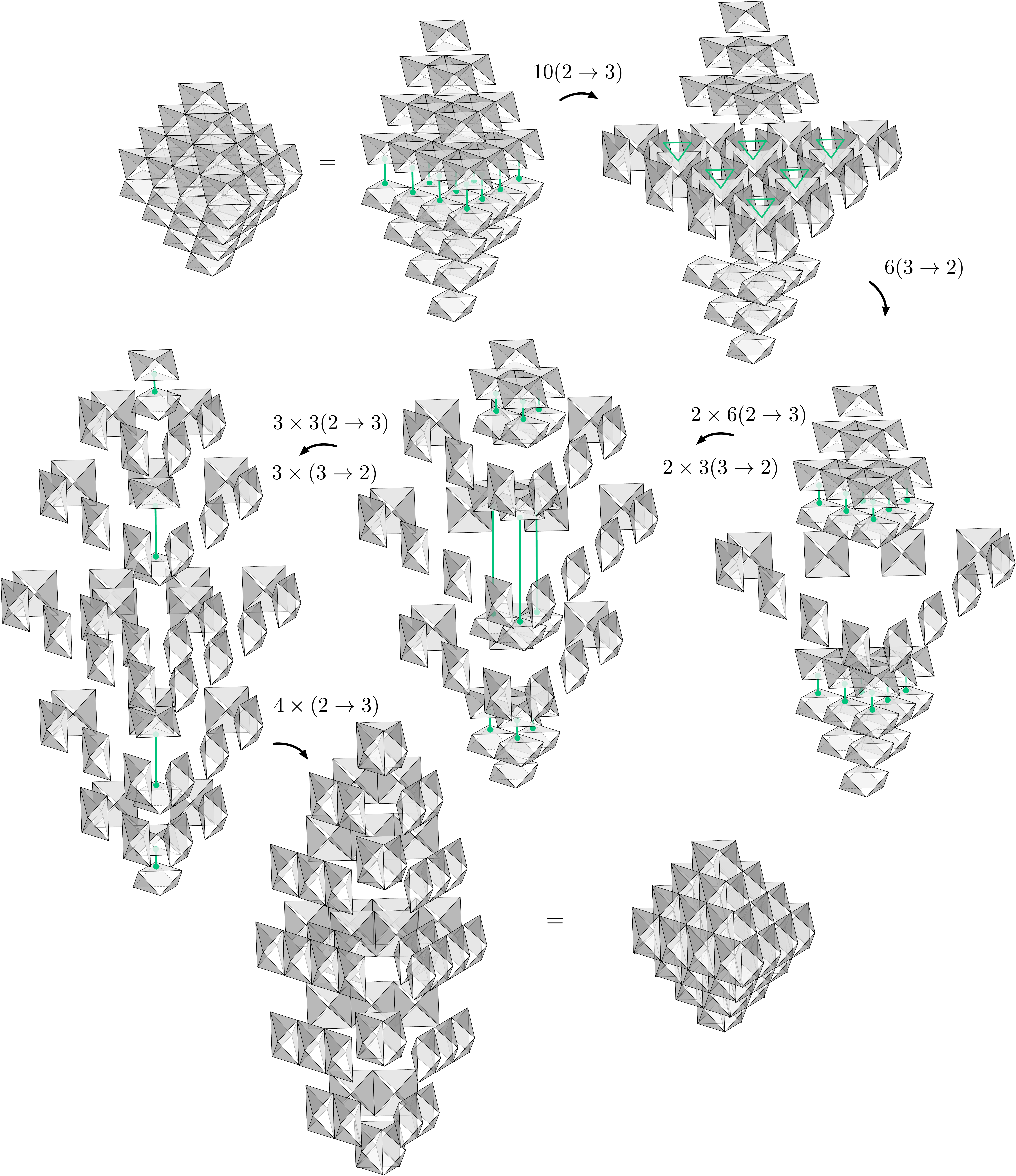}
\caption{The sequence of elementary 2--3 moves for $K=5$.}
\label{fig:23-5}
\end{figure}

The order in which the 2--3 elementary moves are performed is important, but not unique. Every choice of 3d slicing of a 4d simplex gives a different ordering. One systematic way to organize the elementary moves is shown in Figure \ref{fig:23-3-abc} for $K=3$ and in Figure \ref{fig:23-5} (below) for $K=5$. It can be described as follows.

We start with a bipyramid made from two tetrahedra. We start on the ``2'' side. In the first step, we take the ${K\choose 2}$ pairs of octahedra that are glued along black faces (at the common face of the two big tetrahedra) and apply the 2--3 move on the left of Figure \ref{fig:23-oct-abc} to each pair. This creates $3\times {K\choose 2}$ octahedra in a single ``equatorial'' plane. Among them, there are ${K-1\choose 2}$ triples of octahedra glued along three white faces and sharing an internal vertex. For the second step, we apply to these triples the move on the right of Figure \ref{fig:23-oct-abc}, in the $3\to 2$ direction. Then we repeat.
At the $i$-th step, for $i$ odd, there should be $(\frac{i+1}{2})\times {K-(i-1)/2\choose 2}$ pairs of octahedra glued along common black faces. They appear in $\frac{i+1}{2}$ horizontal planes. We apply the $2\to 3$ move on the left of Figure \ref{fig:23-oct-abc} to them. At the $i$-th step for $i$ even, there should be $\frac i2\times{K-i/2\choose 2}$ triples of octahedra glued along three white faces, with an internal trivalent vertex. They appear in $i/2$ horizontal planes. We apply the $3\to 2$ move on the right of Figure \ref{fig:23-oct-abc} to them. After $2K-3$ steps we are done: we arrive at a big bipyramid triangulated into three ideal tetrahedra.

\subsubsection{Coordinates}

Geometrically, the decomposed 2--3 move gives us many different coordinate systems for a generic configuration of five flags in $V_K$. They are the five flags $(A,B,C,D,E)$ at the vertices of the bipyramid, or (better yet) the five flags at the vertices of a 4d simplex. Then every full slice of the 4d simplex --- \ie\ every collection of octahedra that fit together into a bipyramid, at some stage in the sequence of 2--3 moves --- provides a complete coordinate system. The octahedra are assigned standard cross-ratio coordinates. 

To demonstrate this, let us label 4d hypersimplices by $5$-tuples in the usual way,
\be \Delta^{1,2}_{abcde} \;\;\leadsto\; \mathfrak P_{abcde}=A^a\cap B^b\cap C^c\cap D^d\cap E^e\,,\quad a+b+c+d+e=K-2\,, \ee
\be
\Delta^{2,1}_{abcde} \;\;\leadsto\; \mathfrak V_{abcde}=A^{a}\cap B^{b}\cap C^{c}\cap D^{d}\cap E^{e}\,,\quad a+b+c+d+e=K-3\,. 
\ee
These hypersimplices are associated with planes $\mathfrak P\subset V_K$ and 3-spaces $\mathfrak V\subset V_K$, built from the flags. Similarly, we may label the $5\times {K+2\choose 4}$ octahedra that are involved in the sequence of 2--3 moves by a 5-tuple $abcde$ ($a+b+c+d+e=K-2$) with one of the five indices ``underlined'' or ``decorated.''
The decoration serves (in part) to encode how octahedra occur as boundaries of hypersimplices: the boundaries of $\Delta^{1,2}$ and $\Delta^{2,1}$ contain
\begin{align} \pd \Delta^{1,2}_{abcde}\;&:\quad \oct_{\underline abcde}\,,\;
 \oct_{a\underline bcde}\,,\; \oct_{ab\underline cde}\,,\; \oct_{abc\underline de}\,,\; \oct_{abcd\underline e}\,; \\[.1cm]
\pd \Delta^{2,1}_{abcde}\;&:\quad \oct_{(\underline{a+1})bcde}\,,\; \oct_{a(\underline{b+1})cde}\,,\; \oct_{ab(\underline{c+1})de}\,,\; \oct_{abc(\underline{d+1})e}\,,\; \oct_{abcd(\underline{e+1})}\,.\;
\end{align}
Notice that the same octahedron may be on the boundary of \emph{both} $\Delta^{1,2}$ and $\Delta^{2,1}$ hypersimplices. On the other hand, the $5\times {K+1\choose 3}$ octahedra that appear at the beginning and end of the full 2--3 move (composing either 2 or 3 complete tetrahedra) are distinguished by the fact that the decorated index vanishes; thus they can only be boundaries of $\Delta^{1,2}$'s.

Every octahedron is associated with a 2-plane, regardless of the decoration, \eg
\be \oct_{a\underline bcde} \;\;\leadsto\; \mathfrak P_{abcde}=A^a\cap B^b\cap C^c\cap D^d\cap E^e\,,\quad a+b+c+d+e=K-2\,.\ee
It is \emph{also} associated with four lines, corresponding to the \emph{un}decorated indices,
\be \label{oct-lines23}
 \oct_{a\underline{b}cde} \quad\leadsto\quad
  \mathfrak L_{(a+1)bcde}\,,\;  \mathfrak L_{ab(c+1)de}\,,\; \mathfrak L_{abc(d+1)e}\,,\; \mathfrak L_{abcd(e+1)}\,. \ee
Then the octahedron inherits standard vertex coordinates $z,z',z''$ formed from cross-ratios of these lines in  $\mathfrak P_{abcde}$. It is easy to check that on any slice of the 4-simplex (at any step in the sequence of 2--3 moves) the product of octahedron parameters at any internal lattice point equals one. That is, \emph{the standard gluing equations are obeyed}. By the usual methods, one checks that the octahedron parameters on a slice are coordinates for the configuration space of five flags.

Moreover, if we isolate a cluster of five octahedra involved in one of the elementary moves of Figure \ref{fig:23-oct-abc}, the products of parameters $z,z',z''$ on the nine external vertices of this cluster are unchanged during the move. The proof follows from analyzing the cross-ratios involved. This is a local version of the universal statement that boundary coordinates are products of bulk coordinates, no matter how the bulk is decomposed.

It should be clear that the first type of elementary move relates coordinates for five lines in $V_2$: the five octahedra on the left of Figure \ref{fig:23-oct-abc} carry subsets of four of the five lines
\be \mathfrak L_{(a+1)bcde}\,,\; \mathfrak L_{a(b+1)cde}\,,\; \mathfrak L_{ab(c+1)de}\,,\; \mathfrak L_{abc(d+1)e}\,,\; \mathfrak L_{abcd(e+1)} \quad\subset \; \mathfrak P_{abcde}\,,\ee
and cross-ratios thereof. On the other hand, the second elementary move relates invariants of the configuration of five planes
\be \qquad  \mathfrak P_{(a+1)bcde}\,,\; \mathfrak P_{a(b+1)cde}\,,\; \mathfrak P_{ab(c+1)de}\,,\; \mathfrak P_{abc(d+1)e}\,,\; \mathfrak P_{abcd(e+1)} \quad\subset \; \mathfrak V_{abcde}\,. \ee
Each of the octahedra in this second move carries the four lines (\cf\ \eqref{oct-lines23}) that are intersections of one of these five planes with the other four! Four example, $\oct_{\underline abcde}$ carries the intersection of $\mathfrak P_{(a+1)bcde}$ with the other planes, and the cross-ratios formed from the resulting lines.

The moduli spaces of generic configurations of five lines in $V_2$ and five planes in $V_3$ 
are canonically isomorphic, or better to say canonically dual to each other (see the end of Section \ref{sec:config-basic}).
This is why the elementary moves on octahedra look so similar.
The ``$K$-decomposition'' of a full 2--3 move into elementary moves on octahedra
is a much refined version of the collection of the ``pentagons'' assigned to a configuration of five flags 
in section \ref{sec:config-basic}.

Finally, we note that the symplectic constructions of phase spaces and Lagrangian submanifolds that we have discussed throughout this section commute beautifully with 2--3 moves --- simply because they commute with the elementary moves on octahedra. For example, once we know that $\CX^{\rm un}_K(\pd M,\mb t)$ is a symplectic reduction of a product phase space $\CP_\times$ for any given triangulation $\mb t_{3d}$, it follows that it can also be obtained by symplectic reduction for any other triangulation $\mb t_{3d}'$, or indeed for any ``partial'' triangulation obtained by doing some series of elementary 2--3 moves on individual octahedra.
 (Note that the 2--3 moves never change the boundary triangulation $\mb t$!)
We will say more about this in Section \ref{sec:23-combi}. Moreover, the elementary 2--3 moves induce birational maps on the spaces $\wt \CL_K(M,\mb t_{\rm 3d})$ and $\CL_K(M,\mb t_{\rm 3d})$. The maps are one-to-one so long as the configurations of five flags involved in the move are generic.

%%%%%%%%%%%%%%%%%%%%%%%%%%%%%%
%%%%%%%%%%%%%%%%%%%%%%%%%%%%%%
%%%%%%%%%%%%%%%%%%%%%%%%%%%%%%
\section{$\mathrm{K}_2$-Lagrangians and the Bloch complex}
\label{sec:Bloch}

In this section and the next, we study the symplectic nature of gluing, and the Lagrangian properties of the moduli space $\CL_K(M)$ of framed flat connections on $\partial M$ that can be extended to $M$.

\subsection{Perspective}
\label{sec:perspective}

The basic result we are after can be phrased in several equivalent ways. Let $M$ be an admissible 3-manifold with a triangulation $\mb t_{3d}$, inducing a triangulation $\mb t$ of the big boundary. Fix $K$, and let $N$ denote the number of octahedra in the $K$-decomposition of $\mb t_{3d}$.
In Section \ref{sec:FFC}, we introduced the three moduli spaces
\be  \wt\CL_K(M,\mb t_{3d}) \to \CL_K(M,\mb t_{3d}) \subset  \CX^{\rm un}_K(\pd M,\mb t)\,,  \ee
and argued that they could be constructed by ``gluing'' the elementary Lagrangian pairs 
\be \begin{array}{ccc} \CL_{\soct_i} &\subset& \CP_{\pd\soct_i} \\
  \rotatebox{90}{=} && \rotatebox{90}{=} \\
\{z_i''+z_i^{-1}=1\} & \subset & \{z_iz_i'z_i''=-1\}\simeq (\C^*)^2 \end{array}
\ee
associated to individual octahedra. Let $2d$ be the dimension of $\CX^{\rm un}_K(\pd M,\mb t)$, so that  $\CX^{\rm un}_K(\pd M,\mb t)\simeq (\C^*)^{2d}$.
 What we have shown so far is that
\begin{itemize}
\item There exists a non-canonical projection $\wt\pi$ from $\prod_{i=1}^N \CP_{\pd\soct_i}\simeq (\C^*)^{2N}$ to $\CX^{\rm un}_K(\pd M,\mb t)$. In particular, the coordinates $x_a$
on $\CX^{\rm un}_K(\pd M,\mb t)$ are monomials of the octahedron parameters $z_i,z_i',z_i''$.

\item $\wt\CL_K(M,\mb t_{3d})$ is the subvariety in the product Lagrangian $\prod_{i=1}^N \CL_{\soct_i}\subset (\C^*)^{2N}$ cut out by 
   gluing equations $\{c_j=1\}_{j=1}^r$, where  $c_j$ are  monomials of the octahedron parameters.

\item There is a canonical projection $\pi:\prod_{i=1}^N \CP_{\pd\soct_i}\cap \{c_j=1\}_{j=1}^r\longrightarrow \CX^{\rm un}_K(\pd M,\mb t)$.

\item $\CL_K(M,\mb t_{3d})\subset \CX^{\rm un}_K(\pd M,\mb t)$ is the image of $\wt\CL_K(M,\mb t_{3d})$ under either the projection $\wt \pi$ or~$\pi$.
\end{itemize}
Taking a union of $\wt\CL_K(M,\mb t_{3d})$ (or $\CL_K(M,\mb t_{3d})$) over finitely many 3d triangulations, or by sufficiently refining the triangulation, we lose the dependence on $\mb t_{3d}$, keeping only the dependence on big-boundary triangulation $\mb t$.

This description of gluing, however, ignores symplectic structures. The spaces $\CP_{\pd \soct_i}$ and $\CX^{\rm un}_K(\pd M,\mb t)$ have natural holomorphic symplectic structures, given by $\Omega_{\pd\soct_i} = d\log z_i\wedge d\log z_i'$  and the  canonical 2-form $\Omega$ on the space of framed flat unipotent connections, respectively. As anticipated in Section \ref{sec:FFC}, we propose that
\begin{conj}[Symplectic Gluing]\label{conj:symp}
The symplectic space $\big(\CX^{\rm un}_K(\pd M,\mb t),\Omega\big)$ is  isomorphic to a holomorphic symplectic quotient of $\big(\prod_{i=1}^N \CP_{\pd\soct_i},\sum_i \Omega_{\pd\soct_i}\big)$ for the Hamiltonian action of the group $(\C^*)^{N-d}$, whose Hamiltonians are  the gluing monomials $\{c_j\}$. Thus
\be \label{sympred} \textstyle \CX^{\rm un}_K(\pd M,\mb t) \,\simeq\, \prod_{i=1}^N \CP_{\pd\soct_i}\big/\!\!\big/(\C^*)^{N-d} = \big[\prod_{i=1}^N \CP_{\pd\soct_i}\cap(c_j=1)\big]\big/(\C^*)^{N-d}\,. \ee
(Sometimes we denote this quotient as $\textstyle \CX^{\rm un}_K(\pd M,\mb t)\simeq \prod_{i=1}^N \CP_{\pd\soct_i}\big/\!\!\big/(c_j=1)$.)
\end{conj}
This is equivalent to saying that there exist monomials $\gamma_j$ in the octahedron parameters (possibly with fractional exponents) such that the holomorphic symplectic forms on $\CX_K(\pd M,\mb t)$ and the octahedron space $\prod_{i=1}^N \CP_{\pd\soct_i}$ are related as
\be \sum_{i=1}^N d\log z_i\wedge d\log z_i'' = \Omega + \sum_j d\log c_j\wedge d\log \gamma_j\,. \label{relateOmega} \ee
In Sections \ref{sec:Bloch-cx} and \ref{sec:K2lift} we consider the Milnor  group $\mathrm{K}_2$ of the fields of functions on $\prod_{i=1}^N \CP_{\pd\soct_i}$ and $\CX^{\rm un}_K(\pd M,\mb t)$, and $\mathrm{K}_2$-avatars $\CW$ 
of the symplectic forms $\Omega$. Relation \eqref{relateOmega} between the symplectic forms is upgraded to a relation between their $\mathrm{K}_2$-avatars:
\be \sum_{i=1}^N  z_i\wedge  z_i'' = \CW  + \sum_j  c_j\wedge  \gamma_j\,. \label{relateK2} \ee

Since the  $\CL(M,\mb t_{3d})$ is  the projection of $\wt\CL(M,\mb t_{3d})= \prod_{i=1}^N \CL_{\soct_i} \cap (c_j=1)$ to $\CX^{\rm un}_K(\pd M,\mb t)$, we  re-interpret $\CL(M,\mb t_{3d})$ as the image of the product Lagrangian $\prod_{i=1}^N \CL_{\soct_i}  \subset \prod_{i=1}^N \CP_{\pd\soct_i}$ under the symplectic quotient \eqref{sympred}. It follows that
\begin{cor}\label{cor:Lag} Conjecture \ref{conj:symp} implies that $\CL(M,\mb t_{3d})$ is a  $\mathrm{K}_2$-Lagrangian, and hence a holomorphic Lagrangian,  subvariety of $\CX^{\rm un}_K(\pd M,\mb t)$. In other words $\CL(M,\mb t_{3d})$ is $d$-dimensional and the restrictions of $\CW$, and hence $\Omega$,   to $\CL(M,\mb t_{3d})$ vanish.
\end{cor}
Since the $\mathrm{K}_2$-class and hence the 
symplectic structures are preserved by 2--3 moves, the refined $\CL(M,\mb t)$ that depends only on big-boundary triangulation is $\mathrm{K}_2$-Lagrangian as well.

In order to quantize $\CL(M,\mb t)$ and to construct the gauge theories $T_K[M]$, we will a slightly generalized version of Conjecture \ref{conj:symp}, involving logarithmic lifts of the various $\C^*$ coordinates. We discuss this in Section \ref{sec:combi}.

We will not prove Conjecture \ref{conj:symp} in full generality in this paper. 
We will, however, prove some special cases, and provide insight into the general case.

In the rest of this section, we assume that the boundary $\pd M$ consists entirely of big boundary, with holes filled in by small discs (no small annuli or tori). In this case, we use the algebraic approach of \cite{G93},  involving a canonical homomorphism from the complex of configurations of decorated flags to the Bloch complex, to prove that $\CL(M,\mb t_{3d})$ is a $\mathrm{K}_2$-isotropic variety of $\CX^{\rm un}_K(\pd M,\mb t)$ --- meaning that the restriction of $\CW$ to $\CL(M,\mb t_{3d})$ vanishes, but that $\CL(M,\mb t_{3d})$ is not necessarily mid-dimensional.    In Section \ref{sec:combi} and Appendix \ref{app:NZ} we will use a combinatorial/topological approach to prove a weaker version of Conjecture \ref{conj:symp} for any admissible $M$, showing that $\CX^{\rm un}_K(\pd M,\mb t)$ is a symplectic reduction of a multiplicative subspace of $\prod_{i=1}^N \CP_{\pd\soct_i}$ --- not necessarily of the entire space; see Proposition \ref{prop:symp} on page \pageref{prop:symp}.%
\footnote{Since the first version of this paper appeared, Garoufalidis and Zickert \cite{GZ-gluing} have adapted the combinatorial approach to prove Conjecture \ref{conj:symp} in the case that $\pd M$ consists entirely of small tori. Similar arguments appeared in \cite{Guilloux-PGL}.}

The canonical homomorphism from the complex of configurations of decorated flags to the Bloch complex, which we are about to review in detail, has the following form:
\be \label{flags2Bloch0}
\begin{array}{ccccccc}
\cdots & \stackrel{\dd}{\lra} &A^{(K)}_4 & \stackrel{\dd}{\lra} 
 &A^{(K)}_3 & \stackrel{\dd}{\lra}  &A^{(K)}_2 \\
&&\downarrow \mbox{\footnotesize $0$} &&\downarrow \mbox{\footnotesize $\alpha_4 $}&&\downarrow \mbox{\footnotesize $\alpha_3 $} \\
\cdots &\lra &0& \lra &B_2(F)& \stackrel{\delta}{\lra} & \wedge^2(F^*)~.
\end{array} 
\ee
Here $\dd$ is the standard simplicial differential. For instance, it maps  a configuration of four flags in $A^{(K)}_3$ (thought of as located at the four vertices of a tetrahedron) to a signed sum over configurations of three flags in $A^{(K)}_2$ (the triangles on the boundary of the tetrahedron).
The map $\delta$ sends an element $\{x\}_2$ of the Bloch group (satisfying the five-term relation) to $(1-x)\wedge x$. 
The maps $\alpha_\bullet$ are easy to describe for $K=2$.
In this case, a flag is just a line and the map $\alpha_4$ is given by the cross-ratio of a configuration of four vectors
\be
\alpha_4 (l_1, l_2, l_3, l_4)  := \frac{\langle l_1\wedge l_2\rangle \langle l_3 \wedge l_4\rangle}{\langle l_2 \wedge l_3\rangle
\langle l_4 \wedge l_1\rangle }~.
\ee
On the other hand, $\alpha_3$ is given by the following element of $\Lambda^2 \C^*$:
\be
\alpha_3(l_1, l_2, l_3):= \langle l_1\wedge l_2\rangle \wedge \langle l_2\wedge l_3\rangle + \langle l_2\wedge l_3\rangle\wedge \langle l_3\wedge l_1\rangle + \langle l_3\wedge l_1\rangle \wedge \langle l_1\wedge l_2\rangle ~.
\ee
(To relate this to \eqref{eq:crossratiox0bc0} or \eqref{crossx1}, set $(l_1,l_2,l_3,l_4)=(a,b,d,c)$.)
The homomorphism of complex $\alpha_\bullet$ for $K>2$ is defined by using the hypersimplicial $K$-decomposition and the related biGrassmannian complex.

If $\pd M$ has just the big boundary, we  use \eqref{flags2Bloch0} to prove that $\CL(M,\mb t_{3d})$ is $\mathrm{K}_2$-isotropic via the following argument. First, the configurations of decorated flags parameterizing generic points of $\wt\CL(M,\mb t_{3d})$ give rise to elements of $A_3^{(K)}$ in \eqref{flags2Bloch0}; while the configurations parametrizing 
generic points of $\CX^{\rm un}_K(\pd M,\mb t)$ relate to elements of $A_2^{(K)}$. The projection of $\wt\CL(M,\mb t_{3d})$ to $\CL(M,\mb t_{3d})\subset \CX^{\rm un}_K(\pd M,\mb t)$ is reflected by the map $d:A_3^{(K)}\to A_2^{(K)}$. Moving to the bottom row of \eqref{flags2Bloch0}, let $F$ be the field of rational functions on $\CX^{\rm un}_K(\pd M,\mb t)$. The group $\mathrm{K}_2(F)$ is the cokernel of $\delta$, \ie\ $\mathrm{K}_2(F) = \wedge^2(F^*)/\delta\big(B_2(F)\big)$.  The $\mathrm{K}_2$-class $\CW$ of the symplectic form on $\CX^{\rm un}_K(\pd M,\mb t)$ was constructed in \cite[Section 15]{FG-Teich}. By the very definition, its pre-image in $\wedge^2(F^*)$ is obtained as follows: we apply the map $\alpha_3$ to the sum, 
over all triangles  in the triangulation ${\bf t}$ of $\pd M$, of the elements of $A_2^{(K)}$ corresponding to the configurations of decorated flags assigned to the triangles. The restriction of $\CW$ to $\CL(M,\mb t_{3d})$ is precisely the image under the composition   $\alpha_3\circ d: A_3^{(K)}\to \wedge^2(F^*)$ 
of the element in $A_3^{(K)}$ describing the generic point of the space $\CX^{\rm un}_K(\pd M,\mb t)$. By the commutativity of the right square in~\eqref{flags2Bloch0}, this can be rewritten as the image under the map $\alpha_3 \circ \dd=\delta \circ \alpha_4$. Since $\mathrm{K}_2(F)$ is the cokernel of $\delta$, this implies that the $\mathrm{K}_2$-class $\CW$ must restrict to zero on $\CL_K(M,\mb t_{3d})$.
Similar considerations also lead to a formula for the differential of the volume in a family of framed flat connections.

Let us now present the details.

%%%%%%%%%
\subsection{From decorated flags to the Bloch complex} \label{sec3.2}

\subsubsection{Decorated flags} \label{sec3.2.2}

Recall that a {\it decorated flag} $\widetilde F^\bullet$ (often called an affine flag) in $V_K$ 
is a flag $F^\bullet$ equipped with additional data: a choice of a non-zero vector $f^i \in F^{i-1}/F^i$ for each $i=1, \ldots, K$. 
 Denote by $ {\cal A}_K$ 
 the space of all decorated flags in $V_K$.   
It is the principal affine space for $GL(K)$. Denote by 
${\rm Conf}^*_{n+1}( {\cal A}_K)$ the space of generic configurations  of $n+1$ decorated flags. 

Denote by ${\rm Conf}^*_m(k)$ 
the space of generic configurations of $m$ vectors in a vector space of dimension $k$. 
Given a partition ${\bf a}$ as in (\ref{pcf}), or, equivalently, a hypersimplex $\Delta^{p,q}_{\bf a}$ 
of the hypersimplicial $K$-decomposition of an $n$-simplex, 
we define a canonical projection
\be
{\cal C}_{\bf a}: {\rm Conf}^*_{n+1}( {\cal A}_K) \lra {\rm Conf}^*_{n+1}({p+1})\,. 
\ee
Namely, take a generic configuration of decorated flags 
$(\widetilde F^\bullet_0, \ldots , \widetilde F^\bullet_n)$ in $V_K$.   
Intersecting the $(p+1)$-dimensional  subspace 
\be
\mathfrak F_{\bf a}:= F^{a_0}_0 \cap \ldots  \cap F^{a_n}_n  
\ee
with the ``previous'' flag subspace in each of the flags, we get a configuration of $n+1$ codimension-one 
subspaces (hyperplanes) in this vector space:  
\be
\mathfrak H_k:= F^{a_k+1}_k\cap \mathfrak F_{\bf a} ~ \subset ~ \mathfrak F_{\bf a}, \qquad k=0, \ldots, n.
\ee
So far we just recovered the configuration of $n+1$ hyperplanes in (\ref{congph}). 
Now we use the decorations to produce a configuration of $n+1$ vectors in $(\mathfrak F_{\bf a})^*$. 
The decoration $f_{a_k}\in F^{a_k}_k/F^{a_k+1}_k$ determines a unique linear functional $l_k$ on 
$F^{a_k}_k/F^{a_k+1}_k$ taking the value $1$ on the vector $f_{a_k}$. It provides 
a linear functional on the $(p+1)$-space 
$\mathfrak F_{\bf a}$ annihilating the hyperplane $\mathfrak H_k$.
Altogether, the linear functionals $l_0, \ldots, l_n \in (\mathfrak F_{\bf a})^*$ provide a configuration 
$(l_0, \ldots , l_n)\in {\rm Conf}^*_{n+1}(p+1)$. 
So we set
\be \label{confvect1}
{\cal C}_{\bf a}(\widetilde F^\bullet_0, \ldots, \widetilde F^\bullet_n):= (l_0, \ldots, l_n).
\ee

\bd \label{confvectors} Given a generic configuration of $n+1$ decorated flags $(\widetilde F^\bullet_{0}, \ldots, \widetilde F^\bullet_{n})$ in $V_K$, 
we assign to a hypersimplex $\Delta^{p,q}_{\bf a}$  the configuration (\ref{confvect1}) of $n+1$ vectors 
in a $(p+1)$-dimensional vector space:
\be \Delta^{p,q}_{\bf a}
 \quad\longmapsto \quad 
{\cal C}_{\bf a}(\widetilde F^\bullet_0, \ldots, \widetilde F^\bullet_n)\; \in\;  {\rm Conf}^*_{n+1}(p+1), \qquad 
{\bf a} = (a_0, \ldots, a_n)~.
\ee
\ed
Its projectivization is the configuration of $n+1$ points 
in a $p$-dimensional projective space from Section~\ref{sec:config-basic}.

\paragraph{Remark:} We use the convention dual to the one in \cite{G93}. 
Notice that a decorated flag in a vector space $V_K$ determines the dual decorated flag in the dual space $V_K^*$.  
To match the two conventions, replace the decorated flags by the dual decorated flags. 
So intersections of flag subspaces here correspond to 
quotients by direct sums of flag subspaces in \cite{G93}.

\subsubsection{BiGrassmannian complex} \label{sec:BiGr}
Our next goal is to relate, when $n$ varies, different configurations of $n+1$ vectors assigned to

\begin{itemize}
\item generic configurations of $(n+1)$ decorated flags in $V_K$, and  
\item the collections of hypersimplices of the $K$-decompositions of an $n$-simplex.
\end{itemize}

Given a configuration $(l_1, \ldots, l_m)$ in $V_n$, we can use two  
operations to get a new configuration 
of $m-1$ vectors:
\begin{enumerate}

\item Forgetting the $i$-th vector $l_i$, we get a map 
\be
f_i: {\rm Conf}^*_m(n) \lra {\rm Conf}^*_{m-1}(n), ~~~~(l_1, \ldots, l_m)\longmapsto 
(l_1, \ldots, \widehat l_i, \ldots, l_m).
\ee

\item Projecting the vectors $(l_1, \ldots, l_m)$ to the quotient $V_n/(l_j)$ along the one-dimensional subspace spanned by $l_j$, we get a map 
\be
p_j: {\rm Conf}^*_m(n) \lra {\rm Conf}^*_{m-1}(n-1), ~~~~(l_1, \ldots, l_m)\longmapsto 
(\overline l_1, \ldots, \widehat l_j, \ldots \overline l_m). 
\ee
Here $\overline l_s$ stands for the projection of the vector $l_s$ to the quotient. 
 \end{enumerate}
Following \cite{G93}, we organize the spaces ${\rm Conf}^*_m(n)$ into a single object, the \emph{biGrassmannian}:  
\be \label{bigrass}
\begin{array}{ccccccccl}
&&&&&&\ldots &\arrowsL &{\bf G}_3^0\\
&&&&&&\arrowsD &&\arrowsD  \\
&&&&\ldots &\arrowsL &{\bf G}_2^1&\arrowsL &{\bf G}_2^0\\
&&&&\arrowsD &&\arrowsD &&\arrowsD\\
&& \ldots &\arrowsL &{\bf G}_1^2 &\arrowsL
 &{\bf G}_1^1 &\arrowsL &{\bf G}_1^0 \\
&& \arrowsD &&\arrowsD&&\arrowsD&&\arrowsD  \\
\hdots &\arrowsL&{\bf G}_0^3&\arrowsL &{\bf G}_0^2&\arrowsL &{\bf G}_0^1&  \arrowsL  & {\bf G}_0^0
\end{array} 
\ee
The name is explained by the canonical isomorphism
\be \label{gale0}
{\rm Conf}^*_{p+q+2}(p+1) = {\bf G}_p^q
\ee
with the Grassmannian ${\bf G}_p^q$ parametrising 
the $(p+1)$-dimensional subspaces in a vector space of dimension $p+q+2$ with 
a given basis $(e_1, \ldots, e_{p+q+2})$, which are in generic position to the coordinate hyperplanes. Namely, a generic 
$(p+1)$-plane $\pi$ gives rise to a configuration of vectors in $\pi^*$ 
given by the restrictions of the coordinate linear functional $x_i$ 
dual to the basis.  
In the diagram (\ref{bigrass}),  given a Grassmannian ${\bf G}_p^q$, there are 
$p+q+2$ horizontal arrows for the maps $f_i$, and $p+q+2$ vertical arrows for the maps $p_j$.  
Notice that the hypersimplex $\Delta^{p,q}$ match the Grassmannian ${\bf G}^q_p$, as the notation suggest. 
The arrows $f_i$ and $p_j$ match the two kinds of boundaries of the hypersimplex $\Delta^{p,q}$ discussed 
in Section \ref{sec:Kdec}. 

There is a canonical isomorphism 
\be \label{gale}
{\bf G}_p^q = {\bf G}_q^p. 
\ee
It assigns to a 
$(p+1)$-dimensional subspace $\pi$ in a coordinate vector space of dimension $p+q+2$ its annihilator $\pi^\perp$ 
 in the dual vector space, equipped with the dual basis.

Given a set $X$, we assign to it a free abelian group $\Z[X]$ with the generators $\{x\}$ parametrized by the elements of $X$. Applying this construction to the sets of complex points of the configuration spaces (identified with Grassmannians), we arrive at abelian groups 
\be
C_m(n):= \Z[{\rm Conf}^*_{m}(n)(\C)].
\ee
They are organized into the {\it Grassmannian bicomplex}  
\be  \label{GrassmBicomplex}
\begin{array}{ccccccccc}
&&&&&&\cdots &\stackrel{f}{\lra} &C_5(4)\\
&&&&&&\downarrow \mbox{\footnotesize $p$} &&\downarrow \mbox{\footnotesize $p$}     \\
&&&&\cdots &\stackrel{f}{\lra} &C_5(3)&\stackrel{f}{\lra} 
&C_4(3)\\
&&&&\downarrow \mbox{\footnotesize $p$} && \downarrow \mbox{\footnotesize $p$} && \downarrow \mbox{\footnotesize $p$} \\
&&\cdots &\stackrel{f}{\lra} &C_5(2) &\stackrel{f}{\lra}
 &C_4(2) &\stackrel{f}{\lra} &C_3(2) \\
&& \downarrow \mbox{\footnotesize $p$}&& \downarrow \mbox{\footnotesize $p$} && \downarrow \mbox{\footnotesize $p$} && \downarrow \mbox{\footnotesize $p$} \\
\cdots &\stackrel{f}{\lra}& C_5(1)&\stackrel{f}{\lra} &C_4(1)&\stackrel{f}{\lra} &C_3(1)& \stackrel{f}{\lra} & C_2(1)
\end{array} 
\ee
Here the maps $f$ and $p$ are the alternating sums of the maps $f_i$, and  $p_j$:
\be
f = \sum_{i=0}^m(-1)^if_i, ~~~~ p  = \sum_{j=0}^m(-1)^jp_j \,.
\ee

Denote by $BC_\ast$ the sum of the groups on the diagonals, shifting the index $m$ by one:
\be
BC_{m-1}:= \bigoplus_{n=1}^{m-1} C_{m}(n)\,.
\ee
As usual, we change the sign of the differentials $f$ in all odd columns, 
and then define the total  differential $D: BC_m \lra BC_{m-1}$ by $D:=f+ p$. 
Then $D^2=0$ (thanks to the sign change). 
We get the {\it biGrassmannian complex}
\be
\cdots \stackrel{D}{\lra} BC_4\stackrel{D}{\lra} BC_3 \stackrel{D}{\lra} BC_2\stackrel{D}{\lra} BC_1\,.
\ee

Finally, consider the complex of generic configurations of decorated flags in $V_K$:
\be
\cdots \lra A^{(K)}_4\lra A^{(K)}_3 \lra A^{(K)}_2\lra A^{(K)}_1, \qquad   A^{(K)}_m:= \Z[{\rm Conf}^*_{m+1}( {\cal A}_K)(\C)]
\ee
with the standard simplicial differential
\be
(\widetilde F_0, \ldots, \widetilde F_n) \lra \sum_{s=0}^n(-1)^s(\widetilde F_0, \ldots, \widehat {\widetilde F_s}, \ldots, 
\widetilde F_n)\,. 
\ee
Let us now recall the main construction of \cite[Section 2]{G93}. 

\bd 
Given a configuration $(\widetilde F_0, \ldots , \widetilde F_n)$ of $n+1$ generic decorated flags in $V_K$,assigned 
to the vertices of the 
$n$-dimensional simplex, 
take the hypersimplicial $K$-decomposition $\Delta^n_K$ of the simplex, assign to \emph{every} hypersimplex 
$\Delta^{p,q}_{\bf a}$ in $\Delta^n_K$ a configuration  
of vectors ${\cal C}_{\bf a}(\widetilde F_0, \ldots, \widetilde F_n)$, thought of as a point of the Grassmannian ${\bf G}^q_p$, and take the sum over all hypersimplices:
\be
c_n: (\widetilde F_0,\ldots, \widetilde F_n) \lra \sum_{{\bf a}}{\cal C}_{\bf a}(\widetilde F_0, \ldots, \widetilde F_n) \in BC_n\,.
\ee
Finally, extend the map to a homomorphism of abelian groups 
\be
c_n: A^{(K)}_n \lra BC_n\,. 
\ee
\ed

The following crucial result was proved in Key Lemma 2.1 from \cite{G93}. 
\begin{theorem}
The collection of maps $c_n$ gives rise to a homomorphism of complexes
\be \label{dfltobigr}
\begin{array}{ccccccccc}
\cdots & \lra &A^{(K)}_4&\lra &A^{(K)}_3& \lra &A^{(K)}_2&\lra &A^{(K)}_1\\
& &\downarrow \mbox{\footnotesize $c_4$} &&\downarrow \mbox{\footnotesize $c_3$}&&\downarrow \mbox{\footnotesize $c_2$}&&\downarrow \mbox{\footnotesize $c_1$}\\
\cdots &\lra &BC_4&\lra &BC_3&\lra &BC_2&\lra &BC_1
\end{array}
\ee
\end{theorem}

{\bf Proof}. We claim that, 
given a configuration of decorated flags $(\widetilde F_0, \ldots, \widetilde F_n)$, 
the map 
\be
{\cal C}: \Delta_{\bf a}^{p,q}\lra {\cal C}_{{\bf a}}(\widetilde F_0, \ldots, \widetilde F_n)
\ee
that assigns to a hypersimplex $\Delta_{\bf a}^{p,q}$ 
the corresponding configuration of vectors intertwines the operation of taking the boundary 
of the hypersimplex 
with the differential in the Grassmannian bicomplex. 
This would immediately imply the theorem since we apply the map ${\cal C}$ to the 
sum of all hypersimplices of the decomposition of a simplex $\Delta^n_K$, 
and so the total boundary is given by the total boundary $d\Delta^n_K$ of the simplex. 
The latter is given by the alternating sum of the $n+1$ boundary $(n-1)$-dimensional simplices, 
which matches the differential in the complex of generic configurations of decorated flags.  

Let us now formulate the claim precisely. 
Recall that a hypersimplex $\Delta^{p,q}$ has two kinds of boundaries, call them $\partial$- and $\partial'$-boundaries, 
one consisting of 
$(n+1)$ hypersimplices of type $\Delta^{p-1,q}$, and another consisting of 
$(n+1)$ hypersimplices of type $\Delta^{p,q-1}$. So, the boundary $d \Delta_{\bf a}^{p,q}$ 
of the hypersimplex $\Delta_{\bf a}^{p,q}$ 
can be written, taking into account the orientations,  as
\be
d [\Delta_{\bf a}^{p,q}] = \sum_{i=0}^n (-1)^i[\Delta_{\partial_i{\bf a}}^{p-1,q}] + \sum_{j=0}^n (-1)^j
[\Delta_{\partial'_j{\bf a}}^{p,q-1}]. 
\ee
Here $\partial_i{\bf a}$ and $\partial'_i{\bf a}$ are the partitions which label the boundary hypersimplices. 
We assert that taking the $\partial_i$-boundary (resp.~$\partial_j'$-boundary) of a hypersimplex $\Delta_{\bf a}^{p,q}$, and then assigning 
to the obtained hypersimplices the configurations of $n$ vectors we get the same result as 
first assigning to the hypersimplex $\Delta_{\bf a}^{p,q}$ the configuration of $(n+1)$ vectors, and then 
applying the map $f_i$  (resp.~$p_j$). Precisely, there are two commutative diagrams: 
\be
\begin{array}{ccccccccccccc}
\Delta_{\bf a}^{p,q}& \stackrel{\partial_i }{\lra} &  \Delta_{\partial_i{\bf a}}^{p-1,q} &&&~~~&&&
&&\Delta_{\bf a}^{p,q}&\stackrel{\partial_j'}{\lra} &  \Delta_{\partial'_j{\bf a}}^{p,q-1}  \vspace{0.3cm}  \\ 
\downarrow \mbox{\footnotesize $\cal C$}  &&\downarrow \mbox{\footnotesize $\cal C$} &&&~~~&&&&&\downarrow \mbox{\footnotesize $\cal C$} &&\downarrow  \mbox{\footnotesize $\cal C$}   \vspace{0.3cm}  \\
{\cal C}_{\bf a}(\widetilde F_0, \ldots, \widetilde F_n)     &\stackrel{ p_i }{\lra} &{\cal C}_{\partial_i{\bf a}}(\widetilde F_0, \ldots, \widetilde F_n)&&& \qquad &&&&
&    {\cal C}_{\bf a}(\widetilde F_0, \ldots, \widetilde F_n)    &\stackrel{ f_j}{\lra} &{\cal C}_{\partial'_j{\bf a}}(\widetilde F_0, \ldots, \widetilde F_n)
\end{array}
\ee
The commutativity of these two diagrams is straightforward to verify, 
and the  theorem follows immediately from this.

\subsubsection{Bloch complex} \label{sec:Bloch-cx}

Let $F$ be a field. Recall first the definition of the {Bloch group} $B_2(F)$ \cite{Sus2}. 
Denote by $\Z[F^*-\{1\}]$ the free abelian group generated by elements $\{z\}$ assigned to $z\in F^*-\{1\}$. 
Let us define a subgroup $R_2(F) \subset \Z[F^*-\{1\}]$ as follows. Take any generic configuration of $5$ points 
$(z_0, \ldots, z_4)$ on ${\mathbb P}^1(F)$. Then the subgroup $R_2(F)$ is generated by the 
five-term relations, where $r$ is the cross-ratio:
\be
\sum_{i=0}^4(-1)^i\{r(z_0, \ldots, \widehat z_i, \ldots, z_4)\}\subset \Z[F^*-\{1\}]\,.
\ee
\bd
The Bloch group $B_2(F)$ is the quotient of the abelian group $\Z[F^*-\{1\}]$ 
by the subgroup of the five-term relations:
\be
B_2(F):= \frac{\Z[F^*-\{1\}]}{R_2(F)}\,.
\ee
\ed
We denote by $\{z\}_2$ the projection of the generator $\{z\}$ to the Bloch group. 

Next, there is a well-known homomorphism of abelian groups
\be
\delta: B_2(F) \longrightarrow F^* \wedge F^*, ~~~~\{x\}_2 \longmapsto (1-x) \wedge x\,. 
\ee
The group $\mathrm{K}_2(F)$ is the cokernel of this map
\be
\mathrm{K}_2(F):= F^* \wedge F^*/\delta (B_2(F))\,. 
\ee

Let us construct a canonical map of complexes \cite[Sections 2.3, 2.4]{G95}\,:
\be \label{bctobloch}
\begin{array}{ccccccc}
\cdots &\lra &{BC}_4 &\lra
 &{BC}_3 &\lra &{BC}_2 \\
&&\downarrow  \mbox{\footnotesize $b_5$} &&\downarrow \mbox{\footnotesize $b_4$}  &&\downarrow \mbox{\footnotesize $b_3$} \\
\cdots &\lra &R_2(\C)& \lra &\Z[F^*-\{1\}]& \stackrel{\delta}{\lra} &\Lambda^2F^*
\end{array} 
\ee
We define it by looking at the Grassmannian bicomplex~(\ref{GrassmBicomplex}), and defining the map row by row. 

\begin{itemize}
\item
First, we send the bottom row $\cdots \to C_3(1)\to C_2(1)$ to zero. 

\item
The map on the next row is the crucial part of the construction. 
It amounts to a construction of the map of complexes 
\be \label{rs}
\begin{array}{ccccccc}
\cdots &\lra &{C}_5(2) &\lra
 &{C}_4(2) &\lra &{C}_3(2) \\
&&\downarrow \mbox{\footnotesize $\gamma_5$} &&\downarrow \mbox{\footnotesize $\gamma_4$} &&\downarrow \mbox{\footnotesize $\gamma_3$} \\
\cdots &\lra &R_2(F)& \lra &\Z[F^*-\{1\}] & \stackrel{\delta}{\lra} &\Lambda^2F^*
\end{array} 
\ee

\begin{itemize}
\item
Given a configuration of vectors $(l_1, l_2, l_3)$ in a two-dimensional space $V_2$, set  
\be
\gamma_3(l_1, l_2, l_3):= \langle l_1\wedge l_2\rangle \wedge \langle l_2\wedge l_3\rangle + \langle l_2\wedge l_3\rangle\wedge \langle l_3\wedge l_1\rangle + \langle l_3\wedge l_1\rangle \wedge \langle l_1\wedge l_2\rangle \in \C^* \wedge \C^*\,.
\ee
Here to define $\langle l_1\wedge l_2\rangle$ we choose a volume form $\Omega \in {\rm det}V_2^*$ and set 
$\langle l_1\wedge l_2\rangle:= ( \Omega,  l_1\wedge l_2)$, as in \eqref{deftrip} or \eqref{eq:crossratiox0bc0}. The invariant $\gamma_3(l_1, l_2, l_3)$ 
does not depend on the choice of volume form $\Omega$. 

\item
The map $\gamma_4$ is given by the cross-ratio: 
\be
\gamma_4 (l_1, l_2, l_3, l_4) = r(l_1, l_2, l_3, l_4):= \frac{\langle l_1\wedge l_2\rangle \langle l_3 \wedge l_4\rangle}{\langle l_2 \wedge l_3\rangle \langle l_4 \wedge l_1\rangle }\,.
\ee

\item
The map $\gamma_5$ assigns to a configuration $(l_1, l_2, l_3, l_4, l_5)$ the corresponding configuration of 
five points on the projective line, which by the very definition is an element of $R_2$. 
\end{itemize}

\item
Finally, we define a map 
$$
C_5(3) \lra R_2
$$
by assigning to a configuration of five vectors in $V_3$ the corresponding 
projective configuration of five points, and then identifying them with five points on a line as in the end of 
\ref{sec:config-basic}. 

\item
All the other maps are set to zero. 

\end{itemize}

To prove that we get a map of complexes one shows (\cite{G95}, Section 2.3) that the right 
square in (\ref{rs}) commutes. The rest is obvious. We arrive at a homomorphism of complexes
\begin{equation} 
\begin{array}{ccccccc}
\ldots &\lra &{BC}_4 &\lra
 &{BC}_3 &\lra &{BC}_2 \\
&&\downarrow  &&\downarrow \mbox{\footnotesize ${\rm b}_4$}  &&\downarrow \mbox{\footnotesize ${\rm b}_3$}\\
\ldots &\lra &0& \lra &B_2(F)& \stackrel{\delta}{\lra} &\Lambda^2F^*
\end{array} 
\ee

\subsubsection{Conclusion: from decorated flags complex to the Bloch complex}
Combining the homomorphism (\ref{dfltobigr}) from the complex of decorated flags to the biGrassmannian complex 
 with the homomorphism  (\ref{bctobloch}) from the biGrassmannian complex to the Bloch complex, we arrive at 
our final output ---  a homomorphism $\alpha_\bullet$ from the complex of decorated flags in $V_K$ to the Bloch complex:   
\be \label{dftobc}
\begin{array}{ccccccc}
\cdots &\lra &A^{(K)}_4 &\lra
 &A^{(K)}_3 &\lra &A^{(K)}_2 \\
&&\downarrow \mbox{\footnotesize $\alpha_5 $}  &&\downarrow \mbox{\footnotesize $\alpha_4 $}&&\downarrow \mbox{\footnotesize $\alpha_3$}\\
\cdots &\lra &R_2(F)& \lra &\Z[F^*-\{1\}]& \stackrel{\delta}{\lra} &\Lambda^2F^*
\end{array} 
\ee
Of course, the complex $R_2(F) \lra \Z[F^*-\{1\}] \stackrel{\delta}{\lra} \Lambda^2F^*$ is quasi-isomorphic to the Bloch complex, so we can cast the map (\ref{dftobc}) in a more traditional form, where we abuse notation slightly by 
denoting the middle map by $\alpha_4$ again: 
\be \label{dftobc1}
\begin{array}{ccccccc}
\cdots &\lra &A^{(K)}_4 &\lra
 &A^{(K)}_3 &\lra &A^{(K)}_2 \\
&&\downarrow \mbox{\footnotesize $0$} &&\downarrow \mbox{\footnotesize $\alpha_4 $}&&\downarrow \mbox{\footnotesize $\alpha_3 $} \\
\cdots &\lra &0& \lra &B_2(F)& \stackrel{\delta}{\lra} &\Lambda^2F^*
\end{array} 
\ee
This resulting map of complexes is the main ingredient 
of the construction of the second motivic Chern class in \cite{G93}, which is a class 
\be \label{mschc}
C_2^{\cal M} \in H^4(BGL_K(\C), \Z_{\cal M}(2))\,.
\ee

In Section \ref{sec:K2lift}--\ref{sec3.3} we show that the $\mathrm{K}_2$-Lagrangian property and the variation formula for the volume 
of a framed flat $PGL(K,\C)$-connection 
on an admissible 3-manifold in the absence of the small toric and annular boundaries 
follow immediately from these homological considerations. 

\subsection{${\cal L}_K(M)$ is a $\mathrm{K}_2$-isotropic subvariety}
\label{sec:K2lift}

The symplectic structure on the moduli space ${\cal X}^{\rm un}_K(\CC)$ 
of unipotent framed flat $PGL(K,\C)$-connections on a punctured surface $\CC$ 
is given by a 2-form $\Omega$,  the  Atiyah-Bott-Goldman form. The form $\Omega$ 
has been identified in \cite[Section 15]{FG-Teich} as the image under the $d\log \wedge d\log$ map 
of a class ${\cal W}$ in $\mathrm{K}_2$ of the space 
 ${\cal X}^{\rm un}_K(\CC)$, called the {\it motivic avatar of the symplectic form}:
\be
\Omega = (d\log \wedge d\log)({\cal W})\,.
\ee
Recall that $\mathrm{K}_2(F)$ is a quotient of $\Lambda^2F^*$. 
To define the class ${\cal W}$ in $\mathrm{K}_2$, a $\Lambda^2F^*$-representative was assigned  to every ideal triangulation 
${\bf t}$ of $\CC$ in~\cite{FG-Teich},
\be
W_K(\CC; {\bf t})\in \C\Bigl({\cal X}^{\rm un}_K(\CC)\Bigr)^* \wedge 
\C\Bigl({\cal X}^{\rm un}_K(\CC)\Bigr)^*\,. 
\ee
Its image in $\mathrm{K}_2$ does not depend on the choice of triangulation, and defines the class ${\cal W}$. 
Specialising it to  a generic unipotent framed flat $PGL(K,\C)$-connection ${\cal V}_{\CC}$ on ${\CC}$ we get an element
\be
W_K({\cal V}_{\CC}; {\bf t})\in \C^*\wedge \C^*\,.
\ee
Let us recall the definition of $W_K({\cal V}_{\CC}; {\bf t})$. 
Pick an invariant 
decorated flag $\widetilde F_v$ near each puncture $v$ on $\CC$. So each triangle $t$ 
of the ideal triangulation ${\bf t}$  with the vertices $v_1(t), v_2(t), v_3(t)$ gives rise to  
a configuration of three decorated flags 
$(\widetilde F_{v_1(t)},\widetilde F_{v_2(t)}, \widetilde F_{v_3(t)})$. 
We get the element $W_K({\cal V}_{\CC}; {\bf t})$ by applying the map $\alpha_3$ from (\ref{dftobc}) to it, and taking the sum over all triangles: 
\be \label{Wclass}
W_K({\cal V}_{\CC}; {\bf t}):= \sum_{t \in {\bf t}}\alpha_3(\widetilde F_{v_1(t)},\widetilde F_{v_2(t)}, \widetilde F_{v_3(t)})\,.
\ee
(The definition given in \cite[Section 15]{FG-Teich} is equivalent to this one, and in fact 
was originally obtained by calculating \eqref{Wclass}.)

Assume that $M$ is an admissible 3-manifold without small torus or annulus components. 
Taking the sum over all big boundary components $\CC_i$ of $\pd M$, we set 
$$
W_K(\pd M; {\bf t}):= \sum_iW_K(\CC_i; {\bf t}). 
$$

Now let us run a similar procedure on an admissible 3-manifold $M$ with an ideal triangulation ${\bf t}_{3d}$. 
The moduli spaces $\CL_K(M)$ and $\widetilde \CL_K(M)$ may have several irreducible components. 

\bd
A framed flat $PGL(K,\C)$-connection ${\cal V}$ on $M$  is generic for a triangulation ${\bf t}_{3d}$ if its restriction to any 
 tetrahedron $\Delta$ of the triangulation ${\bf t}_{3d}$ provides a generic configuration of four flags 
$(F_{1}(\Delta), \ldots, F_{4}(\Delta))$. 

The ${\cal V}$ is generic if there exists a triangulation ${\bf t}_{3d}$ of $M$ for which it is generic. 

A  component of   $\widetilde \CL_K(M)$ is  generic if it contains
 a generic framed flat connection.
 \ed
 
  Being generic is an open condition: framed flat $PGL(K,\C)$-connections close to a generic one are also generic. 
 This is why it is enough to have a single generic one in a component.  
 
  \paragraph{Examples.}  1. Let $M$ be a polyhedron. So it is a ball whose boundary is a sphere, punctured in  $n$ points. Then the moduli space 
    $\widetilde \CL_K(M)$ has a single irreducible component, and this component is generic. Indeed, by the very definition it parametrizes 
  arbitrary configurations of $n$ flags in a $K$-dimensional vector space. And  evidently $\widetilde \CL_K(M) = \CL_K(M)$.

  2. Let $M$ be a hyperbolic threefold with big boundary. Then it has an Epstein-Penner triangulation into ideal tetrahedra. 
  Notice that a configuration of four points on the projective line is generic if all the points are distinct. So an ideal tetrahedron with distinct vertices, even a flat one, 
  provides a generic configuration of points on ${\Bbb C}P^1$. Therefore the framed flat $PGL(2, \C)$-bundle describing the hyperbolic structure is generic. 
  So the component of $\widetilde \CL_K(M)$  containing the hyperbolic structures on $M$ is generic. 
  We call it the hyperbolic component. 
  
  3. Take the standard embedding $PGL(2) \lra PGL(K)$ induced by the linear representation  of $SL(2)$ in polynomials in two variables of degree $K-1$. 
  Then any hyperbolic threefold $M$ gives rise to a framed flat $PGL(K, \C)$-bundle on $M$: it is obtained by taking the framed flat $PGL(2, \C)$-connection on $M$ 
  assigned to the hyperbolic structure, and using the standard embedding above. The component of  $\widetilde \CL_K(M)$ 
  containing this representation is generic.   Here is a proof. A flag in a two-dimensional vector space $V_2$, that is
   a linear subspace $L \subset V_2$, determines a flag $F_L$ in $S^{K-1}V_2$, 
   given by the nested subspaces $\{L^mS^{K-m-1}(V_2)\}$, $m=K-1, ..., 0$.   It is the unique flag fixed by the Borel subgroup 
   of $PGL_2$ stabilizing the $L$.   Then four distinct subspaces $(L_1, ..., L_4)$ give rise to a generic configuration of four 
   flags $(F_{L_1}, ..., F_{L_4})$.  To see this, one can argue geometrically. 
   For example, if $K=3$, consider a conic in $P^2 = P(S^2(V_2))$ given by the projectivisation of the set 
   of decomposable  vectors $v^2$. Take a  point $l$ on the conic. Then the flag given by the point $l$ and the tangent line to 
  the conic at the point $l$ is the flag $F_L$. Evidently, given four distinct points $(l_1, ..., l_4)$ on the conic, the associated configuration of flags  
  $(F_{L_1}, ..., F_{L_4})$ is generic.

     \vskip 3mm
Take a {\it generic} framed flat $PGL(K,\C)$-connection ${\cal V}$ on $M$. Given a tetrahedron $\Delta$ from a triangulation ${\bf t}_{3d}$ of $M$, and 
 applying the map $\alpha_4$ to the configuration of four flags assigned to $\Delta$, we get 
\be \label{octp}
\alpha_4(F_1(\Delta), \ldots, F_4(\Delta))= \sum_{\bf a}\{z_{\bf a}\}_2 \in B_2(\C)\,. 
\ee
Here $\{z_{\bf a}\}$ are the octahedron parameters. Taking the sum over all tetrahedra $\Delta_i$ of the triangulation, 
we get a $B_2(\C)$-valued invariant: 
\be \label{motvol1}
\alpha_4({\cal V}; {\bf t}_{3d}):= \sum_i \alpha_4(F_{1}(\Delta_i), \ldots, F_{4}(\Delta_i)) \in B_2(\C)\,.
\ee

\begin{theorem}\label{thm:K2}
Let $M$ be an admissible 3-fold without small torus or annulus components. 
Let ${\cal V}\in \widetilde {\cal L}_K(M)$ be a generic framed flat $PGL_K(\C)$-connection on $M$, and 
${\rm Res}_{\pd M}({\cal V})$ its restriction to the boundary.

i) If a triangulation ${\bf t}$ of $\pd M$ is the boundary of a triangulation ${\bf t}_{3d}$ of ${M}$, then 
\be \label{formulaW}
W_K({\rm Res}_{\pd M}({\cal V}); {\bf t}) = \delta\circ \alpha_4({\cal V}; {\bf t}_{3d}). 
\ee
So the restriction of the $\mathrm{K}_2$-class $W_K(\pd M)$ to the component of ${\cal L}_K(M)$ containing ${\cal V}$ is zero.
 Therefore the component of ${\cal L}_K(M)$ containing ${\cal V}$ is $\mathrm{K}_2$-isotropic. 

ii) \textit{If $M$ is a convex polyhedron, then $\CL_K(M)$ is a $\mathrm{K}_2$-Lagrangian subvariety of $\CX^{\rm un}_K(\pd M)$.}

iii) The element $\alpha_4({\cal V}; {\bf t}_{3d}) \in B_2(\C)$ 
does not depend  on  the choice of a triangulation  ${\bf t}_{3d}$, although it does depend 
on  the boundary triangulation ${\bf t}$. 
Therefore we get  an invariant of $({\cal V}, {\bf t})$, which we call the motivic volume invariant: \
\be \label{motvol}
\alpha_4({\cal V}; {\bf t})  \in B_2(\C).
\ee
\end{theorem}

 \paragraph{Proof:} 
i) The second claim  follows from the first, since $\mathrm{K}_2$ is the cokernel of $\delta$. The first 
 follows from the commutativity of the right square in (\ref{dftobc}). 
Indeed, calculating $\delta \circ \alpha_4({\cal V})$ we pick a tetrahedron of the ideal triangulation 
of $M$, and calculate the map $\alpha_4$ followed by $\delta$ for the corresponding 
configuration of four decorated flags. This means taking these four flags and going down and to the right 
in (\ref{dftobc}). Thanks to the commutativity of the right square in (\ref{dftobc}), 
we get the same result by applying the differential to the configuration of four decorated flags (going to the right in (\ref{dftobc})), and then apply the map $\alpha_3$ (going down). After taking the sum 
over all tetrahedra of the ideal triangulation of $M$, the contributions of the 
internal big faces of the tetrahedra cancel pairwise, and the contribution of the other faces 
match the definition of $W_K({\rm Res}_{\pd M}({\cal V}); {\bf t})$ given by (\ref{Wclass}).

  ii) This was proved right after formulation of the Theorem \ref{thm:K2} in the Introduction. 
  
iii) It is clear from (\ref{dftobc}) that the element $\alpha_4({\cal V}; {\bf t}_{3d}) \in B_2(\C)$ 
does not change under the 2--3 moves. Indeed, the change is described by the composition 
$A^{(K)}_4 \lra A_3^{(K)} \lra B_2(\C)$, which is zero by (\ref{dftobc1}). 
So it does not depend on the choice of the 3d triangulation ${\bf t}_{3d}$ of $M$. 
 
The theorem is proved.

\paragraph{Small boundaries:}
Here is how to extend formula (\ref{formulaW}) when $M$ has small toric boundary components. 
On each of the toric components $T^2_i$ of the boundary, the symplectic 
form is naturally lifted to a $\Lambda^2\C^*$-invariant $W_K(T^2_i)$. Namely, 
choose a pair of generators $A, B$  of $H_1(T^2, \Z)$ with intersection number $\langle A, B\rangle=1$.
Present the diagonal parts of the holonomies 
of a flat $PGL(K,\C)$-connection ${\cal V}$ on the torus around the A- and B-cycles as in (\ref{ABholonomies}), 
using  the values  $\ell_a$, $m_b$, where $a,b=1, \ldots K-1$, of the 
positive simple roots on the Cartan group. 
Then 
\be \label{torusW}
W_K(T^2)({\cal V}):= \sum_{a,b=1}^{K-1} (\kappa^{-1})_{ab}\,
\ell_a \wedge m_b \in \C^*\wedge \C^*\,. 
\ee
Here $\kappa_{ab}$ 
is the Cartan matrix. See Sections \ref{sec:torusbdy} and \ref{sec:sympsum}, where we elaborate on the Poisson structure 
for the symplectic form defined by this class.

The Poisson brackets for eigenvalues derived combinatorially in Appendix \ref{app:NZ} are almost sufficient to imply \eqref{formulaW} for toric boundary. 
See Section \ref{sec:sympsum} for further remarks on this. Formula \eqref{formulaW} can also be generalized to admissible 3-manifolds with small annular boundaries (or, equivalently, laminations), but this will be discussed elsewhere (\cf\ \cite{DGV-hybrid} for $K=2$).

\paragraph{Remark:} Another way to get Theorem \ref{thm:K2} is to use the result of \cite[Section 10]{FG-Teich} that a flip 
of a 2d triangulation on a surface $\CC$ amounts to a cluster transformation on the moduli space 
${\cal A}_{SL(K), \CC}$. Indeed, it was shown in \cite[Section 6]{FG-cluster},  that 
any cluster transformation preserves the class ${\cal W}_{\cal A}$ in $\mathrm{K}_2$ of a cluster 
${\cal A}$-variety, and that the difference of the $\Lambda^2\C^*$-invariants 
${\cal W}_{\cal A}({\bf s}_1) - {\cal W}_{\cal A}({\bf s}_2)$ assigned to two seeds 
${\bf s}_1$ and ${\bf s}_2$  related by a cluster transformation is 
$$
{\cal W}_{\cal A}({\bf s}_1) - {\cal W}_{\cal A}({\bf s}_2) = \sum_{i=1}^n(1-z_i)\wedge z_i.
$$
Here the cluster transformation ${\bf s}_1 \to {\bf s}_2$ is written as a composition of $n$ mutations, 
and the parameters $z_1, \ldots, z_n$ are the ${\cal X}$-coordinates related to these mutations. 
In the case of a flip of triangulation, these are the octahedron parameters $\{z_{\bf a}\}$ in (\ref{octp}). 
These two approaches have the same 
origin: the homomorphism of complexes (\ref{bctobloch}), which was in fact used in \cite{FG-Teich} 
to obtain the cluster picture just outlined. However, the 
map $\alpha_4$ has $S_4$ skew-symmetry built in its definition. 
To see it by decomposing a flip into a composition of mutations is 
a non-trivial problem.

%%%%%%%%%%%%%%%%%%%%
\subsection{Differential of the volume of a framed flat 
$PGL(K,\C)$-connection}\label{sec3.3}

Let $G$ be a complex semi-simple Lie group. Let $M$ be a closed manifold
 and ${\cal V}$ a $G$-bundle with flat connection  on $M$.  
The flat $G$-bundle is described by a map
to the classifying space $BG$ of the group $G$ made discrete:
\be
\varphi_{\cal V}: M \lra BG^\delta\,. 
\ee
It is well defined modulo homotopy. 
It maps the fundamental class $[M]$ of $M$ to a class $\varphi_{\cal V}[M] \in H_3(BG^\delta, \Z)$. 
Combined with the regulator map 
\be
r_2: H_3(BG^\delta, \Z) \lra \C/(2\pi i)^2\Z
\ee
we get an invariant of a $G$-bundle with flat connection ${\cal V}$ on $M$:
\be
c_2({\cal V}):= r_2(\varphi_{\cal V}[M])\in \C/(2\pi i)^2\Z\,.
\ee 
In particular, taking the imaginary part of $c_2({\cal V})$ we recover the volume invariant of ${\cal V}$.
The name is suggested by the representations to $PGL(2,\C)$ provided by  
hyperbolic 3-manifolds $M$, when it is the hyperbolic volume. All this is well-known. 

Now let $M$ have a boundary. 
Then, although the fundamental class of $M$ does not exist, for a unipotent $G$-vector 
bundle with flat connection ${\cal V}$ 
on $M$ one can still define an invariant 
\be
c_2({\cal V})\in \C/(2\pi i)^2\Z
\ee
using the ``relative fundamental class'' construction from \cite[Section 3]{G96}, or \cite{Zickert-rep}.  

For generic \emph{non-unipotent} local systems on an admissible 3-manifold $M$, the $B_2(\C)$-valued 
invariant (\ref{motvol})  is a motivic refinement of the volume invariant. 
In fact it amounts to the pullback of the explicit construction of the second motivic Chern class \eqref{mschc}.

To recover the $\R$-valued volume one applies to (\ref{motvol}) the Bloch-Wigner dilogarithm 
\be \label{defBW}
{\cal L}i_2(z):= {\rm Im}\big(\Li_2(z) + \log(1-z)\log|z|\big)\,. 
\ee
It satisfies the five-term relation, which just means that it provides a homomorphism
\be
{\cal L}i_2: B_2(\C) \longmapsto \R, ~~~ \{z\}_2 \longmapsto {\cal L}i_2(z)\,. 
\ee
\vskip 2mm
Notice that even in the simplest possible case when $M$ is the tetrahedron and 
thus the flat connection itself is trivial, the volume is a nontrivial 
invariant. In this particular case it is nothing else but the measurable $3$-cocycle 
of $PGL(K, \C)$ defined in \cite{G93b}: given a flag $F$, 
\be
(g_1, \ldots, g_4) \in PGL(K, \C)^4 \longmapsto {\cal L}i_2(\alpha_4(g_1 F, \ldots, g_4 F))\,.
\ee
\vskip 2mm

Let us show now how formula (\ref{formulaW}) leads to a variation of the volume formula. 

Given a family ${\cal V}(t)$ of framed flat $PGL(K,\C)$-connections 
on an admissible 3-manifold $M$, depending on a parameter $t$, 
 we would like to calculate the differential of the $PGL(K,\C)$-volume ${\rm Vol}_K$ 
 in this family.

Let $X$ be a complex algebraic variety. There is a homomorphism to the space ${\cal A}^1(X)$ of real valued $1$-forms, which are smooth at the generic point of $X$:
\be
\log d\arg: \C(X)^* \wedge \C(X)^* \longrightarrow {\cal A}^1(X), ~~~~ f\wedge g 
\longmapsto \log |f|\, d\arg g - \log |g|\, d\arg f\,.
\ee
The differential of the Bloch-Wigner dilogarithm can be written as a composition of the map $\delta$ with the map 
$\log d\arg$:
\be
d{\cal L}i_2(z):= \log |1-z| d\arg z - \log |z| d\arg (1-z) = \log d\arg \circ ~\delta\{z\}_2\,.
\ee
This shows that to calculate $\sum_id{\cal L}i_2(z_i)$ one needs to calculate 
$\delta \sum_i\{z_i\}_2 =\sum_i (1-z_i) \wedge z_i$.

Let us return to the calculation of the differential of the volume. 
Assume that $M$ is as in Theorem \ref{thm:K2}. 
Choose an ideal  triangulation ${\bf t}$ of the big boundary.
As explained above, 
\be
d{\rm Vol}_K({\cal V}(t); {\bf t})    = d {\cal L}i_2 (\alpha_4({\cal V}(t); {\bf t})) = \log d\arg \circ ~\delta \circ\alpha_4 ({\cal V}(t); {\bf t})\,. 
\ee 
Applying the $\log d\arg$ map to the right hand side of (\ref{formulaW}), 
we get the differential of the volume:
\be
d{\rm Vol}_K({\cal V}(t); {\bf t})    = \log d\arg \Bigl(W_K({\rm Res}_{\pd M}({\cal V}(t)); {\bf t})\Bigr)\,. 
\ee

So to calculate the differential $d {\rm Vol}_K({\cal V}(t))$, 
we restrict the family   ${\cal V}(t)$ to the boundary $\pd M$, and then 
evaluate the universal $1$-form $\log d\arg \Bigl(W_K({\rm Res}_{\pd M}({\cal V}(t)); {\bf t})\Bigr)$ 
on the resulting family of the framed flat connections 
on the boundary. 

We conclude that although the motivic volume of a framed flat $PGL(K, \C)$-connection on $M$ is determined by 
the connection on $M$, the differential of the volume is calculated using the class ${\cal W}$ only. 

We emphasize that the motivic volume invariant does depend on the choice of the boundary triangulation ${\bf t}$ of $M$. 
In some situations there is a preferred triangulation, e.g. in the case when $M$ is a hyperbolic manifold with big geodesic boundary. 
So the motivic volume of a generic framed flat connection on a hyperbolic manifold is an invariant.

%%%%%%%%%%%%%%%%%%%%%%%%%%%%%%%%%%%%%%%%%%%%%%%%%%%%%%%%%%%%%%%%%%%%%%%%%%%%%
%%%%%%%%%%%%%%%%%%%%%%%%%%%%%%%%%%%%%%%%%%%%%%%%%%%%%%%%%%%%%%%%%%%%%%%%%%%%%
%%%%%%%%%%%%%%%%%%%%%%%%%%%%%%%%%%%%%%%%%%%%%%%%%%%%%%%%%%%%%%%%%%%%%%%%%%%%%

\section{Combinatorics of $K$-decompositions}
\label{sec:combi}

In this section we introduce a complementary graphical approach for describing the combinatorics of coordinates for framed flat $PGL(K)$-connections that were developed in Section~\ref{sec:FFC}. We will use an algebra of paths on slices in the $K$-decomposition of a 3-manifold $M$ to encode the relationship between octahedron parameters on one hand, and gluing functions and boundary coordinates on the other. This generalizes the work of \cite{Neumann-combinatorics} in the context of hyperbolic geometry. The approach is especially useful for describing holonomy eigenvalues around small torus boundaries and analyzing their Poisson brackets. In turn, the Poisson brackets lead to a proof of a weak version of the Symplectic Gluing Conjecture \ref{conj:symp} (Proposition \ref{prop:symp}, page \pageref{prop:symp}). It seems promising that with additional effort the combinatorial analysis here could be upgraded to produce a full proof of the Conjecture.

The algebra of paths also has a second important advantage: it allows us define a consistent logarithmic lift of all the phase space coordinates, promoting them to additive $\C$-valued (rather than $\C^*$-valued) functions.
Using $\C^*$-coordinates as in Section \ref{sec:FFC}, we know that all gluing functions $c_k$ and boundary coordinates are monomials in the octahedron parameters $z_i,z_i',z_i''$. Assuming Conjecture \ref{conj:symp}, we could encode the exponents of these monomials in a symplectic matrix $g$, which holds all the necessary information for expressing a patch of $\CX_K^{\rm un}(\pd M)$ as a symplectic reduction of the product $\CP_\times$ of octahedron phase spaces, as well as reproducing the Lagrangian submanifold $\CL_K(M)\subset \CX_K^{\rm un}(\pd M)$. However, this information is not sufficient for quantization or for defining theories $T_K[M]$. A consistent quantization (\eg\ in the sense of \cite{Dimofte-QRS}) requires keeping track of an extra vector of integers $\sigma$ that translate to $q$-corrections in the quantum Lagrangian operators $\hat \CL_K(M)$, as well as in any quantum wavefunctions. In terms of 3d $\CN=2$ theories $T_K[M]$, the extra vector of integers encodes a consistent UV R-charge assignment, which is necessary for compactification of $T_K[M]$ on any curved backgrounds. The formal algebra of path coordinates in this section is used to specify, unambiguously, the vector $\sigma$.

The consistent definition of path coordinates fundamentally requires all small boundary components of $M$ to admit a Euclidean 2d structure. Indeed, the path coordinates will look like Euclidean angle structures on slices of $M$ parallel to the small boundary. This consideration motivated our original definition of admissible manifolds, back in Section \ref{sec:admissible}.

Ultimately, we will also want to choose polarizations $\Pi$ and $\Pi_\times$ for the logarithmic lifts of phase spaces $\CX_K^{\rm un}({\pd M})$ and the product octahedron space $\CP_\times$. Then the gluing data $g$ and $\sigma$ should be combined into an affine symplectic transformation $(g,\sigma)\in ISp(2N,\Z)$, or sometimes $ISp(2N,\Q)$ (where $N$ is the number of octahedra), which relates the two polarizations and is the fundamental piece of information needed to define $T_K[M]$ and any quantum wavefunctions associated with $M$.

Much of this section can be read without knowledge of the framed flat connections of Sections \ref{sec:basics}--\ref{sec:FFC}. In this sense it is a combinatorial guide to producing the data for quantization and for theories $T_K[M]$. \\

In this section, we will suppress the dependence of various moduli spaces (coordinate phase spaces and Lagrangians) on 2d triangulations $\mb t$ and 3d triangulations $\mb t_{\rm 3d}$.
We assume throughout that $\mb t$ and $\mb t_{3d}$ are fixed. We use the notation $\CP_K(\pd M)$ to denote the logarithmic version of coordinate phase spaces $\CX_K^{\rm un}(\pd M,\mb t)$, and use uppercase variables (\eg\ $Z, Z',X,\ldots$) to denote the logarithmic $\C$-valued lifts of the corresponding $\C^*$-valued parameters (\eg\ $z=\exp Z,\,z'=\exp Z',\,x=\exp X$,\ldots). (We abuse notation, continuing to use $\CP_{\pd\soct}$ for the canonical logarithmic lifts of octahedron phase spaces; we also will not distinguish between Lagrangian submanifolds $\CL_K(M,\mb t_{\rm 3d})\subset \CX_K^{\rm un}(\pd M,\mb t)$ and their preimages in $\CP_K(\pd M)$.)

\subsection{Octahedra and slices}
\label{sec:octslice}

Recall that an admissible 3-manifold $M$ can be decomposed into ideal tetrahedra, each of which has a further $K$-decomposition. In particular, the $K$-decomposition cuts an ideal tetrahedron into ${K+1\choose 3}=\frac16 K(K^2-1)$ octahedra $\oct_i$, which are our central players.

\begin{figure}[htb]
\centering
\includegraphics[width=5in]{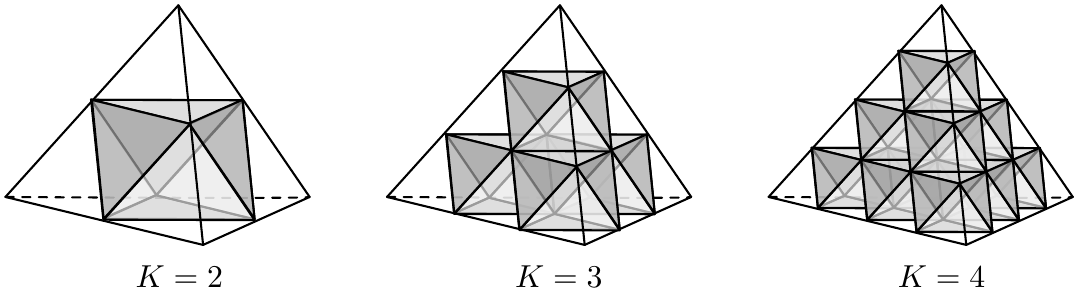}
\caption{$K$-decompositions of an ideal tetrahedron, for $K=2,3,4$.}
\label{fig:oct-K234}
\end{figure}

\begin{wrapfigure}[8]{l}{1.2in}
\centering
\includegraphics[width=1.1in]{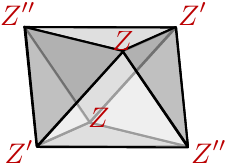}
\caption{Labeling of vertices.}
\label{fig:octZ}
\end{wrapfigure}

Each octahedron is oriented and its faces can be canonically assigned two distinct colors, according to whether they align with big faces of a tetrahedron (black) or with the truncations of vertices (white).
The colors alternate on neighboring faces. We label the six vertices of every octahedron with formal complex parameters $Z_i,Z_i',Z_i''$, equal on opposite vertices and occurring in a clockwise order around every black face. (Thus there are three distinct ways to label any octahedron, related by a cyclic permutation $Z_i\to Z_i'\to Z_i''\to Z_i$.)

\begin{figure}[htb]
\centering
\includegraphics[width=6in]{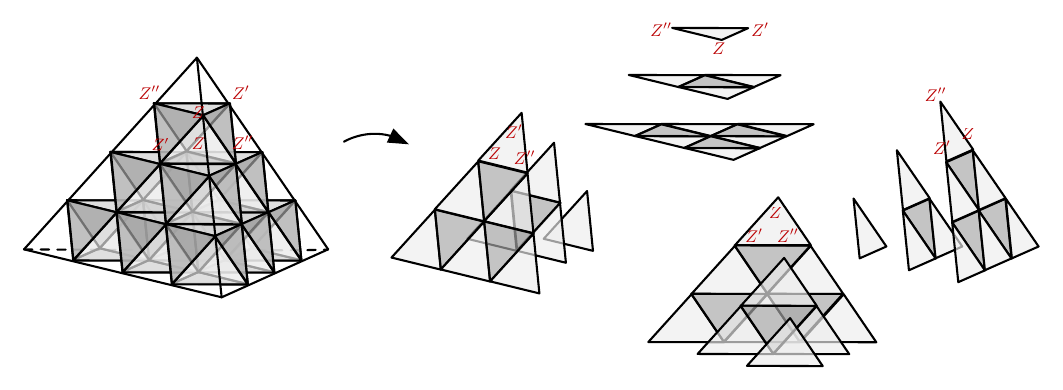}
\caption{Four parallel sets of slices in the $K=4$ decomposition.}
\label{fig:slices}
\end{figure}

Notice that the decomposition into octahedra naturally identifies four families of $K-1$ parallel \emph{slices} of every tetrahedron $\Delta$, as shown in Figure \ref{fig:slices}. Each family of slices is centered around one of the four vertices (a small boundary component). Moreover, every slice is densely tiled by black and white faces of octahedra: white faces from octahedra below the slice and black faces from octahedra above the slice. The three dihedral angles of these small black and white triangles can all be labeled by the vertex parameters $Z_i,Z_i',Z_i''$ of the octahedra that they came from. We have attempted to indicate this in Figure \ref{fig:slices}, for the four triangles associated with the top octahedron.

\begin{figure}[htb]
\centering
\includegraphics[width=5.5in]{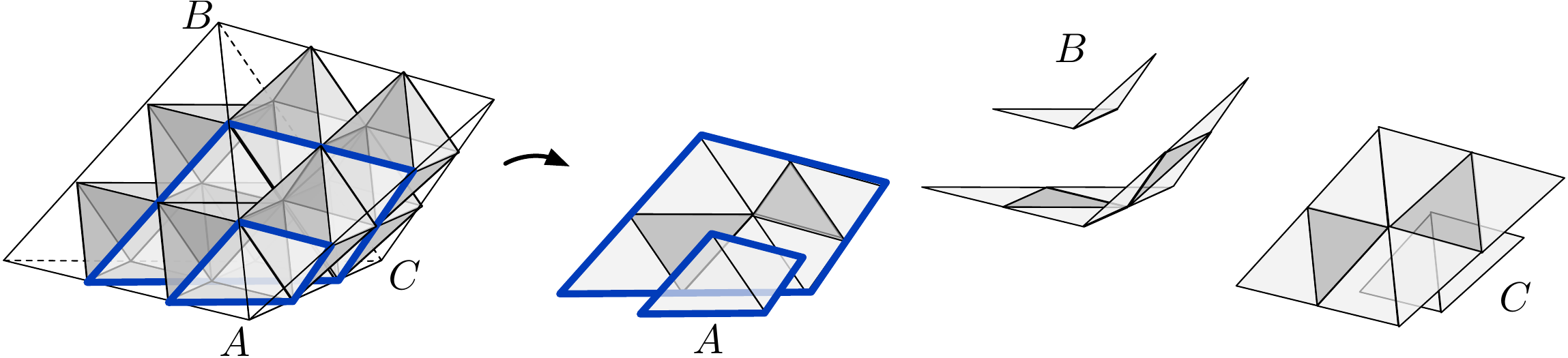}
\caption{Continuing slices from one tetrahedron to another; the sets of slices on the right here are associated with vertices $A$, $B$, and $C$, respectively.}
\label{fig:slice-continue}
\end{figure}

When ideal tetrahedra are glued together to form $M$, the slices can be continued from one tetrahedron to the next, as indicated in Figure \ref{fig:slice-continue}. Indeed, every small boundary component of $\pd M$ acquires a family of $K-1$ global slices throughout $M$ centered around that small boundary and having the same topology. For example, a small torus boundary comes with $K-1$ toroidal slices in the $K$-decomposition.

\subsection{Coordinates on the product phase space}
\label{sec:path-coords}

We now want to revisit the symplectic gluing of elementary Lagrangian pairs for octahedra to form the Lagrangian pair for an admissible 3-manifold $M$. We will assume that the boundary $\pd M$ does not contain small annuli (\ie\ it only has small tori and the small discs that fill in holes on the big boundary). The extension to small annuli follows in a straightforward way be combining results of \cite{DGV-hybrid} with the present analysis.

Suppose that $M$ is triangulated into $N$ ideal tetrahedra, so that the $K$-decomposition has $\frac 16 NK(K^2-1)$ total octahedra. To each octahedron $\oct_i$ we associate a formal logarithmic phase space
\be \label{Poct}
 \CP_{\pd \soct_i} = \big\{(Z_i,Z_i',Z_i'')\,\big|\,Z_i+Z_i'+Z_i''=i\pi\big\}\,, \ee
with holomorphic symplectic form $\Omega_{\pd\soct_i}=dZ_i\wedge dZ_i'$, or equivalently with Poisson bracket
\be \{Z_i,Z_i'\}=\{Z_i',Z_i''\}=\{Z_i'',Z_i\}=1\,.\ee
These coordinates $Z_i,Z_i',Z_i''$ are the octahedron vertex parameters, logarithmic lifts of the cross-ratios of Section \ref{sec:bulkflat}. We also associate with $\oct_i$ the canonical Lagrangian submanifold
\be \label{Loct}
 \CL_{\soct_i}\,=\,\{z_i''+z_i^{-1}-1=0\}\;\subset\,\CP_{\pd\soct_i}\,,\ee
with $z_i=\exp Z_i$, etc.;
and we recall that both the phase space and Lagrangian are invariant under cyclic permutations $Z_i\to Z_i'\to Z_i''\to Z_i$.

For the entire manifold $M$, we build a product phase space
\be \CP_{\times} := \prod_{i=1}^{\frac16 NK(K^2-1)} \CP_{\pd \soct_i}\,,\ee
with independent Poisson brackets for $(Z_i,Z_i',Z_i'')$ and $(Z_j,Z_j',Z_j'')$ when $i\neq j$. This space contains the canonical product Lagrangian
\be \CL_\times := \prod_{i=1}^{\frac16 NK(K^2-1)} \CL_{\soct_i}\;\subset\, \CP_\times\,.\ee
We conjectured in Section \ref{sec:perspective} that the phase space $\CP_K(\pd M)$ and Lagrangian $\CL_K(M)$ associated with the glued manifold $M$ itself are symplectic reductions of $\CP_\times$, $\CL_\times$. Let us describe how this works logarithmically.

\subsubsection{Paths on slices}
\label{sec:pathcoord}

It is extremely useful to encode various affine linear functions on $\CP_\times$ in terms of paths on the slices of the octahedral decomposition of a manifold $M$.

\begin{wrapfigure}{L}{2.6in}
\centering
\includegraphics[width=2.5in]{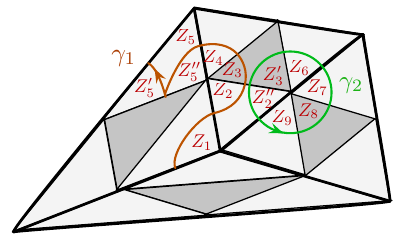}
\caption{Open ($\gamma_1$) and closed ($\gamma_2$) paths on a slice in $M$.}
\vspace{-5pt}
\label{fig:paths}
\end{wrapfigure}

Recall that there are $K-1$ global slices associated with every small boundary component of $M$, which extend from one tetrahedron to the next. We always view the slices from \emph{above}, that is, from the point of view of the small boundary. On any one of these slices, we consider oriented paths made of segments that begin at the midpoint of a side of a small black or white triangle, curve to the left or right inside the triangle, and end at the midpoint of another side. Two examples are shown in Figure~\ref{fig:paths}. The entire path may either be closed or open; if open we require it to begin and end on faces of big tetrahedra.

To any such oriented path $\gamma$, we can assign an affine linear function on the product phase space $\CP_\times$ by
\begin{enumerate}

\item adding octahedron vertex parameters $(Z_i,Z_i',Z_i'')$ for the dihedral angles subtended counterclockwise by segments of $\gamma$;

\item subtracting vertex parameters for dihedral angles subtended clockwise;

\item adding (resp., subtracting) $i\pi$ for a 180-degree ``bounce'' in a counterclockwise (resp., clockwise) direction.

\end{enumerate}

\begin{wrapfigure}[7]{r}{1.8in}
\centering
\vspace{-20pt}
\includegraphics[width=1.8in]{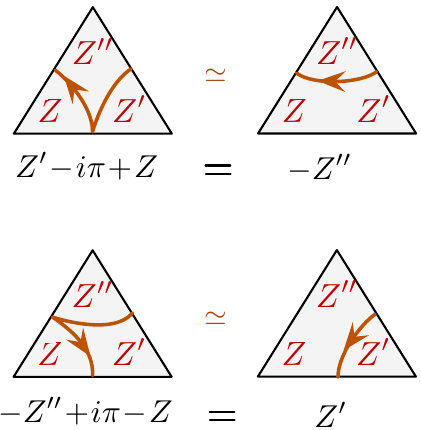}
\vspace{-25pt}
\caption{Path homotopy.}
\vspace{-25pt}
\label{fig:pathH1}
\end{wrapfigure}

\noindent We will call these functions path-coordinates. For example, the open path $\gamma_1$ in Figure \ref{fig:paths} has path-coordinate $-Z_1+Z_2+Z_3+Z_4+Z_5''-i\pi+Z_5'$, while the closed path $\gamma_2$ has path-coordinate $Z_9+Z_8+Z_7+Z_6+Z_5+Z_3'+Z_2''$.

The constraint that $Z_i+Z_i'+Z_i''=i\pi$ (for all octahedra $\oct_i$) in the product phase space implies that path-coordinates are invariant under homotopies inside the small black and white triangles, as shown in Figure~\ref{fig:pathH1}.

\subsubsection{Gluing functions}
\label{sec:C-combi}

Now we proceed to define the new coordinates on $\CP_\times$.

For every internal lattice point $v_k$ in the octahedral decomposition of $M$, we define an affine linear function $C_k$ on the product phase space to be the sum of octahedron parameters ($Z_i$, $Z_i'$, or $Z_i''$) assigned to octahedron vertices identified with $v_k$, minus an overall factor of $2\pi i$. These \emph{gluing functions} $C_k$ include:
\begin{itemize}

\item ${K-1\choose3}=\frac16 (K-1)(K-2)(K-3)$ points inside every ideal tetrahedron (for example, there is one internal point in the $K=4$ tetrahedron of Figure \ref{fig:oct-K234});

\item ${K-1\choose2}=\frac12 (K-1)(K-2)$ points on every pair of big faces of ideal tetrahedra that are glued together (\ie\ on every internal tetrahedron face);

\item ${K-1\choose1}=K-1$ points on every internal tetrahedron edge.

\end{itemize}
For example, for the two $K=3$ tetrahedra glued together in Figure \ref{fig:slice-continue}, there are no points inside tetrahedra, and there are no internal tetrahedron edges; but there is one lattice point in the center of the internal face.

In general, suppose that the big boundary of $M$ (tiled by unglued faces of tetrahedra) has genus $g$ and $h$ holes. Let us define the quantity $d_K(g,h) = 2(K^2-1)(g-1)+K(K-1)h$ as in \eqref{dKgh}, which is the expected complex dimension of the phase space associated with that boundary. Then a straightforward Euler-character argument shows that the total number of internal points in the octahedral decomposition of $M$ is exactly
\be \label{Cdep}
 \text{$\#$ $C_k$'s} \;=\;  \frac16 NK(K^2-1)- \frac12d_K(g,h)\,.
\ee
(If there are multiple disjoint big boundaries, each contributes $-\frac12 d_K(g_i,h_i)$ to \eqref{Cdep}.)
In addition, for every small torus boundary of $M$ there are $K-1$ linear relations among the $C_k$'s, described momentarily. Thus, if there are $t$ torus boundaries one \emph{expects}
\be \label{Cindep}
 \text{$\#$ independent $C_k$'s} \;=\; \frac16 NK(K^2-1)- \frac12d_K(g,h) - (K-1)t\,.
\ee
%

%\begin{wrapfigure}{l}{3in}
\begin{figure}
\centering
\includegraphics[width=3in]{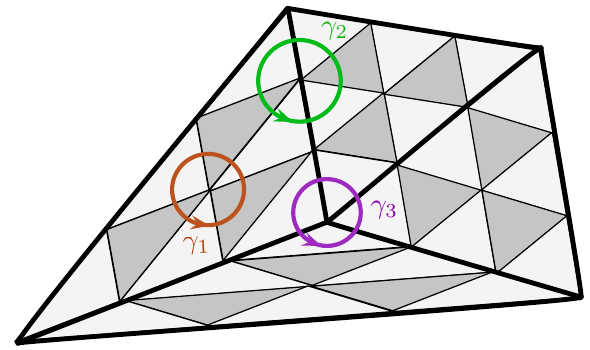}
\caption{Gluing functions from loops on a slice of $M$.}
\label{fig:sliceC}
\end{figure}

Every function $C_k$ is the path-coordinate (minus $2\pi i$) for a small closed loop surrounding a lattice point on a slice of $M$. The three types of vertices described above correspond to the three types of closed loops shown in Figure \ref{fig:sliceC}: $\gamma_1$ surrounds a point inside a tetrahedron, $\gamma_2$ surrounds a point on the face between tetrahedra, and $\gamma_3$ surrounds a point on a shared tetrahedron edge. Conversely, every path-coordinate (minus $2\pi i$) for such a small closed loop on a slice defines a $C_k$,
though the same $C_k$ may arise from multiple loops on different slices.

\begin{wrapfigure}{r}{1.5in}
\centering
\vspace{-5pt}
\includegraphics[width=1.5in]{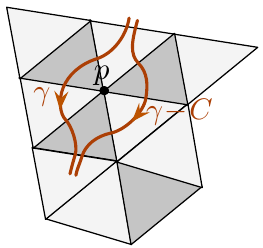}
\caption{Homotopy through a vertex $p$.}
\vspace{-5pt}
\label{fig:pathH2}
\end{wrapfigure}

When homotoping a path through a lattice point, as in Figure~\ref{fig:pathH2}, its path-coordinate simply picks up a contribution $\pm C$ for the gluing function assigned to that point. A nice application of this fact is to illustrate the $K-1$ linear relations among the $C_k$ coming from every small torus boundary of $M$. Consider any of the $K-1$ global slices of $M$ associated with a small torus boundary --- each having the topology of a torus. Choose two small adjacent triangles on this slice, as in Figure \ref{fig:cusp-constraint}, and draw a closed path $\gamma_1$ as shown, so that everywhere outside the two triangles segments of $\gamma_1$ always come in pairs with opposite orientations. Due to cancellations from these segments, the path-coordinate for $\gamma_1$ is just $2\pi i$. We can then homotope $\gamma_1$ to a second path $\gamma_2$ that just surrounds a single vertex $p$, as shown on the left of Figure \ref{fig:cusp-constraint}.
\begin{figure}[htb]
\centering
\includegraphics[width=5.2in]{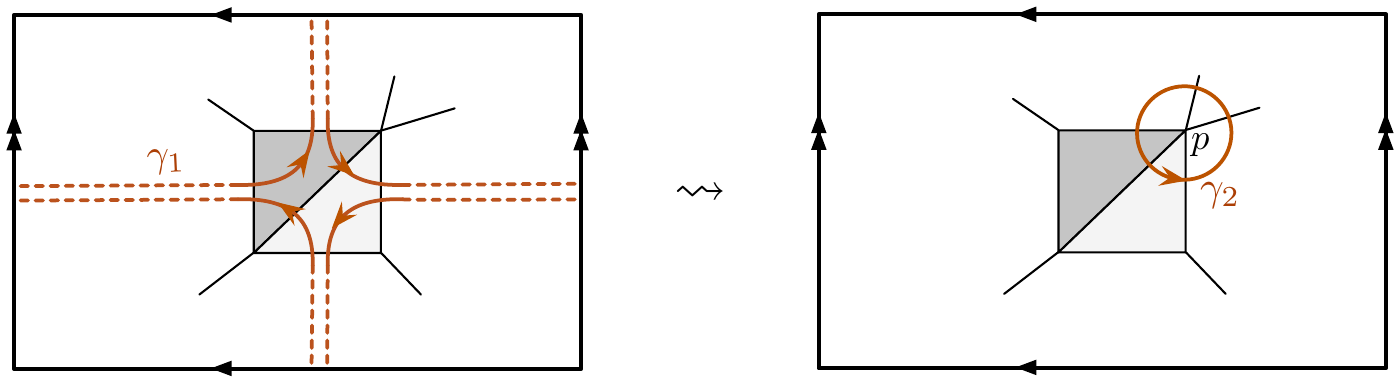}
\caption{Using path homotopy to obtain relations among $C_k$'s.}
\label{fig:cusp-constraint}
\end{figure}
The difference in the path-coordinates for $\gamma_1$ and $\gamma_2$ is the sum of $C_k$'s corresponding to every lattice point on this slice except for $p$. The path-coordinate for $\gamma_2$ is $C_p+2\pi i$ (where $C_p$ corresponds to $p$). Therefore, we find that \emph{the sum of $C_k$'s for all the vertices lying on this slice (including $C_p$) must be zero.} Repeating the argument for the $K-1$ slices parallel to any small torus boundary produces $K-1$ linear constraints, which modify the counting in~\eqref{Cindep}.

\subsubsection{Big boundary coordinates}

Recall that the ideal triangulation of $M$ induces a 2d ideal triangulation of the big boundary. Moreover, every big triangle (the face of a tetrahedron) on the big boundary acquires a 2d $K$-triangulation.

For every lattice point $p_j$ in the $K$-triangulation of the big boundary we define the function $\CX_j$ to be the sum of octahedron parameters $(Z_i,Z_i',Z_i'')$ at octahedron vertices incident to $p_j$. This includes $\frac12 (K-1)(K-2)$ functions for every big tetrahedron face on the boundary, and $K-1$ functions for every edge on the boundary. These are logarithmic versions of the boundary triple-ratios and cross-ratios of Section \ref{sec:bdyflat}.

\begin{figure}[htb]
\centering
\includegraphics[width=4.5in]{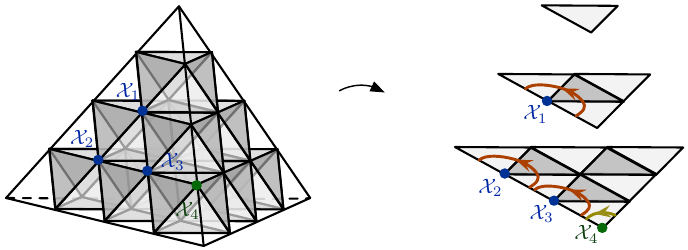}
\caption{The big boundary functions $\CX_j$ as path-coordinates on slices.}
\label{fig:sliceX}
\end{figure}

The functions $\CX_j$ can be interpreted as path-coordinates for open paths on the slices associated with small disc boundaries of $M$. This is illustrated in Figure \ref{fig:sliceX}, in the case of three vertices on a face and one on the edge of a single tetrahedron. Just as in the case of gluing functions $C_k$, a single $\CX_j$ may by represented by multiple paths on multiple slices.

\subsubsection{Small torus coordinates}
\label{sec:UandVcoords}

Finally, we define $2(K-1)$ functions on $\CP_\times$ for every small torus boundary of $M$. Given a torus boundary, we begin by choosing a basis of A and B homology cycles for it, with intersection
%
%\begin{wrapfigure}{l}{2in}
\begin{figure}[tbh]
\centering
\includegraphics[width=2in]{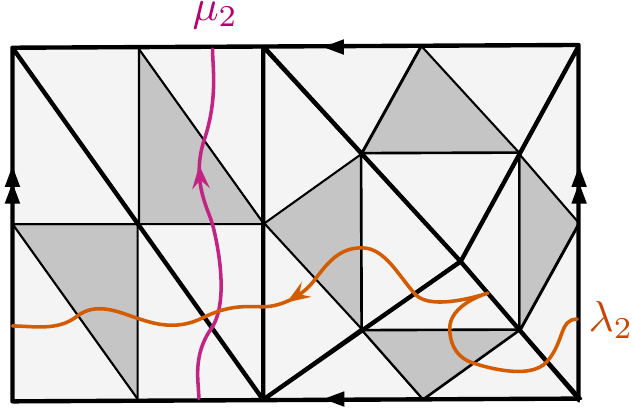}
\caption{Longitude and meridian paths on the second slice of a torus boundary.}
\label{fig:LM3}
\end{figure}
%\end{wrapfigure}
%
number $\langle A,B\rangle=1$.
If $M$ is a knot complement in $S^3$, there is a distinguished basis given by the \emph{meridian} and \emph{longitude} of the torus boundary, and we will sometimes use this terminology. If $M$ is not a knot complement, the basis of cycles can be chosen at will.

We consider the $K-1$ torus slices in the octahedral decomposition of $M$ that are associated with the small torus boundary. Let us number the slices in order of their proximity to the torus boundary --- so that inside a given tetrahedron $\Delta$ the $a$-th slice is tiled by $a^2$ small triangles. On each slice, we draw two non-self-intersecting paths $\mu_a$ and $\lambda_a$ such that their projections to the actual torus boundary are in the homology classes of the A and B cycles, respectively. An example of such paths is shown in Figure~\ref{fig:LM3}.
The path-coordinates $U_a$ and $V_a$ of the paths $\mu_a$ and $\lambda_a$ provide the desired functions on $\CP_\times$.

Notice that the definition of $U_a$ and $V_a$ depends on precisely which paths $\mu_a$ and $\lambda_a$ are chosen to represent a given homology class. However, changing the choice of paths can only modify $U_a$ and $V_a$ by multiples of the gluing functions $C_k$, since any two paths in a fixed homology class are homotopic (using the moves in Figures~\ref{fig:pathH1} and \ref{fig:pathH2}).

After setting $C_k=0$, the path coordinates $U_a$ and $V_a$ are completely unambiguous. They are logarithms of the holonomy eigenvalues called $m_a$ and $\ell_a$ in Section \ref{sec:torusbdy}\,! It is quite easy to show this by using 3d snakes to compute the holonomies, as depicted in Figure \ref{fig:torus-snake}. A snake rotating around a torus boundary simultaneously sweeps out $K-1$ paths (say) $\mu_a$ on the $K-1$ slices parallel to that boundary, and the octahedron parameters that the snake moves pick up are exactly the exponentiated versions of the corresponding path coordinates.

\subsection{Poisson brackets and the quotient}
\label{sec:sympsum}

For a triangulated manifold $M$, whose tetrahedra are decomposed into octahedra, we have defined three sets of affine linear functions on $\CP_\times$: the gluing functions $C_k$ associated with internal vertices in the octahedral decomposition; the functions $\CX_j$ associated with external vertices (on big boundaries); and the $2(K-1)$ functions $U^{(\nu)}_a$ and $V^{(\nu)}_a$ associated with cycles on every small torus boundary. Here we've indexed the torus boundaries by $\nu=1,\ldots,t$.

These functions on the product phase space $\CP_\times$ satisfy very special Poisson brackets. We can summarize the result as follows.
 First, the gluing functions all commute with each other and with the boundary functions:
\begin{subequations} \label{comm}
\be \label{commC}
\{C_k,C_{k'}\} = \{C_k,\CX_j\} = \{C_k,U^{(\nu)}_a\} = \{C_k,V^{(\nu)}_a\} = 0\,.
\ee
Second, the Poisson bracket of two functions $\CX_j$ and $\CX_{j'}$ associated with vertices $p_j$ and $p_{j'}$ on the big boundary is equal to the number of octahedron edges going from $p_j$ to $p_{j'}$, counted with orientation:
\begin{align} \label{commX}
\{\CX_j,\CX_{j'}\} &= \text{$\#$ edges from $p_j$ to $p_{j'}$}  -\text{$\#$ edges from $p_{j'}$ to $p_{j}$}\,.
\end{align}
The orientation is clockwise around every black octahedron face. Thus, on every external tetrahedron face, the RHS of \eqref{commX} can be represented by arrows drawn as in Figure \ref{fig:Xbracket} (\cf\ Section \ref{sec:bdyflat}). 
%\begin{wrapfigure}{r}{1.6in}
\begin{figure}[tbh]
\centering
\includegraphics[width=1.6in]{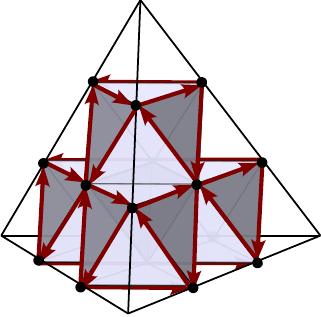}
\caption{Arrows representing the Poisson bracket on big boundaries.}
\label{fig:Xbracket}
\end{figure}
Note that the $\CX_j$ coordinates on two disjoint big boundaries necessarily commute. 
Third, the big boundary coordinates commute with torus coordinates,
\be \{\CX_j,U^{(\nu)}_a\}=\{\CX_j,V^{(\nu)}_a\}=0\,,\ee
and the torus coordinates satisfy
\be \label{commUV}
 \big\{V^{(\nu)}_a,U^{(\nu')}_{a'}\big\} = \delta^{\nu\nu'} \kappa_{aa'} ~,\ee
\be \big\{V^{(\nu)}_a,V^{(\nu')}_{a'}\big\} = \big\{U^{(\nu)}_a,U^{(\nu')}_{a'}\big\} = 0\, ,\ee
\end{subequations}
where
\be \kappa_{aa'} = \begin{cases} 2 & a=a' \\ -1 & |a-a'|=1 \\ 0 &\text{otherwise}\,,\end{cases}
\ee
is the Cartan matrix of $SL(K)$.
Namely, the logarithmic A and B cycle coordinates on different torus boundaries commute; while on the slices corresponding to a single boundary the Poisson bracket is proportional to the intersection number of the cycles, and vanishes if parallel slices are more than one layer apart. These Poisson brackets generalize a fundamental theorem of Neumann and Zagier \cite{NZ} for $K=2$.%

All of the brackets above can be understood intuitively as statements about path-coordi\-nates. In particular, the coordinate of a closed contractible path $\gamma$ commutes with the coordinate of any other open or closed path $\gamma'$, so long as $\gamma'$ (if open) begins and ends at the boundary of $M$. The coordinates of two general paths that intersect have a bracket (roughly) proportional to their intersection number.

We have already proved that the $C_k$ commute with each other and with all other coordinates in Section \ref{sec:bulkflat}. We also proved there that big-boundary coordinates $\CX_j$ have the expected Poisson bracket \eqref{commX}.
The trickiest commutation relations to show are the ones for the A and B cycle holonomies around small tori; we present a full proof in Appendix~\ref{app:NZ}.

We may recognize $\CX_j,U_a^{(\nu)},V_a^{(\nu)}$ as coordinates on the universal cover $\CP_K(\pd M)=\CP_K(\pd M,\mb t)$ of the space $\CX_K^{\rm un}(\pd M,\mb t)$ of framed flat $PGL(K)$-connections on the boundary.
Thus, on one hand, we have a projection
\be \CP_\times \to  \CP_K(\pd M)\,.\ee
On the other hand, the Poisson brackets \eqref{commX}--\eqref{commUV} precisely match the Atiyah-Bott-Goldman bracket on $\CX_K^{\rm un}(\pd M)$, \cf\ \eqref{torusS}, \eqref{flatPoisson}.
This means that we may (non-canonically) \emph{embed}
\be \CP_K(\pd M) \,\subset\, \CP_\times\,.\ee
as a complex symplectic subspace. 
Moreover, the brackets $\{C_k,C_{k'}\}=0$ among gluing functions allow us to define the holomorphic symplectic quotient $\CP_\times \big/\!\!\big/(C_k)$, using the $C_k$'s as a collection of commuting moment maps. Since our space $\CP_\times$ is affine-linear, the quotient is rather trivial to describe: each $C_k$ generates a $\C$ translation action; we quotient out by these translations and then set $C_k=0$. Since the $C_k$ commute with all boundary coordinates $\CX_j,U_a^{(\nu)},V_a^{(\nu)}$, the boundary coordinates descend to well-defined functions on the symplectic quotient,%
\footnote{Note in particular that on the quotient $\CP_\times \big/\!\!\big/(C_k)$, the functions $U^{(\nu)}_a$ and $V^{(\nu)}_a$ no longer have any dependence on the choice of paths drawn on slices, since the $C_k$ are all set to zero. These functions depend only on the homology classes of the paths.} %
and we obtain
\begin{prop}\label{prop:symp} The boundary phase space $\CP_K(\pd M)$ can be identified (non-canonically) as a subspace of the symplectic quotient
\be \CP_K(\pd M) \,\subset\, \CP_\times \big/\!\!\big/(C_k)\,. \label{PdM} \ee
Equivalently, we may factor $\CP_\times \simeq \CP_K(\pd M)\times \CP_C\times \CP_C^\perp$, with symplectic form $\Omega=\sum_i dZ\wedge dZ' = \Omega_{\rm WP} + \sum_k dC_k\wedge d(...) + \Omega^\perp$ on the three factors. The quotient above equals $\CP_\times \big/\!\!\big/(C_k) \simeq \CP_K(\pd M)\times \CP_C^\perp$.
\end{prop}

The Symplectic Gluing Conjecture \ref{conj:symp} amounts to the statement that $\CP_K(\pd M) \simeq \CP_\times \big/\!\!\big/(C_k)$, or in other words that $\CP_C^\perp$ is trivial. If we \emph{assume} that the number of independent gluing functions $C_k$ is exactly the expected number \eqref{Cindep}, then the triviality of $\CP_C^\perp$ follows from an elementary counting of dimensions. Namely, the boundary space $\CP_K(\pd M)$ must have complex dimension
\be \dim_\C \CP_K(\pd M) = 2(K-1)t+\sum_{i=1}^b d_K(g_i,h_i)\,,\ee
where $t$ is the number of small torus boundaries, $g_i,h_i$ are the genus and number of holes for each of the $b$ big boundaries, and $d_K(g,h) = 2(K^2-1)(g-1)+K(K-1)h$ as in \eqref{dKgh}. Then from \eqref{Cindep} we find
\be \dim_\C \CP_K(\pd M) = \dim_\C\CP_\times - 2(\text{$\#$ independent $C_k$'s})\,, \ee
leaving no room for $\CP_C^\perp$.

Corollary \ref{cor:Lag} would then tell us that the image of $\CL_\times$ under the quotient \eqref{PdM} defines the Lagrangian submanifold  $\CL_K(M) \subset \CP_K(\pd M)$. This involves projecting with respect to the flows of the moment maps $C_k$, then intersecting with $C_k=0$. (Technically speaking, $\CL_K(M)$ is the logarithmic lift of the space we have previously denoted $\CL_K(M,\mb t_{3d})$.)

Unfortunately, the combinatorics required to prove that the exact number of independent gluing functions agrees with \eqref{Cindep} are rather complicated for general admissible $M$. Neumann succeeded in proving this for $K=2$ in \cite{Neumann-combinatorics}. After the first version of our paper, Garoufalidis-Zickert proposed a proof for general $K\geq 2$ \cite{GZ-gluing}.

In the case of hyperbolic 3-manifolds and $PGL(2,\C)$-connections, the independence of the gluing functions was originally proved by \cite{NZ} by (in our language) analyzing the images of Lagrangian submanifolds under symplectic reduction, and using Mostow rigidity.
The argument could easily be generalized here, if one knew about rigidity for framed flat $PGL(K,\C)$-connections. However, the statement about independence of the $C_k$'s is ultimately topological in nature, and should hold independent of any rigidity properties of connections in the bulk.

\subsubsection{Unipotent conditions}

The functions $U^{(\nu)}_a$ and $V^{(\nu)}_a$ are independent on $\CP_K(\pd M)$, but the $\CX_j$ are not --- we know from Section \ref{sec:bdyflat} (\cf\ Figure \ref{fig:tetHolonomy}) that requiring unipotent holonomy around holes on the big boundary leads to constraints among the $x_j=\exp(\CX_j)$. It is easy to understand these constraints in terms of paths and slices. We briefly explain how this works.

\begin{figure}[htb]
\centering
\includegraphics[width=2.2in]{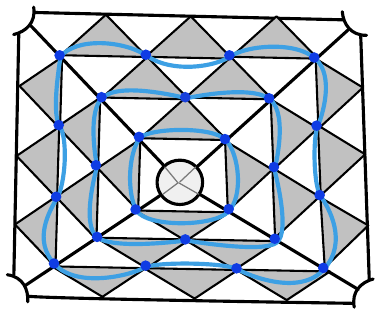}\hspace{.5in}\raisebox{.8in}{$\leadsto$}\hspace{.5in}
\raisebox{.25in}{\includegraphics[width=1.4in]{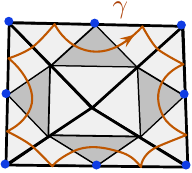}}
\caption{Loops surrounding a hole on the big boundary (left), and a contractible path on the $2^{\rm nd}$ slice parallel to the hole that demonstrates the constraint on eight $X_j$ functions (right).}
\label{fig:holeloops}
\end{figure}

For every hole in the big boundary of $M$ (filled in by a small disc) we expect $K-1$ central relations among the $\CX_j$. The hole in the big boundary is filled in by a small disc, and in the octahedral decomposition of $M$ there are $K-1$ slices parallel to this disc. The path-coordinate for a path on the periphery of such a slice, as on the right of Figure~\ref{fig:holeloops}, is a sum of functions $(\CX_j-i\pi)$; but since the path is a contractible clockwise loop its total coordinate (after setting $C_k=0$) must equal $-2\pi i$. Therefore, we find that the sum of functions $(\CX_j-i\pi)$ around each of the $K-1$ loops surrounding a hole (as on the left of Figure \ref{fig:holeloops}) must equal $-2\pi i$,
\be \label{Xconst}
\sum_{\text{$p_j$ around a hole}} (\CX_j-i\pi) = -2\pi i\,.
\ee
Exponentiating these relations, we recover the constraints introduced in Section \ref{sec:bdyflat}.

\subsubsection{Symplectic data}
\label{sec:Spdata}

Finally, let us summarize the combinatorial data needed to define a class-$\CR$ theory $T_K[M]$ associated with $M$ (as well as any quantum wavefunctions). This requires making a few more choices. Most importantly, we must choose a \emph{polarization} for $\CP(\pd M)$.

For the part of $\CP(\pd M)$ coming from the big boundary, a choice of polarization amounts to a splitting of the $\CX_j$ into linearly independent ``positions'' $X_j$ and ``momenta'' $ P_j$ that obey canonical Poisson brackets
\be \{  P_j, X_{j'}\} = \delta_{jj'}\,.\ee

For the part of $\CP(\pd M)$ coming from small tori, the choice of meridian and longitude (or A and B) cycles already amounts to a polarization. In addition, it is useful to choose (rational) linear combinations $\wt V_a$ of the $V_a$'s that have canonical Poisson brackets with the meridian coordinates,
\be \{ \wt V_a, U_{a'}\} = \delta_{aa'}\,.\ee
Given \eqref{commUV}, it suffices to take $\wt V_a = \sum_b (\kappa^{-1})_{ab}V_b$.

Altogether, we call these choices a polarization $\Pi$ for the boundary $\pd M$. The theory $T_K[M,\Pi]$ depends on $\Pi$.

We also make two inconsequential choices: we choose a maximal independent set of gluing functions $C_k$, and a set of affine linear functions $\Gamma_k$ on $\CP_\times$ that are canonically conjugate to these independent $C_k$, \ie\ $\{\Gamma_k, C_{k'}\}=\delta_{kk'}$.

If we assume the Symplectic Gluing Conjecture \ref{conj:symp}, then the independent $C_k$, along with the $\Gamma_k$ and the functions $ X_j,\, P_j,\, U^{(\nu)}_a,\wt V^{(\nu)}_a$, define a new set of symplectic coordinates on $\CP_\times$. Their relation to the original coordinates $(Z_i,Z_i',Z_i'')$ can be encoded in the affine symplectic transformation
\be \label{gs-main}
\begin{pmatrix}  X_j \\ U^{(\nu)}_a \\ C_k \\
  P_j \\ \wt V^{(\nu)}_a \\ \Gamma_k \end{pmatrix} \;=\; g\cdot \begin{pmatrix} Z_i \\ Z_i'' \end{pmatrix} -i\pi\,\sigma\,,
\ee
where $g$ is a symplectic matrix in $Sp\big(\frac13 NK(K^2-1),\Q\big)$, and $\sigma$ is a $\frac13 NK(K^2-1)$-dimensional vector of rational numbers. (Recall that $N$ is the number of tetrahedra in the triangulation of $M$.)

The rows of $g$ and $\sigma$ corresponding to the gluing functions $C_k$ are quite special; their properties are discussed further in Appendix \ref{app:symp}. Here it is at least clear that they only contain integer entries. Often (but not always) it is possible to choose polarization wisely so that the entire $g$ and $\sigma$ contain only integer entries, \ie\  $(g,\sigma)\in ISp\big(\frac13 NK(K^2\!-\!1),\Z\big)$. For example, this seems possible for any knot complement in $S^3$, for any $K$.

Given the symplectic data $(g,\sigma)$ corresponding to the triangulation of $M$ and a choice of polarization, one can attempt to define a gauge theory $T_K[M,\Pi]$ as well as various quantum wavefunctions associated with $M$, following \cite{Dimofte-QRS, DGG, DGG-index, KashAnd, AK-new, AK-complexCS, D-levelk}, etc. Each of these objects requires that the triangulation satisfy a few extra properties in order to be well-defined.

\subsection{Remarks on quantization}
\label{sec:qremarks}

We have claimed that constructing a logarithmic lift of functions on $\CP_\times$ and $\CP_K(\pd M)$ using path coordinates is necessary for quantization. We may briefly illustrate this idea, in terms of quantizing the Lagrangian $\CL_K(M) = \CL_K(M,\mb t_{3d})$ --- \ie\ promoting $\CL_K(M,\mb t_{3d})$ to a left ideal for the quantized algebra of functions on $\CX_K(\pd M,\mb t)$. We follow the formalism of \cite{Dimofte-QRS}.

(We note that, while constructing theories $T_K[M, \Pi]$ and wavefunctions requires a choice of polarization, the quantization of the Lagrangian $\CL_K(M)$ as a left ideal is independent of~$\Pi$. Thus, the quantization of $\CL_K(M)$ provides the simplest possible example of quantization and the necessity of logarithmic lifts.)

For every octahedron, quantization promotes the coordinates $Z_i,Z_i',Z_i''$ to operators with commutation relations
\be [\hat Z_i,\hat Z_i'] = [\hat Z_i',\hat Z_i''] =[\hat Z_i'',\hat Z_i] = \hbar\,, \ee
and a central constraint
\be \hat Z_i + \hat Z_i'+\hat Z_i'' = i\pi +\hbar/2\,. \label{qPoct}\ee
We may also exponentiate, setting $\hat z_i=\exp \hat Z_i$, etc., as well as $q=\exp \hbar$. Then the commutation relations are $\hat z_i\hat z_i' = q\hat z_i'\hat z_i$, etc., and the central constraint is $\hat z_i\hat z_i'\hat z_i''=-q$. The octahedron Lagrangian becomes an operator
\be \hat \CL_{\soct_i}:\quad \big(\hat z_i''+\hat z_i^{-1}-1\big)\,, \label{qLoct}\ee
which generates a left ideal in the $q$-Weyl algebra of exponentiated operators. Notice that the left ideal is invariant under cyclic permutations $\hat z_i\to \hat z_i'\to \hat z_i''\to\hat z_i$, due to the $\hbar$ correction in the central constraint \eqref{qPoct}.

Now we can promote the rules for path coordinates of Section \ref{sec:path-coords} to define new linear operators $\hat C_k,\hat \CX_j,\hat U_a,\hat V_a$ in the algebra generated by the $\hat Z_i,\hat Z_i',\hat Z_i''$ for all octahedra $\oct_i$. Along a given path, we simply add octahedron parameters with the appropriate signs, and modify every contribution of $\pm i\pi$ to $\pm (i\pi+\hbar/2)$. Moreover, we set $\hat C_k$ to be the sum of octahedron parameters around a closed loop, minus $(2\pi i+\hbar)$. The consistent modification of factors of $i\pi$ to $i\pi+\hbar/2$ leads to a consistent prescription for quantum path coordinates that are invariant under path homotopy.

In order to glue the octahedra together, we perform the quantum-mechanical version of symplectic reduction. We form the product left ideal $\hat \CL_\times =(\hat z_i''+\hat z_i^{-1}-1)_{i=1}^{\text{\#\,octs}}$, and take the centralizer of this ideal with respect to the quantum gluing functions $\hat C_k$. This leads to an ideal that involves only the exponentiated operators $\hat c_k=\exp\hat C_k$, $\hat x_j = \exp\hat \CX_j$, $\hat m_a=\exp\hat U_a$, and $\hat \ell_a=\exp \hat V_a$. In the centralizer, we can unambiguously set $\hat c_k=1$, and obtain a quantization of the Lagrangian $\hat \CL_K(M)$. By construction, upon setting $q\to1$, this Lagrangian should reduce to the classical ideal $\CL_K(M)$.

This procedure illustrates that the purpose of logarithmic path coordinates was to properly keep track of powers of $q$. The proposed construction of $\hat \CL_K(M)$ here suffers from a some technical difficulties (for example, the operation of taking the centralizer of $\hat \CL_\times$ with respect to the $\hat C_k$ is not guaranteed to succeed in general). However, modulo these difficulties, it was shown in \cite{Dimofte-QRS} that the prescription for assigning $q$-corrections here is absolutely unique. It is the only prescription that allows cyclic invariance of the \eqref{qLoct} and allows $\hat \CL_K(M)$ to be invariant under 2--3 moves.

\subsection{2--3 moves and path coordinates}
\label{sec:23-combi}

We now return briefly to the 2--3 moves of Section \ref{sec:23}. We demonstrate how our logarithmic path coordinates change under elementary 2--3 moves, and why they can be consistently defined as the triangulation of an admissible manifold $M$ changes. In the process, we explicitly provide the birational map between varieties $\CL_K(M,\mb t_{3d})$, $\CL_K(M,\mb t_{3d}')$ associated to different triangulations.

(The discussion here is parallels the treatment of 2--3 moves and their quantization in \cite[Section 5.4]{Dimofte-QRS}; we simply replace $K=2$ tetrahedra replaced by octahedra.)

\subsubsection{Elementary moves}
\label{sec:23-elem}

Recall the two elementary moves on octahedra from Section \ref{sec:23}. We show them again in Figure \ref{fig:23-oct}. From a combinatorial perspective, they are absolutely identical.

\begin{figure}[htb]
\centering
\includegraphics[width=6in]{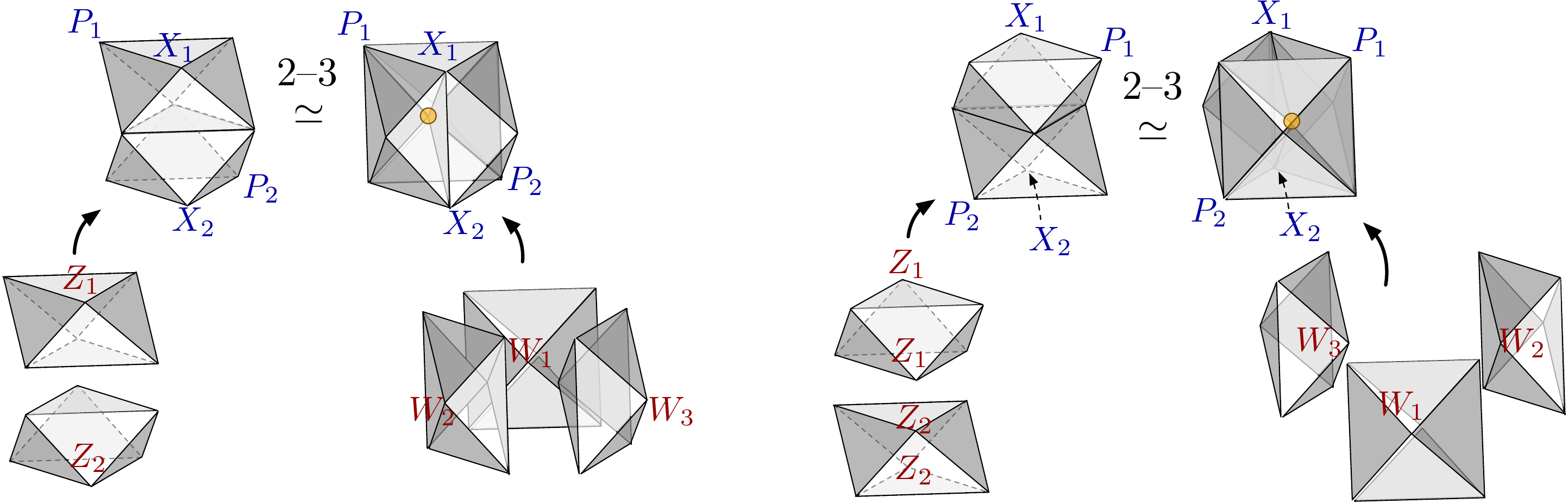}
\caption{The two 2--3 moves on octahedra. The second is equivalent to the first, viewed from behind, with white and black faces exchanged.}
\label{fig:23-oct}
\end{figure}

Let us assign logarithmic parameters $Z_i,Z_i',Z_i''$ to the octahedra on the ``2'' side of the move, and parameters $W_i,W_i',W_i''$ to the octahedra on the ``3'' side. We found previously that the coordinates associated with the nine points on the ``boundary'' of the clump of octahedra involved in the move did not change under the move --- they were products of octahedron cross-ratios before and after. Logarithmically, this leads us to the nine relations
\be \begin{array}{l@{\quad}l@{\quad}l} Z_1=W_2''+W_3',&Z_1'=W_3''+W_1',&Z_1''=W_1''+W_2', \\
 Z_2=W_2'+W_3'',& Z_2'=W_1'+W_2'',&Z_2''=W_3'+W_1''\, , \\
 W_1=Z_1+Z_2,&W_2=Z_1'+Z_2'',& W_3=Z_1''+Z_2'\,. \end{array} \label{P23rels} 
\ee
On the ``3'' side, there is also an internal trivalent vertex, giving a gluing function $C = W_1+W_2+W_3-2\pi i$. Therefore, the dimension of the boundary phase space associated with this clump of octahedra is 4 on both sides. For example, if we choose a polarization $(X_1,X_2;P_1,P_2)=(Z_1,Z_2;Z_1'',Z_2'')$ for the product phase space $\CP_{\pd\soct}^{(Z_1)}\!\times\! \CP_{\pd\soct}^{(Z_2)}$ on the ``2'' side and a consistent polarization $(X_1,X_2,C;P_1,P_2,\Gamma)=(W_2''+W_3',W_2'+W_3'',W_1+W_2+W_3-2\pi i;W_1''+W_2',W_3'+W_1'', W_1'')$ for the product phase space $\CP_{\pd\soct}^{(W_1)}\!\times\! \CP_{\pd\soct}^{(W_2)}\!\times\! \CP_{\pd\soct}^{(W_3)}$ on the ``3'' side, then it is clear that
\be \CP_{\pd(\text{clump})}  \simeq \CP_{\pd\soct}^{(Z_1)}\!\times\! \CP_{\pd\soct}^{(Z_2)} = \{X_1,X_2,P_1,P_2\} = \big(\CP_{\pd\soct}^{(W_1)}\!\times\! \CP_{\pd\soct}^{(W_2)}\!\times\! \CP_{\pd\soct}^{(W_3)}\big)\big/\!\!\big/(C=0)\,. \ee

We can also compute the Lagrangians corresponding to the clump, and check invariance. Let us set $x_i=e^{X_i},\,p_i=e^{P_i}$ as usual. On the ``2'' side we trivially find
\be \CL_{\text{clump},2} =  \{p_1+x_1^{-1}-1=0\,,\, p_2+x_2^{-1}-1=0\}\,,
\label{Lag2} \ee
while the Lagrangian on the ``3'' side is given, after eliminating $e^\Gamma$ and setting $e^C=1$, by
\be \CL_{\text{clump},3} = \{(1-x_1 x_2)(p_1+x_1^{-1}-1)=0\,,\,(1-x_1 x_2)(p_2+x_2^{-1}-1)=0\}\,. \label{Lag3} \ee
The ideals \eqref{Lag2} and \eqref{Lag3} are equivalent, provided that $x_1 x_2=e^{W_1}\neq 1$. This is precisely the condition that octahedra are non-degenerate, \ie\ their cross-ratio parameters take values in $\C^*\bs\{1\}$, which is ensured if classical configurations of framing flats are generic. As we already know, in general, the map of Lagrangians is birational.

\begin{figure}[htb]
\hspace{-.5in}\includegraphics[width=7in]{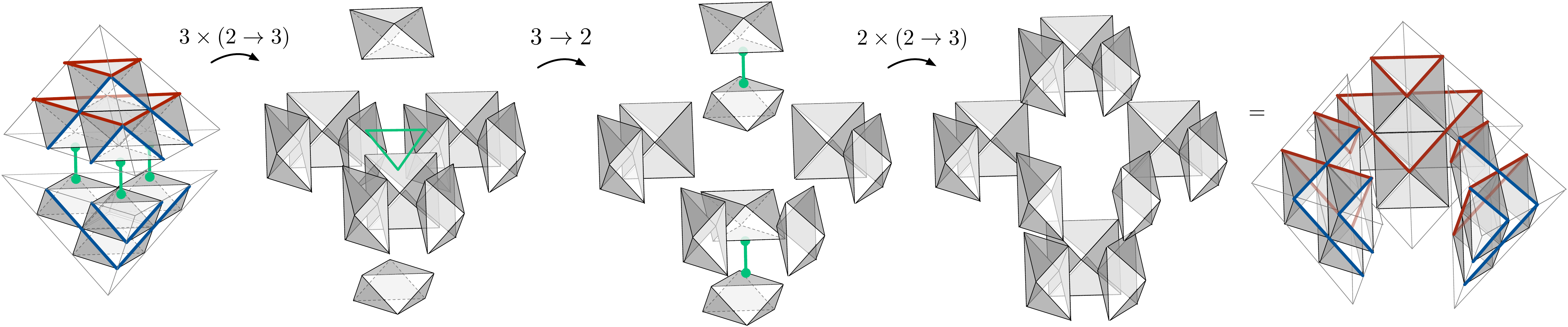}
\caption{The sequence of moves on octahedra that realizes a full 2--3 move for $K=3$.}
\label{fig:23-3}
\end{figure}

\subsubsection{Composing elementary moves}

In order to perform a full 2--3 move on ideal tetrahedra that have $K$-decompositions, we iterate the elementary 2--3 moves. We recall in Figure \ref{fig:23-3} how this worked for $K=3$.

\begin{figure}[htb]
\centering
\includegraphics[width=2.5in]{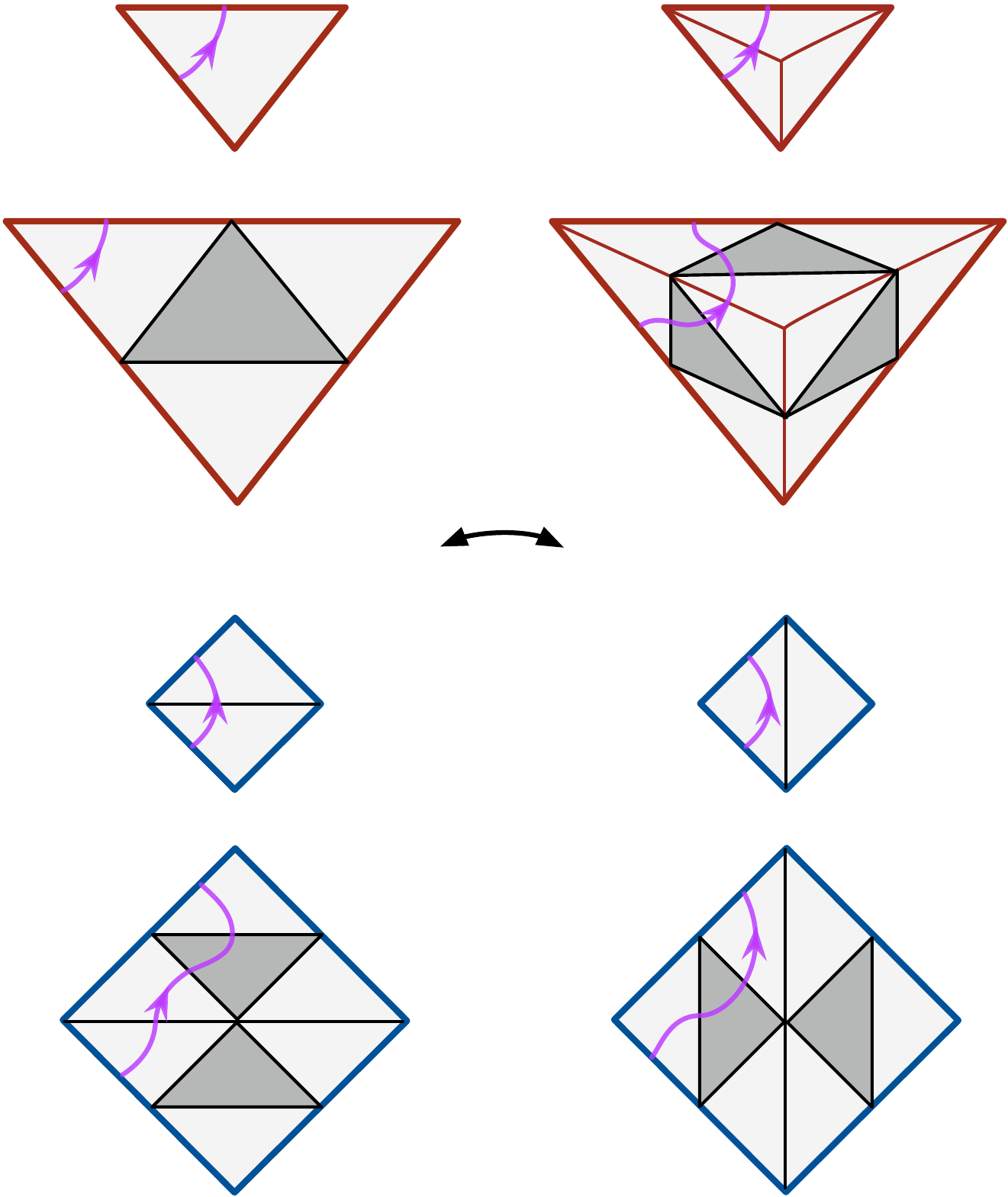}
\caption{Transforming paths during the 2--3 move}
\label{fig:23-3-slices}
\end{figure}

%\begin{wrapfigure}{r}{2.5in}
%%\vspace{-20pt}
%\includegraphics[width=2.5in]{23-3-slices}
%%\vspace{-25pt}
%\caption{Transforming paths during the 2--3 move}
%%\vspace{-25pt}
%\label{fig:23-3-slices}
%\end{wrapfigure}

Notice that path-coordinates as discussed in Section \ref{sec:pathcoord} can be translated from one side of the full 2--3 move to the other. Recall that the path-coordinates are affine linear combinations of octahedron parameters. Then, at each step in the sequence of elementary 2--3 moves, the relations \eqref{P23rels} provide a dictionary for transforming the path-coordinates. These transformations are well-defined, modulo the addition of gluing constraints $C_k$. Graphically, homotopy classes of paths on slices on one side of the 2--3 move are mapped to homotopy classes of paths on the other side. In Figure \ref{fig:23-3-slices} we give an example of how this works on two sets of slices of the $K=3$ bipyramid, outlined in red and in blue in Figure \ref{fig:23-3}. Note, however, that slices of the bipyramid are \emph{not} uniquely defined during the intermediate steps of the full 2--3 move; they are only canonically defined at the beginning and at the end.

%%%%%%%%%%%%%%%%%%%%%%%%%%%%%%%%%%%%%%%%%%%%%%%%%%%%%%%%%%%%%%%%%%%%%%%%%%%%%
%%%%%%%%%%%%%%%%%%%%%%%%%%%%%%%%%%%%%%%%%%%%%%%%%%%%%%%%%%%%%%%%%%%%%%%%%%%%%
%%%%%%%%%%%%%%%%%%%%%%%%%%%%%%%%%%%%%%%%%%%%%%%%%%%%%%%%%%%%%%%%%%%%%%%%%%%%%

\section{Tetrahedron and polyhedron theories}
\label{sec:Tpoly}

In this section we present the first and simplest examples of $\CN=2$ gauge theories $T_K[M]$ associated with 3-manifolds with $K>2$.
We will also compute some of their observables and the corresponding geometric quantities for the 3-manifolds, such as the supersymmetric parameter space $\CL_{\rm SUSY} \simeq \CL_K(M)$.

We start with a survey of tetrahedron theories $T_K[{\Delta,\Pi}]$ for $K=2,3,4,5$. These beautifully illustrate the types of complexities that arise at higher $K$ and how they can be handled. With a judicious choice of boundary polarization $\Pi$, the theories can be described as
\be \label{TKthys}
\begin{array}{c|c|c}
 K & T_K[\Delta] & \text{flavor sym.} \\\hline
 2 & \text{1 free chiral} & U(1) \\
 3 & \text{4 free chirals} & U(1)^{4} \\
 4 & \text{10 chirals + degree-six superpotential} & U(1)^9 \\
 5 & \;\text{20 chirals, $U(1)^2$ gauge group, $W=\sum$(four monopole ops)}\; & U(1)^{16}
\end{array} \ee
In general, the class-$\CR$ methods of this paper construct $T_K[\Delta]$ as a theory with manifest flavor symmetry%
\footnote{With a proper choice of background Chern-Simons terms (part of the polarization data), these flavor symmetries may occasionally be enhanced to non-abelian groups of the same rank. We will only write down the maximal abelian part of the flavor symmetry group.} %
$U(1)^{(K-1)^2}$, where $2(K-1)^2$ is the dimension of the phase space $\CP_K[{\pd \Delta}]$. The theory has $\frac16 K(K^2-1)$ chirals (the number of octahedra in the $K$-decomposition) and $\frac16(K-1)(K-2)(K-3)$ operators added to its superpotential (the number of internal lattice points). The dynamical gauge group depends on the choice of boundary polarization, though for $K\geq 5$ there will always be a nontrivial gauge group; likewise, for $K\geq 5$ the superpotential will always contain some monopole operators.

We also consider the bipyramid theory for $K=3$. The triangular bipyramid can be triangulated in two different ways, leading to two different theories $T_3[{\text{bip},\Pi}]$ that are mirror symmetric. With a suitable choice of polarization, one theory (coming from two tetrahedra) can be described as a collection of eight chirals with a single degree-six superpotential; while the other (coming from three tetrahedra) is a $U(1)^6$ gauge theory with twelve chirals and five operators in its superpotential, including three monopole operators. The mirror symmetry is quite non-trivial! Nevertheless, we know mirror symmetry must hold, and it can systematically be broken down into a sequence of fundamental XYZ $\leftrightarrow$ SQED mirror symmetries just by looking at the geometric decomposition of a 2--3 move into octahedra, as in Figure \ref{fig:23-3}. This is a nice example of how a complicated physical duality is encoded by relatively simple geometry.

The polyhedron theories $T_K[{M,\Pi}]$ in this section all depend on a choice of polarization, and in general there is no canonical choice.%
\footnote{Put differently, in the spirit of \cite{CCV}, there is a great deal of freedom in choosing asymptotic boundary conditions for $K$ M5-branes wrapped on a polyhedron.} %
This makes them somewhat ill-suited for a study of large-$K$ behavior, despite the nice hint of $K^3$ scaling for degrees of freedom that comes from counting octahedra. On the other hand, we should keep in mind that in many circumstances the choice of polarization is not important. For example, the choice dissolves once the theories are re-interpreted as boundary conditions for $U(1)$ gauge fields in four dimensions (\ie\ by weakly gauging $U(1)$ flavor symmetries in a 4d bulk). This was explained in \cite{DGG}. In three dimensions, it is natural to think of theories $T_K[{M,\Pi}]$ as coming in families or equivalence classes whose elements are related by $Sp(2N,\Z)$ transformations. While some observables of a theory (such as partition functions or indices) depend on a choice of representative in the class, and transform under $Sp(2N,\Z)$, others (such as $\CL_{\rm SUSY}\simeq \CL_K(M)$) do not. \\

In this section, we will usually work with the logarithmic lifts $\CP_K(\pd M)$ of phase spaces $\CX_K(\pd M)$, as in Section \ref{sec:combi}. We alternatively use uppercase logarithmic coordinates (\eg\ $Z_i$) and the lowercase exponentiated ones (\eg\ $z_i = \exp Z_i$), as appropriate.

A summary of the systematic rules used to obtain a 3d gauge theory in class $\CR$ from the symplectic gluing data $(g,\sigma)$ of an octahedral decomposition (assuming the Symplectic Gluing Conjecture \ref{conj:symp}) is given in Appendix \ref{app:symp}. We use these rules implicitly to derive gauge theories below, but focus the discussion on more intuitive aspects of the geometry-physics correspondence.

\subsection{Review: the octahedron theory}
\label{sec:Toct}

To begin, let us review the theory $T_\soct$ associated with an octahedron --- or, equivalently a tetrahedron for $K=2$. It was constructed in \cite{DGG}.

Recall that the vertices of an octahedron are labeled by a triplet of parameters $(Z,Z',Z'')$, occurring in the same clockwise cyclic order around every black face. Classically they satisfy $Z+Z'+Z''=i\pi$ and the Poisson brackets $\{Z,Z'\}=\{Z',Z''\}=\{Z'',Z\}=1$. Once we fix a labeling, the octahedron has a canonical initial polarization with $Z$ a position and $Z''$ a momentum. Typically we will also choose the final polarization $\Pi_Z=(X;P) = (Z;Z'')$ to agree with the initial one. Then $T_{\soct} := T_2[{\oct,\Pi_Z}]$ is the theory of a free $\CN=2$ chiral multiplet.

Specifically, we define
\be \label{Toct}
 T_{\soct}:\left\{\begin{array}{l}
 \text{Free chiral $\phi$\,,}\\\text{$U(1)_F$ flavor symmetry with complex mass parameter $Z$\,;} \\[.2cm]
 \text{charges:}\;\;\begin{array}{c|c}
  & \phi \\\hline
  F & 1 \\
  R & 0 \end{array} \qquad\qquad \text{CS levels:}\;\;\begin{array}{c|cc}
   & F  & R \\\hline
   F & -\frac12 & \frac 12 \\
   R &  \frac12 & *
   \end{array}\, \end{array}\right.
   \quad\raisebox{-.4in}{\includegraphics[width=1.2in]{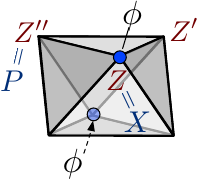}}
\ee
The theory has a $U(1)_F$ flavor symmetry rotating the phase of the chiral, and we add $-1/2$ units of Chern-Simons coupling%
\footnote{As usual, chiral matter in three dimensions produces one-loop half-integer contributions to effective Chern-Simons levels for gauge or flavor symmetries, \cf\ \cite{Redlich-parity, Redlich-parity2, AHISS}. Thus bare Chern-Simons must sometimes take half-integer values, as in \eqref{Toct}, in order for the full theory to be gauge invariant.} %
for a background vector multiplet associated with this flavor symmetry. This is a Chern-Simons contact term, as in \cite{CDFKS, CDFKS-CS}.
The theory also has a $U(1)_R$ R-symmetry. We specify an R-charge assignment, as well as mixed background Chern-Simons couplings between R-symmetry and flavor-symmetry fields. These are irrelevant in flat space, but become important when one compactifies the theory on $S^1$, $S^3$, $S^2\times S^1$, etc.
Similarly, in flat 3d space the mass parameter associated with the flavor symmetry (a scalar in the vector multiplet) is real, but gets complexified in any of these compactifications. This mass parameter is identified with $Z$.
On the 3-manifold side, it is the logarithmic lift of phase-space coordinates that lets us consistently keep track of R-charge assignments and complexified masses $Z$ in the compactifications. (These comments apply not just to the octahedron theory, but to all 3-manifold theories obtained from triangulations.)

Notice that $T_\soct$ has a chiral operator $\phi$. Geometrically, it can be associated with both of the vertices with final position coordinate $X=Z$ in \eqref{Toct}. We are also at liberty to choose \emph{different} final polarizations for the octahedron theory, related to $\Pi_Z$ by an $Sp(2,\Z)$ transformation. In any given polarization $\Pi=(X;P)$, the octahedron vertices commuting with the position coordinate $X$ (if any exist) will be assigned chiral operators. 

For example, if we choose polarization $\Pi_{Z'} = (Z';Z)$ with $Z'$ as a position, the theory becomes $T_{\soct}' := T_2[{\oct,\Pi_{Z'}}]$
\be \label{Toct1}
 T_{\soct}':\left\{\begin{array}{l}
 \text{$U(1)_G$ gauge theory w/ a chiral $\phi'$\,,}\\\text{$U(1)_J$ topological flavor symmetry w/ complex mass $Z'$\,;} \\[.2cm]
 \text{charges:}\;\;\begin{array}{c|c|c}
  & \phi' & V_+ \\\hline
  G & 1 & 0 \\\hline
  J & 0 & 1 \\
  R & 0 & 0  \end{array} \qquad \text{CS levels:}\;\;\begin{array}{c|c|cc}
   & G & J  & R \\\hline
   G & \frac12 & 1 & -\frac12 \\\hline
   J &1 & 0 & 0 \\
   R & -\frac12 & 0 &  * %-\frac12
   \end{array} \end{array}\right.
   \quad\raisebox{-.3in}{\includegraphics[width=1.1in]{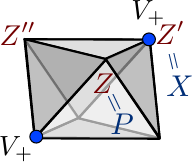}}
\ee
It has a gauge-invariant monopole operator $V_+$ ``sitting'' on the $Z'$ vertices. Similarly, in polarization $\Pi_{Z''}=(Z'';Z')$ we have $T_{\soct}'' := T_2[{\oct,\Pi_{Z''}}]$
\be \label{Toct2}
 T_{\soct}'':\left\{\begin{array}{l}
 \text{$U(1)_G$ gauge theory w/ a chiral $\phi''$\,,}\\\text{$U(1)_J$ topological flavor symmetry w/ complex mass $Z''$\,;} \\[.2cm]
 \text{charges:}\;\;\begin{array}{c|c|c}
  & \phi'' & V_- \\\hline
  G & 1 & 0 \\\hline
  J & 0 & 1 \\
  R & 0 & 0  \end{array} \qquad \text{CS levels:}\;\;\begin{array}{c|c|cc}
   & G & J  & R \\\hline
   G & -\frac12 & -1 & \frac12 \\\hline
   J &-1 & -1 & 1 \\
   R & \frac12 & 1 &  * %-\frac12
   \end{array} \end{array}\right.
  \quad \raisebox{-.3in}{\includegraphics[width=1.2in]{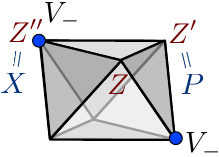}}
\ee
with an anti-monopole operator $V_-$ on the $Z''$ vertices.

The three theories above are actually all mirror-symmetric:
\be T_{\soct}\simeq T_{\soct}'\simeq T_{\soct}''\,. \ee
The flavor symmetry $U(1)_F$ of $T_{\soct}$ matches the topological $U(1)_J$ symmetries of $T_{\soct}'$ and $T_{\soct}''$. Moreover, the respective monopole ($V_+$) and anti-monopole ($V_-$) operators of $T_{\soct}'$ and $T_{\soct}''$
match the chiral operator $\phi$ of $T_{\soct}$.
This special mirror symmetry shows that in a given 3-manifold $M$, where final polarizations have been fixed, it makes no difference how the octahedra are labeled by $(Z,Z',Z'')$ as long as the cyclic order around black faces is preserved --- there are three equivalent choices for every octahedron.

In \emph{any} polarization, the octahedron theory will have a supersymmetric parameter space%
\footnote{The Lagrangian $\CL_{\rm SUSY}$ is the supersymmetric parameter space of a theory on $\C\times S^1$. See \cite{DG-Sdual, DGG, DGG-index} and Appendix \ref{app:assoc} for a brief review.} %
isomorphic to $\CL_{\rm SUSY}\simeq \CL_{\soct}=\{p+x^{-1}-1=0\}$. Changing the polarization just acts as an $Sp(2,\Z)$ transformation on $(X;P)$, with $x=\exp X$ and $p=\exp P$ as usual.

\subsection{Warmup: elementary 2--3 move theories}
\label{sec:23warmup}

As a warmup exercise, we derive the theories associated with the clusters of either 2 or 3 octahedra that enter elementary 2--3 moves. Alternatively, these are theories $T_2[{\rm bip}]$ for a $K=2$ bipyramid, as investigated in \cite{DGG}.
Versions of these theories appear as building blocks in almost all more interesting 3-manifold theories. They also illustrate the fundamental mirror symmetry associated with the 2--3 move on octahedra, which (putatively) is used to build up all the mirror symmetries in class $\CR$.

In all cases, the theories have $U(1)^2$ flavor symmetry. Put differently, the dimension of the phase spaces $\CP_{\pd(\text{cluster})}$ corresponding to these clumps of octahedra is always four. We label octahedra by parameters $(z_1,z_2)$ on the 2-side and $(w_1,w_2,w_3)$ on the 3-side, as shown in Figures \ref{fig:SQED4hyper}--\ref{fig:SQEDXYZ} (just as in Section \ref{sec:23-combi}). Then we have
\be \CP_{\pd(\text{cluster})} \simeq \CP_{\pd\soct z_1}\times \CP_{\pd \soct z_2}
\simeq \big(\CP_{\pd\soct w_1}\times \CP_{\pd \soct w_2}\times \CP_{\pd \soct w_3}\big)\big/\!\!\big/(C=0)\,,
\ee
where $C = W_1+W_2+W_3-2\pi i$ is the gluing function for the internal point on the 3-side. There is no canonical final polarization on $\CP_{\pd(\text{cluster})}$, but there are two somewhat natural choices: a ``longitudinal'' polarization (Figure \ref{fig:SQED4hyper}) and an ``equatorial'' polarization (Figure \ref{fig:SQEDXYZ}). The two choices give slightly different descriptions of the mirror pairs of theories.

\begin{figure}[htb]
\centering
\includegraphics[width=5.4in]{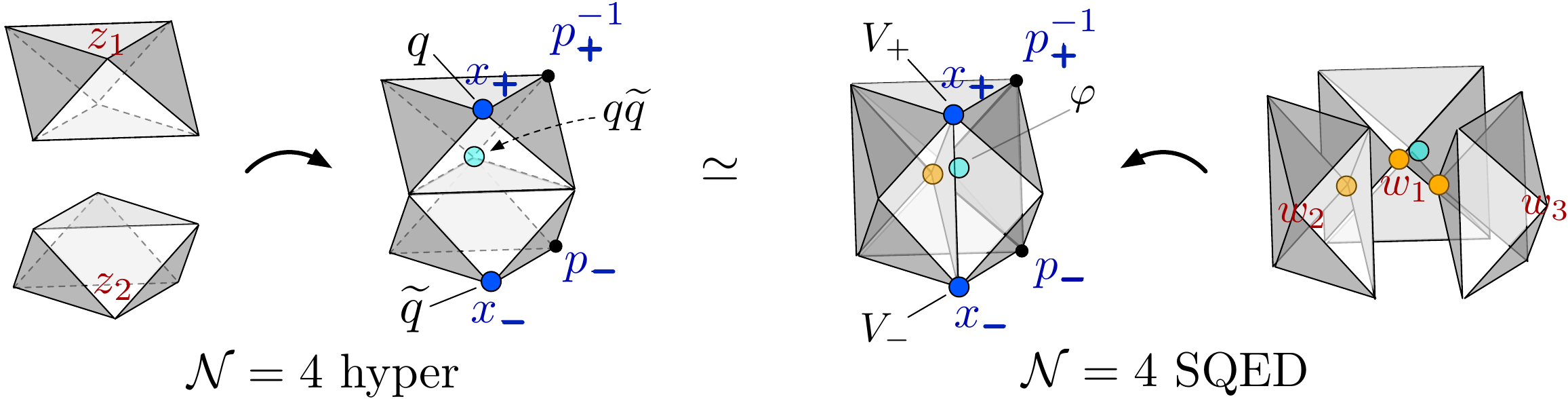}
\caption{Longitudinal polarization on mirror-symmetric clumps of octahedra.}
\label{fig:SQED4hyper}
\end{figure}

In the longitudinal polarization, the cluster of two octahedra is very simple. We take $\Pi_{\rm long} = (X_+,X_-;P_+,P_-) := (Z_1,Z_2;-Z_1',Z_2'')$. Since the final positions $X_\pm$ match the initial positions $Z_{1,2}$, the theory is just a product of two octahedron theories, \emph{a.k.a.} an $\CN=4$ hypermultiplet:
\be \label{Thyper}
T_{\rm hyper}:\;\left\{ \begin{array}{l}
 \text{Two chirals $q,\wt q$, with $U(1)_+\times U(1)_-$ flavor symmetry}\\[.1cm]
 \text{charges:}\;\begin{array}{c|cc}
  & q & \wt q \\\hline
  X_+ & 1 & 0 \\
  X_- & 0 & 1 \\
  R & 0 & 0 \end{array} \qquad
 \text{CS levels:}\;\begin{array}{c|ccc}
  & X_+ & X_- & R \\\hline
   X_+ & \frac12 & 0 & -\frac12 \\
   X_- & 0 & -\frac12 & \frac12 \\
   R & -\frac12 & \frac12 & * %-1
   \end{array}
 \end{array}\right.
\ee
In the charge matrix, we have used the positions $X_\pm$ to label the $U(1)_\pm$ symmetries.
The background Chern-Simons levels follow from the choice of momenta $P_\pm$.
Notice that the cluster of octahedra has three external coordinates $Z_1,Z_2,$ and $Z_1+Z_2$ that commute with the positions $X_\pm$; correspondingly, $T_{\rm hyper}$ has three chiral operators $q$, $\wt q$, and $q\wt q$ transforming under the flavor symmetries at these coordinates.

If instead we build the cluster with three octahedra, we now have a non-trivial $Sp(6,\Z)$ transformation acting on the product phase space:
\be \label{SQED4mx}
\begin{pmatrix} X_+\\X_-\\C\\P_+\\P_-\\\Gamma\end{pmatrix}
= \begin{pmatrix} 0 & 0 & -1 & 0 & 1 & -1 \\
 0 & -1 & 0 & 0 & -1 & 1 \\
 1 & 1 & 1 & 0 & 0 & 0 \\
 1 & 0 & 0 & 1 & 0 & -1 \\
 0 & 0 & -1 & 1 & 0 & -1  \\
 0 & 0 & -1 & 1 & 1 & -1 \end{pmatrix}
 \begin{pmatrix} W_1\\W_2\\W_3\\W_1''\\W_2''\\W_3'' \end{pmatrix}
 -i\pi \begin{pmatrix} -1\\-1\\2\\1\\-1\\ 0 \end{pmatrix}
\ee
The fact that there are initial momenta $(W_2'',W_3'')$ mixed into the final positions $X_\pm$ means that we must have a non-trivial gauging. Indeed, we find that the gauge theory is now described as $\CN=4$ SQED. In $\CN=2$ language:
\be \label{TSQED4}
 T_{\rm SQED}^{\CN=4}: \left\{\begin{array}{l}
 \text{Dynamical $U(1)$ theory with three chirals and a superpotential $W=\wt \phi \varphi\phi$;} \\
 \text{flavor symmetry $U(1)_+\times U(1)_-$ corresponding to (axial$\pm$topological)} \\[.2cm]
 \text{charges:}\quad\begin{array}{c|ccc|cc}
  & \phi & \wt\phi & \varphi & V_+ & V_- \\\hline
 G & 1 & -1 & 0 & 0 & 0 \\\hline
 X_+ & -1 & 0 & 1 & 1 & 0 \\
 X_- & -1 & 0 & 1 & 0 & 1 \\
 R & 2 & 0 & 0 & 0 & 0
 \end{array}\qquad \text{CS levels:}\quad
 \begin{array}{c|c|ccc}
  & G & X_+ & X_- & R \\\hline
  G&0 & \frac{1}{2} & -\frac{1}{2} & 0 \\\hline
 X_+&\frac{1}{2} & 0 & 0 & 0 \\
 X_-&-\frac{1}{2} & 0 & 0 & 0 \\
 R&0 & 0 & 0 & * \end{array}
\end{array}\right.
\ee
Each chiral $\phi,\wt\phi,\varphi$ corresponds to one of the original octahedra, and their product forms an operator $\CO_C$ that must be added to the superpotential to enforce the internal gluing constraint. The remaining flavor symmetry is an axial $U(1)_A$ and a topological $U(1)_J$, which are re-grouped into $U(1)_\pm$.
It can be checked (following, \eg, rules of \cite{AHISS}) that the theory has both a gauge-invariant monopole operator $V_+$ and an anti-monopole $V_-$, with charges as shown. The operators $V_\pm$ and the neutral chiral $\varphi$ now sit at the three external vertices of the clump that commute with $X_\pm$. The duality
\be T_{\rm hyper} \simeq T_{\rm SQED}^{\CN=4} \label{MS4} \ee
is one of the classic examples of 3d mirror symmetry \cite{IS, dBHOY}.

In the equatorial polarization, the final positions $X_\pm$ now involve initial momenta on the $2$-side but not on the $3$-side. Thus, the 2-side gives a dynamical $U(1)$ gauge theory while the 3-side has three ungauged chirals with a superpotential.

\begin{figure}[htb]
\centering
\includegraphics[width=5.4in]{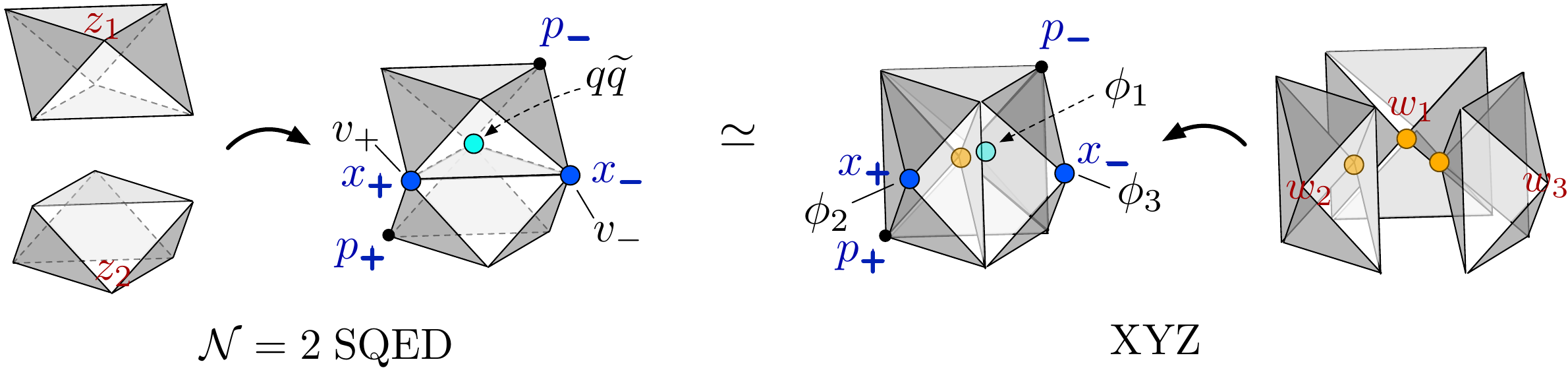}
\caption{Equatorial polarization on mirror-symmetric clumps of octahedra.}
\label{fig:SQEDXYZ}
\end{figure}

Explicitly, on the 2-side, we set $(X_+,X_-;P_+,P_-)=(Z_1'+Z_2'',Z_1''+Z_2';Z_2',Z_1')$, or
\be \label{SQED2mx}
\begin{pmatrix} X_+ \\ X_- \\ P_+ \\ P_- \end{pmatrix} = \begin{pmatrix} 
  -1&0&-1&1 \\
  0&-1&1&-1 \\
  0&-1 & 0 & -1\\
  -1&0&-1&0 \end{pmatrix} \begin{pmatrix}Z_1\\Z_2\\Z_1''\\Z_2''\end{pmatrix}
 -i\pi \begin{pmatrix} -1\\-1\\-1\\-1\end{pmatrix}\,.
\ee
This results in $\CN=2$ SQED,
\be \label{TSQED2}
 T_{\rm SQED}^{\CN=2}: \left\{\begin{array}{l}
 \text{Dynamical $U(1)$ theory with two chirals $q,\,\wt q$;} \\
 \text{flavor symmetry $U(1)_+\times U(1)_-$ corresponding to (axial\,$\pm$\,topological)} \\[.2cm]
 \text{charges:}\quad\begin{array}{c|cc|cc}
  & q & \wt q & v_+ & v_- \\\hline
 G & 1 & -1  & 0 & 0 \\\hline
 X_+ & -1 & 0 & 1 & 0 \\
 X_- & -1 & 0 & 0 & 1 \\
 R & 2 & 0 & 0 & 0
 \end{array}\qquad \text{CS levels:}\quad
 \begin{array}{c|c|ccc}
  & G & X_+ & X_- & R \\\hline
  G&0 & \frac{1}{2} & -\frac{1}{2} & 0 \\\hline
 X_+ &\frac{1}{2} & -\frac{1}{2} & \frac{1}{2} & \frac{1}{2} \\
 X_- & -\frac{1}{2} & \frac{1}{2} & \frac{1}{2} & -\frac{1}{2} \\
 R& 0 & \frac{1}{2} & -\frac{1}{2} & * \end{array}
\end{array}\right.
\ee
This is nothing but $\CN=4$ SQED without the neutral chiral and the superpotential; indeed, geometrically, if we remove the third octahedron from the right side of Figure \ref{fig:SQED4hyper} and rotate the picture by ninety degrees, we recover the left side of Figure \ref{fig:SQEDXYZ}. The theory $T_{\rm SQED}^{\CN=2}$ has axial and topological $U(1)$ symmetries. Its gauge-invariant operators are a monopole/anti-monopole $v_\pm$ and a meson $q\wt q$, which sit at three vertices on the boundary of the cluster.

Finally, if we use the equatorial polarization and build the cluster from three octahedra we obtain the ``XYZ'' model:
\be \label{TXYZ}
T_{\rm XYZ}:\quad \left\{\begin{array}{l}
 \text{Three chirals $\phi_1,\phi_2,\phi_3$ with a superpotential $W=\phi_1\phi_2\phi_3$,}\\[.1cm]
 \text{unbroken flavor symmetry $U(1)_+ \times U(1)_-$} \\[.1cm]
 \text{charges:}\;\begin{array}{c|ccc}
  & \phi_1 & \phi_2 & \phi_3 \\\hline
  X_+ & -1 & 1 & 0 \\
  X_- & -1 & 0 & 1 \\
  R & 2 & 0 & 0
  \end{array} \qquad
  \text{CS levels:}\; \begin{array}{c|ccc}
   & X_+ & X_- & R \\\hline
  X_+ & 0 & \frac12 & 0 \\
  X_- & \frac12 & 0 & 0 \\
  R & 0 & 0 & * \end{array} \end{array}\right.
\ee
The chiral operators $\phi_1,\phi_2,\phi_3$ sit at external vertices of the cluster.

The mirror symmetry
\be T_{\rm SQED}^{\CN=2} \simeq T_{\rm XYZ} \label{MS2} \ee
is a classic example of $\CN=2$ mirror symmetry \cite{AHISS, dBHO}. It is \emph{equivalent} to the $\CN=4$ mirror symmetry \eqref{MS4}. We can simply act with an $Sp(4,\Z)$ transformation on both sides of \eqref{MS4} to change the boundary polarization from longitudinal to equatorial, recovering \eqref{MS2}. In physical language: if we simultaneously gauge a vector $U(1)$ in the free hyper theory and a corresponding topological $U(1)_J$ in $\CN=4$ SQED, we recover $\CN=2$ SQED and the XYZ model, respectively.%
\footnote{To see that we recover the XYZ model this way, we must remember that gauging a $U(1)$ and subsequently gauging its $U(1)_J$ is equivalent in the IR to gauging nothing at all --- it is the statement that $S^2=-I$ in the symplectic group \cite{Witten-SL2}.} %

The twisted superpotentials $\wt W_{(2)}(x,y)$ and $\wt W_{(3)}(x,y)$ discussed back in Section \ref{sec:intro3d} of the Introduction are associated to the theories $T_{\rm SQED}^{\CN=2}$ and $T_{\rm XYZ}$ respectively. The variables $x$ and $y$ correspond to real masses (exponentiated and complexified) for the flavor symmetries $U(1)_+$ and $U(1)_-$, respectively.

\subsection{Tetrahedra at higher $K$}
\label{sec:tet-theories}

We now proceed to describe the ideal-tetrahedron theories $T_K[\Delta]$ for $K>2$. They were summarized in \eqref{TKthys}, and we want to demonstrate how they are constructed.

\subsubsection{$K=3$}

\begin{wrapfigure}{r}{2in}
\vspace{-1cm}
\includegraphics[width=2in]{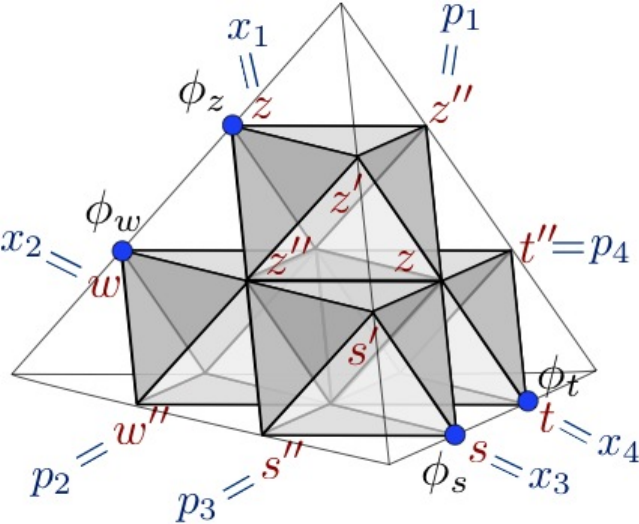}
\caption{Octahedra and final polarization for $K=3$.}
\label{fig:SL3-thy}
\end{wrapfigure}

The four octahedra in a $K=3$ tetrahedron can be assigned triples of parameters $(z,w,s,t)$ as in Figure \ref{fig:SL3-thy}. Since there are no internal lattice points, we expect that the phase space $\CP_3({\pd \Delta})$ and Lagrangian $\CL_3({\Delta})$ are just products of four copies of the octahedron phase space and Lagrangian. Similarly, in suitable polarizations, the theory $T_K[{\Delta}] \simeq  T_{\soct z}\otimes  T_{\soct w} \otimes  T_{\soct s} \otimes  T_{\soct t}$ should just be a product of four octahedron theories.

Explicitly, the phase space $\CP_3({\pd \Delta})$ --- an open patch of the space of framed flat $PGL(3,\C)$-connections on the boundary with unipotent holonomy at the punctures --- is parametrized by the twelve edge coordinates and four face coordinates on the boundary of the tetrahedron, subject to eight constraints from the vertices. The vertex constraints, imposing unipotent holonomy, say that sums of logarithmic coordinates around loops encircling the vertices equal $-2\pi i$, \cf\ \eqref{Xconst} and Figure \ref{fig:tetHolonomy}. A convenient choice of linearly independent boundary coordinates is given by the polarization
\be \label{polD3}
\Pi = (X_1,X_2,X_3,X_4;P_1,P_2,P_3,P_4) = (Z,W,S,T;Z'',W'',S'',T'')\,, \ee
a simple product of octahedron polarizations. Then the Lagrangian becomes the product
\be \CL_3(\Delta) = \{p_i+x_i^{-1}-1=0\}_{i=1}^4\,. \ee
It parametrizes an open patch of the configuration space of four flags in $\C^3$ --- the invariant flags on the tetrahedron's vertices.

Using the polarization \eqref{polD3} we find that the theory is exactly
\be T_3[{\Delta,\Pi}] = T_{\soct z}\otimes T_{\soct w}\otimes T_{\soct s}\otimes T_{\soct t}\,, \ee
\ie\ a theory of four free chirals.
It has $U(1)^4$ flavor symmetry. The chiral operators $\phi_z,\phi_w,\phi_s,\phi_t$ are associated with four external points on the $K=3$ tetrahedron, as indicated in Figure \ref{fig:SL3-thy}. Only these external points have coordinates that commute with the positions in $\Pi$.

\subsubsection{$K=4$}

For $K=4$, the tetrahedron is decomposed into ten octahedra, as in Figure \ref{fig:SL4-thy}. The novelty is that there is an internal lattice point in the middle of the tetrahedron. It leads to a gluing constraint and to a superpotential coupling.

\begin{figure}[htb]
\centering
\includegraphics[width=6in]{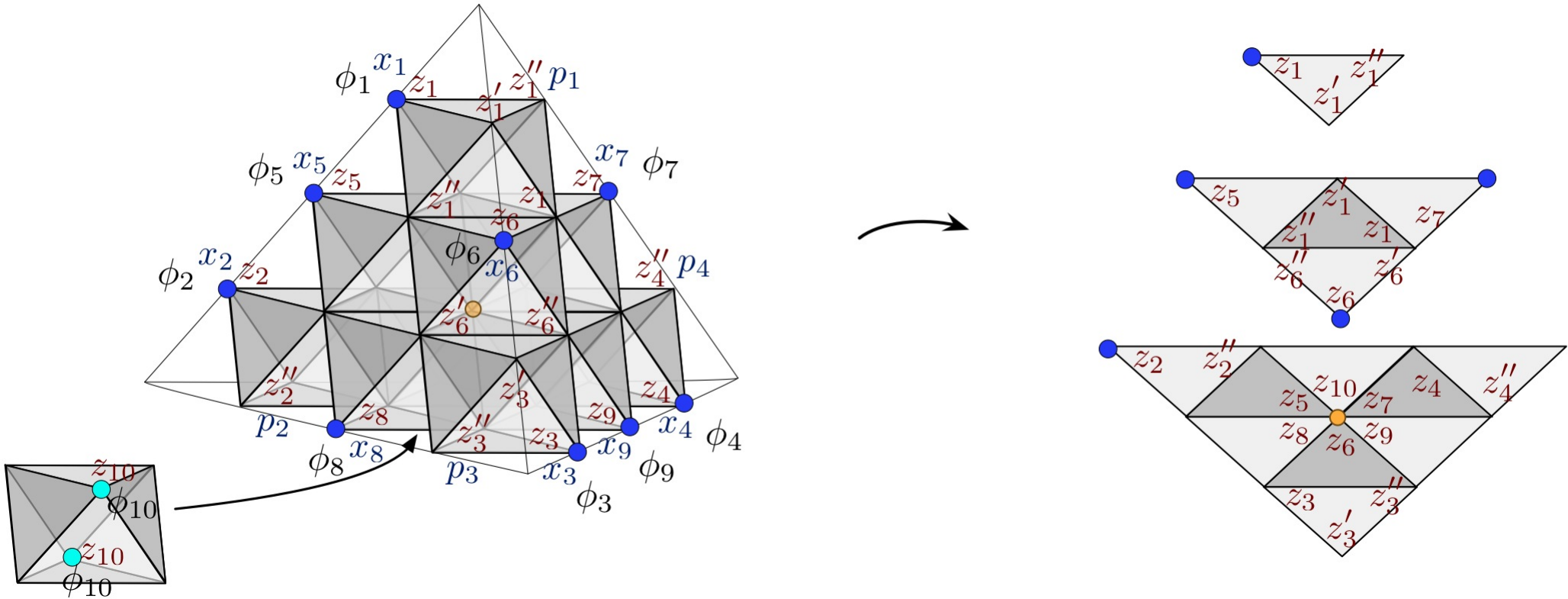}
\caption{Octahedra, final polarization, and operators for $K=4$. The octahedron in the middle of the back edge is drawn separately; note that its parameter is not part of the polarization. On the right we show the three slices around the top vertex.}
\label{fig:SL4-thy}
\end{figure}

A convenient choice of octahedron labelings is shown in the figure. We arrange that all six octahedron vertices surrounding the internal point are labeled by un-primed coordinates, so that the classical gluing constraint takes the form
\be C=Z_5+Z_6+Z_7+Z_8+Z_9+Z_{10}-2\pi i\;\to 0\,. \ee
Then in the gauge theory the corresponding operator will easily be built as a product of six fundamental chirals. Moreover, we notice that all ten of the initial positions $Z_i$ appear as coordinates on the edges of the boundary. Therefore, we can choose an independent set of nine of them (independent modulo the gluing constraint) as positions in the final polarization $\Pi$ as well; for example, we take
\be X_i = Z_i\,,\qquad i=1,\ldots,9\,. \ee
Then the theory $T_4[{\Delta,\Pi}]$ involves no gauging at all. It is basically a product of octahedron theories:
\be  T_4[{\Delta,\Pi}]:\;\left\{\begin{array}{l}
\text{ten chirals $\phi_i$ with a superpotential $W=\phi_5\phi_6\phi_7\phi_8\phi_9\phi_{10}$} \\[.1cm]
\text{$U(1)^9$ flavor symmetry left unbroken}
\end{array}\right.
\ee
As dictated by this polarization, the first four $U(1)_i$ flavor symmetries rotate the first four $\phi_i$'s; while the for $5\leq i\leq 9$ the field $\phi_i$ has charge $+1$ and $\phi_{10}$ has charge $-1$. Moreover, the first nine fields have R-charge zero and $\phi_{10}$ has R-charge 2. The background Chern-Simons levels can be determined after a precise choice of momenta $P_i$ in $\Pi$ is made. (In general, changing the momenta only effects background couplings.)

We might notice that the construction of $T_4[{\Delta,\Pi}]$ is a simple generalization of the construction of the XYZ model in Section \ref{sec:23warmup}.

The phase space $\CP_4({\pd \Delta})$ for the boundary of the $K=4$ tetrahedron is the symplectic quotient of a product of ten octahedron spaces, using the gluing coordinate $C$ as a moment map. Thus it has complex dimension $18$.  Recall that $C$ should be thought of as a coordinate on the product $\CP_\times = \prod_{i=1}^{10}\CP_{\pd\soct i}$, with product polarization $\Pi_\times = (Z_i;Z_i'')$.
We can describe $\CP_4({\pd \Delta})$ more explicitly by choosing new symplectic coordinates on $\CP_\times$ that include the 18 positions and momenta $(X_i,P_i)$ of a final boundary polarization, as well as $C$ and a canonical conjugate $\Gamma$ so that $\{\Gamma,C\}=1$. For example, it suffices to choose
\be \Gamma = Z_{10}''\,,
\ee
together with $X_i=Z_i$ $(1\leq i\leq 9)$ as well as $P_i=Z_i''$ for $1\leq i\leq 4$ and $P_i=Z_i''-\Gamma$ for $5\leq i \leq 9$. Then
\be \CP_4({\pd \Delta}) = \Big(\prod_{i=1}^{10}\CP_{\pd\soct i}\Big)\big/(\Gamma\sim \Gamma+t)\big|_{C=0} = \{X_i,P_i\}_{i=1}^9 \simeq \C^{18}\,.
\ee
Note that all the $P_i$ here can be written as linear combinations of boundary coordinates on the tetrahedron. This must always be the case, due to the Poisson brackets of Section \ref{sec:sympsum}.

The nine-dimensional Lagrangian submanifold is obtained by eliminating $\gamma = z_{10}''$ from $\CL_\times = \{z_i''+z_i^{-1}-1=0\}_{i=1}^{10}$, setting $c=z_5z_6z_7z_8z_9z_{10}=1$, and re-writing the equations in terms of $x_i$ and $p_i$:
\be \CL_4(\Delta):\quad \{p_i-x_i^{-1}-1=0\}_{i=1}^4\,\cap\, \{(1-x_5x_6x_7x_8x_9)p_i+x_i^{-1}-1=0\}_{i=5}^9\,.\ee
In fact, it is also easy to quantize the Lagrangian using the prescription of \cite{Dimofte-QRS}, outlined in Section \ref{sec:qremarks}. We find the left ideal
\be \CL_4(\Delta)=  \big(\hat p_i-\hat x_i^{-1}-1\big)_{i=1}^4\cap \big((1-\hat x_5\hat x_6\hat x_7\hat x_8\hat x_9)\hat p_i+\hat x_i^{-1}-1\big)_{i=5}^9\,.\ee
These are the line operators that annihilate partition functions of $T_4[{\Delta,\Pi}]$.

\subsubsection{$K=5$}
\label{sec:SL5tet}

The $K=5$ tetrahedron is a good example of the complexity that arises for general $K$. It gets decomposed into twenty octahedra (Figure \ref{fig:SL5-assemble}), with four internal lattice points. No matter how we label the octahedra, the gluing functions at these lattice points can never simultaneously be written as sums of initial positions $Z_i$. Correspondingly, the four operators added to the superpotential of the theory $T_5[\Delta]$ can never simultaneously be written as products of elementary chiral fields --- monopole operators are unavoidable. 

\begin{wrapfigure}[11]{r}{1.7in}
\vspace{-.5cm}
\includegraphics[width=1.7in]{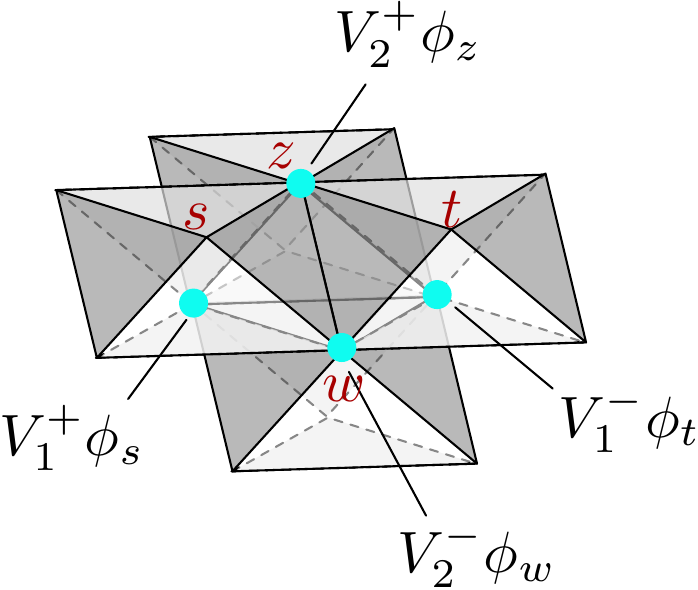}
\caption{The central cluster of the $K=5$ tetrahedron.}
\label{fig:SL5-cluster}
\end{wrapfigure}

It is instructive to build the tetrahedron and the corresponding gauge theory in pieces. We start with a cluster of four octahedra, which have a small upright tetrahedron ($\Delta^{0,2}$ simplex) in their center (Figure \ref{fig:SL5-cluster}). We label octahedra with parameters $(z,w,s,t)$ as shown, and choose a boundary polarization $\Pi_{\rm cluster} =(X_i;P_i)_{i=1}^4$ for the cluster so that the four light-blue boundary points in the figure commute with the positions of $\Pi_{\rm cluster}$. That means that there will be gauge-invariant operators associated with them.

Explicitly, we set $\{X_1= Z+W ,\,X_2=S+T,\,X_3=Z+S'+T'',\,X_4= S+Z'+W''\}$. The positions $X_1,X_2$ are combinations of initial octahedron positions, while $X_3,X_4$ involve initial momenta; thus we expect something like a $U(1)^2$ gauge theory.
In matrix form,
\be  \label{SL5mx}
\begin{pmatrix} X_1\\X_2\\X_3\\X_4 \\ P_1\\P_2\\P_3\\P_4\end{pmatrix}= \left(
\begin{array}{cccccccc}
 1 & 1 & 0 & 0 & 0 & 0 & 0 & 0 \\
 0 & 0 & 1 & 1 & 0 & 0 & 0 & 0 \\
 1 & 0 & -1 & 0 & 0 & 0 & -1 & 1 \\
 -1 & 0 & 1 & 0 & -1 & 1 & 0 & 0 \\
 0 & 0 & 0 & 0 & 0 & 1 & 0 & 0 \\
 0 & 0 & 0 & 0 & 0 & 0 & 0 & 1 \\
 0 & 0 & 1 & 0 & 0 & 0 & 0 & 0 \\
 1 & 0 & 0 & 0 & 0 & 0 & 0 & 0
\end{array}
\right) \begin{pmatrix} Z\\W\\S\\T \\Z''\\W''\\S''\\T''\end{pmatrix} -i\pi \begin{pmatrix} 0\\0\\-1\\-1\\0\\0\\0\\0 \end{pmatrix}\,.
\ee
By comparing the form of this matrix with \eqref{SQED2mx}, we find that the cluster theory is basically two copies of $\CN=2$ SQED. More accurately, it is two copies of SQED whose $U(1)$ gauge fields are coupled by a mixed Chern-Simons term:
\be \label{Tcluster}
\hspace{-.5in}T_{\rm cluster}:\left\{
\begin{array}{l} \text{$U(1)_1\times U(1)_2$ gauge theory coupled to four chirals, no superpotential}\\\text{$U(1)_A^2\times U(1)_J^2$ flavor symmetry (with $(A_1,A_2,J_1,J_2)\sim (X_1,X_2,X_4,X_3)$ above)} \\[.2cm]
\text{charges:}\quad
 \begin{array}{c|cccc|cccc} & \phi_z & \phi_w & \phi_s & \phi_t & V_1^+ & V_1^- & V_2^+ & V_2^- \\\hline
 G_1 & 1 & -1 & 0 & 0 & 0 & 0 & -1 & 1 \\
 G_2 & 0 & 0 & 1 & -1 & -1 & 1 & 0 & 0 \\\hline
 A_1 & 0 & 1 & 0 & 0 & 0 & -1 & 0 & 0 \\
 A_2 & 0 & 0 & 0 & 1 & 0 & 0 & 0 & -1 \\
 J_1 & 0 & 0 & 0 & 0 & 1 & -1 & 0 & 0 \\
 J_2 & 0 & 0 & 0 & 0 & 0 & 0 & 1 & -1 \\
 R & 0 & 0 & 0 & 0 & 0 & 2 & 0 & 2 \end{array} \qquad
 \text{CS levels:} \quad
 \begin{array}{c|cc|ccccc}
  & G_1 & G_2 & A_1 & A_2 & J_1 & J_2 & R \\\hline
 G_1 & 0 & -1 & \frac{1}{2} & 0 & 1 & 0 & -1 \\
 G_2 & -1 & 0 & 0 & \frac{1}{2} & 0 & 1 & -1 \\\hline
 A_1 & \frac{1}{2} & 0 & -\frac{1}{2} & 0 & 0 & 0 & \frac{1}{2} \\
 A_2 & 0 & \frac{1}{2} & 0 & -\frac{1}{2} & 0 & 0 & \frac{1}{2} \\
 J_1 & 1 & 0 & 0 & 0 & 0 & 0 & 0 \\
 J_2 &0 & 1 & 0 & 0 & 0 & 0 & 0 \\
 R & -1 & -1 & \frac{1}{2} & \frac{1}{2} & 0 & 0 & -2\end{array}
 \end{array}\right.
\ee
Notice that each $U(1)_i$ gauge group has both monopole and anti-monopole operators, denoted by $V_i^\pm$. However, due to the dynamical mixed Chern-Simons term, the monopoles of $U(1)_1$ are not gauge invariant under $U(1)_2$, and vice versa. We fix this by ``dressing'' the monopoles with chiral fields. We then find the four gauge-invariant operators
\be \begin{array}{c|c@{\;\;}c@{\;\;}c@{\;\;}c} &  V_1^+\phi_s & V_1^-\phi_t & V_2^+\phi_z &  V_2^-\phi_w \\\hline
A_1 & 0 & -1 & 0 & 1 \\
A_2 & 0 & 1 & 0 & -1 \\
J_1 & 1 & -1 & 0 & 0 \\
J_2 & 0 & 0 & 1 & -1 \\
R & 0 & 2 & 0 & 2 \end{array}\,.
\ee
with flavor charges as indicated. These operators correspond to the light-blue dots in Figure~\ref{fig:SL5-cluster}.

\begin{figure}[htb]
\hspace{-.5in}\includegraphics[width=7in]{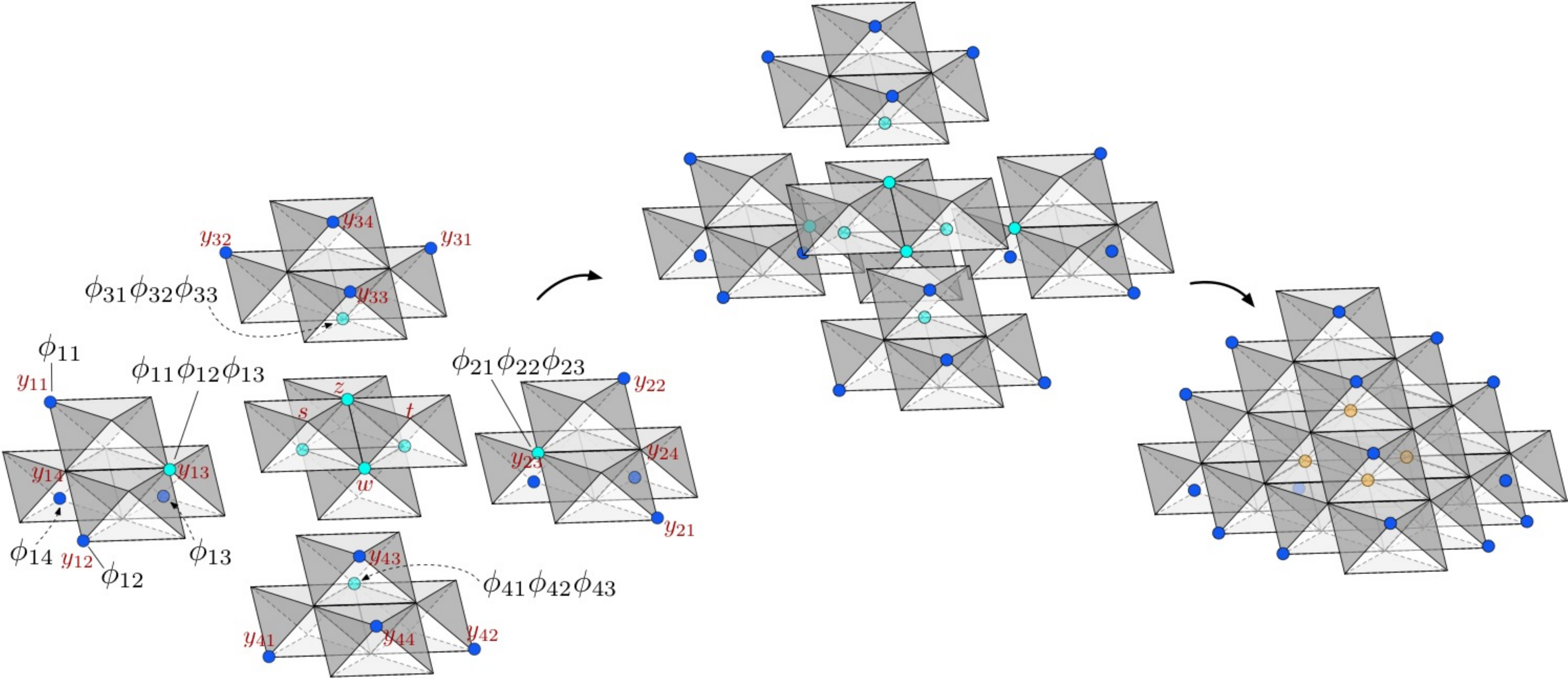}
\caption{Assembling the $SL(5)$ tetrahedron from a central cluster and 16 peripheral octahedra.}
\label{fig:SL5-assemble}
\end{figure}

To complete the $SL(5)$ tetrahedron, we couple the central cluster to sixteen additional octahedra on the periphery, as shown in Figure \ref{fig:SL5-assemble}. We label the new octahedra with parameters $\{y_{ij}\}_{i,j=1}^4$ as shown. The precise gauge theory we obtain depends on the final boundary polarization for the whole ensemble.

The simplest final polarization keeps $X_1$ and $X_2$ from $\Pi_{\rm cluster}$ as position coordinates. We can re-write each of them as a sum of four coordinates on the final boundary; for example $X_1 = Z+W = (Z+Y_{13}'+Y_{23}'')+(W+Y_{13}''+Y_{23}')+Y_{13}+Y_{23}-2\pi i$ and $X_2 = S+T = (S+Y_{33}'+Y_{43}'')+(T+Y_{33}''+Y_{43}')+Y_{33}+Y_{43}-2\pi i$.
For the fourteen remaining positions we take a subset of fourteen $Y_{ij}$ coordinates from octahedra on the periphery (due to the internal gluing constraints, these boundary coordinates are not all independent).
 Then the final gauge theory is simply the cluster theory \eqref{Tcluster}, coupled to 16 $T_\soct$ theories with chirals $\phi_{ij}$ ($1\leq i,j\leq 4$) via a superpotential,
\be T_5[{\Delta}] :\left\{
\begin{array}{c} \text{$T_{\rm cluster}\otimes (T_\soct)^{16}$, with $U(1)^{16}$ flavor symmetry preserved by the superpotential} \\[.2cm]
W = V_1^+\phi_s\phi_{11}\phi_{12}\phi_{13}+ V_1^-\phi_t \phi_{21}\phi_{22}\phi_{23} + V_2^+\phi_z\phi_{31}\phi_{32}\phi_{33}+ V_2^-\phi_w \phi_{41}\phi_{42}\phi_{43}\,.
\end{array} \right.
\ee
Notice that the four products of chirals $\phi_{i1}\phi_{i2}\phi_{i3}$ are associated with vertices on the groups of peripheral octahedra that become identified with the four distinguished vertices of the central cluster during the gluing. The four superpotential couplings combine the operators on the two sides of the gluing.

\subsection{The $K=3$ bipyramid}

We now come to the theory of the bipyramid for $K=3$. We expect that by splitting the bipyramid into ideal tetrahedra two different ways we will find two different theories $T_3[{\text{bip-2}}]$ and $T_3[{\text{bip-3}}]$ that are mirror symmetric, generalizing the $K=2$ examples of Section \ref{sec:23warmup}. We will check that this is the case by relating $T_3[{\text{bip-2}}]$ and $T_3[{\text{bip-3}}]$ via a sequence of fundamental mirror symmetries for octahedra.

\begin{figure}[htb]
\centering
\includegraphics[width=4in]{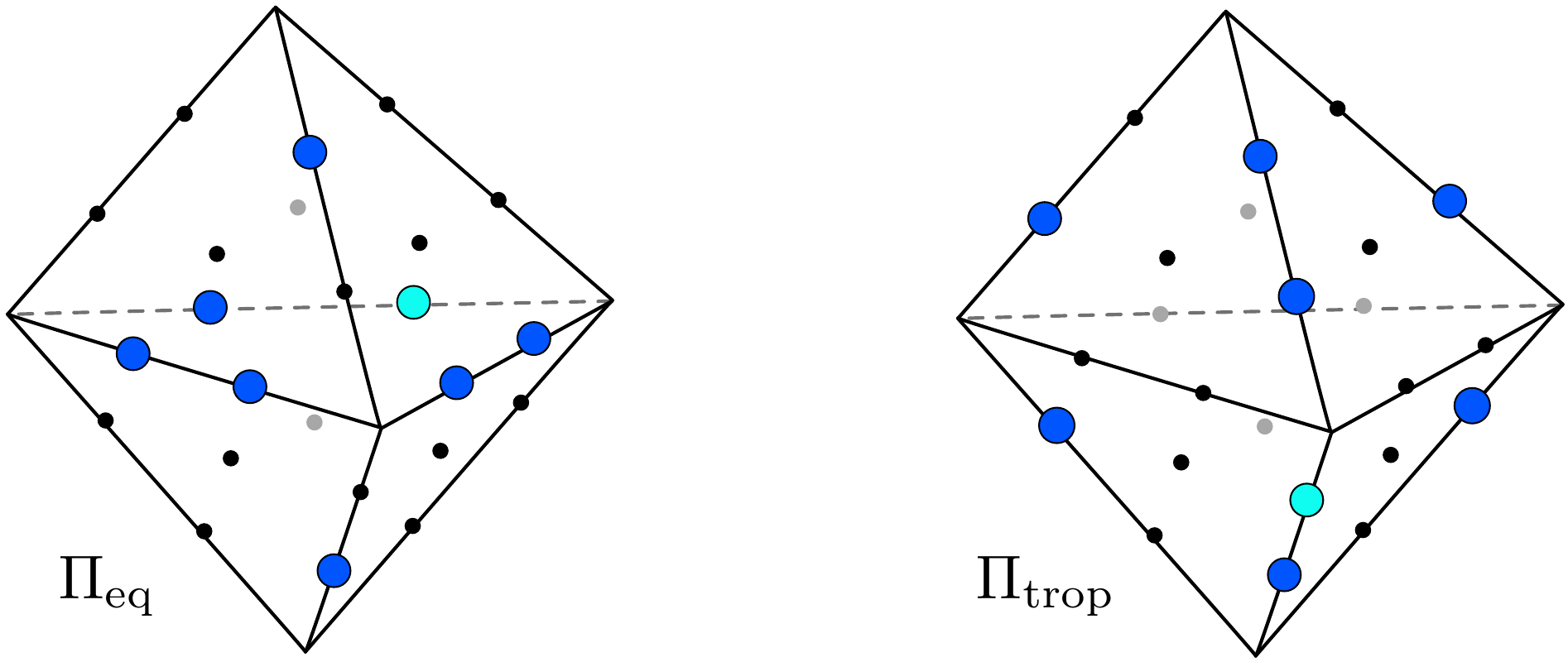}
\caption{The equatorial (left) and tropical (right) polarizations for the $K=3$ bipyramid. We only specify the seven position coordinates, indicating them by dark-blue dots on the $K$-triangulation of the boundary. The light-blue dots are lattice points/coordinates that commute with the positions (and so are assigned operators as well).}
\label{fig:23-3-pol}
\end{figure}

Of course, the $K=3$ bipyramid theories depend on a choice of polarization. There are two somewhat natural options, shown in Figure \ref{fig:23-3-pol}, and designated ``equatorial'' and ``tropical''. Some of the main features of the theories in their two polarizations can be summarized as:
\be \begin{array}{c|cccc|cccc} \text{pol.} & \multicolumn{4}{c|}{\text{2 tetrahedra, $T_3[{\text{bip-2}}]$}} & \multicolumn{4}{c}{\text{3 tetrahedra, $T_3[{\text{bip-3}}]$}} \\\hline
& \text{\;gauge\;} & \text{\;flavor\;} & \text{\# chirals\;} & \text{\;\# $W$ ops} 
& \text{\;gauge\;} & \text{\;flavor\;} & \text{\# chirals\;} & \text{\;\# $W$ ops}  \\\hline
\Pi_{\rm eq} & U(1)^3 & U(1)^7 & 8 & 1 & U(1)^3 & U(1)^7 & 12 & 5 \\
\Pi_{\rm trop} & \text{---} & U(1)^7 & 8 & 1 & U(1)^6 & U(1)^7 & 12 & 5 \\
\end{array}
\ee
Just like in the $K=2$ case, we switch from one polarization to the other by performing an $Sp(14,\Z)$ transformation simultaneously on both sides --- in this case, gauging three $U(1)$ flavor symmetries. In \emph{any} polarization, we will always have mirror symmetry
\be \label{MSbip3} T_3[{\text{bip-2}}]\simeq T_3[{\text{bip-3}}]\,. \ee

We begin by describing the theories on the two sides, and then explain how the mirror symmetry works.

\subsubsection{Two tetrahedra}

\begin{wrapfigure}{r}{1.9in}
%\begin{figure}
%\centering
\vspace{-.5cm}
\includegraphics[width=1.9in]{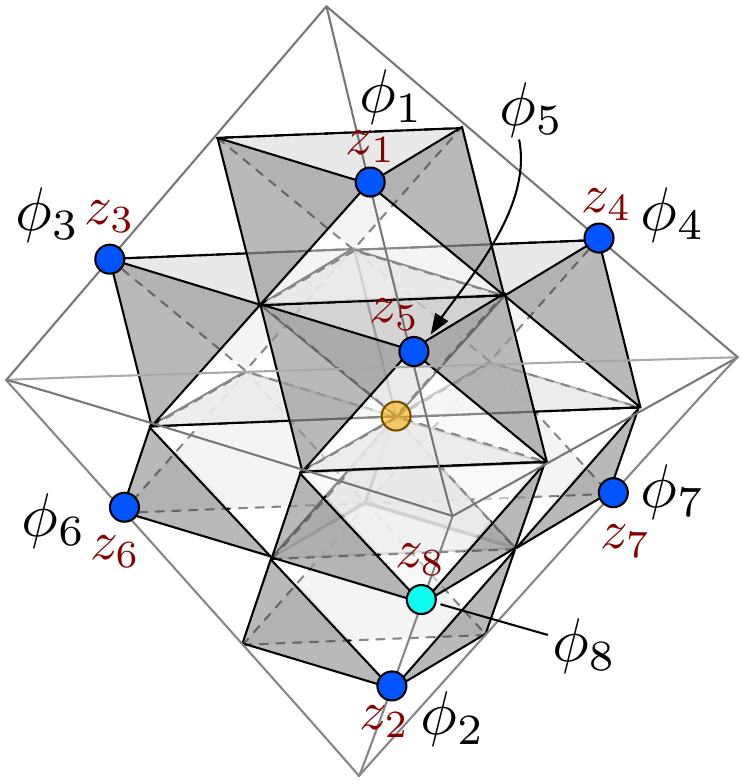}
\caption{Decomposition into eight octahedra, with $\Pi_{\rm trop}$.}
\label{fig:23-3-trop}
%\end{figure}
\end{wrapfigure}

Let us focus on the tropical polarization, since it makes the two-tetrahedron theory especially simple. The decomposition of the bipyramid into two tetrahedra, and subsequently into eight octahedra, is shown in Figure \ref{fig:23-3-trop}. It has a single internal lattice point, lying in the center of the identified faces of the octahedra.
Quite conveniently, we can label the octahedra so that their initial positions $Z_i$ coincide with the final positions $X_i$ in the polarization. Specifically, we can set $X_i=Z_i$ for $1\leq i \leq 7$. Moreover, the central gluing function $C=Z_3+Z_4+Z_5+Z_6+Z_7+Z_8-2\pi i$ is made of six octahedron positions (with no octahedron momenta).

This situation is exactly analogous to the XYZ model of Section \ref{sec:23warmup} or to the $K=4$ tetrahedron theory of Section \ref{sec:tet-theories}. We immediately recognize
\be \label{T-bip2}
T_3[{\text{bip-2},\Pi_{\rm trop}}]:\;\left\{
\begin{array}{l}
\text{Theory of eight chirals $\phi_i$ with a superpotential $W=\phi_3\phi_4\phi_5\phi_6\phi_7\phi_8$\,,}\\[.1cm]
\text{preserving $U(1)^7$ flavor symmetry.}
\end{array}\right.
\ee
The first two flavor symmetries $U(1)_1,U(1)_2$ rotate the decoupled chirals $\phi_1,\phi_2$; while the remaining five symmetries $\{U(1)_i\}_{i=3}^7$ rotate $\phi_i$ and $\phi_8$ with charges $(+1,-1)$.

We could proceed further and specify seven momenta $P_i$ for the polarization $\Pi_{\rm trop}$. They must be some linear combinations of boundary coordinates. However, the choice only affects background Chern-Simons levels, of which we won't keep track in this example.

If we used the equatorial polarization rather than the tropical polarization, the three pairs of octahedra that are glued together along a face would become three copies of $\CN=2$ SQED, as on the left of Figure \ref{fig:SQEDXYZ}. (In the tropical polarization, we could have said that these pairs of octahedra were three $\CN=4$ hypermultiplets, as on the left of Figure \ref{fig:SQED4hyper}.) Nothing happens with the two octahedra at the poles. Then we would get
\be \label{T-bip2-eq}
T_3[{\text{bip-2},\Pi_{\rm eq}}]:\;\left\{
\begin{array}{l}
\text{Three copies of SQED with chirals $\{q_i,\wt q_i\}_{i=1}^{3}$}\,,\\[.1cm]
\text{plus two additional decoupled chirals $\phi_1,\phi_2$\,;}\\[.1cm]
\text{$W=(q_1\wt q_1)(q_2\wt q_2)(q_3\wt q_3)$, preserving $U(1)^7$ flavor sym.}
\end{array}\right.
\ee
The superpotential is a product of the three mesons.

\subsubsection{Three tetrahedra}
\label{sec:23theory-3}

\begin{wrapfigure}{r}{1.8in}
\vspace{-25pt}
\hspace{.1in}\includegraphics[width=1.8in]{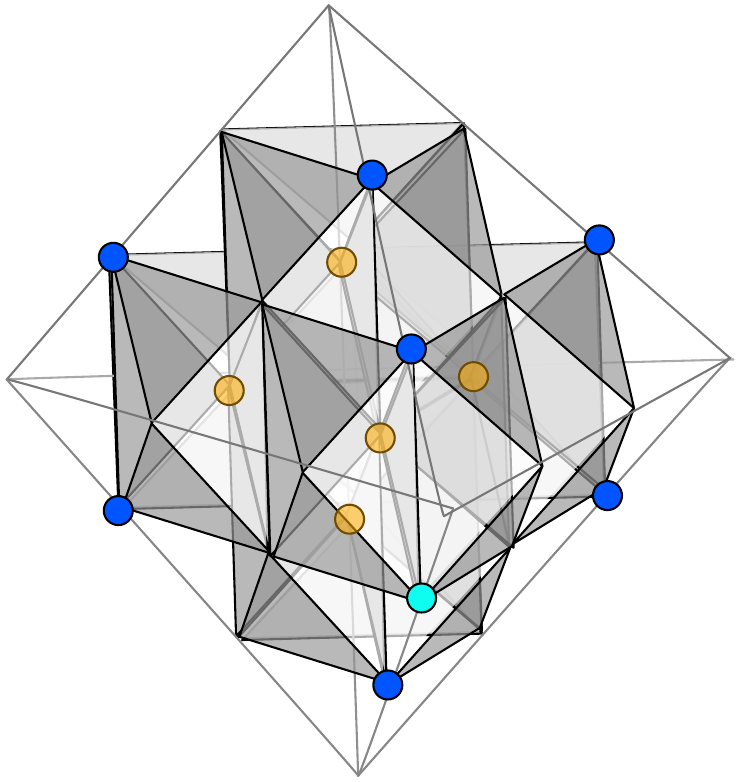}
\vspace{-25pt}
\caption{Decomposition into 12 octahedra, with $\Pi_{\rm trop}$.}
\vspace{-10pt}
\label{fig:23-3-trop3}
\end{wrapfigure}

The decomposition of the bipyramid into three tetrahedra, and subsequently into twelve octahedra, is shown in Figure \ref{fig:23-3-trop3}. Now there are five gluing constraints: three on the glued pairs of tetrahedron faces, and two on the internal edge.
The theory can be assembled in a few steps, much like the $K=5$ tetrahedron theory of Section \ref{sec:tet-theories}.
We will first build a central ``core'' out of six octahedra surrounding the central edge. Then we will couple the core to a halo of six octahedra around the equator (see Figure \ref{fig:23assemble}).

To construct the core, we put together two copies of $\CN=4$ SQED, exactly as in \eqref{TSQED4}, corresponding to the right side of~Figure \ref{fig:SQED4hyper}. We call the copies $T_{\rm SQED}^{\CN=4}$ and $T_{\rm SQED}^{\CN=4}{}'$. Each of them already has a cubic superpotential enforcing a gluing constraint, so there are no additional constraints in assembling them. However, we should change the polarization in order to allow more boundary points on the core to carry gauge-invariant operators, as indicated in Figure~\ref{fig:23Core}. 
We choose two new positions $\wt X_1 = X_-+X_+'$ and $\wt X_2 = P_--P_+'$ (and conjugate momenta $\wt P_1=P_-,\, \wt P_2= X_+'$) in addition to the old positions $X_+$ and $X_-'$ (and conjugate momenta $P_+,\,P_-'$). In matrix form, we have
\be \begin{pmatrix} \wt X_1 \\\wt X_2 \\ \wt P_1 \\ \wt P_2 \end{pmatrix} = \begin{pmatrix} 1 & 1 & 0 & 0 \\
   0 & 0 & 1 & -1 \\
    0 & 0 & 1 & 0 \\
    0 & 1 & 0 & 0
   \end{pmatrix}
\begin{pmatrix} X_- \\ X_+' \\ P_- \\ P_+' \end{pmatrix} =: g\cdot \begin{pmatrix} X_- \\ X_+' \\ P_- \\ P_+' \end{pmatrix}\,.
\ee
This amounts to gauging the anti-diagonal $U(1)_{\ol d}$ of the global symmetry group $U(1)_-\times U(1)_+'$. The operators $V_+$ and $V_-'$ are clearly uncharged under $U(1)_{\ol d}$, as is the product $\eta = V_-V_+'$, so they survive the gauging. In addition, we gain a new pair $\mu_\pm$ of monopole/anti-monopole operators%
\footnote{The anti-monopole $\mu_-$ for $U(1)_{\ol d}$ turns out not to be invariant under the old gauge symmetries of the two copies of SQED, and must be ``dressed'' by fundamental chirals to fix this, $\mu_-\to \mu_-\phi\phi'$. 
The situation is similar to that for monopoles of the $SL(5)$ tetrahedron theory in Section \ref{sec:SL5tet}. In the tables below, and in the figure, it is the dressed, gauge-invariant monopole that is denoted $\mu_-$.} %
for $U(1)_{\ol d}$, charged under a new topological symmetry $U(1)_J$. They are associated with the boundary coordinates shown in Figure \ref{fig:23Core}. In summary, the core theory theory is two copies of SQED with a new polarization, $T_{\rm core} = g\circ( T_{\rm SQED}^{\CN=4}\otimes T_{\rm SQED}^{\CN=4}{}')$, and explicitly
\be \label{T-core}
\hspace{-.3in} T_{\rm core}: \left\{\begin{array}{l}
 \text{$U(1)^3$ gauge theory, with $U(1)_+\times U(1)_-'\times U(1)_d\times U(1)_J$ symmetry,} \\
 \text{six chirals, and superpotential $W = \wt\phi\varphi\phi + \wt\phi'\varphi'\phi'$}\\[.2cm]
 \text{charges:}\;\begin{array}{c|cccccc|ccccc}
 & \phi & \wt\phi & \varphi & \phi' & \wt\phi' & \varphi'
   & V_+ & V_-' & \eta & \mu_+ & \mu_- \\\hline
 G & 1 & -1 & 0 & 0 & 0 & 0 & 0 & 0 & 0 & 0 & 0 \\
 G' & 0 & 0 & 0 & 1 & -1 & 0 & 0 & 0 & 0 & 0 & 0 \\
 G_{\ol d} & 1 & 0 & -1 & -1 & 0 & 1 & 0 & 0 & 0 & 0 & 0  \\\hline
 X_+ & -1 & 0 & 1 & 0 & 0 & 0 & 1 & 0 & 0 & 0 & -1 \\
 X_-'& 0 & 0 & 0 & -1 & 0 & 1 & 0 & 1 & 0 & 0 & -1 \\
 D & -1 & 0 & 1 & 0 & 0 & 0 & 0 & 0 & 1 & 0 & -1 \\
 J & 0 & 0 & 0 & 0 & 0 & 0 & 0 & 0 & 0 & 1 & -1 \\
 R & 2 & 0 & 0 & 2 & 0 & 0 & 0 & 0 & 0 & 0 & 2 
 \end{array} \quad
 \text{CS:}\; \begin{array}{c|ccc|ccccc}
 & G & G' & G_{\ol d} & X_+ & X_-' & D & J & R \\\hline
 G& 0 & 0 & \frac{1}{2} & \frac{1}{2} & 0 & -\frac{1}{2} & 0 & 0 \\
 G' &0 & 0 & \frac{1}{2} & 0 & -\frac{1}{2} & 0 & 0 & 0 \\
 G_{\ol d} &\frac{1}{2} & \frac{1}{2} & 0 & 0 & 0 & 0 & 1 & 0 \\\hline
 X_+&\frac{1}{2} & 0 & 0 & 0 & 0 & 0 & 0 & 0 \\
X_-' & 0 & -\frac{1}{2} & 0 & 0 & 0 & 0 & 0 & 0 \\
D & -\frac{1}{2} & 0 & 0 & 0 & 0 & 0 & 0 & 0 \\
 J &0 & 0 & 1 & 0 & 0 & 0 & 0 & 0 \\
 R &0 & 0 & 0 & 0 & 0 & 0 & 0 & * \end{array}
\end{array}\right.
\ee
Here we have denoted by $U(1)_d$ (or `$D$') the diagonal of $U(1)_-\times U(1)_+'$ that is preserved as a flavor symmetry.

\begin{figure}[tbh]
\centering
\includegraphics[width=2.8in]{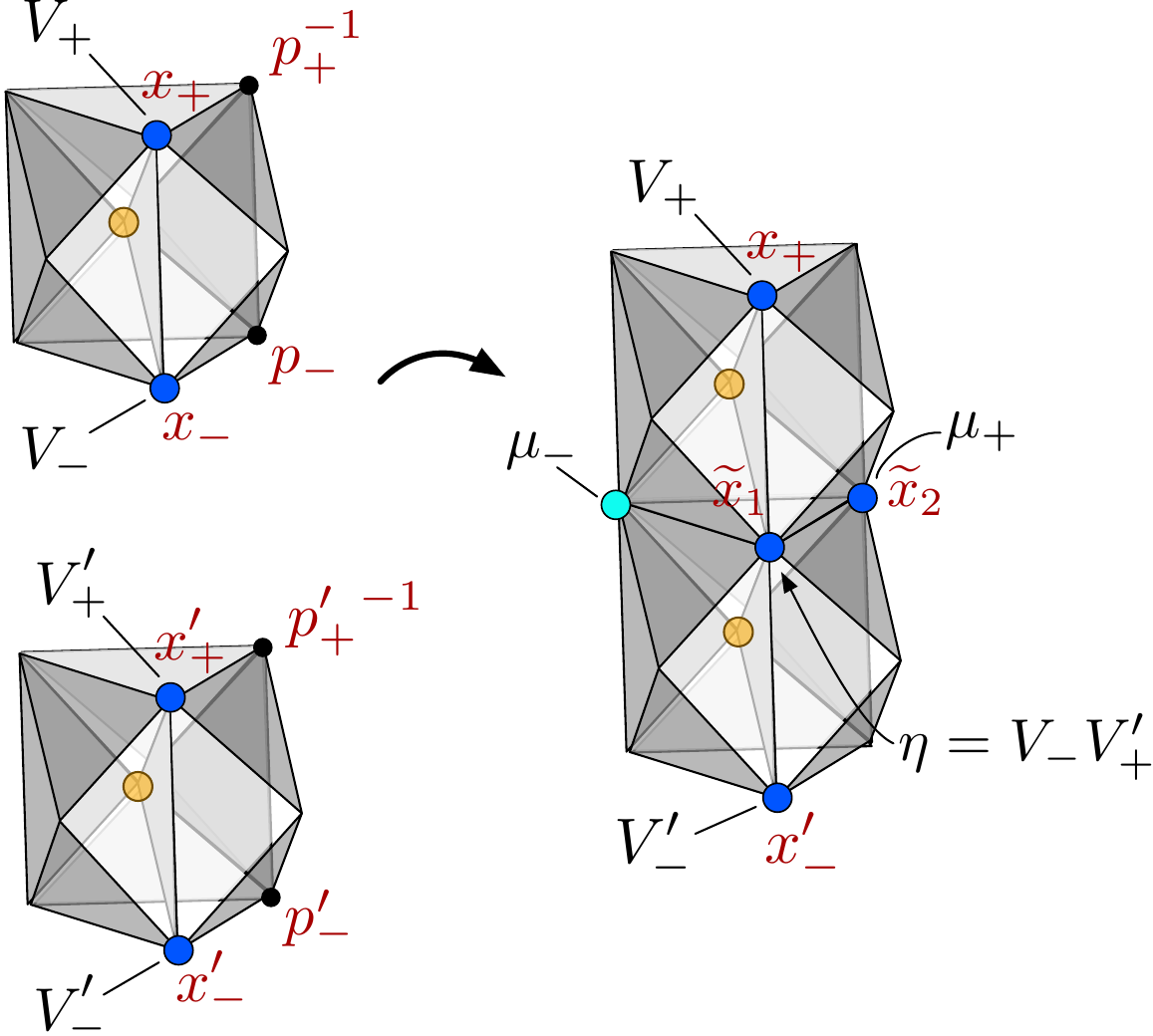}
\caption{The core of the $SL(3)$ bypyramid, built from two copies of $\CN=4$ SQED.}
\label{fig:23Core}
\end{figure}

In order to complete the bipyramid, we must add a ``halo'' of six octahedra to the core of Figure \ref{fig:23Core}. This is shown in Figure \ref{fig:23assemble}. Three new gluing constraints arise, and each of the operators $\eta,\mu_+,\mu_-$ from the core participates in one of three new superpotential couplings.

\begin{figure}[htb]
\centering
\includegraphics[width=5.4in]{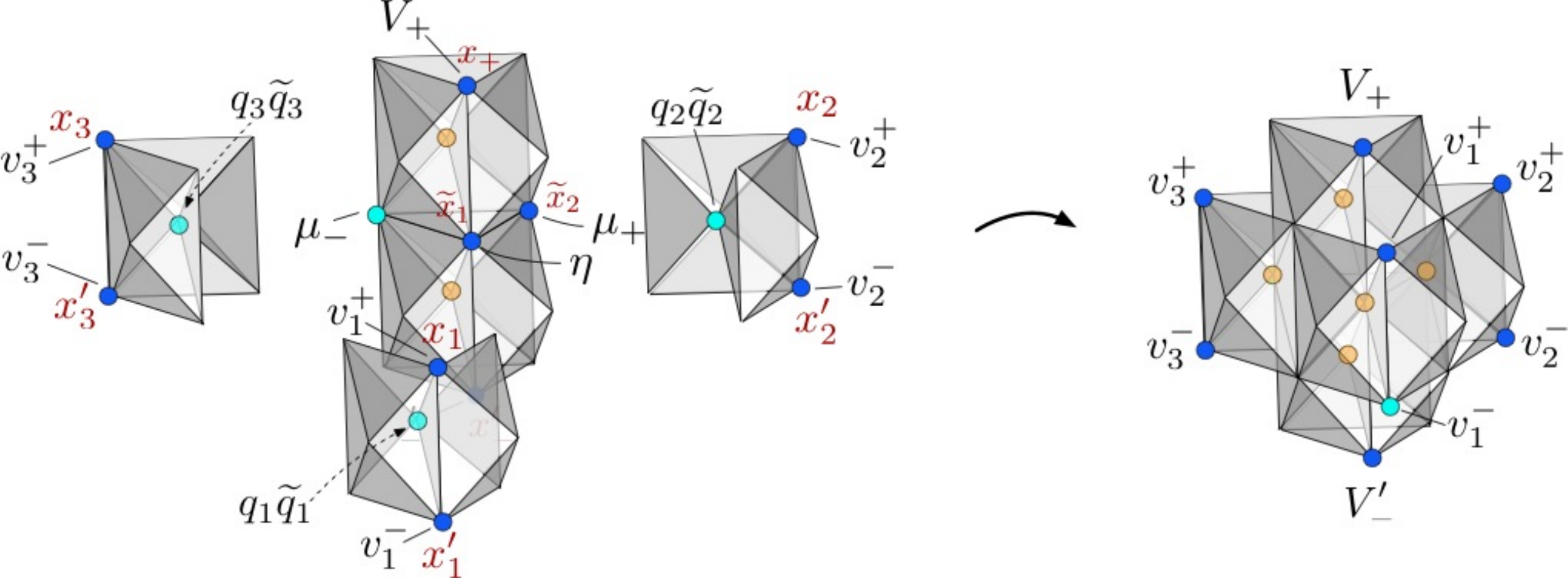}
\caption{Assembling the $SL(3)$ bipyramid from a core and a halo.}
\label{fig:23assemble}
\end{figure}

For the tropical polarization on the final boundary, we should represent the halo as three copies of $\CN=2$ SQED (\cf\ \eqref{TSQED2} and Figure \ref{fig:SQEDXYZ}). The meson operators $q_i\wt q_i$ in the three copies of SQED couple with the three operators $\eta,\mu_+,\mu_-$ of the core in a superpotential. The theory becomes
\be \label{T-bip3}
T_3[{\text{bip-3}}]: \left\{\begin{array}{l}
\text{$T_{\rm core}\otimes T_{\rm halo} \simeq \big[g\circ(T_{\rm SQED}^{\CN=4}\otimes T_{\rm SQED}^{\CN=4}{}\!')\big]\otimes (T_{\rm SQED}^{\CN=2})^3$,}\\[.2cm]
\text{$U(1)^6$ gauge theory with 12 chirals and the $U(1)^7$ flavor symmetry preserved by} \\[.1cm]
\text{\qquad $W = \wt\phi\varphi\phi + \wt\phi'\varphi'\phi' + q_1\wt q_1\eta + q_2\wt q_2\mu_++ q_3\wt q_3\mu_-$\,.}
\end{array}\right.
\ee
The $U(1)^7$ flavor symmetry comes from $U(1)^4$ from the core and $U(1)^6$ from the halo, minus the three symmetries broken by the core-halo couplings. The positions in the final tropical polarization consist of $X_+,X_-'$ from the core and any five of the six $X_i,X_i'$ from the halo (after gluing, the six halo positions are not all independent). The theory has eight gauge-invariant operators associated with the external light- and dark-blue dots on the bipyramid: the monopoles $V_+,V_-'$ of the core and the monopoles $v^+_i,v^-_i$ ($i=1,2,3$) for the three copies of SQED in the halo.

If instead we wanted to reproduce the equatorial polarization on the final boundary, we would just use a different polarization for the halo. In fact, in the equatorial polarization, the halo just becomes six free chirals (or three $\CN=4$ hypers). Three pairs of these chirals combine with the three core operators $\eta,\mu_+,\mu_-$ to form three cubic superpotential couplings. We obtain a $U(1)^3$ gauge theory with a total of 12 fundamental chirals, five superpotential couplings, and a preserved $U(1)^7$ flavor symmetry.

\subsubsection{Mirror symmetry}

We have found an extremely simple bipyramid theory $T_3[{\text{bip-2}}]$, and a somewhat more complicated one $T_3[{\text{bip-3}}]$.
To finish, let us proceed through the chain of fundamental mirror symmetries that link these two theories, demonstrating their IR equivalence. We simply follow the chain of 2--3 moves on octahedra shown in Figure \ref{fig:23-3}. We will be somewhat loose on the details, since the construction (ultimately) is completely systematic.

\begin{figure}[htb]
\hspace{-.5in}
\includegraphics[width=7in]{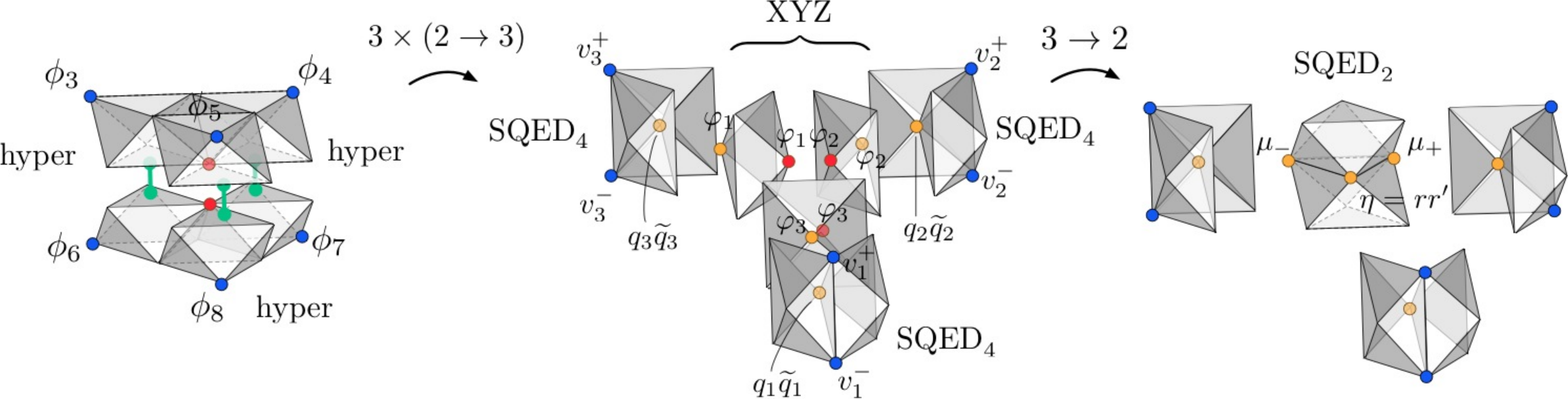}
\caption{The first two steps in going from two $SL(3)$ tetrahedra to three.}
\label{fig:23theorymovesI}
\end{figure}

We work in the tropical polarization, and start with the theory $T_3[{\text{bip-2}}]$ \eqref{T-bip2} built from two tetrahedra and eight octahedra. Consider the three pairs of chirals $(\phi_3,\phi_6)$, $(\phi_4,\phi_7)$, $(\phi_5,\phi_8)$ corresponding to the three pairs of octahedra that touch across the big internal face.  They represent three $\CN=4$ hypers, coupled by an $\CN=2$ superpotential
\be W = \phi_3\phi_4\phi_5\phi_6\phi_7\phi_8\,. \label{W6}\ee
The first set of three $2\to 3$ moves dualizes the hypers to three copies of $\CN=4$ SQED (Figure~\ref{fig:23theorymovesI}). Let us denote the charged and neutral chirals in the $\CN=4$ SQED's as $(q_i,\wt q_i,\varphi_i)_{i=1}^3$.
 Each hyper is replaced with a pair of SQED monopole/anti-monopole operators, \eg\ $(\phi_3,\phi_6)\to (v_3^+,v_3^-)$; while the three quadratic operators $(\phi_3\phi_6,\phi_4\phi_7,\phi_5\phi_8)$ are replaced by the neutral chirals $(\varphi_1,\varphi_2,\varphi_3)$. Thus the superpotential becomes
\be W = \varphi_1\varphi_2\varphi_3 + \sum_{i=1}^3 \wt q_i\varphi_iq_i\,. \label{W4} \ee
(We have added three $\CN=4$ SQED superpotentials to \eqref{W6}.)

Now we notice that there is a copy of the XYZ model embedded in the theory: it involves the three neutral chirals. We dualize it to a copy of $\CN=2$ SQED, performing the next $3\to 2$ move (also in Figure \ref{fig:23theorymovesI}). Let us call the charged chirals of this SQED $r$ and $r'$. One of the XYZ fields (say $\varphi_3$) is replaced by the meson $\eta=rr'$, while $\varphi_1$ and $\varphi_2$ are replaced by a monopole/anti-monopole pair $\mu_+,\mu_-$. The superpotential is now
\be W = \wt q_1q_1\eta + \wt q_2q_2\mu_+ + \wt q_3q_3\mu_-\,.\ee
(The XYZ superpotential has been eliminated in the duality.) The gauge group is $U(1)_{\ol d}$, acting on $(r,r')$.

\begin{figure}[htb]
\centering
\includegraphics[width=5in]{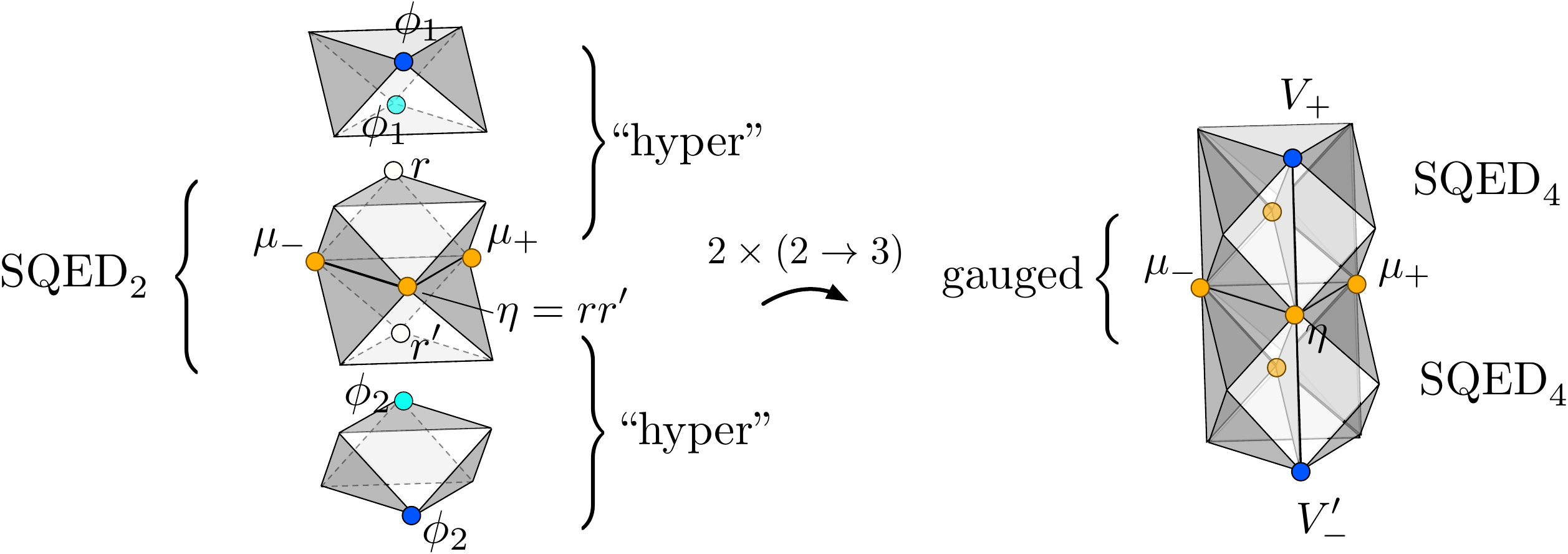}
\caption{The final step in going from two $SL(3)$ tetrahedra to three: forming the core.}
\label{fig:23theorymovesII}
\end{figure}

The halo of six octahedra is now complete, and we just need to form the core. It requires the last set of two $2\to 3$ moves, shown in Figure \ref{fig:23theorymovesII}. We bring in the two octahedra at the poles of the bipyramid, and think of each pair $(r,\phi_1)$ and $(r,\phi_2)$ as an $\CN=4$ hyper. We want to dualize these pairs to $\CN=4$ SQED. This is subtle: the chirals $r,r'$ are \emph{not} gauge-invariant, and we are effectively assuming that it is ok to interchange an order of limits (of gaugings) as we flow to the IR. With this assumption, the result of the duality is the core theory of \eqref{T-core}: two copies of $\CN=4$ SQED, for which a combination of axial and topological symmetries --- identified with $U(1)_{\ol d}$ --- has been subsequently gauged.

The operators $\mu_+,\mu_-,\eta$ all survive as monopoles of the core theory, as described in Section \ref{sec:23theory-3}. In addition, there are two new monopoles $V_+$ and $V_-'$ coming from the copies of $\CN=4$ SQED. The gauge group has become $U(1)^3$, comprising the two $\CN=4$ SQED gauge groups and $U(1)_{\ol d}$. The superpotential gets two cubic $\CN=4$ SQED operators added to it:
\be W =  \wt q_1q_1\eta + \wt q_2q_2\mu_+ + \wt q_3q_3\mu_- + \wt\phi\varphi\phi+\wt\phi'\varphi'\phi'\,.\ee
Altogether, we recover the coupled core-halo system that describes $T^{(3)}_{\text{bip-3}}$.

%%%%%%%%%%%%%%%%%%%%%%%%%%%%%%%%%%%%%%%%%%%%%%%%%%%%%%%%%%%%%%%%%%%%%%%%%%%%
%%%%%%%%%%%%%%%%%%%%%%%%%%%%%%%%%%%%%%%%%%%%%%%%%%%%%%%%%%%%%%%%%%%%%%%%%%%%
%%%%%%%%%%%%%%%%%%%%%%%%%%%%%%%%%%%%%%%%%%%%%%%%%%%%%%%%%%%%%%%%%%%%%%%%%%%%

\section{Knot complement theories}
\label{sec:Tcusp}

We have finally arrived at the theories $T_K[M]$ associated with knot complements $M=S^3\bs \CK$. More generally, we can take $M=\ol M\bs \CK$, where $\ol M$ is any closed 3-manifold. Recall from Section \ref{sec:combi} that in the ideal triangulation of a knot complement, all big faces of tetrahedra are completely glued, leaving behind a small torus boundary on which all the tetrahedron vertices sit.

Physically, the theory $T_K[M]$ should correspond to wrapping $K$ M5-branes on $\ol M\times \R^3$, in the ambient geometry $T^*\ol M\times \R^3\times \C$. We then add $K$ additional ``probe'' branes along $N^*\CK\times\R^3$, where $N^*\CK \subset T^*M$ is the conormal bundle of the knot.%
\footnote{Constructions of this type, which use intersecting branes to embed a knot-complement geometry in M-theory, are quite common. They were introduced in \cite{OV}, and have made many recent appearances, including \cite{Wfiveknots} as well as \cite{DGG,DGG-index}.} %
This is the M-theory version of creating a small torus boundary --- specifically one with boundary conditions relevant for parametrizing irreducible flat connections on $M$.
The setup preserves four supercharges and flows to an effective $\CN=2$ theory on $\R^3$ in the infrared. Alternatively, we may start with the $A_{K-1}$ (2,0) theory in six dimensions and compactify it on $\ol M$ (with a suitable topological twist), adding a codimension-two defect along $\CK$. In the language of \cite{Gaiotto-dualities} (see also \cite{GW-surface, GW-Sduality}), the codimension-two defect is of the type corresponding to a ``full'' or ``maximal'' puncture. From either construction, it is straightforward to deduce that $T_K[M]$ should be an SCFT with $SU(K)$ flavor symmetry.%
\footnote{The relation between theories $T[M]$, compactifications of the 6d (2,0) theory, and M-theory is further explored in \cite{DGV-hybrid}.
} %

As discussed in Section \ref{sec:combi} (and Appendix \ref{app:symp}), a knot-complement theory can systematically be defined by extracting symplectic gluing data $(g,\sigma)$ from an ideal triangulation of $M$, and the $K$-decomposition thereof. We will give some examples below in Sections \ref{sec:trefoil}--\ref{sec:figure-eight}. The complexity of the theory does grow quite quickly with $K$. With $N$ tetrahedra, our description of $T_K[M]$ will contain $\frac16 NK(K^2-1)$ chirals, and potentially a very large abelian gauge group. The superpotential will \emph{also} contain $\frac16 NK(K^2-1)$ operators --- corresponding to the internal lattice points of the octahedral decomposition --- of which $K-1$ do not break flavor symmetries. Nevertheless, despite this apparent complexity, we will be able to use combinatorics to deduce (and/or verify) several interesting general properties of knot-complement theories. For example, we can see immediately from the dimension of the boundary phase space $\CP_K({T^2})$ that $T_K[M]$ has $U(1)^{K-1}$ flavor symmetry, which may be enhanced to $SU(K)$.

One particularly interesting question to ask is how theories $T_K[M]$ depend on $K$ as $K$ becomes large. Simply counting chiral multiplets in a UV description leads to the naive guess that degrees of freedom scale like $K(K^2-1)$. We can make this somewhat more precise by (conjecturally) relating degrees of freedom to the volume of flat $PGL(K,\C)$-connections on $M$ --- as given by the imaginary part of the Chern-Simons functional, $\Im \int_M \Tr\big(\CA d\CA+\frac23\CA^3\big)$. In turn, when $M$ is hyperbolic, we will be able to see combinatorially that there is a flat $PGL(K,\C)$-connection whose volume is exactly $\frac16 K(K^2-1)$ times the hyperbolic volume of $M$. This adds some further evidence that the naive guess for the scaling is correct.

Let us observe that this scaling of degrees of freedom appears to be in very good agreement with results from the AdS/CFT correspondence.
If $M$ is a closed manifold, the gravity solutions dual to theories $T_K[M]$ should be given by the AdS$_4$ compactifications of M-theory that were obtained in~\cite{GKW-5branes} by considering the near-horizon limit of M5-branes wrapping a Special Lagrangian 3-cycle. 
One explicit solution is known, in which the internal 7-manifold is an $S^4$-fibration over a closed hyperbolic 3-manifold $M$
(this is in fact the 11d uplift of a 7d supergravity solution originally found in~\cite{PerniciSezgin}).
The free energy $\CF = - \log |\CZ |$ can be determined from the computation of the effective 4d Newton constant using standard AdS/CFT techniques \cite{HenningsonSkenderis, EJM-surface} and is expected to be given essentially by $K^3\, \text{Vol}(M)$, where $\text{Vol}(M)$ is the hyperbolic volume.
In the particular case of knot complements, one would have to enhance the setup slightly by introducing the additional probe branes that intersect $M$ along the knot $\CK$, much as was done for 4d $\CN=2$ theories of class $\CS$ in \cite{GaiottoMaldacena}. However, as long as the knot complement $M$ has nonzero hyperbolic volume, the $K^3\text{Vol}(M)$ scaling should remain robust. 
Tantalizingly, a subleading correction to this result modifying $K^3\to K(K^2-1)$ was predicted by the authors of \cite{HMM-anomalies} from the requirement of anomaly cancellation for M5-branes.
This appears to match the exact scaling of volumes of flat connections mentioned above.

We remark that all the ideas and constructions in this section apply in a completely straightforward way to ``link complements,'' or manifolds $M$ with multiple small torus boundaries. We focus on knot complements for simplicity.

%%%%%%%%%%%%%%%%%%%%%%%%%%%%%%%%%%%%%%%%%%%%%%%%%%%%%%%%%%%%%%%%%%%%%%%%%%%%%
%%%%%%%%%%%%%%%%%%%%%%%%%%%%%%%%%%%%%%%%%%%%%%%%%%%%%%%%%%%%%%%%%%%%%%%%%%%%%
%%%%%%%%%%%%%%%%%%%%%%%%%%%%%%%%%%%%%%%%%%%%%%%%%%%%%%%%%%%%%%%%%%%%%%%%%%%%%

\subsection{Flavor symmetry and marginal operators of $T_K[M]$}
\label{sec:marginal}

A knot complement theory $T_K[M]$ that is constructed combinatorially from an ideal triangulation of $M$ will always have flavor symmetry that is at least $U(1)^{K-1}$. This is clear from the dimension of the boundary phase space, $\dim_\C \CP_K({T^2})=2(K-1)$ (Sections \ref{sec:sympsum} and \ref{sec:torusbdy}). We can say quite a bit more about $T_K[M]$, however, by looking at how the combinatorial gluing actually happens.

Recall from Sections \ref{sec:torusbdy} and \ref{sec:path-coords} that an ideal triangulation of a knot complement $M$ with $N$ big tetrahedra, has exactly $N$ internal edges as well. There are also $2N$ distinct internal faces. Therefore, after decomposing further into octahedra, the total number of internal gluing constraints $C_k$ (\ie\ lattice points) will be
\be (K-1)(\text{\# edges})+ {K-1\choose 2}(\text{\# faces})+{K-1\choose 3}(\text{\# faces}) = \frac16 N K(K^2-1)\,.\ee
This is equal to the number of octahedra. However, not all of these gluing constraints are independent. As discussed in Section \ref{sec:C-combi} (\cf\ Figure \ref{fig:cusp-constraint}), $K-1$ linear combinations of them will vanish automatically. Therefore, the final boundary phase space $\CP_K({T^2})$ can be written as a symplectic reduction of the $\frac13 N K(K^2-1)$-dimensional product octahedron phase space using any independent subset of the $C_k$ as moment maps.

Now consider what this means in terms of gauge theory. We build $T_K[M]$ by starting with $\frac16 N K(K^2-1)$ octahedron theories $T_\soct$, applying a symplectic transformation $(g,\sigma)$ to obtain a theory $\wt T$ with $U(1)^{\frac16 N K(K^2-1)}$ flavor symmetry, and then adding operators $\CO_k$ to the superpotential to break flavor symmetries. We should add one $\CO_k$ for every gluing constraint $C_k$. However, we find that after adding operators $\CO_k$ for any independent subset of $\frac16 N K(K^2-1)-(K-1)$ gluing constraints, we are done. The remaining $(K-1)$ operators can be added, but they do not break any additional flavor symmetries. Moreover, their R-charge is necessarily set equal to two. Thus, they are marginal operators that do not break any flavor symmetry, and by the arguments of \cite{GKSTW} they must be exactly marginal.

This set of $(K-1)$ exactly marginal operators provides a strong indication that the remaining $U(1)^{K-1}$ flavor symmetry of $T_K[M]$ can be enhanced to $SU(K)$ at some special point $p^*$ in the space of marginal deformations. Indeed, at such a point $p^*$, an additional $K(K-1)$ marginal operators must arise, paired with the currents of $SU(K)/U(1)^{K-1}$ that are restored \cite{GKSTW} (see also \cite{DG-E7} for similar arguments). Altogether, the $K(K-1)+(K-1) = K^2-1$ marginal operators at $p^*$ beautifully fill out the irreducible adjoint representation of $SU(K)$.

Both the marginal operators that are lifted off of $p^*$ and the $K-1$ exactly marginal operators that survive can be understood from an M-theory perspective. The M-theory explanation appears in \cite{DGV-hybrid}.

We can also address the issue of polarization. As usual, for any $M$ with a boundary, the theory $T_K[M]$ will depend on a polarization $\Pi$. In the case of a knot complement, however, the choice is basically canonical. Indeed, by representing a manifold $M$ with torus boundary as the complement of a knot $\CK$ in a specific closed manifold $\ol M$ (which is certainly necessary from the M-theory perspective, where the $\CK$ arises from intersecting branes or defects), we automatically identify a ``meridian'' cycle $\mu$ on the torus boundary. It is the cycle that forms an infinitesimally small loop linking $\CK$ in $\ol M$, and would be contractible if we removed $\CK$. The eigenvalues of the meridian holonomy then become canonical position coordinates for $\Pi$.

If moreover $M$ is a knot complement in $\ol M=S^3$ or any homology sphere, then there is also a canonical longitude cycle $\lambda$ dual to the meridian on the boundary. It is the unique cycle that is null-homologous in $M$. The  eigenvalues of the longitude holonomy become canonical momentum coordinates, as explained in Sections \ref{sec:torusbdy} and \ref{sec:sympsum}. However, we often do not require a canonical choice of momenta, since they only affect background Chern-Simons couplings.

Note that the $K-1$ real mass deformations of $T_K[M]$ coming from the $U(1)^{K-1}$ flavor symmetry correspond directly to the meridian eigenvalues of a flat connection on $M$ --- \ie\ the position coordinates for the boundary phase space, in the canonical polarization. More precisely, after compactifying $T_K[M]$ on any combination of circles or spheres, the real masses get complexified, and the complex masses correspond to the complex meridian eigenvalues. If we want to keep $T_K[M]$ at a conformal point (that is, on the conformal manifold parametrized by marginal deformations), we should turn all mass deformations off. This means making all eigenvalues trivial, setting
\be \label{hol-parab}
{\rm Hol}(\mu) \sim \begin{pmatrix}
   1&0&0& &0&0\\
   1&1&0& \cdots &0&0\\
   0&1&1&&0&0 \\
   \vdots &&&\ddots&& \\
   %0&0&0&&1&0\\
   0&0&0&&1&1
\end{pmatrix}
\ee
up to conjugation.
This is not a trivial boundary condition: in order for the holonomy of a flat connection to be irreducible (\eg\ as discussed in Section \ref{sec:torusbdy}) the meridian holonomy must contain a single Jordan block. The matrix \eqref{hol-parab} is called the maximal parabolic element of $PGL(K,\C)$, and can be described nicely as $\rho_{[K]}\cdot\left(\begin{smallmatrix}1&0\\1&1 \end{smallmatrix}\right)$, where $\rho_{[K]}:PGL(2,\C)\to PGL(K,\C)$ is the $K$-dimensional irreducible representation, or principal embedding.

%%%%%%%%%%%%%%%%%%%%%%%%%%%%%%%%%%%%%%%%%%%%%%%%%%%%%%%%%%%%%%%%%%%%%%%%%%%%%
%%%%%%%%%%%%%%%%%%%%%%%%%%%%%%%%%%%%%%%%%%%%%%%%%%%%%%%%%%%%%%%%%%%%%%%%%%%%%
%%%%%%%%%%%%%%%%%%%%%%%%%%%%%%%%%%%%%%%%%%%%%%%%%%%%%%%%%%%%%%%%%%%%%%%%%%%%%

\subsection{$K^3$ scaling in $T_K[M]$}

We would like to sketch an argument indicating that the degrees of freedom of $T_K[M]$ at a conformal point scale like $\frac16K(K^2-1)$ when $K$ is large, and $M$ is hyperbolic. The argument has two parts. First we observe that, when $M$ is hyperbolic, there is a flat $PGL(K,\C)$-connection on $M$ with boundary conditions \eqref{hol-parab} whose volume is $\frac 16K(K^2-1)$ times the hyperbolic volume ${\rm Vol}(M)$. Second, we try to relate the free energy of $T_K[M]$ on a sphere to the volume of this flat $PGL(K,\C)$-connection. The second part of the argument involves several nontrivial assumptions that we hope to investigate in subsequent work.

\subsubsection{$PGL(K)$ volume from octahedra}

The complex volume of a framed flat $PGL(K,\C)$-connection $\CA$ on a knot complement $M$ was discussed in Section \ref{sec3.3}. It can be interpreted as the value of the holomorphic Chern-Simons functional,
\be \label{VolCS}
\CV(\CA) = i({\rm Vol}(\CA)+i{\rm CS}(\CA)) \sim \int_{M}\Tr\big[\CA d\CA+\tfrac23\CA^3\big]\,,\ee
with appropriate boundary conditions. When a connection has unipotent boundary holonomy, the real part is defined modulo $(2\pi i)^2\Z$ while the imaginary part is a well-defined real ``volume.'' When $M$ is a hyperbolic 3-manifold, the unique complete hyperbolic metric can be re-written as a flat $PGL(2,\C)$-connection $\CA_{\rm geom}$. This geometric flat connection has parabolic holonomy at the meridian, ${\rm Hol}(\mu) \sim \left(\begin{smallmatrix}1&0\\1&1 \end{smallmatrix}\right)$, and its real volume equals the hyperbolic volume of $M$,
\be {\rm Vol}(\CA_{\rm geom}) = {\rm Vol}(M)\,. \ee
In general there are a finite number of flat $PGL(2,\C)$-connections with parabolic holonomy, but the geometric one is distinguished as the connection with the \emph{largest} volume.%
\footnote{For physics discussions of the relation between hyperbolic geometry and complex Chern-Simons theory, see \cite{Witten-gravCS, gukov-2003, DGLZ}. Classic mathematics references are \cite{thurston-1980, NZ, yoshida-1985}.}

On a hyperbolic manifold $M$ there is also a distinguished conjugate connection $\CA_{\rm conj}=\ol\CA_{\rm geom}$. Its real volume is clearly $-{\rm Vol}(\CA_{\rm geom})$, and it has the \emph{smallest} volume of all connections with parabolic holonomy.

Given the geometric $PGL(2,\C)$-connection $\CA_{\rm geom}$, we can always construct an irreducible $PGL(K,\C)$-connection $\CA_{\rm geom}^{(K)}$ on $M$ by using the irreducible $K$-dimensional representation $\rho_{[K]}$ to embed $PGL(2,\C)$ into $PGL(K,\C)$. As explained above, $\rho_{[K]}$ simply maps one parabolic element to another; therefore, $\CA_{\rm geom}^{(K)}$ will have parabolic holonomy \eqref{hol-parab} around the meridian. Moreover, the complex volume of $\CA_{\rm geom}^{(K)}$ is exactly $\frac16K(K^2-1)$ times $\CV(\CA_{\rm geom})$; in particular,
\be {\rm Vol}(\CA_{\rm geom}^{(K)}) = \frac16K(K^2-1) {\rm Vol}(\CA_{\rm geom}) =  \frac16K(K^2-1) {\rm Vol}(M)\,. \label{VKV2} \ee
This fact was proved in \cite{GTZ-slN}. It can be intuited from the RHS of \eqref{VolCS}. Momentarily we will give a simple, constructive combinatorial argument for the validity of \eqref{VKV2}, similar to that used by \cite{GTZ-slN}.

In general, for a hyperbolic manifold $M$ there may be (finitely) many flat $PGL(K,\C)$-connections $\CA_\alpha$ with parabolic meridian holonomy. They may or may not have anything to do with lower-rank embeddings. Explicitly, they can be described as the solutions to the classical gluing equations of an octahedral decomposition that have meridian eigenvalues $m_a=1$; or as the points of intersection of the Lagrangian submanifold $\CL_K(M)$ with the Lagrangian $\{m_a = 1\}$. Interestingly, however, the geometric connection $\CA_{\rm geom}^{(K)}$ seems to have the largest volume of all the $\CA_\alpha$ (and its conjugate $\CA^{(K)}_{\rm conj}=\ol \CA_{\rm geom}^{(K)}$ the smallest).%
\footnote{This was recently conjectured by S. Garoufalidis, M. Goerner, and C. Zickert, after computing volumes of boundary-parabolic flat connections for hundreds of hyperbolic knots.}

In order to demonstrate \eqref{VKV2} using the combinatorics of this paper, we proceed as follows. Let $M$ be a hyperbolic knot complement. Promote $\CA_{\rm geom}$ to a framed flat connection (the choice of invariant flag at the torus boundary is unique); and choose an ideal triangulation $\{\Delta_i\}_{i=1}^N$ of $M$ such that $\CA_{\rm geom}$ induces well-defined cross-ratio coordinates for the octahedra of the $2$-decomposition. Almost any triangulation will do. Recall that in the 2-decompositions, there is a single octahedron per tetrahedron; so we can label the octahedron parameters $\{z_i,z_i',z_i''\}_{i=1}^N$, with $z_iz_i'z_i''=-1$ and $z_i''+z_i^{-1}-1=0$. The hyperbolic volume of $M$ is computed as
\be {\rm Vol}(M) = \sum_{i=1}^N {\rm Vol}(\oct_i) = \sum_{i=1}^N {\CL i}(z_i)\,,\ee
where $\CL i(z)$ is the Bloch-Wigner dilogarithm function \eqref{defBW}.

Now consider the $K$-decomposition of the same triangulation. Suppose that every octahedron in a big tetrahedron $\Delta_i$ is labeled with the same triple of parameters $\{z_i,z_i',z_i''\}_{i=1}^N$ that encoded $\CA_{\rm geom}$ above. We align all the primes and double-primes to coincide with their former locations in the 2-decomposition. Then every gluing constraint at a lattice point inside a tetrahedron $\Delta_i$ takes the form $z_iz_i'z_i''z_iz_i'z_i''=1$ and is automatically satisfied. Every face gluing constraint takes the form $z_iz_i'z_i''z_jz_j'z_j''=1$, which is automatic as well; and every edge gluing constraint is a copy of the corresponding edge constraint for $K=2$, which is already satisfied. Since this uniform assignment of octahedron parameters satisfies all of the octahedron gluing constraints, it defines a framed flat $PGL(K,\C)$-connection on $M$. Moreover, it is straightforward to check from the snake rules of Section \ref{sec:projbases} that the holonomy around any closed cycle in $M$ is simply the image of the $\CA_{\rm geom}$ holonomy under $\rho_{[K]}$. In particular, the meridian holonomy at the boundary is parabolic.

Therefore, this assignment of octahedron parameters at general $K$ simply parametrizes the flat connection $\CA_{\rm geom}^{(K)}$. Its volume is easy to calculate, since every octahedron in a big tetrahedron gives the same contribution:
\be {\rm Vol}(\CA_{\rm geom}^{(K)}) \,=\, \sum_{\text{all octahedra}}{\rm Vol}(\oct) \,=\, \frac16 K(K^2-1)\sum_{i=1}^N \CL i_2(z_i) \,=\, \frac16 K(K^2-1){\rm Vol}(M)\,.\ee

\subsubsection{From free energy to volumes}
\label{sec:energy}

Now we come to the more nontrivial part of the scaling argument. A standard measure for the number of degrees of freedom in a three-dimensional theory is the $S^3$ partition function $\CZ[T]$. The free energy $\CF = -\log|\CZ[T]|$ is extensive in the number of degrees of freedom, and has several extremely nice properties; for example, for an $\CN=2$ theory, it is maximized at the superconformal values of R-charges \cite{Jafferis-Zmin, JKPS-Fthm}. Unfortunately, although the $S^3$ partition function can be computed systematically for theories $T_K[M]$ in class $\CR$ \cite{Kapustin-3dloc, DGG}, it is not immediately clear how it relates to the volumes of flat connections discussed above.

There does exist a closely related measure of degrees of freedom for $\CN=2$ theories that \emph{is} related to volumes: the partition function $\CZ_b[T]$ on an ellipsoid $S^3_b$. The ellipsoid deformation stretches out the three-sphere, so that metrically it can be described by the embedding $b^2|z|^2+b^{-2}|w|^2=1$ in flat $\C^2\simeq\R^4$; for $b=1$ we recover round $S^3$. The partition functions of class-$\CR$ theories on $S^3_b$ can also be systematically calculated \cite{HHL, DGG}. For a 3-manifold theory, $\CZ_b[T_K[M]]$ agrees with the analytically continued partition function of $SL(K,\R)$ Chern-Simons theory on $M$, with appropriate boundary conditions at $\pd M$. Moreover, it follows from results of \cite{DGG} (based on \cite{gukov-2003, hikami-2006, DGLZ, Dimofte-QRS}) that in the $b\to 0$ limit,
\be \CF_b[T_K[M]]=-\log\big|\CZ_b[T_K[M]]\big| \overset{b\to 0}\sim -\frac 1{2\pi b}{\rm Vol}(\CA_\alpha) + O(1)\,, \label{FVol}\ee
where $\CA_\alpha$ is the volume of one of the flat $PGL(K,\C)$-connections on $M$ with parabolic meridian holonomy. Note that the holographic free energy on $S^3_b$ is simply related to the free energy on $S^3$ by $\CF_b = (b+1/b)^2 \CF/4$~\cite{MPS-S3b}.

In order to prove the desired large-$K$ scaling from \eqref{FVol}, we must make two assumptions --- both of which are rather interesting and will be elucidated elsewhere. First, we must assume that \eqref{FVol} accurately measures the degrees of freedom of $T_K[M]$ at a superconformal fixed point, and in particular with the correct superconformal R-symmetry. Second, we need to know that the flat connection $\CA_\alpha$ entering the RHS of \eqref{FVol} is the conjugate flat connection $\CA_{\rm conj}^{(K)}=\ol \CA_{\rm geom}^{(K)}$ so that the volume becomes $-{\rm Vol}(\CA_{\rm conj}^{(K)}) = +\frac16K(K^2-1){\rm Vol}(M)$.
The appearance of $\CA_{\rm conj}^{(K)}$ on the RHS of \eqref{FVol} means that this connection dominates the $SL(K,\R)$ Chern-Simons partition function. There have been hints in \cite{KashAnd} that this does happen for $K=2$, though more is needed to understand the general story.%
\footnote{The question of which complex critical points dominate a real partition function is a tricky and subtle one, and closely related to the (still open) Volume Conjecture for compact Chern-Simons theory \cite{kashaev-1997, Mur-Mur}. The Volume Conjecture could be proved physically if it were known that geometric flat connections dominated $SU(2)$ partition functions \cite{gukov-2003, Wit-anal}.}

%%%%%%%%%%%%%%%%%%%%%%%%%%%%%%%%%%%%%%%%%%%%%%%%%%%%%%%%%%%%%%%%%%%%%%%%%%%%%
%%%%%%%%%%%%%%%%%%%%%%%%%%%%%%%%%%%%%%%%%%%%%%%%%%%%%%%%%%%%%%%%%%%%%%%%%%%%%
%%%%%%%%%%%%%%%%%%%%%%%%%%%%%%%%%%%%%%%%%%%%%%%%%%%%%%%%%%%%%%%%%%%%%%%%%%%%%

\subsection{Trefoil knot invariants}
\label{sec:trefoil}

We conclude this section with two examples of knot complements and the combinatorics of their $K=3$ triangulations.

To keep things simple, we will use minimal triangulations of the trefoil and the figure-eight knot complements, into two tetrahedra (and eight octahedra) each. These triangulations are not quite sufficiently refined to define theories $T_3[M]$ due to a technical constraint discussed in \cite[Sec. 4.1]{DGG} and Appendix \ref{app:symp}: in order to construct superpotential operators $\CO_k$ corresponding to gluing constraints $C_k$, the constraints must be ``easy.'' This means that the functions $C_k$ never contain a sum of two non-commuting octahedron parameters, such as $Z_i+2Z_i''$. For $K\geq 4$ this is never a problem, but for $K=2,3$ the triangulations must be further refined. Nevertheless, even the unrefined triangulations are good enough for calculating observables of the theories --- such as $\CL_{\rm SUSY}\simeq \CL_K(M)$, volumes, wavefunctions, etc. --- which is what we focus on here.

\subsubsection{SL(2) invariants}

Let us briefly review the structure of trefoil invariants for $K=2$.

The moduli space of flat $SL(2,\C)$-connections on the trefoil knot complement $M=S^3\bs \mb{3_1}$ has two irreducible components of dimension one, corresponding (respectively) to flat connections whose holonomy is abelian and to flat connections that are genuinely nonabelian and irreducible. Each component projects to a one-dimensional Lagrangian submanifold in the boundary phase space $\CP_2({\pd\mb{3_1}})$. If we parameterize the meridian and longitude eigenvalues as
\be \mu:\quad \begin{pmatrix} m & \\ & m^{-1} \end{pmatrix}\,,\qquad \lambda:\quad \begin{pmatrix} \ell & \\ & \ell^{-1} \end{pmatrix}\,, \ee
so that (in exponentiated coordinates) $\CP_2({\pd\mb{3_1}})=\{(m,\ell)\}\simeq \C^*\times \C^*$, one finds that the two Lagrangians are
\begin{align} \CL_{\rm abelian}:\quad& \ell-1=0 \\ \label{Lag231}
\CL_2({\mb{3_1}})=\CL_{\text{non-abelian}}: \quad& \ell+m^6=0\,. \notag \end{align}
The union of these two components is the well-known A-polynomial of the trefoil \cite{cooper-1994}.
Notice that the Lagrangians have Weyl symmetry, \ie\ they are invariant under $(\ell,m)\to (\ell^{-1},m^{-1})$. However, if we think of $\CP_2({\pd\mb{3_1}})$ as a moduli space of \emph{framed} flat connections we do not quotient out by the Weyl group.

Only the non-abelian flat connections can be parameterized by cross-ratio coordinates in an ideal triangulation. Indeed, the Lagrangian $\CL_2({\mb{3_1}})$ obtained from an ideal triangulation agrees with $\CL_{\text{non-abelian}}$.

In order to discuss $PGL(2,\C)$ flat connections rather than $SL(2,\C)$ flat connections, we would simply re-write the above equations in terms of the invariant squared eigenvalues $m^2$ and $\ell^2$. It is useful to note that the Lagrangians are \emph{already} written in terms of $m^2$. Conversely, in order to lift from $PGL(2,\C)$ to $SL(2,\C)$, one only needs to take an appropriate square root of $\ell^2$, and not $m^2$. The pattern for general $K$ is that the equations for $SL(K,\C)$ Lagrangians involve the individual longitude eigenvalues but only the $K$-th powers of the meridian eigenvalues. This is true for any knot complement in a homology sphere.%
\footnote{It is true because for knot complements in homology sphere one can always tensor a flat $\C^K$ bundle with a flat complex line bundle whose holonomy around the meridian is any $K$-th root of unity.} %
Therefore, to lift from $PGL(K,\C)$ to $SL(K,\C)$ one only needs to ``take a $K$-th root of the $PGL(K,\C)$ longitude.''

The full Lagrangian $(\ell-1)(m^6\ell+1)$ has been quantized in the context of $SU(2)$ Chern-Simons theory \cite{garoufalidis-2004, gukov-2003, Gar-Le}. Using combinatorics of ideal triangulations, one can quantize $\CL_{\mb{3_1}}^{(2)}$, obtaining $\hat\ell+q^{\frac 32}\hat m^6 \simeq 0$. One can then proceed to calculate analytically continued Chern-Simons partition functions, \emph{a.k.a.} sphere partition functions and indices of the trefoil-knot theory. However, they are all extremely simple. For example, the index just looks like
\be \CI_{\mb{3_1}}^{(2)}(m,\zeta;q) = \zeta^{3m}\,. \ee
in the conventions of \cite{DGG-index}. This ultimately reflects the fact that the trefoil is \emph{not} a hyperbolic knot, and we do not expect $T_2[{\bm{3_1}}]$ (or for that matter $T_K[{\bm{3_1}}]$) to contain many interesting degrees of freedom in the infrared.

\subsubsection{The SL(3) character variety}

Now consider the full moduli space of flat $SL(3,\C)$-connections. A straightforward but extremely lengthy computation%
\footnote{A first-principles computation of this moduli space involves looking at representations of $\pi_1(M)$ into $SL(3,\C)$, and imposing the relations in $\pi_1(M)$ as algebraic constraints on elements of $SL(3,\C)$ matrices.} %
shows that the moduli space has five irreducible components of dimension two on which the two meridian eigenvalues are allowed to vary freely. If we parametrize the boundary eigenvalues as
\be \label{MLtref}
 \mu:\quad \begin{pmatrix} m_1 && \\
 & m_2 & \\
 && \frac{1}{m_1m_2} \end{pmatrix},\qquad
 \lambda:\quad \begin{pmatrix} \ell_1 && \\
 & \ell_2 & \\
 && \frac{1}{\ell_1\ell_2} \end{pmatrix},
\ee
then the five components project to the Lagrangian submanifolds
\be   \label{Lag331}
\begin{array}{rll} (1):\quad & \{\;\ell_1-1=0,\; & \ell_2-1=0\;\}  \\
   (2_1):\quad & \{\;\ell_1-1=0,\;& \ell_2+m_1^3m_2^6=0\;\}  \\
   (2_2):\quad & \{\;\ell_1+m_1^6m_2^3=0,\;\;&\ell_2-1=0\;\}  \\
   (2_3):\quad & \{\;m_2^3\ell_1+m_1^3=0,\;\;&m_1^3\ell_2+m_2^3=0\;\}  \\
   (3):\quad & \{\;\ell_1-m_1^6=0,\;&\ell_2-m_2^6=0\;\}\quad = \CL_3({\bm{3_1}}) \,. \end{array}
\ee
Branch $(1)$ contains the $SL(3,\C)$ flat connections with abelian holonomy; branches $(2_i)$ contain the flat connections whose holonomy reduces to an $SL(2,\C)\times U(1)$ subgroup, and branch $(3)$ contains the connections whose holonomy is irreducible. Thus, using the combinatorics of this paper, we expect to be able to reproduce (and a associate a 3d theory to) branch (3). Branches (1) and (3) are invariant under the $S_3$ Weyl symmetry of $SL(3,\C)$, while the three branches $(2_i)$ are permuted among themselves.

Note that each of these branches contains embeddings of $SL(2,\C)$ flat connections along one-dimensional slices. For example, if we set $\ell_1=\ell_2$ and $m_1=m_2$ in branch (3), we obtain the principal embedding (using the representation $\rho_{[3]}$) of the nonabelian $SL(2,\C)$-connections.

We may also observe that, as predicted, the Lagrangian equations only depend on the \emph{cubes} of the meridian eigenvalues. To reduce to $PGL(3,\C)$ flat connections we could rewrite the equations in terms of the invariants $\ell_1^3$ and $\ell_2^3$ as well.

\subsubsection{Symplectic gluing data}

Let us see how we can extract symplectic gluing data from the trefoil triangulation at $K=3$, in a (relatively) much more efficient manner.

\begin{figure}[tbh] \center 
\includegraphics[scale=.63]{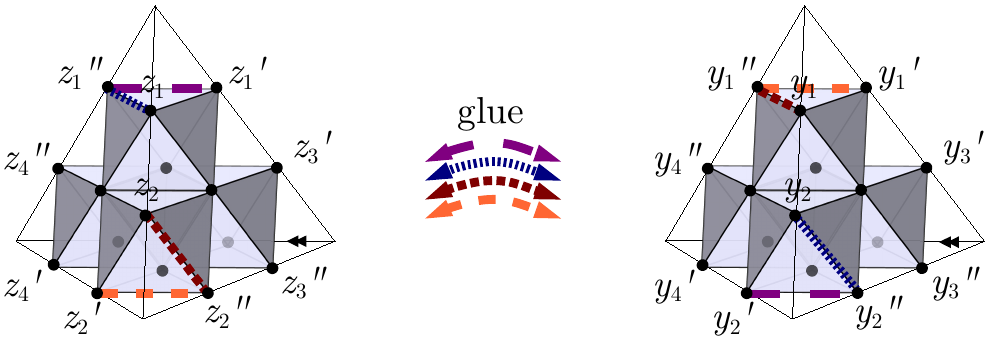}
\caption{Construction of the trefoil knot complement from two $SL(3)$ tetrahedra. Octahedron edges with matching colors are glued together.}
\label{fig:trefoilSL3}
\end{figure}

The complement of the trefoil knot can be decomposed into two tetrahedra, as shown in the Figure~\ref{fig:trefoilSL3}.
There are eight gluing functions, corresponding to lattice points on internal faces and edges, which we write in exponentiated form as
\bea
&c_1 = z_3 y_3~, \qquad &\qquad   c_5 = z_1 z_4'z_3'' y_2 y_3'y_4''~,  \\
&c_2 = z_4 y_4~, \qquad &\qquad c_6 = z_2 z_3'z_4'' y_1 y_4'y_3''~,  \\
&c_3 = z_1 z_1' z_2'' z_4'z_4'' y_1 y_1' y_2'' y_4'y_4''  ~, \qquad &\qquad  c_7 = z_3 z_2'z_1'' y_4 y_1'y_2''~,  \\
&c_4 = z_1'' z_2 z_2' z_3' z_3'' y_1'' y_2 y_2' y_3' y_3''   ~,  \qquad &\qquad c_8 = z_4 z_1'z_2'' y_3 y_2'y_1''~.  
\eea
(In logarithmic form, each of these is a sum of octahedron parameters, minus $2\pi i$; \eg, $C_1=Z_3+Y_3-2\pi i$.)
Notice that there are two automatic relations $c_1c_2c_3c_4=1$ and $c_5c_6c_7c_8=1$.
Thus, as we just emphasized in Section \ref{sec:marginal}, only six gluing constraints are independent.

\begin{figure}[tbh] \center 
\includegraphics[scale=.63]{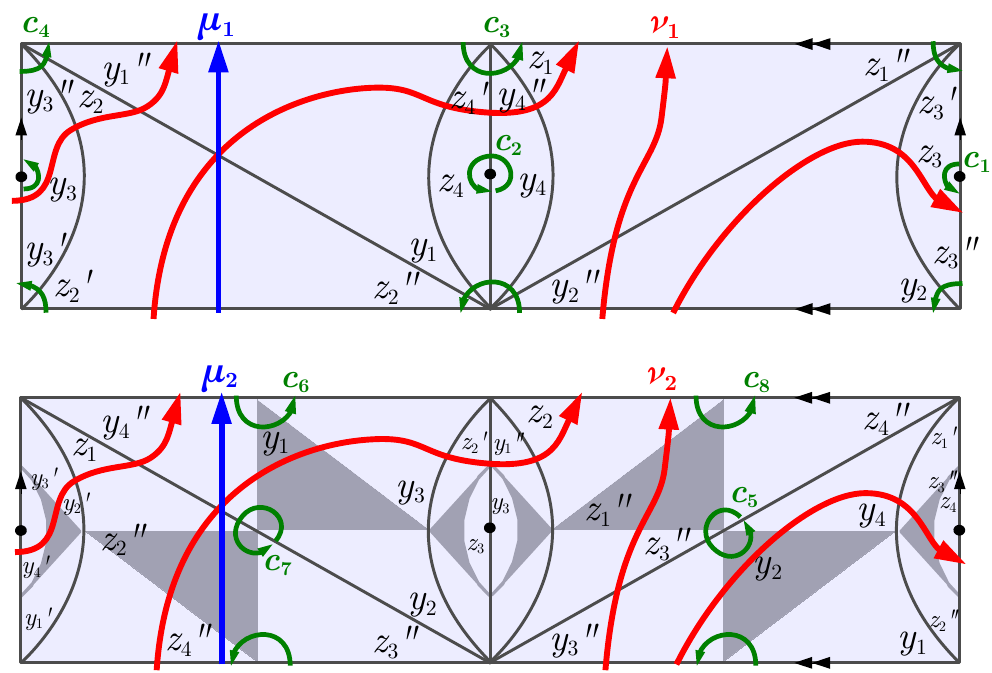}
\caption{First and second slices of the torus boundary for the trefoil knot. The paths $\mu_a$ and $\nu_a$ represent the meridian and longitude, while the small circular paths drawn in green correspond to the gluing constraints.}
\label{fig:trefoillevel2}
\end{figure}

As explained in Section~\ref{sec:UandVcoords}, one can obtain the $U_a$ and $V_a$ 
coordinates that parametrize meridian and longitude eigenvalues by taking two global slices parallel to the small torus boundary. The slices are drawn in Figure \ref{fig:trefoillevel2}, along with meridian and longitude paths on each slice. (We also see the eight gluing functions as coming from small circular paths, and we can check that each of the relations $c_1c_2c_3c_4=1$, and $c_1c_2c_3c_4c_5c_6c_7c_8=1$ involve the gluing functions on a single slice.) In exponentiated form, we find the meridian and longitude eigenvalues to be
\bea
e^{U_1} = \frac{y_1''}{z_2''}~,\quad \qquad && \qquad 
e^{V_1} = \frac{z_1z_2z_4'y_1''y_2''y_4''}{z_1''z_2''z_3'' y_1y_2y_3'}~,\\
e^{U_2} =  \frac{z_2'' y_4''}{z_1''z_4''}~, \qquad && \qquad 
e^{V_2} =  \frac{z_1 y_2 (y_4'')^2}{z_1'' z_2''(z_4'')^2y_1'y_2'y_3^3y_4} ~.
\eea

In order to have canonical Poisson brackets, we redefine the longitude coordinates as in section~\ref{sec:Spdata}:
\begin{subequations}\label{31-V}
\bea
\wt V_1 &:= & \frac13 (2V_1 + V_2 + C_1 -C_2 + C_6)  ~, \\
\wt V_2 &:= & \frac13 (V_1 + 2 V_2 - C_1 + C_2 - C_6)  ~.
\eea
\end{subequations}
One can check that now $\{\wt V_a,U_b\}=\delta_{ab}$. The division by $K=3$ here (\ie\ by the determinant of the Cartan matrix) is a convenient way of lifting from $PGL(3,\C)$ representations to $SL(3,\C)$ representations. Moreover, the judicious addition of gluing constraints (which are ultimately set to zero) in \eqref{31-V} allows $\wt V_1,\wt V_2$ to be \emph{integer} linear combinations of octahedron parameters after the division.

Note that holonomy eigenvalues are now parametrized in several different ways. For the meridian, the cubes of the eigenvalues in \eqref{MLtref} are related to the path-coordinates as
\be m_1^3 = \exp(-2U_1-U_2)\,,\quad m_2^3 = \exp(U_1-U_2)\,,\quad \frac{1}{m_1^3m_2^3} = \exp(U_1+2U_2)\,;\ee
while for the longitude the eigenvalues themselves are
\be \ell_1 = \exp(-\wt V_1)\,,\quad \ell_2 = \exp(\wt V_1-\wt V_2)\,,\quad \frac{1}{\ell_1\ell_2} = \exp \wt V_2\,. \ee

We obtain the classical Lagrangian submanifold corresponding to the space of flat, framed, irreducible $SL(3,\C)$-connections on the trefoil knot complement by rewriting the product Lagrangian $\CL_\times = \big\{ z_i''+z_i^{-1}-1=0,  y_i''+y_i^{-1}-1=0\big\}_{i=1}^{4}$ in terms of the holonomy eigenvalues $m_a$ and $\ell_a$
and making use of the classical gluing constraints $c_k = 1$. 
The result is
\be \CL_3({\mb{3_1}}) = \big\{ \ell_1 - m_1^6=0 ,\; \ell_2 - m_2^6=0 \big\}~. \ee
It is precisely branch (3) of \eqref{Lag331}.

Finally, we can summarize the gluing data in the affine-symplectic $ISp(16,\Z)$ transformation $(g,\sigma)$ that transforms the initial polarization with positions $(Z_1'',Z_2'',Z_3,Z_4'', Y_1'',Y_2'',Y_3,Y_4'')$ and momenta $(Z_1',Z_2',Z_3'',Z_4',Y_1',Y_2',Y_3'',Y_4')$ to the final (canonical!) polarization 
\be \Pi=(U_1,U_2, C_1, C_2, C_3, C_5, C_6, C_7; \tilde V_1,\tilde V_2, \Gamma_1, \Gamma_2, \Gamma_3, \Gamma_5, \Gamma_6, \Gamma_7)~.  \ee
We obtain
\be g_3^{\mb{3_1}}  =  \left(
\begin{array}{cccccccc|cccccccc}
 0 & -1 & 0 & 0 & 1 & 0 & 0 & 0 & 0 & 0 & 0 & 0 & 0 & 0 & 0 & 0 \\
 -1 & 1 & 0 & -1 & 0 & 0 & 0 & 1 & 0 & 0 & 0 & 0 & 0 & 0 & 0 & 0 \\
 0 & 0 & 1 & 0 & 0 & 0 & 1 & 0 & 0 & 0 & 0 & 0 & 0 & 0 & 0 & 0 \\
 0 & 0 & 0 & -1 & 0 & 0 & 0 & -1 & 0 & 0 & 0 & -1 & 0 & 0 & 0 & -1 \\
 -1 & 1 & 0 & 1 & -1 & 1 & 0 & 1 & 0 & 0 & 0 & 1 & 0 & 0 & 0 & 1 \\
 -1 & 0 & 0 & 0 & 0 & -1 & -1 & 1 & -1 & 0 & 1 & 1 & 0 & -1 & -1 & 0 \\
 0 & -1 & -1 & 1 & -1 & 0 & 0 & 0 & 0 & -1 & -1 & 0 & -1 & 0 & 1 & 1 \\
 1 & 0 & 1 & 0 & 0 & 1 & 0 & -1 & 0 & 1 & 0 & 0 & 1 & 0 & 0 & -1 \\ \hline
 -2 & -2 & 0 & 0 & 1 & 1 & 0 & 2 & -1 & -1 & -1 & 1 & 0 & 0 & 1 & 1 \\
 -2 & -1 & 0 & -2 & 1 & 0 & -2 & 2 & -1 & 0 & 0 & 0 & 0 & -1 & 0 & 0 \\
 0 & 0 & 0 & 0 & 1 & 0 & 0 & 0 & 0 & 0 & 1 & 0 & 0 & 0 & 0 & 0 \\
 0 & 0 & 0 & 1 & 0 & 0 & 0 & 1 & -1 & 0 & 0 & 1 & 0 & 0 & 0 & 0 \\
 0 & -1 & 0 & 1 & 0 & 0 & 0 & 0 & -1 & 0 & 0 & 1 & 0 & 0 & 0 & 0 \\
 -1 & 1 & 0 & -1 & 0 & 1 & 0 & 1 & 0 & 0 & 0 & 0 & 0 & 0 & 0 & 0 \\
 -1 & 0 & 0 & -1 & 1 & 1 & -1 & 1 & 0 & 0 & 0 & 0 & 0 & 0 & 0 & 0 \\
 -1 & 0 & 0 & -1 & 0 & 1 & -1 & 1 & 0 & 0 & 0 & 0 & 0 & 0 & 0 & 0
\end{array}
\right)  ~, \qquad 
\sigma_3^{\mb{3_1}}  =  \left(
\begin{array}{c}
 0 \\
 2 \\
 0 \\
 -2 \\
 -2 \\
 -3 \\
 -3 \\
 -1 \\ \hline 
 2 \\
 4 \\
 0 \\
 0 \\
 0 \\
 0 \\
 0 \\
 0
\end{array}
\right) ~.
\ee

\subsection{Figure-eight knot gluing}
\label{sec:figure-eight}

Since the figure-eight knot is the simplest hyperbolic knot and a fairly standard example in knot theory, we briefly demonstrate the construction of $K=3$ gluing data for it as well.

Like for the trefoil, the complement of the figure-eight knot can be triangulated into two tetrahedra (Figure \ref{fig:fig8SL3}). 
One of the most interesting properties of the figure-eight knot is that the geometric flat $PGL(2,\C)$-connection $\CA_{\rm geom}$ with parabolic holonomy is represented by two regular ideal tetrahedra --- \emph{\ie} two octahedra that have all coordinates $z_i=z_i'=z_i''\equiv \exp \frac{i\pi}{3}$. This value of $z_i$ maximizes the volume function $\CL i_2(z_i)$. For general $K$, the complement of the figure-eight knot can similarly be decomposed into $2\times \frac16K(K^2-1)$ ``regular'' octahedra, again with all $z_i=z_i'=z_i'' \equiv\exp \frac{i\pi}{3}$. (One can check explicitly that this is a solution to the gluing equations below.)

\begin{figure}[tbh] \center
\includegraphics[scale=.63]{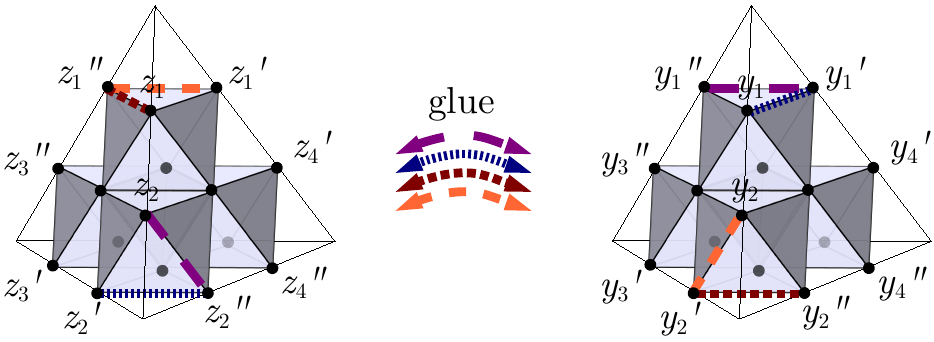}
\caption{Construction of the figure-eight knot complement from two $SL(3)$ tetrahedra.}
\label{fig:fig8SL3}
\end{figure}

Using the $K=3$ decomposition into tetrahedra and octahedra shown in Figure \ref{fig:fig8SL3}, we find that the $K=3$ gluing constraints are
\bea
c_1 = z_1 z_1'z_3'y_2'y_4y_4'~,\qquad && \qquad  c_5 =   z_2 z_4'z_3'' y_4y_2'y_1''~, \\
c_2 = z_2 z_2'z_4' y_1'y_3y_3' ~, \qquad && \qquad c_6 =    z_1 z_3'z_4'' y_3y_1'y_2''~, \\
c_3 = z_2''z_3z_3''y_1y_1''y_4'' ~, \qquad && \qquad c_7 =   z_4 z_2'z_1'' y_1y_3'y_4''~, \\
c_4 = z_1''z_4z_4''y_2y_2''y_3'' ~, \qquad && \qquad c_8 =    z_3 z_1'z_2'' y_2y_4'y_3''~,
\eea
and satisfy the relations $c_1c_2c_3c_4=1$ and $(c_1c_2c_3c_4)c_5c_6c_7c_8=1$, one for each slice.

\begin{figure}[tbh] \center 
\includegraphics[scale=.63]{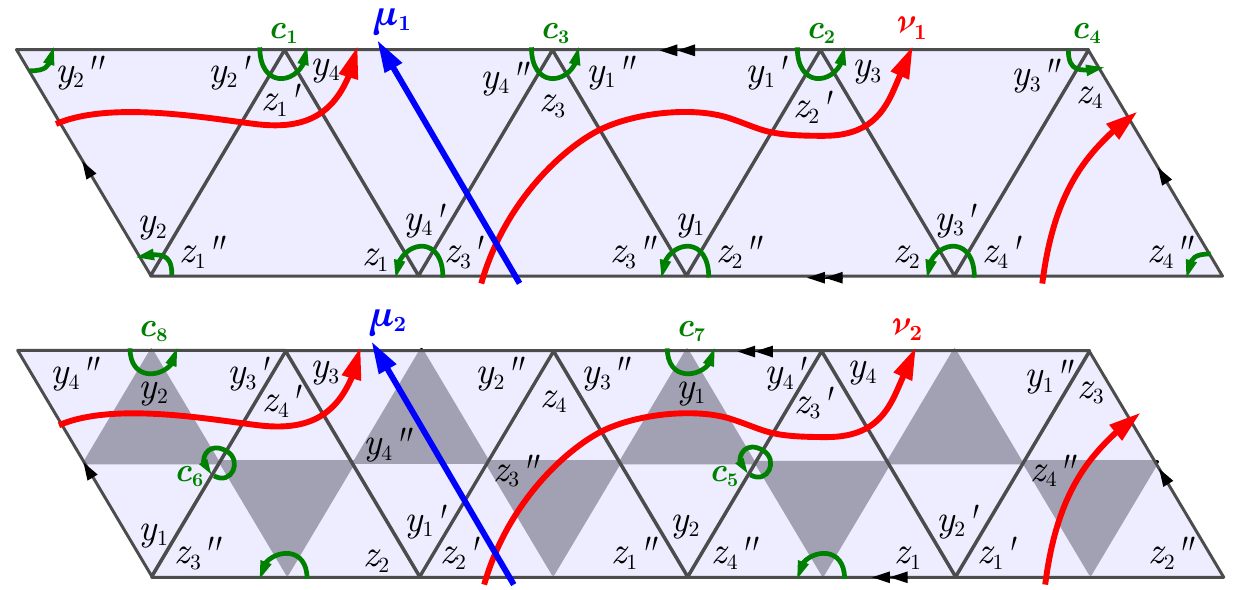}
\caption{First and second slices of the torus boundary for the figure-eight knot.}
\label{fig:fig8level2}
\end{figure}
The coordinates $u_a$ and $v_a$ for the meridian and longitude paths drawn in Figure~\ref{fig:fig8level2} are
\bea
u_1 = \frac{z_3'}{y_4''} ~,  \qquad && \qquad    u_2    = \frac{z_2' y_4'' }{y_1''y_3''} ~, \\
v_1 = \frac{z_1' z_2' y_3 y_4}{z_3'' z_4'' y_1 y_2 }~,   \qquad && \qquad     v_2 =   \frac{z_3'  z_4'  y_1 y_2 }{z_1'' z_2'' y_3 y_4 }~.
\eea
We redefine the longitude coordinates to have canonical Poisson brackets:
\bea
\tilde V_1 &\equiv & \frac13 (2V_1 + V_2 - C_5 - C_6 )  ~, \\
\tilde V_2 &\equiv & \frac13 (V_1 + 2 V_2 + C_5 +  C_6)  ~.
\eea

Taking the initial polarization with positions $\{ Z_1'',Z_2'',Z_3'',Z_4'',Y_1',Y_2',Y_3',Y_4' \}$, we 
find the following symplectic gluing matrix, with all integer coefficients:
\be
g_3^{\mb{4_1}}  = \left(
\begin{array}{cccccccc|cccccccc}
 0 & 0 & 0 & 0 & 0 & 0 & 0 & 1 & 0 & 0 & 1 & 0 & 0 & 0 & 0 & 1 \\
 0 & 0 & 0 & 0 & 1 & 0 & 1 & -1 & 0 & 1 & 0 & 0 & 1 & 0 & 1 & -1 \\
 -1 & 0 & 0 & 0 & 0 & 1 & 0 & 1 & 0 & 0 & 1 & 0 & 0 & 0 & 0 & 1 \\
 0 & -1 & 0 & 0 & 1 & 0 & 1 & 0 & 0 & 0 & 0 & 1 & 0 & 0 & 1 & 0 \\
 0 & 1 & 0 & 0 & -1 & 0 & 0 & -1 & 0 & 0 & -1 & 0 & 0 & 0 & 0 & -1 \\
 0 & -1 & 1 & 0 & -1 & 1 & 0 & 0 & 0 & -1 & 0 & 1 & -1 & 0 & 0 & 1 \\
 -1 & 0 & 0 & 1 & 1 & -1 & 0 & 0 & -1 & 0 & 1 & 0 & 0 & -1 & 1 & 0 \\
 1 & 0 & 0 & -1 & 0 & 0 & 1 & -1 & 0 & 1 & 0 & -1 & 1 & 0 & 0 & -1 \\ \hline
 0 & 0 & -1 & -1 & 0 & 0 & 0 & 0 & 1 & 1 & 0 & 0 & 0 & 0 & 0 & 0 \\
 -1 & -1 & 0 & 0 & 0 & 0 & 0 & 0 & 0 & 0 & 1 & 1 & 0 & 0 & 0 & 0 \\
 -1 & 2 & 0 & 1 & -1 & 0 & -1 & 0 & -1 & 0 & 0 & -1 & 0 & 0 & 0 & 0 \\
 1 & 1 & 0 & 0 & 0 & -1 & -1 & 0 & 0 & 0 & -1 & -1 & 0 & 0 & 0 & 0 \\
 0 & 1 & 0 & -1 & -1 & 1 & 0 & 0 & 0 & 1 & 0 & -1 & 0 & 0 & 0 & 0 \\
 -1 & 0 & 0 & 1 & 0 & 0 & -1 & 0 & 0 & -1 & 0 & 1 & -1 & 0 & 0 & 0 \\
 0 & 0 & 0 & 0 & 0 & 1 & 0 & 0 & 0 & 0 & 0 & 0 & 0 & 0 & 0 & 0 \\
 -1 & 2 & -1 & 2 & 0 & -1 & -2 & 1 & 0 & -1 & -1 & 0 & -1 & 0 & 0 & 0
\end{array}
\right) ~, \qquad 
\sigma_3^{\mb{4_1}}  =  \left(
\begin{array}{c}
 1 \\
 3 \\
 -1 \\
 -1 \\
 -3 \\
 -2 \\
 -2 \\
 -2 \\ \hline 
 4 \\
 4 \\
 0 \\
 0 \\
 0 \\
 0 \\
 0 \\
 0
\end{array}
\right) ~.
\ee

\section*{Acknowledgements}

We would like to thank Christopher Beem, Nicolas Bergeron, Elisha Falbel, Davide Gaiotto, Stavros Garoufalidis, Antonin Guilloux, Sergei Gukov, Lotte Hollands, Daniel Jafferis, Juan Maldacena, Greg Moore, Andrew Neitzke, Nathan Seiberg, Roland van der Veen, Edward Witten, and Don Zagier for helpful comments and discussions. 
We would especially like to thank Dan Xie and Masahito Yamazaki for collaboration at an early stage of this project.

All the authors would like to thank the Simons Center for Geometry and Physics for hospitality during the completion of this work, both during June 2012 and during the 2012 Summer Workshop in Mathematics and Physics. TD is also grateful to the Max Planck Institute for Mathematics, Bonn, for hospitality during September, 2012; and MG would like to thank the Institute for Advanced Studies for hospitality.
The work of TD is supported by a William D. Loughlin Membership
at the Institute for Advanced Study, with additional support from DOE grant DE-FG02-90ER-40542. 
The work of MG is supported by the ERC Starting Independent Researcher Grant 259133.
The work of AG was supported by NSF grants DMS-1059129 and DMS-1301776.

%%%%%%%%%%%%%%%%%%%%%%%%%%%%%%%%%%%%%%%%%%%%%%%%%%%%%%%%%%%%%%%%%%%%%%%%%%%%%
%%%%%%%%%%%%%%%%%%%%%%%%%%%%%%%%%%%%%%%%%%%%%%%%%%%%%%%%%%%%%%%%%%%%%%%%%%%%%
%%%%%%%%%%%%%%%%%%%%%%%%%%%%%%%%%%%%%%%%%%%%%%%%%%%%%%%%%%%%%%%%%%%%%%%%%%%%%
%%%%%%%%%%%%%%%%%%%%%%%%%%%%%%%%%%%%%%%%%%%%%%%%%%%%%%%%%%%%%%%%%%%%%%%%%%%%%
%%%%%%%%%%%%%%%%%%%%%%%%%%%%%%%%%%%%%%%%%%%%%%%%%%%%%%%%%%%%%%%%%%%%%%%%%%%%%

\appendix

\section{Symplectic data and class $\CR$}
\label{app:symp}

Section \ref{sec:combi} of this paper was devoted to describing how the $K$-triangulation of 3-manifolds defines symplectic gluing data $(g,\sigma)\in ISp(2N,\Z)$ or $(g,\sigma)\in ISp(2N,\Q)$ with certain properties, if one assumes the Symplectic Gluing Conjecture \ref{conj:symp}. In this appendix we review how such abstract data $(g,\sigma)$ --- whether it arises from a 3-manifold or not --- subsequently defines 3d SCFT's $T_{(g,\sigma)}$ in class $\CR$ and various associated invariants. We largely follow \cite{DGG}.

Of course, multiple pairs $(g,\sigma)$ can lead to the same SCFT $T_{(g,\sigma)}$. Thus it will be natural to define a set $\IR$ of equivalence classes of symplectic data whose elements are in one-to-one correspondence with 3d theories. Conjecturally, all nontrivial equivalences $(g,\sigma)\sim (g',\sigma')$ are generated by the algebraic analogues of 2--3 moves and octahedron rotations. (Physically, we would like to say that all non-trivial mirror symmetries of 3d abelian Chern-Simons matter theories are generated by the basic XYZ\,$\leftrightarrow$\,SQED duality.)

\subsection{The symplectic data}
\label{sec:data}

To be more concrete, we recall that elements of the affine symplectic group%
\footnote{Here we will focus on $ISp(2N,\Z)$ rather than $ISp(2N,\Q)$. The generalization to rational transformations is straightforward, but requires extra technical care.} %
$ISp(2N,\Z) \simeq Sp(2N,\Z)\ltimes \Z^{2N}$ can be represented as pairs $(g,\sigma)$, where $g \in Sp(2N,\Z)$ and $\sigma$ is a vector of $2N$ integers. We define a \emph{decoration} for $(g,\sigma)$ to be an integer $0\leq d\leq N$, and distinguish the first $d$ rows of $g$ (and the first $d$ elements of $\sigma$). We then let $\wh \IR$ be the set of such decorated elements $(g,\sigma)$, for all $d,N\geq 0$.

We further represent an element $(g,\sigma)\in\wh \IR$ as a transformation on a complex symplectic phase space $\CP_\times = \C^{2N}$:
\be \label{gsymp}
\begin{pmatrix} X_1 \\ \vdots \\ X_d \\ C_1 \\ \vdots\\ C_{N-d} \\\hline
 P_1 \\ \vdots \\ P_d \\ \Gamma_1 \\\vdots \\ \Gamma_{N-d} \end{pmatrix}
 =
 \begin{matrix} \vspace{.4cm} \\ \left(\begin{array}{@{\qquad}c@{\qquad}|@{\qquad}c@{\qquad}} \\  \mb A  &\mb B \\ \\\hline \\ \mb C & \mb D \\ \\
 \end{array}\right) \\
 \underbrace{\hspace{1.6in}}_{\ds g}
 \end{matrix}
  \cdot
 \begin{pmatrix} Z_1\\\vdots \\Z_N \\\hline Z_1'' \\\vdots\\ Z_N'' \end{pmatrix} - i\pi \begin{matrix} \vspace{.4cm} \\
 \begin{pmatrix} \\ \sigma_X \\ \\\hline \\ \sigma_P \\ \\ \end{pmatrix}\\
 \underbrace{\hspace{1cm}}_{\ds \sigma}
 \end{matrix}\,,
\ee
where $\mb A,\,\mb B,\,\mb C,\,\mb D$ are $N\times N$ blocks and $\sigma_X=(\sigma_1,\ldots,\sigma_N)^T$ and $\sigma_P=(\sigma_{N+1},\ldots,\sigma_{2N})^T$ are $N$-vectors of integers. The transformation preserves the complex symplectic form
\be \Omega_\times  = \sum_{i=1}^N dZ_i''\wedge dZ_i = \sum_{j=1}^d dP_j\wedge dX_j + \sum_{k=1}^{N-d} d\Gamma_k\wedge dC_k\,,\ee
and can be thought of as a change of polarization. It sends $\Pi_\times=(Z_i;Z_i'')$ (with $Z_i$ as positions and $Z_i''$ as conjugate momenta) to $\wt \Pi = (X_j,C_k;P_j,\Gamma_k)$, where $X_j,C_k$ are positions and $P_j,\Gamma_k$ are conjugate momenta.
The positions $X_j$ label the $d$ decorated rows.

When $(g,\sigma)$ comes from a $K$-triangulation of a 3-manifold $M$, as described in Section \ref{sec:combi}, the $X_j$ and $P_j$ are coordinates for  a boundary phase space $\CP_K({\pd M})$; while the $C_k$ are the gluing functions.

\subsection{Constraints on the data}

We impose two constraints on the elements of $\wh \IR$, which seem to be necessary for defining 3d theories and other invariants.

The first constraint is satisfied by every decorated $(g,\sigma)$ that arises from geometry. We call it the \emph{Angle Constraint}. To describe it most symmetrically, we introduce $N$ auxiliary coordinates
\be Z_i' = i\pi - Z_i-Z_i'' \ee
on the phase space $\CP_\times$. The Angle Constraint requires that every undecorated coordinate $C_k$ can be written in the form 
\be \label{Cangle}
C_k = \sum_{i=1}^N \big(\alpha_{ki}Z_i + \alpha_{ki}'Z_i'+\alpha_{ki}''Z_i''\big)-2\pi i\,,
\ee
where $\alpha_{ki},\,\alpha_{ki}',\,\alpha_{ki}''$ (with $1\leq k \leq N-d$ and $1\leq i\leq N$) are three matrices of non-negative integers that satisfy
\be \label{alphasum}
\sum_{j=1}^{N-d} \alpha_{ki} \leq 2\,,\quad \sum_{j=1}^{N-d} \alpha_{ki}'\leq 2\,,\quad \sum_{j=1}^{N-d} \alpha_{ki}'' \leq 2\qquad \forall\;\, 1\leq i \leq N\,.
\ee
(It is easy to see that geometric gluing functions $C_k$ all satisfy the Angle Constraint, since they are positive sums of octahedron parameters, minus $2\pi i$, and every octahedron parameter occurs at most twice.) The Angle Constraint implies that the undecorated entries of the blocks $\mb A$ and $\mb B$ in \eqref{gsymp} obey $|\mb A_{d+k,i}|\leq 2,\, |\mb B_{d+k,i}|\leq 2,$ and $|\mb A_{d+k,i}-\mb B_{d+k,i}|\leq 2$.

In order to define gauge theories, we need to supplement the Angle Constraint with a stronger \emph{Superpotential Constraint}. It requires that the undecorated coordinates can be written in the form \eqref{Cangle} where for each $i$ and $k$ only one of $\alpha_{ki},\alpha_{ki}',\alpha_{ki}''$ is nonzero. Alternatively, we could say that for any fixed $k$ there exist suitable cyclic rotations of the initial coordinates $Z_i\to Z_i'\to Z_i''\to Z_i$ (independent for each $i$) so that after these rotations 
\be \label{Csuper}
C_k = \sum_{i=1}^N \alpha_{ki}Z_i -2\pi i\,.\ee
This will ultimately allow a construction of chiral operators $\CO_k$ associated with the $C_k$.

\subsection{The gauge theory}
\label{sec:gauge}

Given an element $(g,\sigma)\in \wh \IR$ that satisfies the Angle and Superpotential constraints, it is possible to define a 3d $\CN=2$ gauge theory $T_{(g,\sigma)}$ in the ultraviolet --- which then flows to a desired SCFT in the infrared. The theory preserves a $U(1)_R$ R-symmetry, and the data $(g,\sigma)$ encodes a specific UV R-charge assignment, which can then be used to compactify $T_{(g,\sigma)}$ on curved manifolds. Moreover, the data $(g,\sigma)$ encodes a specific choice of Chern-Simons couplings for the background gauge fields associated with $U(1)$ flavor symmetries and with the R-symmetry.%
\footnote{More precisely, $(g,\sigma)$ encodes flavor-flavor and mixed flavor-R Chern-Simons couplings. It does not uniquely specify R-R couplings. Such background couplings and their effect on partition functions and dualities were recently discussed in \cite{CDFKS, CDFKS-CS}.}

The gauge theory $T_{(g,\sigma)}$ is built in the following steps.

\begin{itemize}

\item[0.] Define $T_\soct \equiv T_2[\oct,\Pi_Z]$ to be the octahedron theory of Section \ref{sec:Toct} (\emph{a.k.a.}~the $K=2$ tetrahedron theory). It is the theory of a free $\CN=2$ chiral multiplet $\phi$, with charge $+1$ under a background $U(1)$ flavor symmetry and R-charge zero, and an additional level $-1/2$ Chern-Simons coupling for difference of flavor and R-charge gauge fields. The Chern-Simons level matrix can be represented as
\be \begin{array}{c|cc}  & F & R \\\hline
  F & -1/2 & 1/2 \\
  R & 1/2 & * \end{array}\;\;.
\ee

\item[1.] Form the product theory $T_\times = T_{\Delta_1}\otimes\cdots\otimes T_{\Delta_N}$. It is a theory of $N$ chirals $\phi_i$, and has maximal abelian flavor symmetry $U(1)^N$. The real masses of the $U(1)$'s (scalars in the background gauge multiplets) can be associated with the real parts of the coordinates $Z_i$ on $\CP_\times$.

\item[2.] Apply the $Sp(2N,\Z)$ symplectic transformation $g$ to the theory $T_\times$ to obtain a theory $\wt T$ with a new $U(1)^N$ flavor symmetry. The $Sp(2N,\Z)$ action on 3d $\CN=2$ gauge theories was defined in \cite{Witten-SL2}, and can be implemented (\eg) by factoring $g$ into generators. Then
\begin{itemize}
\item[a.] ``U-type'' generators of the form $\left(\begin{smallmatrix} \mb U & 0 \\ 0 & \mb U^{-1T} \end{smallmatrix}\right)$, where $\mb U\in GL(N,\Z)$, act by changing the basis of $U(1)^N$ flavor currents. If $V=(V_1,\ldots,V_N)^T$ is a vector of background gauge multiplets corresponding to the $U(1)^N$ symmetry, we send $V\mapsto \mb U^{-1}V$.
\item[b.] ``T-type'' generators of the form $\left(\begin{smallmatrix} \mb I & 0 \\ \mb k & \mb I \end{smallmatrix}\right)$, where $\mb I$ is the $N\times N$ identity and $\mb k$ is $N\times N$ symmetric, act by adding supersymmetric Chern-Simons terms for the background $U(1)^N$ flavor symmetry with (mixed) level matrix $\mb k$.
\item[c.] ``S-type'' generators of the form $\left(\begin{smallmatrix} \mb I-\mb E_i & -\mb E_i \\ \mb E_i & \mb I-\mb E_i\end{smallmatrix}\right)$, where $E_i$ is an $N\times N$ matrix with entry $1$ at the $i$-th spot on the diagonal and zeroes everywhere else, act by gauging the $i$-th $U(1)_i$ symmetry and replacing it with a new topological $U(1)_J$ flavor symmetry. A background gauge multiplet for $U(1)_J$ is coupled to the gauge multiplet of the gauged $U(1)_i$ via a supersymemtric FI term, which is the same thing as a mixed Chern-Simons term with level matrix $\left(\begin{smallmatrix}0&1\\1&0\end{smallmatrix}\right)$.
\end{itemize}

This $Sp(2N,\Z)$ action can be extended to rational matrices $Sp(2N,\Q)$ if one allows rescaling of the charge lattice of $T_\times$.

\item[3.] Now the theory $\wt T$ from Step 2 has a $U(1)^N$ flavor symmetry with real masses associated with the top $N$ rows of $g$, \ie\ to the coordinates $X_j$ and $C_k$.

Change the R-charge assignment of the theory $\wt T$ by adding $\sigma_i$ ($1\leq i \leq N$) multiples of each $U(1)_i$ flavor charge to the R-charge. 

For each component $\sigma_{N+i}$ in the bottom half of the shift vector, $1\leq i \leq N$, add $-\sigma_{N+i}$ units
of mixed Chern-Simons coupling between the background gauge field for the $U(1)_i$ flavor symmetry and the background gauge field for the R-symmetry.

This interpretation of affine shifts defines an action of the full affine symplectic group $ISp(2N,\Z)$ on 3d SCFT's. In particular, the full relations of $ISp(2N,\Z)$ are satisfied (after flowing to the IR), up to an integer ambiguity in the background R-R Chern-Simons level.

\item[4.] Finally, break $N-d$ flavor symmetries $U(1)_k$ associated with the undecorated coordinates $C_k$. This is done by adding $N-d$ operators $\CO_k$ to the superpotential, charged under each respective $U(1)_k$. We'll review the construction of the operators below. The breaking can be thought of as setting the respective mass parameters $C_k\to 0$.

\end{itemize}
In the end, this produces a theory $T_{(g,\sigma)}$ with manifest $U(1)^d$ flavor symmetry. The mass parameters of the $d$ unbroken $U(1)$'s are associated with the coordinates $X_j$ (\ie\ with the decorated rows of $(g,\sigma)$).

The rank of the gauge group of $T_{(g,\sigma)}$ depends on exactly how the symplectic transformation $g$ is applied in Step 2. Different decompositions into generators lead to slightly different UV descriptions that flow to the same thing in the IR. (In the IR the relations of the symplectic group hold.) It is fairly easy to see that a lower bound on the rank of the UV gauge group is given by ${\rm rank}(\mb B)$. Often the bound can be realized --- \ie\ there exists a decomposition of $g$ that involves exactly ${\rm rank}(\mb B)$
$S$-type generators.

Alternatively, if $\mb B$ has maximal rank and is unimodular, there exists a way to apply the affine-symplectic transformations of Steps 0--3 all at once, \cf\ \cite{DG-quantumNZ}. We define a theory $\wt T$ by starting with $N$ chirals $\phi_i$, each charged under an independent dynamical gauge symmetry $U(1)_i$ (and having zero R-charge). Thus the gauge group is $U(1)^N$. The flavor group is also $U(1)^N$, since each dynamical $U(1)_i$ has an associated withpological $U(1)_{J,i}$. We then specify the full Chern-Simons coupling matrix to be
\be \label{CS-allin1}
 \begin{array}{c|@{\;}c@{\;\;}c@{\;}c}
  & G & J & R \\\hline
 G & \;\;\mb B^{-1}\mb A-\frac12 \mb I_{N\times N} & -\mb B^{-1} & \frac12-\mb B^{-1}\sigma_X \\[.05cm]
 J & -\mb B^{-1\,T} & \mb D\mb B^{-1} & \mb D\mb B^{-1}\sigma_X-\sigma_P \\[.05cm]
 R & \frac12-(\mb B^{-1}\sigma_X)^T & (\mb D\mb B^{-1}\sigma_X-\sigma_P)^T & *
 \end{array}\,.
\ee
(Here `$G$' stands for the $N$ dynamical gauge fields, `$J$' for the $N$ background flavor gauge fields, and `$R$' for the background $U(1)_R$ field. For instance, the $G$--$J$ blocks encode FI terms. The $G$--$G$ blocks encodes dynamical CS terms that couple the otherwise independent gauge fields.) The $N$ flavor symmetries of $\wt T$ correspond directly to the coordinates $(X_j,C_k)$.  To obtain $T_{(g,\sigma)}$ from $\wt T$ we ``simply'' have to apply Step 4, adding appropriate operators to the superpotential to break the $C_k$ symmetries.

\subsection{Equivalences}
\label{sec:equiv}

Two elements $(g,\sigma),(g',\sigma')\in \wh\IR$ can sometimes define theories that are equivalent after flowing to the IR, $T_{(g,\sigma)}\simeq T_{(g',\sigma')}$. There are three basic ways in which this can happen. (In the following, we will \emph{assume} that both $T_{(g,\sigma)}$ and $T_{(g',\sigma')}$ possess the chiral operators needed to break flavor symmetries in Step 4 of the gauge theory construction; we return to this in Section \ref{sec:defO}.)

First, there are some trivial equivalences. We'll describe them in terms of initial and final coordinates on a phase space $\CP_\times$, as in \eqref{gsymp}. We can permute pairs of initial coordinates $(Z_i,Z_i'')\leftrightarrow(Z_{i'},Z_{i'}'')$ (which corresponds to permuting the $N$ chirals of $T_{(g,\sigma)}$). Similarly, we can permute the final coordinates $C_k$. Moreover, we can redefine the $X_j$ and $P_j$ coordinates by any integer linear combination of the $C_k$, as long as we simultaneously redefine the $\Gamma_k$ so that $g$ remains symplectic. We can also redefine the $\Gamma_k$ by integer linear combinations of the $C_k$ and any multiples of $i\pi$. In gauge theory, these latter redefinitions of coordinates all correspond to redefining the flavor currents (or background CS levels) for flavor symmetries $U(1)_k$ that will ultimately be broken.

We might also allow ourselves redefine the $P_j$ by integer linear combinations of the $X_j$ and by multiples of $i\pi$, while keeping $g$ symplectic. This changes background Chern-Simons levels for the flavor and R-symmetries of $T_{(g,\sigma)}$, which one is sometimes interested in keeping track of and sometimes not.

The second type of equivalence is an ``octahedron rotation.'' In terms of phase-space coordinates, it corresponds to a cyclic permutation $Z_i\to Z_i'\to Z_i''\to Z_i$ for any fixed $i$\,; in other words,
\be \begin{pmatrix} Z_i \\ Z_i''\end{pmatrix} \mapsto \begin{pmatrix} i\pi - Z_i-Z_i'' \\ Z_i \end{pmatrix}\,.\ee
This is an operation on two columns of $g$, and on $\sigma$. In terms of gauge theory, the cyclic rotation corresponds to taking a chiral $\phi_i$ of $T_{(g,\sigma)}$ and replacing it by a $U(1)$ gauge theory coupled to another chiral $\phi_i'$ to obtain $T_{(g',\sigma')}$, as discussed in detail in Section \ref{sec:Toct}. This should be an IR duality.%
\footnote{As usual, this statement must be taken with a grain of salt. We are assuming that the mirror symmetry between a free chiral $T_{\soct,\Pi_Z}$ and a free ``vortex'' $T_{\soct,\Pi_{Z'}}$, as in \eqref{Toct}--\eqref{Toct1}, implies IR duality between any gauge theories $T_{(g,\sigma)}, T_{(g',\sigma')}$ in which the free chiral and the free vortex (respectively) are embedded. Arguing this carefully may require a subtle interchange of limits of IR flow. The same comments apply to the 2--3 moves below.}

The third type of equivalence is the algebraic version of an elementary ``2--3 move'' on octahedra. It replaces $(g,\sigma)$ of rank $N$ and $d$ decorated rows with $(g',\sigma)$ of rank $N'=N+1$ and $d'=d$. To implement it in the $2\to 3$ direction, we select any two pairs of initial coordinates, say $(Z_1,Z_1'')$ and $(Z_2,Z_2'')$, and replace them with new initial coordinates $(W_i,W_i'')_{i=1}^3$, where
\be Z_1=W_2''+W_3'\,,\quad \quad Z_1''=W_1''+W_2'\,,\quad  Z_2=W_2'+W_3''\,,\quad Z_2''=W_3'+W_1''\,,\ee
as in Section \ref{sec:23-elem}, with $W_i'=i\pi-W_i-W_i''$. We also add a new undecorated pair of rows $(C_0,\Gamma_0)=(W_1+W_2+W_3-2\pi i,W_1'')$. The new $(g',\sigma')$ can be worked out in a straightforward manner. In the opposite direction, a $3\to 2$ move can only be applied to an element $(g',\sigma')$ that is already the image of a $2\to 3$ move. In particular, an undecorated row of $(g',\sigma')$ must already define a coordinate of the form $C_0=W_1+W_2+W_3-2\pi i$.

The interpretation of the 2--3 move in gauge theory was reviewed in Section \ref{sec:23warmup}. To apply a $2\to 3$ move, we take any pair of chirals $\phi_1,\phi_2$, think of them as an embedded hypermultiplet theory \eqref{Thyper}, and replace the hypermultiplet theory with a copy of $\CN=4$ SQED \eqref{TSQED4}. In the opposite direction, we isolate any three chirals coupled by a cubic superpotential, treat them as a copy of the XYZ model \eqref{TXYZ}, and replace them with a copy of $\CN=2$ SQED \eqref{TSQED2}. We must be careful, though: these applications of a 2--3 move do \emph{not} necessarily preserve all of the operators $\CO_k$ associated with additional broken symmetries. In order to hope that $T_{(g,\sigma)}$ and its putative image $T_{(g',\sigma')}$ are truly IR dual, both of them must independently contain all the necessary $\CO_k$'s.

Algebraically, we note that the Angle Constraint \eqref{Cangle} is preserved by all three types equivalences we have just described. In contrast, the Superpotential Constraint is preserved by the first and second types, but not  by all 2--3 moves. A preliminary guess for a set that might be in one-to-one correspondence with duality classes of 3d $\CN=2$ SCFT's (that flow from abelian theories in the UV) is
\begin{align} \label{RR}
\IR &:= \begin{array}{rr}\big\{ (g,\sigma)\in \wh\IR\,\big|\, \text{$(g,\sigma)$ satisfy the Angle and Superpotential Constraints} \big\}\Big/ \\
\text{modulo equivalences} \\
 \rotatebox{-90}{\!\!\!\!\!$\simeq$}\;\;?\hspace{2.5in}
\end{array} \\
\CR &:= \begin{array}{ll}\big\{\text{3d $\CN=2$ abelian Chern-Simons-matter theories} \\ \hspace{.5in}\text{with any $U(1)_R$--preserving superpotentials}\big\}\big/\text{modulo IR duality}\,. \end{array}\notag
\end{align}
In defining $\IR$, we only quotient out by equivalences (including 2--3 moves) that do preserve the Superpotential Constraint. The potential correspondence $\IR\simeq \CR$ deserves much further study.

\subsection{The existence of operators $\CO_k$}
\label{sec:defO}

For completeness, let us briefly describe the construction of the operators $\CO_k$. We assume that we have followed Steps 0--3 of the gauge theory construction to build a theory $\wt T$, associated with symplectic data $(g,\sigma)$, and we want to apply Step 4.

If $(g,\sigma)$ satisfies the Angle and Superpotential Constraints, then for every symmetry $U(1)_k$ that must be broken there is another pair $(g,\sigma)_k$, related to $(g,\sigma)$ by ``octahedron rotations,'' such that the corresponding coordinate $C_k$ takes the form \eqref{Csuper}. The pair $(g,\sigma)_k$ defines a theory $\wt T_k$ after Step 3 that is mirror symmetric to $\wt T$. In $\wt T_k$ one defines an operator
\be  \label{Wdef}
\hat \CO_k = \phi_1^{\alpha_{k1}}\phi_2^{\alpha_{k2}}\cdots \phi_N^{\alpha_{kN}}\,,
\ee
a simple product of elementary chiral fields of the theory. (Recall that the $\alpha_{ki}$ must be nonnegative integers!) Because of the affine term $-2\pi i$ in \eqref{Csuper}, this operator is guaranteed to have UV R-charge 2. By using the mirror symmetry between $\wt T$ and $\wt T_k$, one can then ``pull back'' the operator $\hat \CO_k$ to an equivalent chiral operator $\CO_k$ in $\wt T$, charged precisely under the desired $U(1)_k$ symmetry. This can be repeated for each of the $N-d$ symmetries that must be broken. Then the final superpotential of $T_{g,\sigma}$ takes the form $\sum_{k=1}^{N-d} \CO_k$.

\subsection{Associated invariants}
\label{app:assoc}

Finally, let us touch upon some mathematical objects associated with an element $(g,\sigma) \in \wh\IR$, and their physical significance.

All of these objects are invariant under the first two types of equivalences presented in Section \ref{sec:equiv}. However, further restrictions are sometimes necessary to ensure invariance under 2--3 moves, such as the non-degeneracy discussed in Section \ref{sec:23-elem}.

\subsubsection*{Phase space and $\mathrm{K}_2$-Lagrangian}

We explained how to build phase spaces and classical Lagrangian submanifolds (and their geometric significance) in Section \ref{sec:combi}. The phase space is a symplectic reduction
\be \CP_{(g,\sigma)} = \CP_\times \big/\!\!\big/(C_k=0) \simeq \{(X_j,P_j)_{j=1}^d\}\,, \ee
and has the holomorphic symplectic structure
\be \Omega = \sum_{j=1}^d dP_j\wedge dX_j\,,\ee
whereas the putative Lagrangian submanifold $\CL_{(g,\sigma)} \subset \CP_{(g,\sigma)}$ is the image of the product $\CL_\times=\{z_i''+z_i^{-1}-1=0\}_{i=1}^N$ (with $z_i=\exp Z_i$, $z_i''=\exp Z_i''$)  under the reduction. Explicitly, this means inverting the transformation \eqref{gsymp} to re-write $z_i,z_i''$ as Laurent monomials in the final coordinates $x_j=\exp X_j$, $c_k=\exp C_k$, $p_j=\exp P_j$, $\gamma_k=\exp \Gamma_k$; then eliminating $\gamma_k$ from the equations and setting $c_j=0$. The result is a Lagrangian submanifold in the exponentiated phase space (with $\C^*$ coordinates $x_j$ and $p_j$), so long as $\CL_\times$ is transverse to the moment maps, and to slices of $\CP_{(g,\sigma)}$ at generic constant $x_j$.

The Lagrangian submanifold is the supersymmetric parameter space for the theory $T_{(g,\sigma)}$ on $\C\times S^1$ \cite{DG-Sdual, DGG, FGSS-AD}. The effective $\CN=(2,2)$ theory obtained from compactifying on $S^1$ at finite radius is governed by a twisted superpotential $\wt W(X_1,\ldots,X_d;\sigma)$. It is the function of complex twisted masses that correspond to the decorated position coordinates of $\CP_{(g,\sigma)}$, as well as dynamical twisted-chiral fields $\sigma$. The Lagrangian equations are
\be \hspace{-.5in} \CL_{(g,\sigma)}\simeq \CL_{\rm SUSY}:\qquad \exp \bigg(\frac{\pd \wt W}{\pd Z_i}\bigg|_{\pd \tilde W/\pd \sigma=0}\bigg) = p_i\,. \ee

We saw in Section \ref{sec:23-elem} that $\CL_{(g,\sigma)}$ is invariant under elementary 2--3 moves so long as the transversality condition is preserved. (In particular, it was necessary that at generic values of $x_j$, the gluing conditions $c_k=1$ never forced $z_i,z_i''$ to take values $0$, $1$, or $\infty$.) In terms of gauge theory, transversality means that $T_{(g,\sigma)}$ has isolated massive vacua on $\C\times S^1$. The vacua correspond to the points $p^{(\alpha)}(x)$ on $\CL_{(g,\sigma)}$ at fixed mass parameters $x$.

\subsubsection*{Volume}

Suppose that the product Lagrangian $\CL_\times$ is transverse to the gluing constraints and to slices at constant $x$, as above. Then by using the equations $z_i''+z_i^{-1}-1=0$ and the exponentiated form of \eqref{gsymp} we can express $z_i$ and $z_i''$ as rational functions on $\CL_{(g,\sigma)}$ --- \ie\ rational functions of $x,p$. (Geometrically, we would be solving for octahedron parameters given fixed boundary coordinates on $\pd M$.) The real volume function associated with $(g,\sigma)$ can be defined as
\be {\rm Vol}_{(g,\sigma)}(x,p) = \sum_{i=1}^N \CL i_2(z_i)\,, \label{vol-abs}\ee
where $\CL i_2(z) = \Im\,\Li_2(z)+\arg(1-z)\log|z|$ is the Bloch-Wigner dilogarithm function {\bf [REF]}. The volume is a function on $\CL_{(g,\sigma)}$, and is invariant under algebraic 2--3 moves (that preserve transversality) due to the 5-term relation for $\CL i_2(z)$.

The data $(g,\sigma) \in ISp(2N,\Z)$ actually allows us to promote \eqref{vol-abs} to a complex volume defined modulo $\frac{\pi^2}{6}\Z$ using methods of \cite{Neumann-combinatorics} (or as in \cite[Sec 5.2]{DG-quantumNZ}).

For the theory $T_{(g,\sigma)}$, the volume ${\rm Vol}_{(g,\sigma)}(x,p^{(\alpha)}(x))$ is the free energy density (the value of $\wt W$) in the vacuum $p^{(\alpha)}(x)$ on $\C\times S^1$. As we already observed in Section \ref{sec:energy}, the volumes of \emph{particular} solutions $p^{(\alpha)}(x)$ dominate the asymptotics of various partition functions of $T_{(g,\sigma)}$.

\subsubsection*{Quantum Lagrangian}

We reviewed in Section \ref{sec:qremarks} how to use the data $(g,\sigma)$ to promote a classical Lagrangian $\CL_{(g,\sigma)}$ to a left ideal of quantum operators $\hat \CL_{(g,\sigma)}$. The basic idea is to promote all logarithmic coordinates to operators satisfying $[\hat Z_i'',\hat Z_i]=\hbar$, etc., and to promote \eqref{gsymp} to linear relations among quantum operators with a canonical correction $i\pi\sigma \to (i\pi+\hbar/2)\sigma$ on the RHS. This then allows the left ideal $\big(\hat z_i''+\hat z_i^{-1}-1\big)_{i=1}^N$ to be rewritten in terms of exponentiated operators $\hat x=\exp \hat X,\,\hat p=\exp \hat P,\,\hat c=\exp \hat C,\,\hat \gamma=\exp\hat \Gamma$. Eliminating $\hat \gamma$ and setting $\hat c=1$ should produce $\hat \CL_{(g,\sigma)}$.%
\footnote{An alternative method of quantizing one-dimensional Lagrangians involves the topological recursion of \cite{EO}, adapted to the current setting in \cite{GS-quant,BorotEynard}. It is expected that the topological recursion is equivalent to the algebraic approach outlined here.} %
The left ideal $\hat \CL_{(g,\sigma)}$ encodes the Ward identities for line operators satisfied by partition functions of $T_{(g,\sigma)}$.

\subsubsection*{Quantum partition functions}

The element $(g,\sigma)$ gives rise to several closely related wavefunctions. Geometrically, they are versions of Chern-Simons partition functions on a 3-manifold $M$; they are also partition functions of $T_{(g,\sigma)}$ on simple curved backgrounds.

\begin{itemize}
\item The supersymmetric index of $T_{(g,\sigma)}$ on $S^2\times S^1$ was described in \cite{DGG-index} following \cite{Kim-index, IY-index, KW-index}. As a wavefunction, it corresponds to quantizing the exponentiated phase space $\CP_{(g,\sigma)}\simeq (\C^*)^{2d}$ with a \emph{real} symplectic form $\omega \sim \hbar^{-1} \Im\, \Omega$. In a particular real polarization, it depends on $d$ integers $m_j$ (magnetic fluxes) and $d$ phases $\zeta_j\in S^1$ (electric fugacities) as well as a formal parameter $q=e^\hbar$. The combinations $q^{m_j/2}\zeta_j$ are identified with the position coordinates $x_j$.
Geometrically, the index matches the partition function of $PGL(K,\C)$ Chern-Simons theory at level $k=1$ on a manifold $M$. The index is well-defined and invariant under 2--3 moves so long as the data $(g,\sigma)$ admits a so-called semi-strict angle structure \cite{Gar-index}.

\item The ellipsoid ($S^3_b$) partition function of $T_{(g,\sigma)}$ can be computed by methods of \cite{Kapustin-3dloc, HHL, DGG}. It corresponds to quantization of $\CP_{(g,\sigma)}$ with respect to a real symplectic form $\Re(\Omega)+s\,\Im(\Omega)$, where $s=i\frac{1-b^2}{1+b^2}$.%
\footnote{Strictly speaking, the symplectic form is real when $|b|=1$. To make sense of other values of $b$, one must analytically continue the answer.} %
As explained in \cite{D-levelk} (after the first version of the current paper appeared), this turns out to be identical to the quantization of the real slice of the phase space $\CP_{(g,\sigma)}$ with $X_j,P_j\in \R$.
Thus, geometrically, one may identify the ellipsoid partition function with either a $PGL(K,\C)$ Chern-Simons partition function at level $k=1$ \emph{or} a $PGL(K,\R)$ Chern-Simons partition function. It depends analytically on $d$ continuous variables $X_j$ (masses).
Geometric partition functions of this type (``state integral models'') first appeared in \cite{hikami-2006}, with later developments including \cite{DGLZ, Dimofte-QRS, KashAnd, AK-new}.

\item The partition function of a 3d $\CN=2$ theory on an ellipsoidally deformed lens space $L(k,1)_b =S^3_b/\Z_k$ was developed and studied in \cite{BNY-Lens, IY-Lens, IMY-fact}.  Geometrically, it is expected to agree with a $PGL(K,\C)$ Chern-Simons partition function at level $k$ \cite{D-levelk, AK-complexCS}.

\item By expanding the formal integrals for ellipsoid partition functions around complex critical points, it is possible to derive a series of perturbative invariants associated with classical solutions $p^{(\alpha)}(x)$ on $\CL_{(g,\sigma)}$ (\ie, geometrically, to flat $PGL(K,\C)$-connections with fixed boundary conditions $x$) \cite{DG-quantumNZ}. After the volume \eqref{vol-abs}, the first subleading invariant is a Reidemeister-Ray-Singer torsion. In the geometric setting, these perturbative invariants are expected to match the asymptotic expansion of colored HOMFLY polynomials on $M$.

\item The partition functions $B^\alpha(x;q)$ of $T_{(g,\sigma)}$ on a twisted (spinning) geometry $\C\times_q S^1$ were recently discussed in \cite{BDP-blocks}. These ``holomorphic blocks'' depend locally on a solution $p^{(\alpha)}(x)$ to the classical Lagrangian equations, but are fully nonperturbative quantum objects. It was conjectured that both the index and the ellipsoid partition functions can be written as sums of products of the same holomorphic blocks. Geometrically, the holomorphic blocks seem to come from nonperturbative completions of analytically continued Chern-Simons theory on a manifold $M$, as in \cite{Wit-anal, Wfiveknots}.

\end{itemize}

\section{The Poisson bracket for eigenvalues}
\label{app:NZ}

Here we prove that the eigenvalues of holonomies around noncontractible cycles on small boundaries (small tori or annuli) have the expected Atiyah-Bott-Goldman Poisson brackets discussed in Sections \ref{sec:torusbdy} and \ref{sec:sympsum}.

We consider an admissible 3-manifold $M$ with an ideal triangulation and a subsequent $K$-decomposition. Let $\CP_\times = \CP_{\pd\soct_i}$ be the linear product phase space of all the octahedra. We work with logarithmic coordinates, as in Section \ref{sec:combi}. For every internal and external point in the $K$-decomposition we define affine-linear functions $C_k$, $\CX_j$ (respectively) on $\CP_\times$ --- the sums of octahedron parameters at vertices that touch those points. For any closed path $\gamma$ on a small boundary, we choose $K-1$ representative paths $\gamma_a$ on the slices parallel to that small boundary; then we use the rules of Section \ref{sec:pathcoord} to define $K-1$ corresponding affine-linear path-coordinates $U_a^\gamma$.

We already argued in Section \ref{sec:tet-symp} (using cancellation of arrows on octahedra) that
\be \{C_k,C_{k'}\}=\{C_k,\CX_j\}=0\,, \ee
and
\be \{\CX_j,\CX_{j'}\} = \text{signed sum of boundary arrows connecting points at $\CX_j$ and $\CX_{j'}$}\,, \ee
which is the expected Poisson structure on the boundary. We now want to show that
\be \{C_k,U_a^\gamma\} = \{\CX_j, U_a^\gamma\} = 0\,, \ee
and
\be \{U_a^\gamma,U_b^\lambda\} = \kappa_{ab}\,\langle\gamma,\lambda\rangle\,,\ee
where $\langle\gamma,\lambda\rangle$ is the signed intersection number of the paths $\gamma,\lambda$ on the small boundary, and
\be \kappa_{ab}= \begin{cases} 2 & a=b \\
 -1 & |a-b|=1 \\
 0 & |a-b|\geq 2 \end{cases}
\ee
is the Cartan matrix of $SL(K)$. This justifies all of the brackets summarized in Section \ref{sec:sympsum}, and generalizes a central result of \cite{NZ} for $K=2$.

Our proof basically extends that of \cite{NZ}, using the framework of slices and path coordinates. There are two basic steps. First we show that the $U_a^\gamma$ commute with all other coordinates. The fact that $U_a^\gamma$ commutes with the $C_k$ immediately implies that the Poisson bracket of $U_a^\gamma$ only depends on the \emph{homotopy class} of the chosen path $\gamma_a$ on the $a$-th slice. In particular, it indicates that $\{U_a^\gamma,U_b^\lambda\}$ should only depend on the intersection number $\langle\gamma,\lambda\rangle$ and some universal function of $a,b$. The second part of the proof fixes the normalization of $\{U_a^\gamma,U_b^\lambda\}$, through what is unfortunately a brute-force, case-by-case computation.

\subsection{$\{C,U\} = \{\CX,U\} = 0$}
\label{sec:UCbracket}

Let $p$ be an internal or external point in the $K$-decomposition of $M$. Let $\CX$ generically denote the sum of octahedron parameters at $p$. Let $U$ be the coordinate for any closed path $\gamma$ (we drop the subscript $a$ in this subsection) on any global slice $S$ in the $K$-decomposition. We want to show that $\{\CX,U\}=0$.

\begin{figure}[htb]
\centering
\includegraphics[width=3.7in]{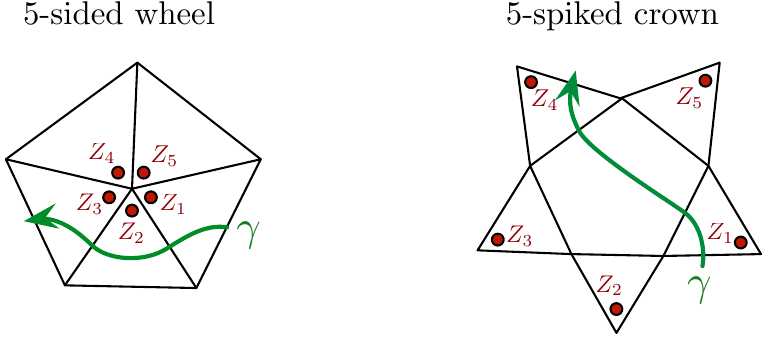}
\caption{The wheel and crown arrangements that can occur on slices.}
\label{fig:crownwheel}
\end{figure}

The approach will be to look at how the octahedron parameters contributing to $\CX$ can appear on the slice $S$. There are two basic cases: either $\CX$ contains a sum of parameters in a \emph{crown-shaped} or \emph{wheel-shaped} arrangement of small triangles on $S$. These two arrangements are shown in Figure \ref{fig:crownwheel}, with the parameters contributing to $\CX$ marked by red dots. Any path disjoint from these arrangements automatically commutes with $\CX$. Any path $\gamma$ that enters and exits these arrangements (without stopping) also has a coordinate $U$ that commutes with the sum of contributions to $\CX$, due to pairwise cancellations in $\{\CX,U\}$.
For example, in Figure \ref{fig:crownwheel}, $\CX$ has contributions $\CX=\sum_{i=1}^5 Z_i+\ldots$. The coordinate $U$ for the path drawn through the wheel equals $Z_1'-Z_2+Z_3''+\ldots$, and the coordinate for the path drawn through the crown equals $Z_1''-Z_4''+\ldots$. In each case there is a pair of $\pm 1$ contributions to $\{\CX,U\}$ that cancel, coming from the entry and exit points.

Now let us go through the various options for locations of $p$ more carefully, and verify that the crown and wheel arrangements arise.

\begin{figure}[htb]
\centering
\includegraphics[width=6in]{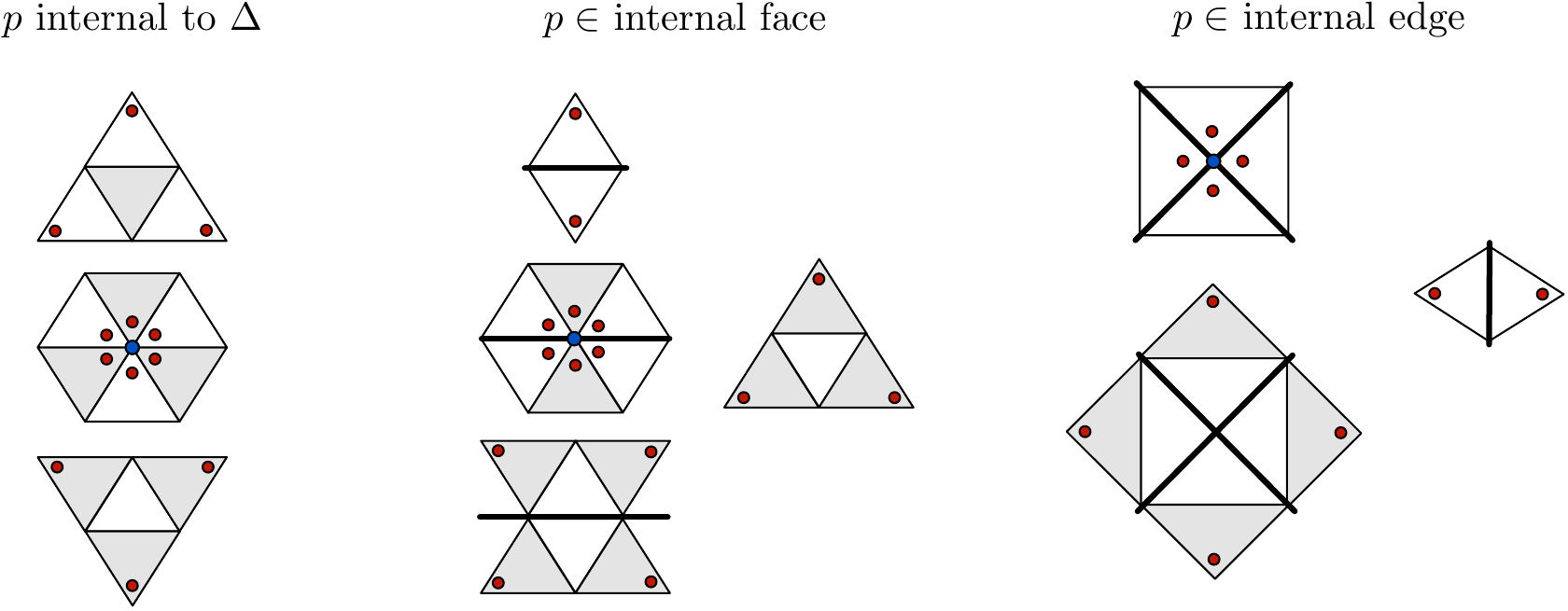}
\caption{Slices in the neighborhood of a point $p$ (blue dot), with octahedron parameters appearing in $\CX$ marked by red dots.}
\label{fig:crown-internal}
\end{figure}

First suppose that $p$ is a point internal to the $K$-decomposition of a tetrahedron. Recall that there are four families of $K-1$ slices through the octahedron (one family is centered around each tetrahedron vertex). In each family there are three slices that contain parameters contributing to $p$: the slice containing $p$ and the slices directly above and below it. On the slice containing $p$, the parameters appear in a six-sided wheel; whereas on the slices above and below the parameters appear in three-spiked crowns (Figure \ref{fig:crown-internal}, left).

If $p$ is on a glued face shared by two tetrahedra, then there are five relevant families of slices --- one centered around each vertex of the big bipyramid that contains $p$. In the three families centered around the vertices of the big glued face, the slices containing $p$ and those above and below $p$ have $\CX$ parameters, in wheels or crowns (Figure \ref{fig:crown-internal}, center). In the two families of slices parallel to the glued face, only the $(K-1)$-st slice has $\CX$ parameters, in a three-spiked crown.

If $p$ is on an internal edge $E$ of the triangulation shared by $N$ tetrahedra (not necessarily distinct) in an $N$-gonal bipyramid, there are two types of families of slices to consider. The two families centered around the ends of $E$ each have two slices with $\CX$ parameters, in an $N$-sided wheel and an $N$-spiked crown (Figure \ref{fig:crown-internal}, right). The $N$ families centered around the ``equatorial'' vertices of the bipyramid each have the $(K-1)$-st slice containing $\CX$ parameters, in a 2-spiked crown.

\begin{figure}[htb]
\centering
\includegraphics[width=3.5in]{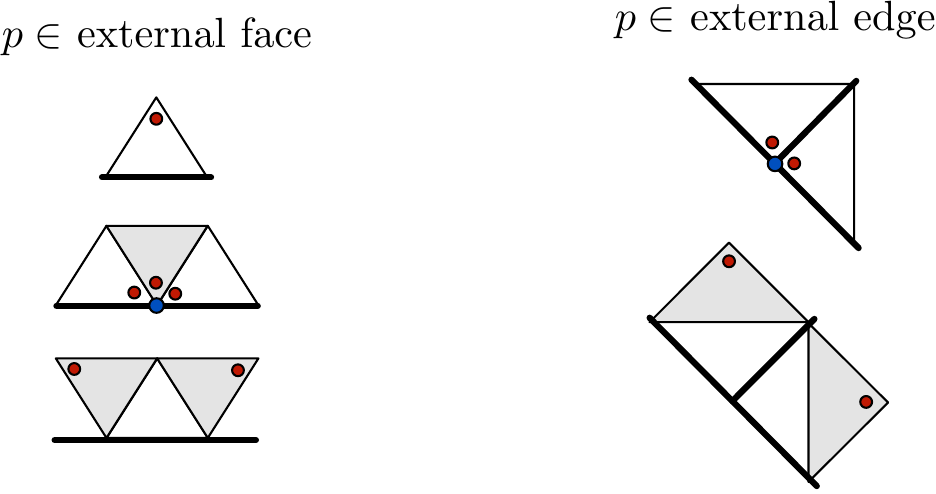}
\caption{Slices in the neighborhood of an external point $p$ (blue dot).}
\label{fig:crown-external}
\end{figure}

Finally, suppose that $p$ is on a big boundary of $M$. Then the slices containing $\CX$ parameters are basically as described above, except some families of slices may get cut in half. For example, if $p$ is on a face of a tetrahedron on the big boundary, the slices centered around the three vertices of that face look like the left side of Figure \ref{fig:crown-external}. If $p$ is on an external edge, then the slices centered around the endpoints of the edge look like the right side of Figure \ref{fig:crown-external}. This does not affect the argument for cancellations above. Any \emph{closed} path $\gamma$ in one of these slices must both enter and exit the ``half-crown'' or ``half-wheel'' regions without stopping, and contributions to $\{\CX,U\}$ will continue to cancel in pairs.

\subsection{$\{U^\gamma,U^\lambda\} = \kappa\cdot\langle \gamma,\lambda\rangle$}
\label{sec:UUbracket}

In the second part of the proof, we consider any two closed paths $\gamma,\lambda$ on slices in the $K$-decomposition of $M$. Again we suppress the subscripts $(a,b)$ denoting the slice number. We want to show that the corresponding path-coordinates $U^\gamma,\,U^\lambda$ have Poisson bracket proportional to the signed intersection number of the projection of the paths to the small boundary; and that the proportionality constant equals $2$ if the paths are on the same slice, $-1$ if the paths are on immediately neighboring slices, and $0$ otherwise.

\begin{wrapfigure}{r}{1.7in}
\includegraphics[width=1.7in]{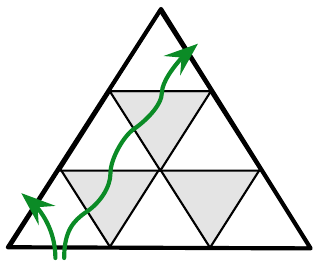}
\caption{Uniformization of paths inside a tetrahedron.}
\label{fig:LRpath}
\end{wrapfigure}

From Section \ref{sec:UCbracket} we know that $\{U^\gamma,U^\lambda\}$ only depends on the homotopy class of $\gamma$ and $\lambda$ (since closed paths commute with all gluing constraints $C_k$). Therefore, we may ``uniformize'' the paths, so that 1) they are smooth (have no bounces); 2) whenever they turn counterclockwise (left) inside a big tetrahedron they stay close to the edge they wind around; and 3) whenever they turn clockwise (right) they stay as far away from the edge as possible (Figure \ref{fig:LRpath}). (It may be useful to recall that slices are always viewed from \emph{above}, from the perspective of a small boundary, in the 3-manifold $M$. Thus clockwise/right and counterclockwise/left orientations are well defined.)

The goal now is to show that $\{U^\gamma,U^\lambda\}$ gets contributions from points of intersection of the paths and nowhere else. We fix $\gamma$ and assume it lies on the $a$-th global slice (parallel to some small boundary component). We will look at how octahedron parameters that contribute to $U^\gamma$ appear in other slices of $M$, and how they interact with other putative paths $\lambda$ on some $b$-th slice. We simply go through an exhaustive case-by-case analysis.

Any putative contribution to $\{U^\gamma,U^\lambda\}$ comes from segments of $\gamma$ and $\lambda$ that run along a common octahedron $\oct_i$ --- so that potentially non-commuting parameters $Z_i,Z_i',Z_i''$ from that octahedron occur in both $U^\gamma$ and $U^\lambda$. Moreover, in order for segments of $\gamma$ and $\lambda$ to share an octahedron, the segments must run along slices in a common big tetrahedron $\Delta$. There are two basic cases to consider: these slices are in the same family within $\Delta$ (\ie\ they are parallel to each other); or they are in different families.

\subsubsection{Parallel slices}

Suppose that segments of $\gamma$ and $\lambda$ run on parallel slices within a single tetrahedron $\Delta$. When exiting $\Delta$ they may continue running on parallel slices in a second tetrahedron $\Delta'$, and into a third, etc. As long as the paths stay together, they simultaneously turn left or right within each successive tetrahedron. We will see in a second that the contributions to $\{U^\gamma,U^\lambda\}$ from such synchronous segments (where they turn together) vanish identically. Then, if the paths continue on this way for their entire length, it follows that $\{U^\gamma,U^\lambda\}=0$, consistent with the fact that $\langle \gamma,\lambda\rangle=0$. Otherwise, there must exist a tetrahedron $\Delta_0$ in which the paths converge, starting their synchronous run, and a tetrahedron $\Delta_1$ in which the paths diverge, ending the synchronous run. We will show that if $\gamma$ and $\lambda$ cross during the run then the contribution to $\{U^\gamma,U^\lambda\}=0$ equals $\pm \kappa_{ab}$, and otherwise vanishes.

\begin{figure}[htb]
\centering
\includegraphics[width=6in]{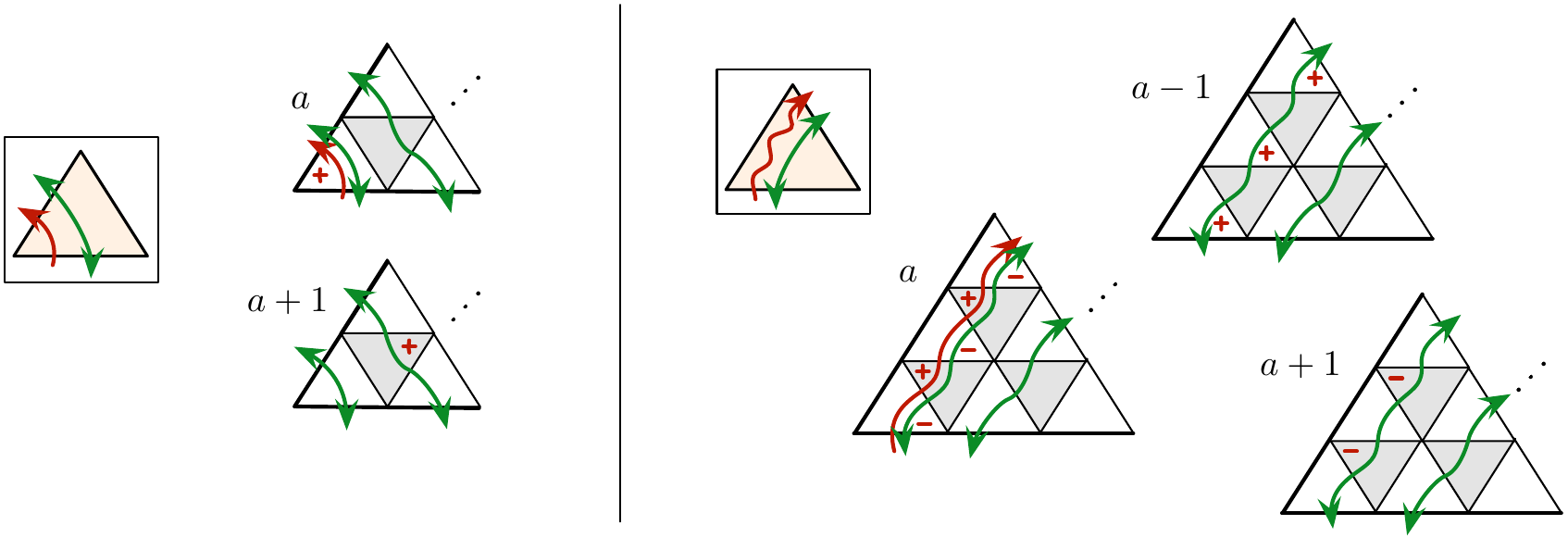}
\caption{Paths turning together within a tetrahedron. $\gamma$ and the octahedron parameters contributing to it are drawn in red, while possible options for $\lambda$ are in green.}
\label{fig:Uparallel-I}
\end{figure}

Thus, let us first consider slices of a tetrahedron $\Delta$ in which the paths turn the same way, modulo orientation (which may be equal or opposite).
The path $\gamma$, on the $a$-th slice, either turns left or right. If it turns left (making a small turn) then the octahedron parameter (call it $Z_i$) contributing to this segment of $\gamma$ also appears on the $(a+1)$-st slice. This is shown on the left of Figure \ref{fig:Uparallel-I} with a red `$+$' sign, indicating that the parameter contributes positively to $U^\gamma$. We also draw possible locations for $\lambda$, in green. Depending on the precise value of $K$ and whether $\lambda$ is oriented equal or opposite to $\gamma$, the path can appear in several different positions. It can also appear on any slice. However, in any position that allows $\lambda$ to pass through small triangles containing $Z_i$ (red $+$'s), the coordinate $U^\lambda$ only picks up $\pm Z_i$ --- never $Z_i'$ or $Z_i''$. Thus, in these parallel slices of $\Delta$, the contribution to $\{U^\gamma,U^\lambda\}$ is zero.

Similarly, if $\gamma$ turns right inside $\Delta$, it picks up a series of (distinct) octahedron parameters with opposite signs, which also appear on the $(a+1)$-st and $(a-1)$-st slices, shown on the left of Figure \ref{fig:Uparallel-I}. In any possible configuration, the path for $\lambda$ at worst picks up these same octahedron parameters. So again the contribution to $\{U^\gamma,U^\lambda\}$ is zero.

Knowing that the intermediate stages of a synchronous run contribute nothing to the Poisson bracket, we can focus on the beginning and end of the run. We might as well assume that the tetrahedra $\Delta_0$, $\Delta_1$ at the beginning and end are immediately adjacent (\ie\ there are no intermediate segments).

\begin{figure}[htb]
\centering
\includegraphics[width=5.5in]{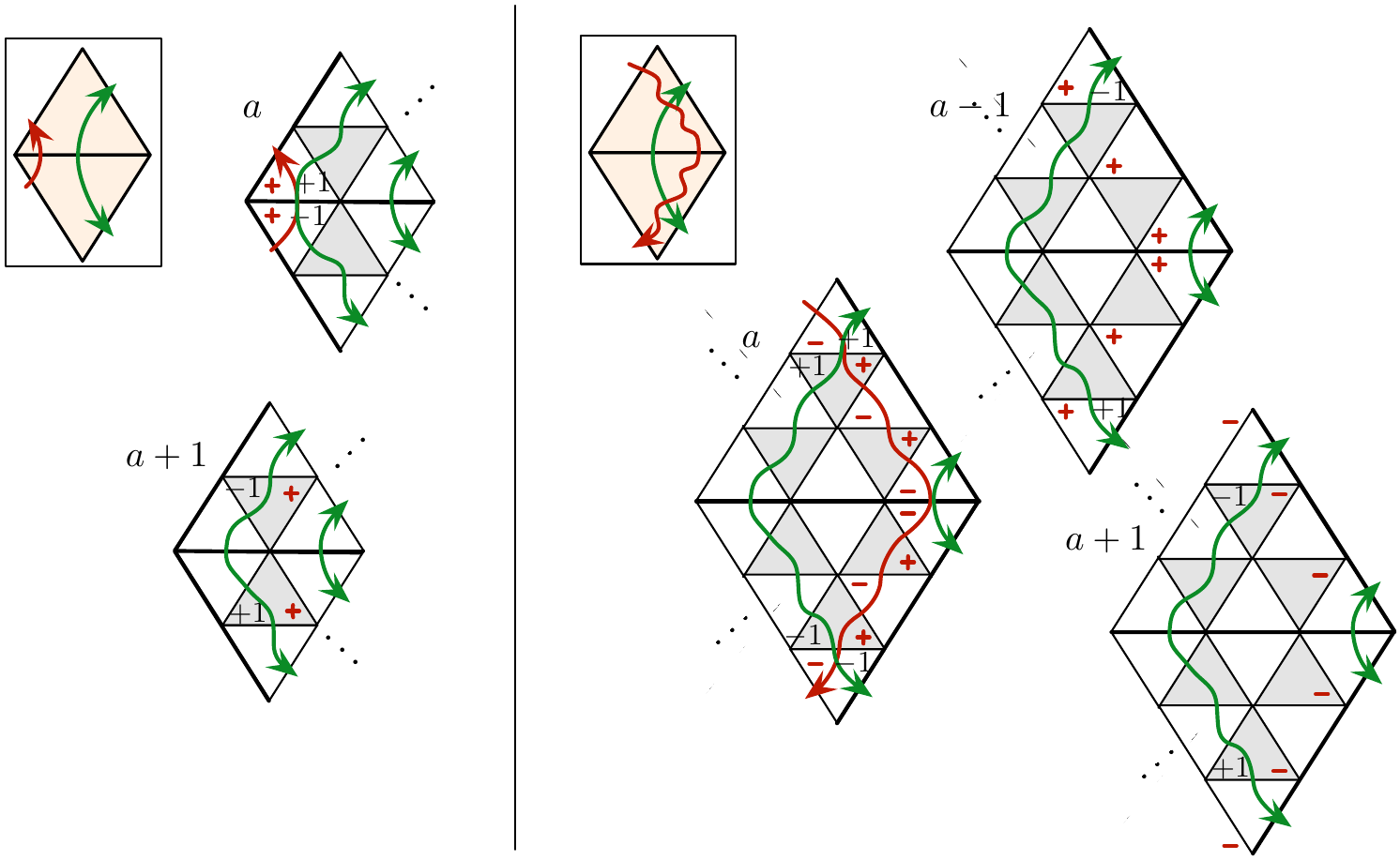}
\caption{Paths converging and then diverging without a net crossing.}
\label{fig:Uparallel-II}
\end{figure}

If the projections of $\gamma$ and $\lambda$ to the common small boundary do \emph{not} have a net crossing during the run, there are two possible cases to analyze, depicted in Figure \ref{fig:Uparallel-II}. On the right, $\gamma$ turns left (making a small turn), and its octahedron parameters appear on two slices. Call the octahedron parameters $Z_0,Z_1$. If $\lambda$ runs along these slices, it can be in several different positions, depending on $K$ and its orientation. In each possible position, any parameter picked up by $U^\lambda$ in $\Delta_0$ that does not commute with $Z_0$ is paired with a parameter in $\Delta_1$ that does not commute with $Z_1$, so that the total contribution to $\{Z_0+Z_1,U^\lambda\}$ cancels. For example, one of the $\lambda$ paths on the $a$ slice picks up $-Z_0''$ in $\Delta_0$ and $-Z_1'$ in $\Delta_1$, with $\{Z_0+Z_1,-Z_0''-Z_1''\}=0$.

Similarly, if $\gamma$ turns to the left, its octahedron parameters appear in slices $a$, $a+1$, and $a-1$. Thus paths $\lambda$ on these slices might have segments that don't commute with $\gamma$. But all non-commuting contributions from $\Delta_0$ cancel with those from $\Delta_1$.

Altogether, with no net crossing, the total contribution to $\{U^\gamma,U^\lambda\}$ from a full synchronous run of the paths vanishes.

\begin{figure}[htb]
\centering
\includegraphics[width=6in]{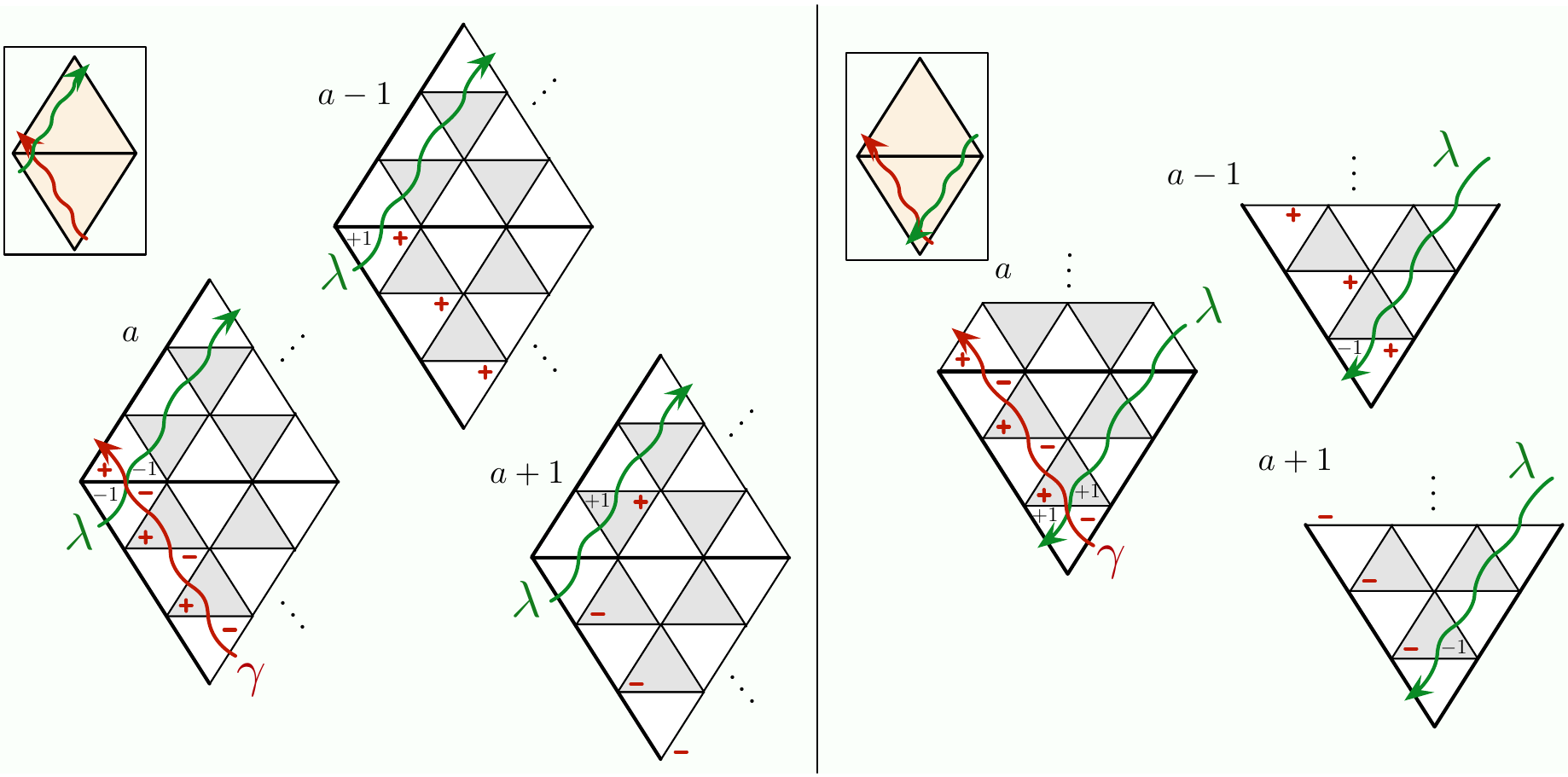}
\caption{Paths converging and then diverging with a crossing. $\langle \gamma,\lambda\rangle=\mp 1$ on the left (right).}
\label{fig:Usameslice}
\end{figure}

Finally, we consider the most interesting case: a nontrivial crossing of $\gamma,\lambda$ going from $\Delta_0$ to $\Delta_1$. There are two possible arrangements. This time we keep careful track of orientations of both $\gamma$ and $\lambda$. The path $\gamma$ must have segments turning both right and left, so in both cases its octahedron parameters appear on the $(a\pm1)$-st slices as well as the $a$-th. On the left (resp., right) of Figure \ref{fig:Usameslice} the intersection number of the projections $\langle \gamma,\lambda\rangle$ equals minus one (resp., plus one).

Correspondingly, on the left of Figure \ref{fig:Usameslice}, we see that if $\lambda$ is in the $a$-th slice then each of the two small triangles around the intersection point of $\gamma$ and $\lambda$ contributes $-1$ to $\{U^\gamma,U^\lambda\}$. If $\lambda$ is on the $(a+1)$-st or $(a-1)$-st slices then $U^\lambda$ picks up a single parameter that doesn't commute with part of $U^\gamma$, contributing $+1$ to $\{U^\gamma,U^\lambda\}$. On the right of Figure \ref{fig:Usameslice}, the same observations hold, with opposite signs. Altogether, we find that the contribution to the bracket from these crossing segments of path is
\be \{U^\gamma,U^\lambda\} = \kappa_{ab}\,\langle \gamma,\lambda\rangle\,, \label{PBproof} \ee
when $\lambda$ lies on the $b$-th slice.

\subsubsection{Slices in different families}

To complete the proof, we need to show that whenever $\gamma$ and $\lambda$ run through the same tetrahedron (or tetrahedra) but along two \emph{different} families of slices (centered locally around two different tetrahedron vertices) the contribution to the Poisson bracket vanishes.

As for parallel slices, is useful to begin with a single tetrahedron $\Delta$ in which the two paths turn the same way --- meaning that they enter and exit in the same two tetrahedron faces. This identifies an edge $E$ of $\Delta$ that both paths turn around. We depict this situation in Figure \ref{fig:Ucomm1}, choosing to place $\gamma$ on the family of slices centered around the ``top'' vertex of $E$, and $\lambda$ around the ``bottom'' vertex.

\begin{figure}[htb]
\centering
\includegraphics[width=6in]{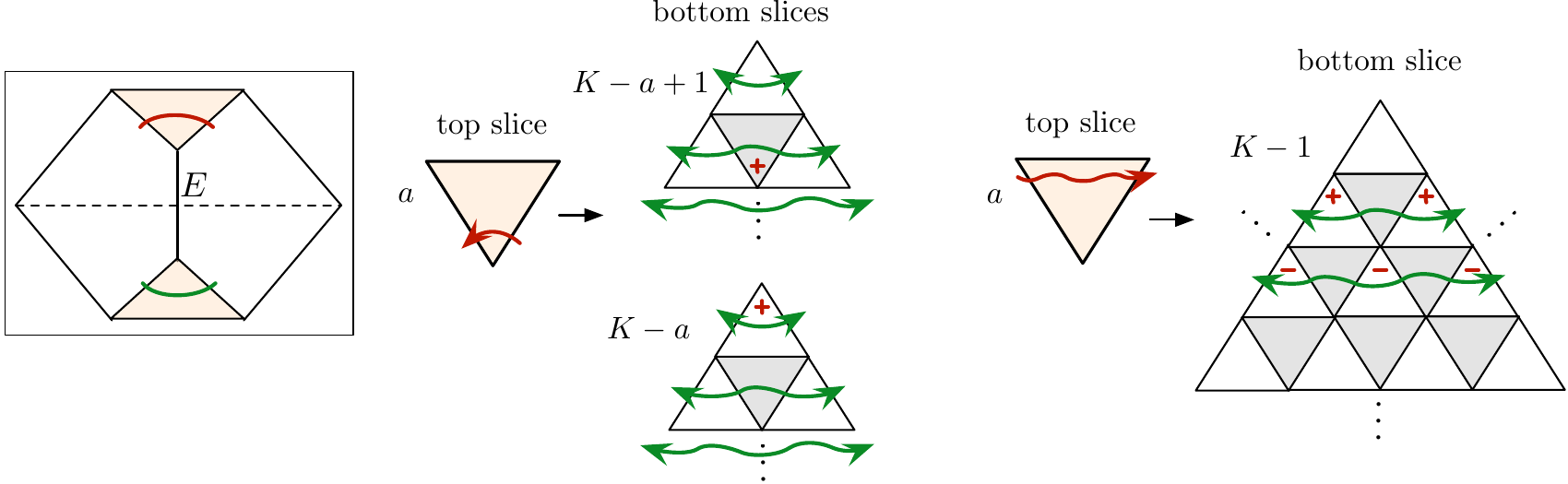}
\caption{Paths entering and exiting a tetrahedron on the same pair of faces.}
\label{fig:Ucomm1}
\end{figure}

There are two possible cases. If $\gamma$ (on the $a$-th slice with respect to the top) turns to the left, then it picks up a single octahedron parameter $Z_i$. This parameter appears on slices $K-a$ and $K-a+1$ with respect to the bottom. Any possible paths $\lambda$ at worst pick up $Z_i$ itself, and so commute with this segment of $\gamma$. Alternatively, if $\gamma$ turns to the right, then its octahedron parameters appear on the $(K-1)$-st (\ie\ the last) slice with respect to the bottom vertex. They appear in two horizontal rows --- the position of these rows depending on $a$. Again, all possible paths $\lambda$ at worst pick up the same octahedron parameters as in $\gamma$, so that the contribution to $\{U^\gamma,U^\lambda\}$ vanishes.

Now we consider the general situation. The paths $\gamma$ and $\lambda$ may run together through a collection of tetrahedra, moving on different families of slices the entire way. In some tetrahedron $\Delta_0$ they must enter on different faces and exit on the same face to start the run. In some other tetrahedron $\Delta_1$ they enter together and exit on different faces to end the run. We may skip intermediate tetrahedra where they move together (because we know the segments in these tetrahedra commute), and simply assume that $\Delta_0$ and $\Delta_1$ are immediately adjacent. There are four basic cases to analyze.

\begin{figure}[htb]
\hspace{-.5cm}
\includegraphics[width=6.5in]{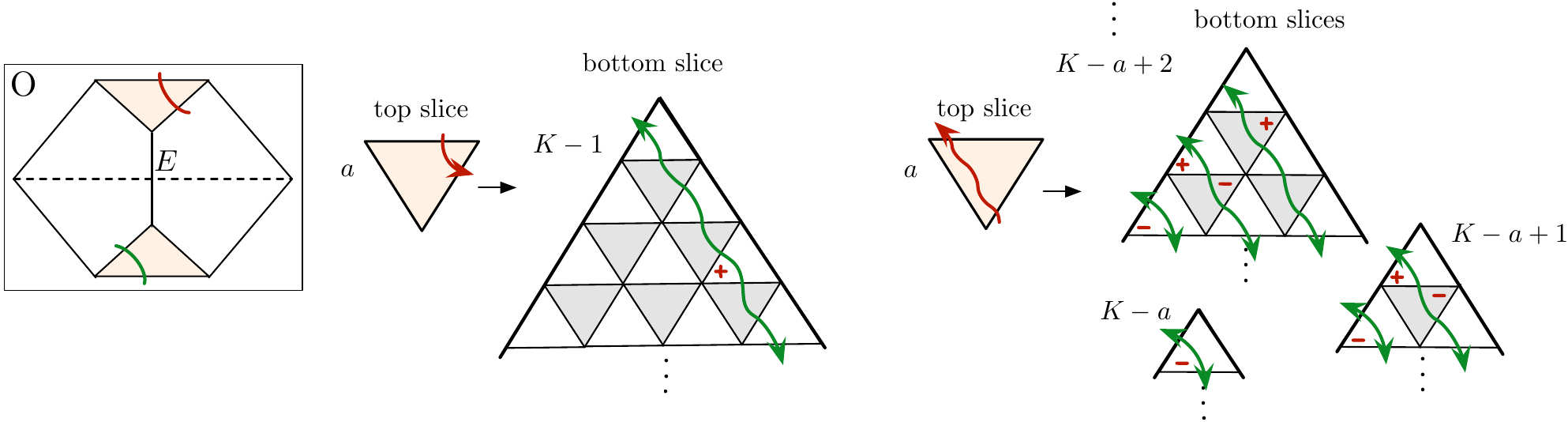}
\caption{Paths entering and exiting a single tetrahedron on four distinct faces.}
\label{fig:Ucomm-0}
\end{figure}

First, it is possible for $\Delta_0$ and $\Delta_1$ to coincide --- \ie\ $\gamma$ and $\lambda$ both enter and exit a single tetrahedron separately. This is depicted in Figure \ref{fig:Ucomm-0}. If $\gamma$ turns to the left, then its single octahedron parameter appears on the last slice with respect to the bottom vertex (at a distance $a$ away from the edge $E$). The only possible $\lambda$ path crossing the small triangle with this parameter just picks up the parameter itself.
If $\gamma$ turns to the right, its $2a-1$ octahedron parameters appear on the first through the $(K-a)$-th slices with respect to the bottom vertex. On the first through the $(K-a+2)$-th slice the octahedron parameters are arranged in two 2-spiked crowns (as in Section \ref{sec:UCbracket}, so every closed path commutes with them. On the $(K-a)$-th and $(K-a+1)$-st slices the possible problematic $\lambda$ paths just pick up the parameters themselves.

Otherwise, we may assume that $\Delta_0$ and $\Delta_1$ are distinct and adjacent. Then there are three remaining arrangements in which $\gamma$ and $\lambda$ may occur. They are depicted in Figures \ref{fig:Ucomm-I}--\ref{fig:Ucomm-III}. Since $\gamma$ and $\lambda$ run along two different families of slices, they distinguish two different vertices common to $\Delta_0$ and $\Delta_1$, which in turn distinguishes a common edge $E$. We choose to position $\gamma$ on slices centered around the ``top'' vertex of $E$ and $\lambda$ on slices centered around the bottom.

\begin{figure}[htb]
\hspace{-.5in}
\includegraphics[width=7in]{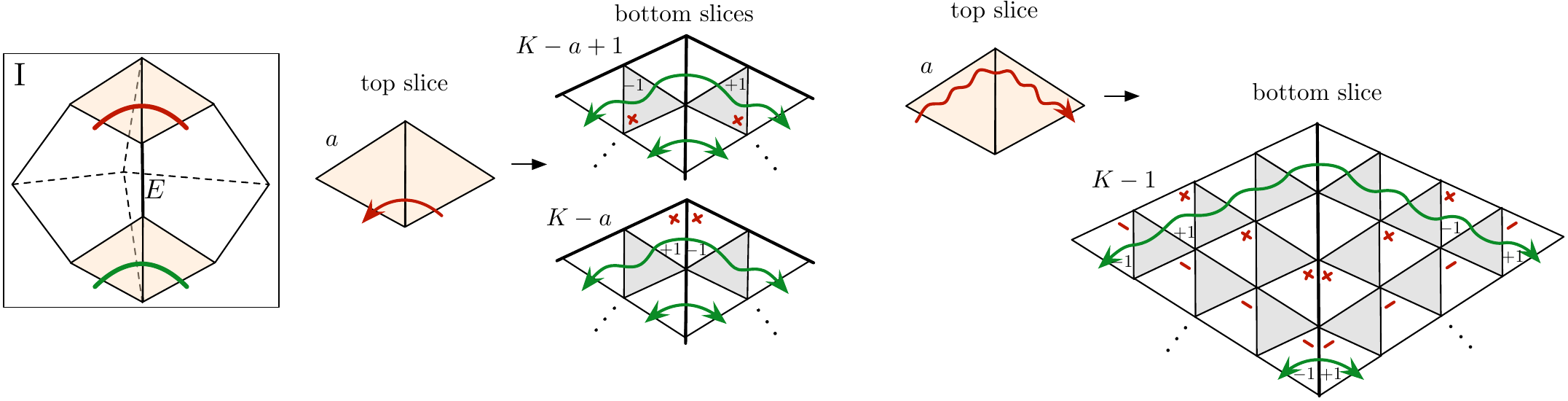}
\caption{Paths entering and exiting a pair of glued tetrahedra on four distinct faces: I}
\label{fig:Ucomm-I}
\end{figure}

\begin{figure}[htb]
\hspace{-.5in}
\includegraphics[width=7in]{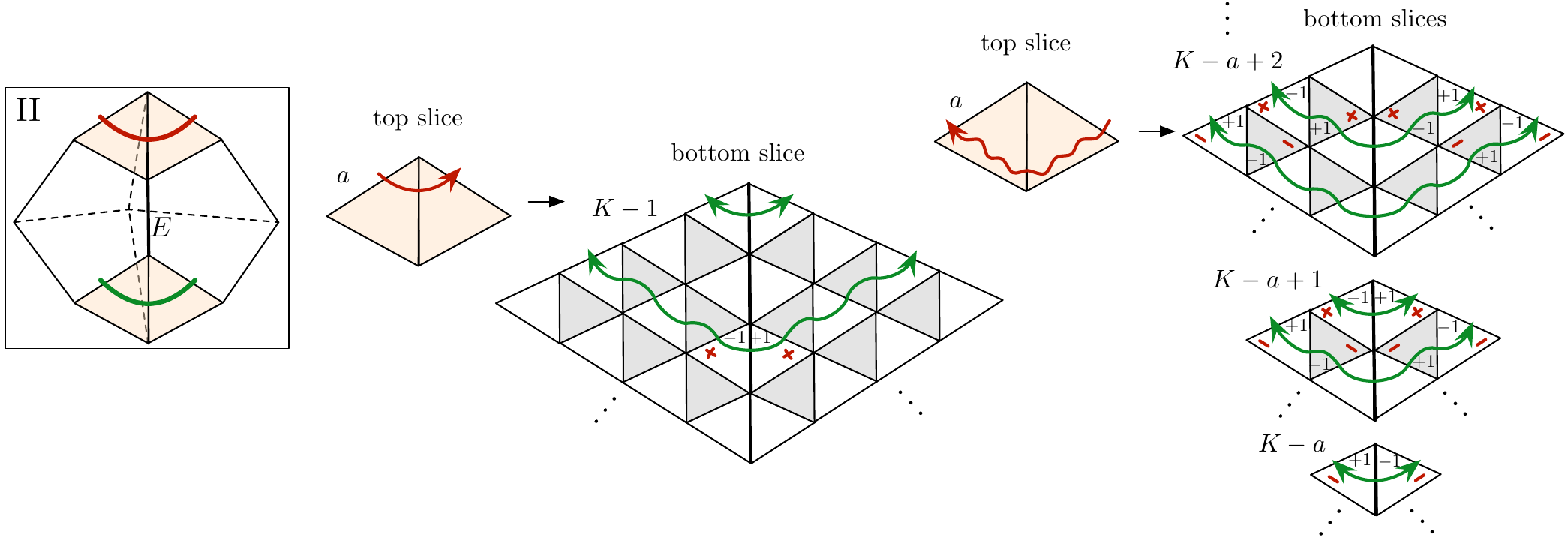}
\caption{Paths entering and exiting a pair of glued tetrahedra on four distinct faces: II}
\label{fig:Ucomm-II}
\end{figure}

\begin{figure}[htb]
\hspace{-.5in}
\includegraphics[width=7in]{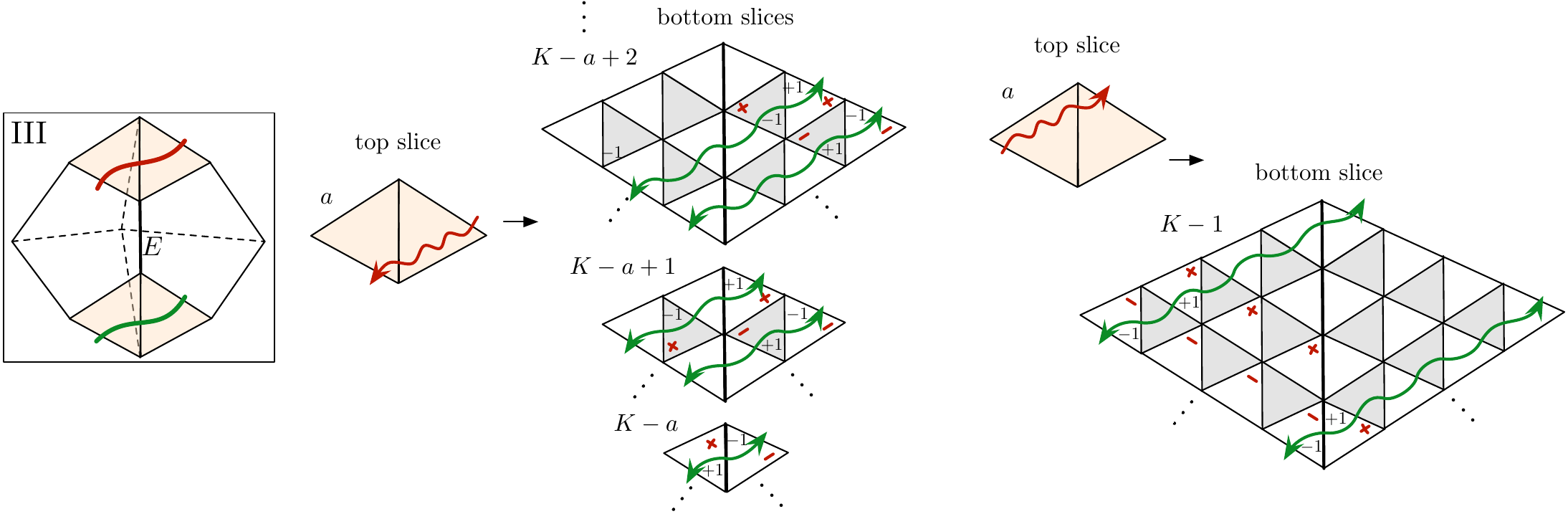}
\caption{Paths entering and exiting a pair of glued tetrahedra on four distinct faces: III}
\label{fig:Ucomm-III}
\end{figure}

In each of these three cases, there are two options for orienting $\gamma$. It is then shown how its octahedron parameters appear on various slices with respect to the bottom vertex of $E$ (the arrangements of $+$'s and $-$'s follow by combining the arrangements in Figures \ref{fig:Ucomm1} and \ref{fig:Ucomm-0}). Now the potential $\lambda$ paths do pick up parameters that don't commute with those in $U^\gamma$. However, the non-commuting contributions always cancel in pairs --- either within $\Delta_0$ and $\Delta_1$ independently, or between $\Delta_0$ and $\Delta_1$. We let the figures speak for themselves.

This finally exhausts all possible ways in which the two paths $\gamma$ and $\lambda$ may come close enough to share an octahedron --- and thus potentially have nontrivial contributions to their Poisson bracket. The \emph{only} contributions that don't vanish identically or cancel in pairs are those that come from paths that cross on adjacent (or identical) parallel slices, as in Figure~\ref{fig:Usameslice}. The contribution \eqref{PBproof} found there is therefore the full Poisson bracket $\{U^\gamma,U^\lambda\}$. This finishes the proof.

\newpage
\bibliographystyle{JHEP_TD}
\bibliography{toolbox}

\providecommand{\href}[2]{#2}\begingroup\raggedright\begin{thebibliography}{100}

\bibitem{DGG}
T.~Dimofte, D.~Gaiotto, and S.~Gukov, {\it Gauge Theories Labelled by
  Three-Manifolds},  {\em Comm. Math. Phys.} {\bf 325} (2014) 367--419,
  [\href{http://xxx.lanl.gov/abs/1108.4389}{{\tt arXiv:1108.4389}}].

\bibitem{Matveev-spines}
S.~V. Matveev, {\it Transformations of special spines, and the Zeeman
  conjecture},  {\em Izv. Akad. Nauk SSSR Ser. Mat.} {\bf 51} (1987), no.~5
  1104--1116, 1119.

\bibitem{Piergallini}
R.~Piergallini, {\it Standard moves for standard polyhedra and spines},  {\em
  Rend. Circ. Mat. Palermo (2) Suppl.} (1988), no.~18 391--414.

\bibitem{AHISS}
O.~Aharony, A.~Hanany, K.~Intriligator, N.~Seiberg, and M.~J. Strassler, {\it
  Aspects of N=2 Supersymmetric Gauge Theories in Three Dimensions},  {\em
  Nucl. Phys.} {\bf B499} (1997), no.~1-2 67--99,
  [\href{http://xxx.lanl.gov/abs/hep-th/9703110v1}{{\tt hep-th/9703110v1}}].

\bibitem{thurston-1980}
W.~Thurston, {\it The Geometry and Topology of Three-Manifolds},  {\em Lecture
  notes at Princeton University} (1980).

\bibitem{Witten-gravCS}
E.~Witten, {\it 2+1 Dimensional Gravity as an Exactly Soluble System},  {\em
  Nucl. Phys.} {\bf B311} (1988), no.~1 46--78.

\bibitem{gukov-2003}
S.~Gukov, {\it Three-Dimensional Quantum Gravity, Chern-Simons Theory, and the
  A-Polynomial},  {\em Commun. Math. Phys.} {\bf 255} (2005), no.~3 577--627,
  [\href{http://xxx.lanl.gov/abs/hep-th/0306165v1}{{\tt hep-th/0306165v1}}].

\bibitem{NZ}
W.~D. Neumann and D.~Zagier, {\it Volumes of hyperbolic three-manifolds},  {\em
  Topology} {\bf 24} (1985), no.~3 307--332.

\bibitem{Dimofte-QRS}
T.~Dimofte, {\it Quantum Riemann Surfaces in Chern-Simons Theory},  {\em Adv.
  Theor. Math. Phys.} {\bf 17} (2013) 479--599,
  [\href{http://xxx.lanl.gov/abs/1102.4847}{{\tt arXiv:1102.4847}}].

\bibitem{hikami-2006}
K.~Hikami, {\it Generalized Volume Conjecture and the A-Polynomials - the
  Neumann-Zagier Potential Function as a Classical Limit of Quantum Invariant},
   {\em J. Geom. Phys.} {\bf 57} (2007), no.~9 1895--1940,
  [\href{http://xxx.lanl.gov/abs/math/0604094v1}{{\tt math/0604094v1}}].

\bibitem{DGLZ}
T.~Dimofte, S.~Gukov, J.~Lenells, and D.~Zagier, {\it Exact Results for
  Perturbative Chern-Simons Theory with Complex Gauge Group},  {\em Comm. Num.
  Thy. and Phys.} {\bf 3} (2009), no.~2 363--443,
  [\href{http://xxx.lanl.gov/abs/0903.2472}{{\tt arXiv:0903.2472}}].

\bibitem{FG-Teich}
V.~V. Fock and A.~B. Goncharov, {\it Moduli spaces of local systems and higher
  Teichmuller theory},  {\em Publ. Math. Inst. Hautes Etudes Sci.} {\bf 103}
  (2006) 1--211, [\href{http://xxx.lanl.gov/abs/math/0311149v4}{{\tt
  math/0311149v4}}].

\bibitem{FG-qdl-cluster}
V.~V. Fock and A.~B. Goncharov, {\it The quantum dilogarithm and
  representations of quantum cluster varieties},  {\em Invent. Math.} {\bf 175}
  (2009), no.~2 223--286, [\href{http://xxx.lanl.gov/abs/math/0702397v6}{{\tt
  math/0702397v6}}].

\bibitem{FG-laminations}
V.~Fock and A.~Goncharov, {\it Symplectic double for moduli spaces of G-local
  systems on surfaces},  \href{http://xxx.lanl.gov/abs/1410.3526}{{\tt
  arXiv:1410.3526}}.

\bibitem{DGV-hybrid}
T.~Dimofte, D.~Gaiotto, and R.~van~der Veen, {\it RG Domain Walls and Hybrid
  Triangulations},  \href{http://xxx.lanl.gov/abs/1304.6721}{{\tt
  arXiv:1304.6721}}.

\bibitem{cooper-1994}
D.~Cooper, M.~Culler, H.~Gillet, D.~Long, and P.~Shalen, {\it Plane Curves
  Associated to Character Varieties of 3-Manifolds},  {\em Invent. Math.} {\bf
  118} (1994), no.~1 47--84.

\bibitem{FG-cluster}
V.~V. Fock and A.~B. Goncharov, {\it Cluster ensembles, quantization and the
  dilogarithm},  {\em Annales Scientifiques L'Ecole Normal Superier, 4-e
  series} {\bf t.42} (2009) 865--930,
  [\href{http://xxx.lanl.gov/abs/math/0311245v5}{{\tt math/0311245v5}}].

\bibitem{Dunfield-mahler}
D.~W. Boyd, F.~Rodriguez-Villegas, and N.~M. Dunfield, {\it Mahler's Measure
  and the Dilogarithm (II)},  {\em Canad. J. Math.} {\bf 54} (2002), no.~3
  468--492, [\href{http://xxx.lanl.gov/abs/math/0308041v2}{{\tt
  math/0308041v2}}].

\bibitem{Champ-hypA}
A.~Champanerkar, {\it A-Polynomial and Bloch Invariants of Hyperbolic
  3-Manifolds},  {\em Ph.D. Thesis, Columbia University} (2003).

\bibitem{GS-quant}
S.~Gukov and P.~Su{\l}kowski, {\it A-polynomial, B-model, and Quantization},
  {\em JHEP} {\bf 1202} (2012) 070,
  [\href{http://xxx.lanl.gov/abs/1108.0002}{{\tt arXiv:1108.0002}}].

\bibitem{Kapustin-Witten}
A.~Kapustin and E.~Witten, {\it Electric-Magnetic Duality And The Geometric
  Langlands Program},  {\em Comm. Num. Th. and Phys.} {\bf 1} (2007) 1--236,
  [\href{http://xxx.lanl.gov/abs/hep-th/0604151v3}{{\tt hep-th/0604151v3}}].

\bibitem{GMN}
D.~Gaiotto, G.~W. Moore, and A.~Neitzke, {\it Four-dimensional wall-crossing
  via three-dimensional field theory},  {\em Comm. Math. Phys.} {\bf 299}
  (2010), no.~1 163--224, [\href{http://xxx.lanl.gov/abs/0807.4723}{{\tt
  arXiv:0807.4723}}].

\bibitem{BFG-sl3}
N.~Bergeron, E.~Falbel, and A.~Guilloux, {\it Tetrahedra of flags, volume and
  homology of SL(3)},  \href{http://xxx.lanl.gov/abs/1101.2742}{{\tt
  arXiv:1101.2742}}.

\bibitem{GGZ-slN}
S.~Garoufalidis, M.~Goerner, and C.~K. Zickert, {\it Gluing equations for
  PGL(n,C)-representations of 3-manifolds},
  \href{http://xxx.lanl.gov/abs/1207.6711}{{\tt arXiv:1207.6711}}.

\bibitem{GTZ-slN}
S.~Garoufalidis, D.~P. Thurston, and C.~K. Zickert, {\it The complex volume of
  SL(n,C)-representations of 3-manifolds},
  \href{http://xxx.lanl.gov/abs/1111.2828}{{\tt arXiv:1111.2828}}.

\bibitem{Zickert-rep}
C.~K. Zickert, {\it The Volume and Chern-Simons Invariant of a Representation},
   {\em Duke Math. J.} {\bf 150} (2009), no.~3 489--532,
  [\href{http://xxx.lanl.gov/abs/0710.2049}{{\tt arXiv:0710.2049}}].

\bibitem{Zickert-sl3}
C.~K. Zickert, {\it The Extended Bloch Group and Algebraic K-Theory},
  \href{http://xxx.lanl.gov/abs/0910.4005}{{\tt arXiv:0910.4005}}.

\bibitem{GMN-spectral}
D.~Gaiotto, G.~W. Moore, and A.~Neitzke, {\it Spectral networks},  {\em Ann.
  Henri Poincare} {\bf 14} (2013), no.~7 1643--1731,
  [\href{http://xxx.lanl.gov/abs/1204.4824}{{\tt arXiv:1204.4824}}].

\bibitem{GMN-snakes}
D.~Gaiotto, G.~W. Moore, and A.~Neitzke, {\it Spectral Networks and Snakes},
  \href{http://xxx.lanl.gov/abs/1209.0866}{{\tt arXiv:1209.0866}}.

\bibitem{CCV}
S.~Cecotti, C.~Cordova, and C.~Vafa, {\it Braids, Walls, and Mirrors},
  \href{http://xxx.lanl.gov/abs/1110.2115}{{\tt arXiv:1110.2115}}.

\bibitem{Cordova-tangles}
C.~Cordova, S.~Espahbodi, B.~Haghighat, A.~Rastogi, and C.~Vafa, {\it Tangles,
  Generalized Reidemeister Moves, and Three-Dimensional Mirror Symmetry},
  \href{http://xxx.lanl.gov/abs/1211.3730}{{\tt arXiv:1211.3730}}.

\bibitem{HenningsonSkenderis}
M.~Henningson and K.~Skenderis, {\it The Holographic Weyl anomaly},  {\em JHEP}
  {\bf 9807} (1998) 023, [\href{http://xxx.lanl.gov/abs/hep-th/9806087v2}{{\tt
  hep-th/9806087v2}}].

\bibitem{HMM-anomalies}
J.~A. Harvey, R.~Minasian, and G.~Moore, {\it Non-abelian Tensor-multiplet
  Anomalies},  {\em JHEP} {\bf 9809} (1998) 004,
  [\href{http://xxx.lanl.gov/abs/hep-th/9808060v1}{{\tt hep-th/9808060v1}}].

\bibitem{FGSS-AD}
H.~Fuji, S.~Gukov, M.~Stosic, and P.~Su{\l}kowski, {\it 3d analogs of
  Argyres-Douglas theories and knot homologies},
  \href{http://xxx.lanl.gov/abs/1209.1416}{{\tt arXiv:1209.1416}}.

\bibitem{CDGS}
H.-J. Chung, T.~Dimofte, S.~Gukov, and P.~Su{\l}kowski, {\it 3d-3d
  Correspondence Revisited},  \href{http://xxx.lanl.gov/abs/1405.3663}{{\tt
  arXiv:1405.3663}}.

\bibitem{GukovPei}
S.~Gukov and D.~Pei, {\it Equivariant Verlinde formula from fivebranes and
  vortices},  \href{http://xxx.lanl.gov/abs/1501.0131}{{\tt arXiv:1501.0131}}.

\bibitem{PeiKe}
D.~Pei and K.~Ye, {\it A 3d-3d appetizer},
  \href{http://xxx.lanl.gov/abs/1503.0480}{{\tt arXiv:1503.0480}}.

\bibitem{DGH}
T.~Dimofte, S.~Gukov, and L.~Hollands, {\it Vortex Counting and Lagrangian
  3-manifolds},  {\em Lett. Math. Phys.} {\bf 98} (2011) 225--287,
  [\href{http://xxx.lanl.gov/abs/1006.0977}{{\tt arXiv:1006.0977}}].

\bibitem{Yamazaki-3d}
Y.~Terashima and M.~Yamazaki, {\it SL(2,R) Chern-Simons, Liouville, and Gauge
  Theory on Duality Walls},  {\em JHEP} {\bf 1108} (2011) 135,
  [\href{http://xxx.lanl.gov/abs/1103.5748}{{\tt arXiv:1103.5748}}].

\bibitem{CJ-S3}
C.~Cordova and D.~L. Jafferis, {\it Complex Chern-Simons from M5-branes on the
  Squashed Three-Sphere},  \href{http://xxx.lanl.gov/abs/1305.2891}{{\tt
  arXiv:1305.2891}}.

\bibitem{LY-S2}
S.~Lee and M.~Yamazaki, {\it 3d Chern-Simons Theory from M5-branes},  {\em
  JHEP} {\bf 1312} (2013) 035, [\href{http://xxx.lanl.gov/abs/1305.2429}{{\tt
  arXiv:1305.2429}}].

\bibitem{D-volume}
T.~Dimofte, {\it 3d Superconformal Theories from 3-Manifolds}, .

\bibitem{Yamazaki-defects}
D.~Gang, N.~Kim, M.~Romo, and M.~Yamazaki, {\it {Aspects of Defects in 3d-3d
  Correspondence}},  \href{http://xxx.lanl.gov/abs/1510.0501}{{\tt
  arXiv:1510.0501}}.

\bibitem{Witten-M}
E.~Witten, {\it Solutions of Four-Dimensional Field Theories Via M Theory},
  {\em Nucl. Phys.} {\bf B500} (Jan, 1997)
  [\href{http://xxx.lanl.gov/abs/hep-th/9703166v1}{{\tt hep-th/9703166v1}}].

\bibitem{Gaiotto-dualities}
D.~Gaiotto, {\it N=2 dualities},  {\em JHEP} {\bf 1208} (2012) 034,
  [\href{http://xxx.lanl.gov/abs/0904.2715}{{\tt arXiv:0904.2715}}].

\bibitem{AGT}
L.~F. Alday, D.~Gaiotto, and Y.~Tachikawa, {\it Liouville Correlation Functions
  from Four-Dimensional Gauge Theories},  {\em Lett. Math. Phys.} {\bf 91}
  (2010), no.~2 167--197, [\href{http://xxx.lanl.gov/abs/0906.3219}{{\tt
  arXiv:0906.3219}}].

\bibitem{GGP-4d}
A.~Gadde, S.~Gukov, and P.~Putrov, {\it Fivebranes and 4-manifolds},
  \href{http://xxx.lanl.gov/abs/1306.4320}{{\tt arXiv:1306.4320}}.

\bibitem{Tachi-discrete}
Y.~Tachikawa, {\it On the 6d origin of discrete additional data of 4d gauge
  theories},  {\em JHEP} {\bf 1405} (2014) 020,
  [\href{http://xxx.lanl.gov/abs/1309.0697}{{\tt arXiv:1309.0697}}].

\bibitem{GMNIII}
D.~Gaiotto, G.~W. Moore, and A.~Neitzke, {\it Framed BPS States},  {\em Adv.
  Theor. Math. Phys.} {\bf 17} (2013) 241--397,
  [\href{http://xxx.lanl.gov/abs/1006.0146}{{\tt arXiv:1006.0146}}].

\bibitem{GW-surface}
S.~Gukov and E.~Witten, {\it Gauge Theory, Ramification, and the Geometric
  Langlands Program},  {\em Curr. Devel. Math.} {\bf 2006} (Dec, 2008) 35--180,
  [\href{http://xxx.lanl.gov/abs/hep-th/0612073v2}{{\tt hep-th/0612073v2}}].

\bibitem{FockChekhov}
L.~Chekhov and V.~V. Fock, {\it Quantum Teichm{\"u}ller Space},  {\em Theoret.
  and Math. Phys.} {\bf 120} (1999), no.~3 1245--1259,
  [\href{http://xxx.lanl.gov/abs/math/9908165v2}{{\tt math/9908165v2}}].

\bibitem{Kash-Teich}
R.~M. Kashaev, {\it Quantization of Teichm{\"u}ller Spaces and the Quantum
  Dilogarithm},  {\em Lett. Math. Phys.} {\bf 43} (1998), no.~2 105--115.

\bibitem{DGG-index}
T.~Dimofte, D.~Gaiotto, and S.~Gukov, {\it 3-Manifolds and 3d Indices},  {\em
  Adv. Theor. Math. Phys.} {\bf 17} (2013) 975--1076,
  [\href{http://xxx.lanl.gov/abs/1112.5179}{{\tt arXiv:1112.5179}}].

\bibitem{D-levelk}
T.~Dimofte, {\it Complex Chern-Simons theory at level k via the 3d-3d
  correspondence},  \href{http://xxx.lanl.gov/abs/1409.0857}{{\tt
  arXiv:1409.0857}}.

\bibitem{NS-I}
N.~A. Nekrasov and S.~L. Shatashvili, {\it Supersymmetric vacua and Bethe
  ansatz},  {\em Nucl. Phys. B, Proc. Suppl.} {\bf 192-193} (2009) 91--112,
  [\href{http://xxx.lanl.gov/abs/0901.4744}{{\tt arXiv:0901.4744}}].

\bibitem{EO}
B.~Eynard and N.~Orantin, {\it Invariants of Algebraic Curves and Topological
  Expansion},  {\em Commun. Number Theory Phys.} {\bf 1} (2007), no.~2
  347--452, [\href{http://xxx.lanl.gov/abs/math-ph/0702045v4}{{\tt
  math-ph/0702045v4}}].

\bibitem{BorotEynard}
G.~Borot and B.~Eynard, {\it All-order asymptotics of hyperbolic knot
  invariants from non-perturbative topological recursion of A-polynomials},
  \href{http://xxx.lanl.gov/abs/1205.2261}{{\tt arXiv:1205.2261}}.

\bibitem{KashAnd}
J.~E. Andersen and R.~Kashaev, {\it A TQFT from quantum Teichm{\"u}ller
  theory},  \href{http://xxx.lanl.gov/abs/1109.6295}{{\tt arXiv:1109.6295}}.

\bibitem{AK-new}
J.~E. Andersen and R.~Kashaev, {\it A new formulation of the Teichm{\"u}ller
  TQFT},  \href{http://xxx.lanl.gov/abs/1305.4291}{{\tt arXiv:1305.4291}}.

\bibitem{Gar-index}
S.~Garoufalidis, {\it The 3D index of an ideal triangulation and angle
  structures},  \href{http://xxx.lanl.gov/abs/1208.1663}{{\tt
  arXiv:1208.1663}}.

\bibitem{GHRS-index}
S.~Garoufalidis, C.~D. Hodgson, J.~H. Rubinstein, and H.~Segerman, {\it
  1-efficient triangulations and the index of a cusped hyperbolic 3-manifold},
  \href{http://xxx.lanl.gov/abs/1303.5278}{{\tt arXiv:1303.5278}}.

\bibitem{Frohman-Gelca}
C.~Frohman, R.~Gelca, and W.~Lofaro, {\it The A-Polynomial From the
  Noncommutative Viewpoint},  {\em Trans. Amer. Math. Soc.} {\bf 354} (2002),
  no.~2 735--747, [\href{http://xxx.lanl.gov/abs/math/9812048v1}{{\tt
  math/9812048v1}}].

\bibitem{Sikora-quant}
A.~Sikora, {\it {Quantizations of Character Varieties and Quantum Knot
  Invariants}},  \href{http://xxx.lanl.gov/abs/0807.0943}{{\tt
  arXiv:0807.0943}}.

\bibitem{AK-complexCS}
J.~E. Andersen and R.~Kashaev, {\it Complex Quantum Chern-Simons},
  \href{http://xxx.lanl.gov/abs/1409.1208}{{\tt arXiv:1409.1208}}.

\bibitem{G93}
A.~B. Goncharov, {\it Explicit construction of characteristic classes},  {\em
  I. M. Gelfand Seminar 169-210, Adv. Soviet Math. 16, Part 11, Amer. Math.
  Soc. (Providence, RI)} (1993).

\bibitem{G95}
A.~B. Goncharov, {\it Geometry of configurations, polylogarithms, and motivic
  cohomology},  {\em Adv. Math.} {\bf 114} (1995), no.~2 197--318.

\bibitem{Bonahon-vol}
F.~Bonahon, {\it A Schlafli-type formula for convex cores of hyperbolic
  3-manifolds},  {\em J. Diff. Geom.} {\bf 50} (1998), no.~1 25--58,
  [\href{http://xxx.lanl.gov/abs/dg-ga/9704017v1}{{\tt dg-ga/9704017v1}}].

\bibitem{Sus1}
A.~A. Suslin, {\it Homology of GL(n), characteristic classes, and Milnor's
  K-theory},  {\em Trudy Mat. Inst. Steklov} {\bf 165} (1984) 188--204.

\bibitem{Neumann-combinatorics}
W.~Neumann, {\it Combinatorics of Triangulations and the Chern-Simons Invariant
  for Hyperbolic 3-Manifolds},  {\em in Topology '90, Ohio State Univ. Math.
  Res. Inst. Publ.} {\bf 1} (1992).

\bibitem{GZ-gluing}
S.~Garoufalidis and C.~K. Zickert, {\it The symplectic properties of the
  PGL(n,C)-gluing equations},  \href{http://xxx.lanl.gov/abs/1310.2497}{{\tt
  arXiv:1310.2497}}.

\bibitem{DV-NZ}
T.~Dimofte and R.~van~der Veen, {\it A Spectral Perspective on Neumann-Zagier},
   \href{http://xxx.lanl.gov/abs/1403.5215}{{\tt arXiv:1403.5215}}.

\bibitem{Kabaya-pants}
Y.~Kabaya, {\it Parametrization of PSL(2,C)-representations of surface groups},
   \href{http://xxx.lanl.gov/abs/1110.6674}{{\tt arXiv:1110.6674}}.

\bibitem{Gar-Le}
S.~Garoufalidis and T.~T. Le, {\it The Colored Jones Function is q-Holonomic},
  {\em Geom. Topol.} {\bf 9} (2005) 1253--1293,
  [\href{http://xxx.lanl.gov/abs/math/0309214v3}{{\tt math/0309214v3}}].

\bibitem{garoufalidis-2004}
S.~Garoufalidis, {\it On the Characteristic and Deformation Varieties of a
  Knot},  {\em Geom. Topol. Monogr.} {\bf 7} (2004) 291--304,
  [\href{http://xxx.lanl.gov/abs/math/0306230v4}{{\tt math/0306230v4}}].

\bibitem{Gar-qSL3}
S.~Garoufalidis and C.~Koutschan, {\it The SL3 Jones polynomial of the trefoil:
  a case study of q-holonomic sequences},
  \href{http://xxx.lanl.gov/abs/1011.6329}{{\tt arXiv:1011.6329}}.

\bibitem{FGS-superA}
H.~Fuji, S.~Gukov, and P.~Su{\l}kowski, {\it Super-A-polynomial for knots and
  BPS states},  \href{http://xxx.lanl.gov/abs/1205.1515}{{\tt
  arXiv:1205.1515}}.

\bibitem{Gar-a}
S.~Garoufalidis, {\it The colored HOMFLY polynomial is q-holonomic},
  \href{http://xxx.lanl.gov/abs/1211.6388}{{\tt arXiv:1211.6388}}.

\bibitem{GGL}
A.~M. Gabrielov, I.~Gelfand, and M.~V. Losik, {\it Combinatorial computation of
  characteristic classes},  {\em Funct. Anal. Appl.} {\bf 9} (1975) 103--115.

\bibitem{Guilloux-PGL}
A.~Guilloux, {\it Representations of 3-manifold groups in PGL(n,C) and their
  restriction to the boundary},  \href{http://xxx.lanl.gov/abs/1310.2907}{{\tt
  arXiv:1310.2907}}.

\bibitem{Sus2}
A.~A. Suslin, {\it K-theory of a field and the Bloch group},  {\em Proc.
  Steklov Inst. Math.} {\bf 4} (1991) 217--239.

\bibitem{G96}
A.~Goncharov, {\it Volumes of hyperbolic manifolds and mixed Tate motives},
  {\em JAMS} {\bf 12} (1996), no.~2 569--619,
  [\href{http://xxx.lanl.gov/abs/alg-geom/9601021v3}{{\tt
  alg-geom/9601021v3}}].

\bibitem{G93b}
A.~B. Goncharov, {\it The classical polylogarithms, algebraic K-theory and
  $\zeta_F(n)$},  {\em Proc. of the Gelfand Seminar, Birkhauser} (1993)
  113--135.

\bibitem{Redlich-parity}
A.~N. Redlich, {\it Gauge noninvariance and parity nonconservation of
  three-dimensional fermions},  {\em Phys. Rev. Lett.} {\bf 52} (1984), no.~1
  18--21.

\bibitem{Redlich-parity2}
A.~N. Redlich, {\it Parity violation and gauge noninvariance of the effective
  gauge field action in three dimensions},  {\em Phys. Rev. D} {\bf 29} (1984),
  no.~10 2366--2374.

\bibitem{CDFKS}
C.~Closset, T.~T. Dumitrescu, G.~Festuccia, Z.~Komargodski, and N.~Seiberg,
  {\it Contact Terms, Unitarity, and F-Maximization in Three-Dimensional
  Superconformal Theories},  {\em JHEP} {\bf 1210} (2012) 053,
  [\href{http://xxx.lanl.gov/abs/1205.4142}{{\tt arXiv:1205.4142}}].

\bibitem{CDFKS-CS}
C.~Closset, T.~T. Dumitrescu, G.~Festuccia, Z.~Komargodski, and N.~Seiberg,
  {\it Comments on Chern-Simons Contact Terms in Three Dimensions},  {\em JHEP}
  {\bf 1209} (2012) 091, [\href{http://xxx.lanl.gov/abs/1206.5218}{{\tt
  arXiv:1206.5218}}].

\bibitem{DG-Sdual}
T.~Dimofte and S.~Gukov, {\it Chern-Simons Theory and S-duality},  {\em JHEP}
  {\bf 1305} (2013) 109, [\href{http://xxx.lanl.gov/abs/1106.4550}{{\tt
  arXiv:1106.4550}}].

\bibitem{IS}
K.~Intriligator and N.~Seiberg, {\it Mirror Symmetry in Three Dimensional Gauge
  Theories},  {\em Phys. Lett.} {\bf B387} (1996) 513--519,
  [\href{http://xxx.lanl.gov/abs/hep-th/9607207v1}{{\tt hep-th/9607207v1}}].

\bibitem{dBHOY}
J.~de~Boer, K.~Hori, H.~Ooguri, Y.~Oz, and Z.~Yin, {\it Mirror Symmetry in
  Three-Dimensional Gauge Theories, SL(2,Z) and D-Brane Moduli Spaces},  {\em
  Nucl. Phys.} {\bf B493} (1996) 148--176,
  [\href{http://xxx.lanl.gov/abs/hep-th/9612131v1}{{\tt hep-th/9612131v1}}].

\bibitem{dBHO}
J.~de~Boer, K.~Hori, and Y.~Oz, {\it Dynamics of N=2 Supersymmetric Gauge
  Theories in Three Dimensions},  {\em Nucl. Phys.} {\bf B500} (1997) 163--191,
  [\href{http://xxx.lanl.gov/abs/hep-th/9703100v3}{{\tt hep-th/9703100v3}}].

\bibitem{Witten-SL2}
E.~Witten, {\it SL(2,Z) Action On Three-Dimensional Conformal Field Theories
  With Abelian Symmetry},  \href{http://xxx.lanl.gov/abs/hep-th/0307041v3}{{\tt
  hep-th/0307041v3}}.

\bibitem{OV}
H.~Ooguri and C.~Vafa, {\it Knot Invariants and Topological Strings},  {\em
  Nucl. Phys.} {\bf B5777} (Jan, 2000) 419--438,
  [\href{http://xxx.lanl.gov/abs/hep-th/9912123v3}{{\tt hep-th/9912123v3}}].

\bibitem{Wfiveknots}
E.~Witten, {\it Fivebranes and Knots},  {\em Quantum Topol.} {\bf 3} (2012),
  no.~1 1--137, [\href{http://xxx.lanl.gov/abs/1101.3216}{{\tt
  arXiv:1101.3216}}].

\bibitem{GW-Sduality}
D.~Gaiotto and E.~Witten, {\it S-Duality of Boundary Conditions In N=4 Super
  Yang-Mills Theory},  {\em Adv. Theor. Math. Phys.} {\bf 13} (2009), no.~2
  721--896, [\href{http://xxx.lanl.gov/abs/0807.3720}{{\tt arXiv:0807.3720}}].

\bibitem{GKW-5branes}
J.~P. Gauntlett, N.~Kim, and D.~Waldram, {\it M-Fivebranes Wrapped on
  Supersymmetric Cycles},  {\em Phys. Rev.} {\bf D63} (2001) 126001,
  [\href{http://xxx.lanl.gov/abs/hep-th/0012195v2}{{\tt hep-th/0012195v2}}].

\bibitem{PerniciSezgin}
M.~Pernici and E.~Sezgin, {\it Spontaneous compactification of
  seven-dimensional supergravity theories},  {\em Class. Quan. Grav.} {\bf 2}
  (1985), no.~5 673--681.

\bibitem{EJM-surface}
R.~Emparan, C.~V. Johnson, and R.~C. Myers, {\it Surface Terms as Counterterms
  in the AdS/CFT Correspondence},  {\em Phys. Rev.} {\bf D60} (1999) 104001,
  [\href{http://xxx.lanl.gov/abs/hep-th/9903238v4}{{\tt hep-th/9903238v4}}].

\bibitem{GaiottoMaldacena}
D.~Gaiotto and J.~Maldacena, {\it The gravity duals of N=2 superconformal field
  theories},  \href{http://xxx.lanl.gov/abs/0904.4466}{{\tt arXiv:0904.4466}}.

\bibitem{GKSTW}
D.~Green, Z.~Komargodski, N.~Seiberg, Y.~Tachikawa, and B.~Wecht, {\it Exactly
  Marginal Deformations and Global Symmetries},  {\em JHEP} {\bf 1006} (2010)
  106, [\href{http://xxx.lanl.gov/abs/1005.3546}{{\tt arXiv:1005.3546}}].

\bibitem{DG-E7}
T.~Dimofte and D.~Gaiotto, {\it An E7 Surprise},  {\em JHEP} {\bf 1210} (2012)
  129, [\href{http://xxx.lanl.gov/abs/1209.1404}{{\tt arXiv:1209.1404}}].

\bibitem{yoshida-1985}
T.~Yoshida, {\it The eta-invariant of hyperbolic 3-manifolds},  {\em Invent.
  Math.} {\bf 81} (Jan, 1985) 473--514.

\bibitem{Jafferis-Zmin}
D.~L. Jafferis, {\it The Exact Superconformal R-Symmetry Extremizes Z},  {\em
  JHEP} {\bf 1205} (2012) 159, [\href{http://xxx.lanl.gov/abs/1012.3210}{{\tt
  arXiv:1012.3210}}].

\bibitem{JKPS-Fthm}
D.~L. Jafferis, I.~R. Klebanov, S.~S. Pufu, and B.~R. Safdi, {\it Towards the
  F-Theorem: N=2 Field Theories on the Three-Sphere},  {\em JHEP} {\bf 1106}
  (2011) 102, [\href{http://xxx.lanl.gov/abs/1103.1181}{{\tt
  arXiv:1103.1181}}].

\bibitem{Kapustin-3dloc}
A.~Kapustin, B.~Willett, and I.~Yaakov, {\it Exact Results for Wilson Loops in
  Superconformal Chern-Simons Theories with Matter},  {\em JHEP} {\bf 1003}
  (2010) 089, [\href{http://xxx.lanl.gov/abs/0909.4559}{{\tt
  arXiv:0909.4559}}]. Published in: JHEP 1003:089,2010 32 pages.

\bibitem{HHL}
N.~Hama, K.~Hosomichi, and S.~Lee, {\it SUSY Gauge Theories on Squashed
  Three-Spheres},  {\em JHEP} {\bf 1105} (2011) 014,
  [\href{http://xxx.lanl.gov/abs/1102.4716}{{\tt arXiv:1102.4716}}].

\bibitem{MPS-S3b}
D.~Martelli, A.~Passias, and J.~Sparks, {\it The gravity dual of supersymmetric
  gauge theories on a squashed three-sphere},  {\em Nucl. Phys.} {\bf B864}
  (2012) 840--868, [\href{http://xxx.lanl.gov/abs/1110.6400}{{\tt
  arXiv:1110.6400}}].

\bibitem{kashaev-1997}
R.~M. Kashaev, {\it The hyperbolic volume of knots from quantum dilogarithm},
  {\em Lett. Math. Phys.} {\bf 39} (1997) 269--265,
  [\href{http://xxx.lanl.gov/abs/q-alg/9601025v2}{{\tt q-alg/9601025v2}}].

\bibitem{Mur-Mur}
H.~Murakami and J.~Murakami, {\it The colored Jones polynomials and the
  simplicial volume of a knot},  {\em Acta Math.} {\bf 186} (Jan, 2001)
  85--104, [\href{http://xxx.lanl.gov/abs/math/9905075v2}{{\tt
  math/9905075v2}}].

\bibitem{Wit-anal}
E.~Witten, {\it Analytic Continuation of Chern-Simons Theory},  {\em
  Chern-Simons gauge theory: 20 Years After (AMS/IP Stud. Adv. Math.)} (2011)
  347--446, [\href{http://xxx.lanl.gov/abs/1001.2933}{{\tt arXiv:1001.2933}}].

\bibitem{DG-quantumNZ}
T.~D. Dimofte and S.~Garoufalidis, {\it The quantum content of the gluing
  equations},  {\em Geom. Topol.} {\bf 17} (2013), no.~3 1253--1315,
  [\href{http://xxx.lanl.gov/abs/1202.6268}{{\tt arXiv:1202.6268}}].

\bibitem{Kim-index}
S.~Kim, {\it The complete superconformal index for N=6 Chern-Simons theory},
  {\em Nucl. Phys.} {\bf B821} (2009) 241--284,
  [\href{http://xxx.lanl.gov/abs/0903.4172}{{\tt arXiv:0903.4172}}].

\bibitem{IY-index}
Y.~Imamura and S.~Yokoyama, {\it Index for three dimensional superconformal
  field theories with general R-charge assignments},  {\em JHEP} {\bf 1104}
  (2011) 007, [\href{http://xxx.lanl.gov/abs/1101.0557}{{\tt
  arXiv:1101.0557}}].

\bibitem{KW-index}
A.~Kapustin and B.~Willett, {\it Generalized Superconformal Index for Three
  Dimensional Field Theories},  \href{http://xxx.lanl.gov/abs/1106.2484}{{\tt
  arXiv:1106.2484}}.

\bibitem{BNY-Lens}
F.~Benini, T.~Nishioka, and M.~Yamazaki, {\it 4d Index to 3d Index and 2d
  TQFT},  {\em Phys. Rev.} {\bf D86} (2012) 065015,
  [\href{http://xxx.lanl.gov/abs/1109.0283}{{\tt arXiv:1109.0283}}].

\bibitem{IY-Lens}
Y.~Imamura and D.~Yokoyama, {\it S3/Zn partition function and dualities},  {\em
  JHEP} {\bf 1211} (2012) 122, [\href{http://xxx.lanl.gov/abs/1208.1404}{{\tt
  arXiv:1208.1404}}].

\bibitem{IMY-fact}
Y.~Imamura, H.~Matsuno, and D.~Yokoyama, {\it Factorization of S3/Zn partition
  function},  {\em Phys. Rev.} {\bf D89} (2014) 085003,
  [\href{http://xxx.lanl.gov/abs/1311.2371}{{\tt arXiv:1311.2371}}].

\bibitem{BDP-blocks}
C.~Beem, T.~Dimofte, and S.~Pasquetti, {\it Holomorphic Blocks in Three
  Dimensions},  \href{http://xxx.lanl.gov/abs/1211.1986}{{\tt
  arXiv:1211.1986}}.

\end{thebibliography}\endgroup

\end{document}